\documentclass[]{tADP2e}
\usepackage{epstopdf}
\usepackage{amsbsy}
\usepackage[english]{babel}
\usepackage{color,xcolor}
\usepackage{hyperref}
\usepackage{amsmath}
\usepackage{amssymb}
\usepackage{mathrsfs}
\usepackage{dsfont}
\usepackage{chngcntr}
\counterwithout{figure}{section}
\counterwithout{table}{section}
\usepackage{subfigure}
\usepackage{multirow}
\usepackage{slashbox}
\usepackage{rotating}

\def\bs#1{\boldsymbol{#1}}
\def \Z {\mathbb{Z}}

\def \H {\mathcal{H}}
\def \A {\mathcal{A}}
\def \F {\mathcal{F}}
\def \K {\hat{\mathcal{K}}}
\def \k {\mathbf{k}}
\def \T {\hat{T}}
\def \C {\hat{C}}
\def \sin {\text{sin}}
\def \cos {\text{cos}}
\clearpage
\phantomsection

\begin{document}
\articletype{REVIEW ARTICLE}
\title{Topological quantum matter with cold atoms}

\author{Dan-Wei Zhang$^{1,3}$
, Yan-Qing Zhu$^{2,3}$,  Y.~X.~Zhao$^{2,3}$, Hui Yan$^{1}$, and Shi-Liang Zhu$^{2,1}$$^{\ast}$\thanks{$^\ast$Corresponding author. Email: slzhu@nju.edu.cn\vspace{6pt}}
\\\vspace{6pt}
$^{1}${\em{Guangdong Provincial Key Laboratory of Quantum Engineering and Quantum Materials,
SPTE, South China Normal University, Guangzhou 510006, China}};\\
$^{2}${\em{National Laboratory of Solid State Microstructures and School of Physics, Nanjing University, Nanjing 210093, China}};\\
$^{3}${\em{Collaborative Innovation Center of Advanced Microstructures, Nanjing 210093, China}};
\\ \received{\today}}

\maketitle
\begin{abstract}
This is an introductory review of the physics of topological quantum matter with cold atoms. Topological quantum phases, originally discovered and investigated in condensed matter physics, have recently been explored in a range of different systems, which produced both fascinating physics findings and exciting opportunities for applications. Among the physical systems that have been considered to realize and probe these intriguing phases, ultracold atoms become promising platforms due to their high flexibility and controllability. Quantum simulation of topological phases with cold atomic gases is a rapidly evolving field, and recent theoretical and experimental developments reveal that some toy models originally proposed in condensed matter physics have been realized with this artificial quantum system. The purpose of this article is to introduce these developments. The article begins with a tutorial review of topological invariants and the methods to control parameters in the Hamiltonians of neutral atoms. Next, topological quantum phases in optical lattices are introduced in some detail, especially several celebrated models, such as the Su-Schrieffer-Heeger model, the Hofstadter-Harper model, the Haldane model and the Kane-Mele model. The theoretical proposals and experimental implementations of these models are discussed. Notably, many of these models cannot be directly realized in conventional solid-state experiments. The newly developed methods for probing the intrinsic properties of the topological phases in cold atom systems are also reviewed. Finally, some topological phases with cold atoms in the continuum and in the presence of interactions are discussed, and an outlook on future work is given.
\end{abstract}

\begin{classcode}
67.85.-d Ultracold gases, trapped gases; 03.75.Ss Degenerate Fermi gases; 73.43.Nq Quantum phase transitions; 71.10.Fd Lattices Fermi models; 03.75.Lm Tunneling Josephson effect, Boson-Einstein condensates in periodic potentials, solitons, vortices, and topological excitations; 03.65.Vf Phases: geometric, dynamic or topological
\end{classcode}


\noindent
\centerline{\bf Contents}\\

\noindent
1. Introduction~~~~~~~~~~~~~~~~~~~~~~~~~~~~~~~~~~~~~~~~~~~~~~~~~~~~~~~~~~~~~~~~~~~~~~~~~~~~~~~~~~~~~~~~~~~~~~3\\

\noindent

2. Topological invariants in momentum space~~~~~~~~~~~~~~~~~~~~~~~~~~~~~~~~~~~~~~~~~~~~~~~~~~~~~~~~~~~~~~~~~~~~~6\\
    \hspace*{7pt} 2.1. Gauge fields in momentum space~~~~~~~~~~~~~~~~~~~~~~~~~~~~~~~~~~~~~~~~~~~~~~~~~~~~~~~~~~~~~~6\\
    \hspace*{7pt} 2.2. Quantized Zak phase~~~~~~~~~~~~~~~~~~~~~~~~~~~~~~~~~~~~~~~~~~~~~~~~~~~~~~~~~~~~~~~~~~~~~~~~~~~~~~7\\
    \hspace*{7pt} 2.3. Chern numbers~~~~~~~~~~~~~~~~~~~~~~~~~~~~~~~~~~~~~~~~~~~~~~~~~~~~~~~~~~~~~~~~~~~~~~~~~~~~~~~~~~~~~8\\\
    \hspace*{7pt} 2.4. Spin Chern number and $\Z_2$ topological invariants~~~~~~~~~~~~~~~~~~~~~~~~~~~~~~~~~~~~~~~~9\\
    \hspace*{7pt} 2.5. The Hopf invariant~~~~~~~~~~~~~~~~~~~~~~~~~~~~~~~~~~~~~~~~~~~~~~~~~~~~~~~~~~~~~~~~~~~~~~~~~~~~~~~~~~9\\

\noindent
3. Engineering the Hamiltonian of atoms~~~~~~~~~~~~~~~~~~~~~~~~~~~~~~~~~~~~~~~~~~~~~~~~~~~~~~~~~~~~~~~~~~~~10\\
    \hspace*{7pt} 3.1. Laser cooling~~~~~~~~~~~~~~~~~~~~~~~~~~~~~~~~~~~~~~~~~~~~~~~~~~~~~~~~~~~~~~~~~~~~~~~~~~~~~~~~~~~~~~10\\
    \hspace*{7pt} 3.2. Effective interactions~~~~~~~~~~~~~~~~~~~~~~~~~~~~~~~~~~~~~~~~~~~~~~~~~~~~~~~~~~~~~~~~~~~~~~~~~~~~~11\\
    \hspace*{7pt} 3.3. Dipole potentials and optical lattices~~~~~~~~~~~~~~~~~~~~~~~~~~~~~~~~~~~~~~~~~~~~~~~~~~~~~~~~~~~~~12\\
    \hspace*{7pt} 3.4. Artificial magnetic fields and spin-orbit couplings~~~~~~~~~~~~~~~~~~~~~~~~~~~~~~~~~~~~~~~~~~~~~~~~17\\
      \hspace*{24pt} 3.4.1. Geometric gauge potentials~~~~~~~~~~~~~~~~~~~~~~~~~~~~~~~~~~~~~~~~~~~~~~~~~~~~~~~~~~~~~~~~~~~~~~~~17\\
      \hspace*{24pt} 3.4.2. Laser-assisted tunneling~~~~~~~~~~~~~~~~~~~~~~~~~~~~~~~~~~~~~~~~~~~~~~~~~~~~~~~~~~~~~~~~~~~~~~~~~~21\\
      \hspace*{24pt} 3.4.3. Periodically driven systems~~~~~~~~~~~~~~~~~~~~~~~~~~~~~~~~~~~~~~~~~~~~~~~~~~~~~~~~~~~~~~~~~~~~~~~23\\

\noindent
4. Topological quantum matter in optical lattices~~~~~~~~~~~~~~~~~~~~~~~~~~~~~~~~~~~~~~~~~~~~~~~~~~~~~~~~~~26\\
    \hspace*{7pt} 4.1. One-dimension~~~~~~~~~~~~~~~~~~~~~~~~~~~~~~~~~~~~~~~~~~~~~~~~~~~~~~~~~~~~~~~~~~~~~~~~~~~~~~~~~~~~~~~~~~27\\
      \hspace*{24pt} 4.1.1. SSH model and Rice-Mele model~~~~~~~~~~~~~~~~~~~~~~~~~~~~~~~~~~~~~~~~~~~~~~~~~~~~~~~~~~~~~~~~~~~~~27\\
      \hspace*{24pt} 4.1.2. Topological pumping~~~~~~~~~~~~~~~~~~~~~~~~~~~~~~~~~~~~~~~~~~~~~~~~~~~~~~~~~~~~~~~~~~~~~~~~~~~~~~~34\\
      \hspace*{24pt} 4.1.3. 1D AIII class topological insulators~~~~~~~~~~~~~~~~~~~~~~~~~~~~~~~~~~~~~~~~~~~~~~~~~~~~~~~~~~~~~~37\\
      \hspace*{24pt} 4.1.4. Creutz ladder model~~~~~~~~~~~~~~~~~~~~~~~~~~~~~~~~~~~~~~~~~~~~~~~~~~~~~~~~~~~~~~~~~~~~~~~~~~~~~~~39\\
      \hspace*{24pt} 4.1.5. Aubry-Andre-Harper model~~~~~~~~~~~~~~~~~~~~~~~~~~~~~~~~~~~~~~~~~~~~~~~~~~~~~~~~~~~~~~~~~~~~~~~~~~40\\
    \hspace*{7pt} 4.2. Two-dimension~~~~~~~~~~~~~~~~~~~~~~~~~~~~~~~~~~~~~~~~~~~~~~~~~~~~~~~~~~~~~~~~~~~~~~~~~~~~~~~~~~~~~~~~~~42\\
      \hspace*{24pt} 4.2.1. Graphene-like physics and Dirac fermions~~~~~~~~~~~~~~~~~~~~~~~~~~~~~~~~~~~~~~~~~~~~~~~~~~~~~~~~~~42\\
      \hspace*{24pt} 4.2.2. Hofstadter model~~~~~~~~~~~~~~~~~~~~~~~~~~~~~~~~~~~~~~~~~~~~~~~~~~~~~~~~~~~~~~~~~~~~~~~~~~~~~~~~~~49\\
      \hspace*{24pt} 4.2.3. Haldane model~~~~~~~~~~~~~~~~~~~~~~~~~~~~~~~~~~~~~~~~~~~~~~~~~~~~~~~~~~~~~~~~~~~~~~~~~~~~~~~~~~~~~54\\
      \hspace*{24pt} 4.2.4. Kane-Mele model~~~~~~~~~~~~~~~~~~~~~~~~~~~~~~~~~~~~~~~~~~~~~~~~~~~~~~~~~~~~~~~~~~~~~~~~~~~~~~~~~~~58\\
    \hspace*{7pt} 4.3. Three-dimension~~~~~~~~~~~~~~~~~~~~~~~~~~~~~~~~~~~~~~~~~~~~~~~~~~~~~~~~~~~~~~~~~~~~~~~~~~~~~~~~~~~~~~~~60\\
      \hspace*{24pt} 4.3.1. 3D Dirac fermions~~~~~~~~~~~~~~~~~~~~~~~~~~~~~~~~~~~~~~~~~~~~~~~~~~~~~~~~~~~~~~~~~~~~~~~~~~~~~~~~~60\\
      \hspace*{24pt} 4.3.2. Weyl semimetals and Weyl fermions~~~~~~~~~~~~~~~~~~~~~~~~~~~~~~~~~~~~~~~~~~~~~~~~~~~~~~~~~~~~~~~~~61\\
      \hspace*{24pt} 4.3.3. Topological nodal-line semimetals~~~~~~~~~~~~~~~~~~~~~~~~~~~~~~~~~~~~~~~~~~~~~~~~~~~~~~~~~~~~~~~~~65\\
      \hspace*{24pt} 4.3.4. 3D $\Z_2$ topological insulators~~~~~~~~~~~~~~~~~~~~~~~~~~~~~~~~~~~~~~~~~~~~~~~~~~~~~~~~~~~~~~~~~~67\\
      \hspace*{24pt} 4.3.5. 3D Chiral topological insulators~~~~~~~~~~~~~~~~~~~~~~~~~~~~~~~~~~~~~~~~~~~~~~~~~~~~~~~~~~~~~~~~~~69\\
      \hspace*{24pt} 4.3.6. Hopf topological insulators~~~~~~~~~~~~~~~~~~~~~~~~~~~~~~~~~~~~~~~~~~~~~~~~~~~~~~~~~~~~~~~~~~~~~~~72\\
      \hspace*{24pt} 4.3.7. Integer quantum Hall effect in 3D~~~~~~~~~~~~~~~~~~~~~~~~~~~~~~~~~~~~~~~~~~~~~~~~~~~~~~~~~~~~~~~~~75\\
    \hspace*{7pt} 4.4. Higher and synthetic dimensions~~~~~~~~~~~~~~~~~~~~~~~~~~~~~~~~~~~~~~~~~~~~~~~~~~~~~~~~~~~~~~~~~~~~~~~~78\\
    \hspace*{7pt} 4.5. Higher-spin topological quasiparticles~~~~~~~~~~~~~~~~~~~~~~~~~~~~~~~~~~~~~~~~~~~~~~~~~~~~~~~~~~~~~~~~~82\\

\noindent
5. Probing methods~~~~~~~~~~~~~~~~~~~~~~~~~~~~~~~~~~~~~~~~~~~~~~~~~~~~~~~~~~~~~~~~~~~~~~~~~~~~~~~~~~~~~~~~~86\\
    \hspace*{7pt} 5.1. Detection of Dirac points and topological transition~~~~~~~~~~~~~~~~~~~~~~~~~~~~~~~~~~~~~~~~~~~~~~~~~~~86\\
    \hspace*{7pt} 5.2. Interferometer in momentum space~~~~~~~~~~~~~~~~~~~~~~~~~~~~~~~~~~~~~~~~~~~~~~~~~~~~~~~~~~~~~~~~~~~~~~~87\\
    \hspace*{7pt} 5.3. Hall drift of accelerated wave packets~~~~~~~~~~~~~~~~~~~~~~~~~~~~~~~~~~~~~~~~~~~~~~~~~~~~~~~~~~~~~~~~~90\\
    \hspace*{7pt} 5.4. Streda formula and density profiles~~~~~~~~~~~~~~~~~~~~~~~~~~~~~~~~~~~~~~~~~~~~~~~~~~~~~~~~~~~~~~~~~~~~92\\
    \hspace*{7pt} 5.5. Tomography of Bloch states~~~~~~~~~~~~~~~~~~~~~~~~~~~~~~~~~~~~~~~~~~~~~~~~~~~~~~~~~~~~~~~~~~~~~~~~~~~~~93\\
    \hspace*{7pt} 5.6. Spin polarization at high symmetry momenta~~~~~~~~~~~~~~~~~~~~~~~~~~~~~~~~~~~~~~~~~~~~~~~~~~~~~~~~~~~~~96\\
    \hspace*{7pt} 5.7. Topological pumping approach~~~~~~~~~~~~~~~~~~~~~~~~~~~~~~~~~~~~~~~~~~~~~~~~~~~~~~~~~~~~~~~~~~~~~~~~~~~98\\
    \hspace*{7pt} 5.8. Detection of topological edge states~~~~~~~~~~~~~~~~~~~~~~~~~~~~~~~~~~~~~~~~~~~~~~~~~~~~~~~~~~~~~~~~~~~99\\

\noindent
6. Topological quantum matter in continuous form~~~~~~~~~~~~~~~~~~~~~~~~~~~~~~~~~~~~~~~~~~~~~~~~~~~~~~~~~~~99\\
    \hspace*{7pt} 6.1. Jackiw-Rebbi model with topological solitons~~~~~~~~~~~~~~~~~~~~~~~~~~~~~~~~~~~~~~~~~~~~~~~~~~~~~~~~~~~99\\
    \hspace*{7pt} 6.2. Topological defects in Bose-Einstein condensates~~~~~~~~~~~~~~~~~~~~~~~~~~~~~~~~~~~~~~~~~~~~~~~~~~~~~~101\\
    \hspace*{7pt} 6.3. Spin Hall effect in atomic gases~~~~~~~~~~~~~~~~~~~~~~~~~~~~~~~~~~~~~~~~~~~~~~~~~~~~~~~~~~~~~~~~~~~~~~105\\

\noindent
7. Topological quantum matter with interactions~~~~~~~~~~~~~~~~~~~~~~~~~~~~~~~~~~~~~~~~~~~~~~~~~~~~~~~~~~~106\\
    \hspace*{7pt} 7.1. Spin chains~~~~~~~~~~~~~~~~~~~~~~~~~~~~~~~~~~~~~~~~~~~~~~~~~~~~~~~~~~~~~~~~~~~~~~~~~~~~~~~~~~~~~~~~~~~106\\
      \hspace*{24pt} 7.1.1. Spin-1/2 chain~~~~~~~~~~~~~~~~~~~~~~~~~~~~~~~~~~~~~~~~~~~~~~~~~~~~~~~~~~~~~~~~~~~~~~~~~~~~~~~~~~~106\\
      \hspace*{24pt} 7.1.2. Spin-1 chain and Haldane phase~~~~~~~~~~~~~~~~~~~~~~~~~~~~~~~~~~~~~~~~~~~~~~~~~~~~~~~~~~~~~~~~~~~109\\
    \hspace*{7pt} 7.2. Kitaev chain model~~~~~~~~~~~~~~~~~~~~~~~~~~~~~~~~~~~~~~~~~~~~~~~~~~~~~~~~~~~~~~~~~~~~~~~~~~~~~~~~~~~~111\\
    \hspace*{7pt} 7.3. 1D Anyon-Hubbard model~~~~~~~~~~~~~~~~~~~~~~~~~~~~~~~~~~~~~~~~~~~~~~~~~~~~~~~~~~~~~~~~~~~~~~~~~~~~~~~~116\\
    \hspace*{7pt} 7.4. Bosonic quantum Hall states~~~~~~~~~~~~~~~~~~~~~~~~~~~~~~~~~~~~~~~~~~~~~~~~~~~~~~~~~~~~~~~~~~~~~~~~~~~118\\
      \hspace*{24pt} 7.4.1. Single-component Bose-Hubbard model~~~~~~~~~~~~~~~~~~~~~~~~~~~~~~~~~~~~~~~~~~~~~~~~~~~~~~~~~~~~~~119\\
      \hspace*{24pt} 7.4.2. Two-component Bose-Hubbard model~~~~~~~~~~~~~~~~~~~~~~~~~~~~~~~~~~~~~~~~~~~~~~~~~~~~~~~~~~~~~~~~~121\\
    \hspace*{7pt} 7.5. Kitaev honeycomb model~~~~~~~~~~~~~~~~~~~~~~~~~~~~~~~~~~~~~~~~~~~~~~~~~~~~~~~~~~~~~~~~~~~~~~~~~~~~~~~~122\\

\noindent
8. Conclusion and outlook~~~~~~~~~~~~~~~~~~~~~~~~~~~~~~~~~~~~~~~~~~~~~~~~~~~~~~~~~~~~~~~~~~~~~~~~~~~~~~~~~125\\
    \hspace*{7pt} 8.1. Unconventional topological bands~~~~~~~~~~~~~~~~~~~~~~~~~~~~~~~~~~~~~~~~~~~~~~~~~~~~~~~~~~~~~~~~~~~~~~125\\
    \hspace*{7pt} 8.2. Other interacting topological phases~~~~~~~~~~~~~~~~~~~~~~~~~~~~~~~~~~~~~~~~~~~~~~~~~~~~~~~~~~~~~~~~~~126\\
    \hspace*{7pt} 8.3. Non-equilibrium dynamics and band topology~~~~~~~~~~~~~~~~~~~~~~~~~~~~~~~~~~~~~~~~~~~~~~~~~~~~~~~~~~~~127\\
    \hspace*{7pt} 8.4. Topological states in open or dissipative systems~~~~~~~~~~~~~~~~~~~~~~~~~~~~~~~~~~~~~~~~~~~~~~~~~~~~~128\\

\noindent
Acknowledgements~~~~~~~~~~~~~~~~~~~~~~~~~~~~~~~~~~~~~~~~~~~~~~~~~~~~~~~~~~~~~~~~~~~~~~~~~~~~~~~~~~~~~~~~~~~~~~~~~129\\
\noindent
Disclosure
\noindent
statement~~~~~~~~~~~~~~~~~~~~~~~~~~~~~~~~~~~~~~~~~~~~~~~~~~~~~~~~~~~~~~~~~~~~~~~~~~~~~~~~~~~~~~~~~~~~~~~~~129\\
Funding~~~~~~~~~~~~~~~~~~~~~~~~~~~~~~~~~~~~~~~~~~~~~~~~~~~~~~~~~~~~~~~~~~~~~~~~~~~~~~~~~~~~~~~~~~~~~~~~~~~~129\\
\noindent
Appendix A. Formulas of topological invariants~~~~~~~~~~~~~~~~~~~~~~~~~~~~~~~~~~~~~~~~~~~~~~~~~~~~~~~~~~~~~~~~~~~129\\

\noindent
References~~~~~~~~~~~~~~~~~~~~~~~~~~~~~~~~~~~~~~~~~~~~~~~~~~~~~~~~~~~~~~~~~~~~~~~~~~~~~~~~~~~~~~~~~~~~~~~~~141 \\

\section{Introduction}

Topology is an important mathematical discipline, starting its prosperity
in the early part of the twentieth century. It is concerned
with the properties of space that are preserved under continuous
deformations, such as stretching, crumpling, and bending, but not
tearing or gluing. Topological methods have recently played increasingly important roles in physics, and it is now difficult to
think of an area of physics where topology does not apply. In early development in this field, Paul
Dirac used topological concepts to show that there are magnetic
monopole solutions to Maxwell's equations \cite{Dirac1931}, and Sir Roger Penrose also
used topological methods to show that singularities are a generic feature of gravitational collapse \cite{Penrose1965}. However, it was not until the 1970's that topology really came to prominence in physics, and that was thanks
to its introduction into gauge theories and condensed matter physics.

What we now know as ``topological quantum states" of condensed
matter  may go back to the Su-Schrieffer-Heeger model for conducting polymers with topological solitons in the 1970's \cite{WPSu1979,Heeger1988,LYu1988} and were encountered around 1980 \cite{Haldane2017}, with
the experimental discovery of the integer \cite{Klitzing1980} and
fractional \cite{Tsui1982} quantum Hall effects (QHE) in the
two-dimensional (2D) electron systems, as well as the theoretical discovery of
the entangled gapped spin-liquid states in quantum integer-spin
chains \cite{Haldane1983a}. Until then, phases of matter have been largely classified based on symmetries
and symmetries breaking known as the Landau paradigm. The discovery of the ``quantum
topological matter" made it clear that the paradigm based on
symmetries is insufficient, as the quantum Hall phases do not break
any symmetry and would seem ``trivial" from the symmetry
standpoint.

A topological phase is an exotic form of matter characterized by non-local properties rather than
local order parameters. An early milestone was the discovery by
David Thouless and collaborators in 1982 of a remarkable formula
[Thouless-Kohmoto-Nightingale-den Nijs (TKNN) formula] for QHE \cite{Thouless1982}, which
was soon recognized by Barry Simon as the first Chern
invariant for the mathematically termed $U(1)$ fiber bundles in topology \cite{Simon1983} with an essential connection to the
geometric phase discovered by Michael Berry \cite{Berry1984}. The
identification of the TKNN formula as a topological invariant
marked the beginning of the recognition that topology would play
an important role in classifying quantum states. The TKNN result
was originally obtained for the band structure of electrons in uniform
magnetic fields. In 1988, F. D. M. Haldane
realized that the necessary condition for a QHE
was not a magnetic field, but broken time-reversal invariance
\cite{Haldane1988}. He investigated a graphene-like tight-binding
toy model (now called the Haldane model) with
next-nearest-neighbor hopping and averaged zero magnetic field,
constructing the first model for the QHE without
Landau levels. The QHE without Landau levels is now known as the quantum anomalous Hall
effect or Chern insulator, and is the first topological insulator
discovered, although it is one with a broken time reversal symmetry (TRS).
D. J. Thouless, J. M. Kostrlitz, and F. D. M. Haldane were awarded
the 2016 Nobel Prize in physics ``for theoretical discoveries of
topological phase transitions and topological phases of matter".

Another major development in this field is the discovery of topological
insulators with TRS in 2-4 dimensions \cite{Hasan2010,XLQi2011}. S.-C. Zhang and J. Hu predicated a kind of four-dimensional QHE, which is characterized by the second Chern number \cite{ZhangSC2001}. It is the first topological insulator with TRS predicted and only recently was experimentally realized with ultracold atoms \cite{Lohse2018}. C. Kane and E. Mele \cite{Kane2005a,Kane2005b}
theoretically combined two conjugate copies of the Haldane model, one for
spin-up electrons for which the valence band has Chern number $\pm
1 $ and one for spin-down electrons where the valence band has the
opposite Chern number $\mp 1$. Since the total Chern number of the
band vanishes, there is no QHE. However, they
discovered that so long as the TRS is unbroken, the system has a previously unexpected $\Z_2$
topological invariant related to Kramers degeneracy. Independently, B. A. Bernevig, T. L. Hughes, and S.-C. Zhang \cite{ Bernevig2006b} predicted the quantum spin Hall effect \cite{Bernevig2006a}  in quantum well structures of HgCdTe, which is known as a state of 2D topological insulators, paving the way to its experimental discovery \cite{Konig2007}. The 3D generalization of this $\Z_2$ invariant was independently and simultaneously predicted in 2007 by three groups \cite{Haldane2017}, which led to the
experimental discovery of the 3D time reversal invariant
topological insulators. 
The discovery of topological insulators signaled the start of a wider search for topological phases of
matter, and this continues to be fertile ground. Since topological quantum numbers are fairly insensitive to local
imperfections and perturbations, topological protection offers fascinating possibilities for applications in quantum technology.

Besides topological insulators, topological phases are generalized to topological (semi)metals, such as Weyl and Dirac semimetals in 3D solids \cite{Wehling2014,Armitage2018}, and new topological materials are being discovered and developed at an impressive rate, the possibilities for creating and probing exotic topological phases would be greatly enhanced if these phases could be realized in systems that are easily tuned. Ultracold atoms with their flexibility could provide such a platform. In particular, some idealized model Hamiltonians for topological quantum matter, which are unrealistic in other quantum systems, can be realized with ultracold atoms in optical lattices (OLs). Below, we briefly summarize the toolbox that has been developed to create and probe topological quantum matter with cold atoms.

i) The lattice structure of a single-particle energy band in a
solid is fundamental for some topological quantum phases. For
instance, both the topological insulators proposed by Handane and Kane
and Mele exist in a honeycomb lattice, while spin liquid states
favor a Kagome lattice. Ultracold atoms can be trapped in the
potential minima formed by the laser beams. By changing the
angles, wavelengths and polizations of the laser beams, one can create different
lattice geometries. OLs with various geometric structures, such as square/cubic, triangular,
honeycomb, and Kagome lattice, and superlattice structures,
have been experimentally realized (see the review on engineering novel OLs \cite{Windpassinger2013}). In addition, OLs
provide convenient ways to control various factors in cold atoms such as the strength
of interatomic interactions, the band structures, the spin composition, and the levels of disorder
more easily than in real crystals.

ii) A necessary condition for the QHE or topological insulators with broken TRS is a
magnetic field (flux). Although atomic gases are neutral
particles, artificial gauge fields can be realized for them \cite{Dalibard2011,Goldman2014}.
Therefore, one can use atomic gases to simulate charged quantum
particles, such as electrons in external electromagnetic fields.
Artificial magnetic fields for atomic gases have been
implemented through several ways: rotating an isotropic 2D
harmonic trap, generating a space-dependent geometric phase by dressing
the atom-light interaction, and suitably shaking an OL. Importantly, the methods based on the atom-light interaction
and shaking lattices are well-suited for implementing an artificial
gauge field in an OL. Artificial gauge fields,
combined with OLs, lead to the realization of several
celebrated toy models proposed but unrealistic in condensed matter
physics. For instance, the Haldane model \cite{Jotzu2014} and the
Hofstadter model \cite{Miyake2013,Aidelsburger2013} have been directly
realized for the first time with ultracold gases.

iii) Spin-orbit coupling (SOC) is a basic ingredient for a $\Z_2$ topological
insulator with TRS. It can also be realized by a non-Abelian geometric phase due to
the atom-laser interaction. To simulate an SOC of spin-1/2 particles, one can
use a configuration where two atomic dressed states form a degenerate
manifold at every point in the laser field. When an atom prepared in a state in the manifold slowly moves along a closed
trajectory, a non-Abelian geometric phase is accumulated in the wave
function, and an SOC is generated if the
non-Abelian geometric phase is space dependent. Recently, one-dimensional (1D) and
2D SOCs for bosonic and fermionic atoms have been experimentally created in the continuum or OLs \cite{YJLin2011b,JYZhang2012,PWang2012,Cheuk2012,LHuang2016,ZWu2016}, which are
the first step towards the simulation of a topological insulator
with TRS.

iv) The concept of synthetic dimensions offers an additional
advantage for the experimental exploration of topological states
in cold gases. One kind of synthetic dimension consists of interpreting a set of addressable
internal states of an atom, e.g. Zeeman sublevels of a hyperfine
state as fictitious lattice sites; this defines an extra spatial
dimension coined synthetic dimension. Therefore, driving
transitions between different internal states corresponds to
inducing hopping processes along the synthetic dimension.
In turn, loading atoms into a real N-dimensional spatial OL potentially allows one to simulate systems of $N+1$
spatial dimensions. Synthetic dimensions were recently realized in 1D OLs for investigating the chiral edge states in the 2D QHE \cite{Mancini2015,Stuhl2015,Livi2016}. Notably a dynamical version of the 4D QHE \cite{ZhangSC2001,YLi2013a,Price2015} has been experimentally achieved with cold atoms in a 2D optical superlattice with two synthetic dimensions \cite{Lohse2018}.

v) Besides the possibility of engineering single particle
Hamiltonians, there are several methods to flexibly tune complex
many-body interactions in cold atoms. Strong correlation plays important
roles for some typical topological quantum matter, such as
fractional quantum Hall states and spin liquids. More recently,
there has been intense interest in the possibility of realizing
fractional quantum Hall states in lattice systems: the fractional
Chern insulators. The tunability of atomic on-site interactions \cite{Chin2010} or long-range
dipole-dipole interactions in ultracold dipolar gases \cite{Baranov2008,Lahaye2009}
opens up the possibility of realizing various new topological states
with strong correlations, including fractional anyonic statistics,
an unambiguous signature of topological phases.

vi) Compared with condensed-matter systems, ultracold atoms
allow detailed studies of the relation between dynamics and
topology as the timescales are experimentally easier to access.
For example, time-dependent OLs constitute a powerful
tool for engineering atomic gases with topological properties.
Recently, the identification of non-equilibrium signatures of
topology in the dynamics of such systems has been reported by
using time- and momentum-resolved full state tomography for
spin-polarized fermionic atoms in driven OLs \cite{TLi2016,Flaschner2016,Flaeschner2017}. These
results pave the way for a deeper understanding of the connection
between topological phases and non-equilibrium dynamics.

vii) Another remarkable advantage for studying topological phases
with cold atoms is that the topological invariants can be directly
detected in this system. For example, the Chern number has been
directly detected by measuring the quantized center-of-mass response \cite{Aidelsburger2014}. It can also be observed through the
Berry-curvature-reconstruction scheme \cite{TLi2016,Flaschner2016} or by measuring the spin
polarization of an atomic cloud at highly-symmetric points of the
Brillouin zone (BZ) \cite{ZWu2016}. Furthermore, nontrivial edge states can be visualized in
real space since the high-resolution addressing techniques offer
the possibility of directly loading atoms into the edge states and
cold atoms can be visualized by imaging the atomic cloud in-situ.
Momentum distributions and band populations can also be obtained
through time-of-flight imaging and band-mapping, respectively.

In this review, we take a closer look at the merger of two fields: topological
quantum matter as discussed in condensed matter physics and ultracold atoms. Both are active fields of research with a large amount of literature. For readers interested in more specialized reviews of quantum simulation with
ultracold atoms, we recommend review articles \cite{Lewenstein2007,Bloch2008,Cooper2008,Dalibard2011,DWZhang2011,Bloch2012,Galitski2013,Goldman2014,HZhai2015,XZhou2015,Goldman2016,Gross2017,Cooper2018}. For readers interested in more dedicated reviews on topological phases in condensed matter, we recommend Refs. \cite{Hasan2010,XLQi2011,Wehling2014,Bansil2016,Chiu2016,Armitage2018}. The aim of
this review is to satisfy the needs of both newcomers and
experts in this interdisciplinary field. To cater to the needs
of newcomers, we devote Sec. \ref{SecII} to a tutorial-style
introduction to topological invariants commonly used in condensed
matter physics, and the more general introductions are put in the Appendix \ref{App}. In Sec. \ref{SecIII}, we describe how the Hamiltonians can be fully engineered
in cold atom systems. A reader new to condensed
matter physics or ultracold atomic physics would find these two sections beneficial. In Sec. \ref{SecIV}, our emphasis is on recent theoretical and experimental developments on how to realize various topological states (models) or phenomena in different OL systems. In Sec. \ref{SecV}, we introduce the developed methods for probing topological invariants and other intrinsic properties of the topological phases in cold atom systems. In Sec. \ref{SecVI} and Sec. \ref{SecVII}, we move beyond single-particle physics of Bloch bands in lattice systems to describe some quantum matter in the continuum and interacting many-body phases that have topologically nontrivial properties. Finally, an outlook on future work and a brief conclusion are given.

\section{Topological invariants in momentum space} \label{SecII}

The purpose of this section is to briefly introduce various topological invariants referenced in the following sections. The more general introduction of topological invariants with the derivations of many formulas in this section are put in the Appendix \ref{App}.

\subsection{Gauge fields in momentum space} \label{gauge-field}

We denote the momentum-space Hamiltonian of an insulator as $\H(\k)$ with $\k$ in the first Brillouin zone (BZ), and assume finite number of bands, namely $\H(\k)$ is a $(M+N)$-dimensional matrix at each $\k$, where $M$ and $N$ are numbers of conduction and valence bands, respectively. At each $\k$, $\H(\k)$ can be diagonalized and the conduction and valence eigenpairs are $(E_{+,a},~|+,\k,a \rangle)$ and $(E_{-,b},~|-,\k,b\rangle)$, respectively, with $a=1,\cdots,M$ and $b=1,\cdots,N$.
At each $\k$, valence states $|-,\k,b\rangle$ span an $N$ dimensional vector space and these vector space spread smoothly over the whole BZ. We can define the Berry connection (gauge potential) as
\begin{equation}
\A_{b,b'}^\mu(\k)=\langle-,\k,b|\frac{\partial}{\partial k_\mu}|-,\k,b'\rangle \label{gauge-potential}
\end{equation}
with $\mu=1,2,\cdots,d$ labeling momentum coordinates. Accordingly, the Berry curvature (gauge field strength) is given by
\begin{equation}
\F^{\mu\nu}=\partial^\mu \A^\nu-\partial^\nu\A^\mu+[\A^\mu,\A^\nu]. \label{gauge-field}
\end{equation}

To have a basic idea of the Berry connection and curvature in momentum space, we take a general two-band model as an example. The Hamiltonian reads
\begin{equation}\label{M-2-b-model}
\mathcal{H}_{2b}(\k)=\mathbf{d}(\k)\cdot\sigma,
\end{equation}
where $\sigma_i$ with $i=1,2,3$ are the Pauli matrices. Strictly speaking, the term $\epsilon(\k)\mathrm{1}_2$ should also be added into Eq.~\eqref{M-2-b-model}. But it is ignored here because it is irrelevant to the topology of the band structure, noticing that it only shifts the energy spectrum and does not affect
eigenstates.  As the spectrum is given by $E_{\pm}(\k)=\pm |\mathbf{d}(\k)|$, for insulator $|\mathbf{d}(\k)|$ is not equal to zero for all $\k$. The valence eigenstates can be represented by $|-,\k\rangle=e^{-i\sigma_3\phi(\k)/2}e^{-i\sigma_2\theta(\k)/2}|\downarrow\rangle$, where $\theta(\k)$ and $\phi(\k)$ are the standard spherical coordinates of $\hat{\mathbf{d}}(\k)\equiv\mathbf{d}(\k)/|\mathbf{d}(\k)|$, and $|\downarrow\rangle$ is the negative eigenstate of $\sigma_3$. The Berry connection can be straightforwardly derived as
\begin{equation}
\A^\mu(\k)=\frac{i}{2}\cos\theta(\k)\,\partial^{\mu}\phi(\k).
\end{equation}
Under the $U(1)$ gauge transformation $|-,\k\rangle\rightarrow e^{i\varphi(\k)}|-,\k\rangle$, the Berry connection $\A^{\mu}(\k)$ is transformed to be $\A^\mu(\k)+i\partial_{k_\mu}\varphi(\k)$. But the Berry curvature is invariant under gauge transformations, and is given from Eq.~\eqref{gauge-field} by $\F^{\mu\nu}(\k)=-\frac{i}{2}\sin\theta(\k)[\partial^\mu\theta(\k)\partial^\nu\phi(\k)-\partial^\nu\theta(\k)\partial^\mu\phi(\k)]$, which can be recast in terms of $\hat{\mathbf{d}}(\k)$ as
\begin{equation}\label{M-Field-2b}
\F^{\mu\nu}(\k)=\frac{1}{2i}\hat{\mathbf{d}}\cdot(\partial^\mu \hat{\mathbf{d}}\times\partial^\nu\hat{\mathbf{d}}).
\end{equation}

\subsection{Quantized Zak phase}

The simplest example of topological invariant in momentum space is the so-called quantized Zak phase.
The Zak phase is a Berry's phase picked up by a particle moving across
a 1D BZ \cite{Zak1989}. For a given Bloch wave $\psi_k(x)$
with quasimomentum $k$, the Zak phase  can be conveniently
expressed through the cell-periodic Bloch function
$u_k(x) = e^{-ikx}\psi_k(x)$:
\begin{equation}
\label{Zak1}
\gamma = i\int_{-G/2}^{G/2} \A_k dk,
\end{equation}
where the gauge potential in Eq.~\eqref{gauge-potential} is given by $\A_k=\langle u_k|\partial_k |u_k\rangle$ and $G = 2\pi/a$ is the reciprocal lattice vector and $a$ is
the lattice period. As $i\partial_k$ in Eq.~\eqref{Zak1} is the position operator, physically $\gamma a/(2\pi)$ is just the center of the Wannier function corresponding to $u_k(x)$. Accordingly, it is noticed that the Zak phase $\gamma $, Eq.~\eqref{Zak1}, is well defined module $2\pi$, because a shift of the lattice origin by $d$, which corresponds to $u_k(x)\rightarrow e^{ikd}u_k(x)$, changes Eq.~\eqref{Zak1} by $2\pi d/a$.
So the Zak phase $\gamma$ can be any real number mod $2\pi$, and therefore is not a topological invariant. However, certain symmetries can quantize it into integers in units of $\pi$. The quantization of Eq.~\eqref{Zak1} was first discussed in 1D band theory by Zak taking into account the inversion symmetry~\cite{Zak1989}. In order to preserve the inversion symmetry, the wanner-function center has to be either concentrated at lattice sites or at the midpoints of lattice sites.

A paradigmatic 1D model with the topological invariant being the Zak phase is provided
by the Su-Schrieffer-Heeger model of polyacetylene \cite{WPSu1979}, which exhibits two topologically distinct
phases. A unit cell in this model has two sites with sublattice symmetry, which quantizes the Zak phase. Accordingly the cell-periodic wave function $u_k$ can be viewed as a two-component spinor $u_k = (\alpha_k, \beta_k)$, and the Zak phase,
Eq.~\eqref{Zak1}, in units of $\pi$ takes an simple form $\nu_{Z} = (i/\pi)\int_{-G/2}^{G/2} (\alpha_k^{*}\partial_k\alpha_k+\beta_k^{*}\partial_k\beta_k) dk$.

\subsection{Chern numbers}

The second example of the topological invariants in momentum space is the famous Chern number, which can be formulated for any even-dimensional spaces. For $2n$ dimensions, the corresponding Chern number is called the $n$th Chern number, and the corresponding integrand is called the $n$th Chern character. We first introduce the first Chern number (conventionally called Chern number), which appears in 2D momentum space $\mathbf{k}=(k_x,k_y)$. The BZ forms a torus $\mathbb{T}^2$
and the  Chern number for a $2$D insulator is given as
\begin{equation}
{C}=\frac{i}{2\pi}\int_{\mathbb{T}^2} d^2k~\mathrm{tr}\F^{xy}. \label{M-First-Chern}
\end{equation}
Noticing that the trace over the commutator in Eq.~\eqref{gauge-field} vanishes, we find that the Chern number essentially comes from the Abelian connection
$a^\mu=\mathrm{tr}\A^\mu$, which is just the sum of the Abelian Berry connection of all valence bands, namely, that $a^j=\sum_{\alpha}\langle \k, \alpha|\partial/\partial k_j| \mathbf{k},\alpha\rangle$ with $\alpha$ labeling the valence bands.
Accordingly the Chern number of Eq.~\eqref{M-First-Chern} can be rewritten in terms of the Abelian connection as
\begin{equation}
\label{Chern-BZ}
{C}=\frac{i}{2\pi}\int \int_{BZ}  f^{xy}\mathbf{(k)}dk_x  dk_y, \ \ \ f^{xy}\mathbf{(k)}=\frac{\partial a^y\mathbf{(k)} }{\partial k_x}-\frac{\partial a^x\mathbf{(k)} }{\partial k_y},
\end{equation}
The Chern number of Eq.~\eqref{M-First-Chern} is also called the Thouless-Kohmoto-Nightingale-den Nijs (TKNN) invariant, which was shown to be the transverse conductance in units of $e^2/h$ using the Kubo formula, and therefore is the topological invariant to characterize the integer quantum Hall effect~\cite{Thouless1982}. A nonvanishing transverse conductance requires the TRS breaking, which is consistent with Eq.~\eqref{M-First-Chern}, since 
$i\F$ is odd under TRS.
For the two-band model of Eq.~\eqref{M-2-b-model}, the Chern number can be expressed explicitly by
\begin{equation}
C=\frac{1}{4\pi}\int\int_{BZ}~\hat{\mathbf{d}}\cdot(\partial_{k_x}\hat{\mathbf{d}}\times\partial_{k_y}\hat{\mathbf{d}})dk_x  dk_y, \label{M-Simplified-Chern}
\end{equation}
which can be derived by directly substituting Eq.~\eqref{M-Field-2b} into Eq.~\eqref{Chern-BZ}.

If $n=2$, the second Chern number for a $4$D insulator is given by
\begin{equation}
{C}_2=-\frac{1}{32\pi^2}\int_{\mathbb{T}^4}d^4k~\epsilon_{\mu\nu\lambda\sigma}\mathrm{tr} \F^{\mu\nu}\F^{\lambda\sigma}. \label{M-Second-Chern}
\end{equation}
For more than one valence bands, the second Chern number, Eq.~\eqref{M-Second-Chern}, cannot be expressed in terms of the Abelian Berry connection $a^\mu(\k)$, which is in contrast to the first Chern number, and therefore is essentially non-Abelian. It was predicted that the second Chern number ${C}_2$ can be used to characterize a quantum Hall effect in 4D space \cite{ZhangSC2001}, which was realized in a recent experiment with ultracold atoms loaded in an optical lattice with synthetic dimensions \cite{Lohse2018}.
Furthermore, in contrast to that all systems with TRS have vanishing first Chern number,  the second Chern number of Eq.~\eqref{M-Second-Chern} can preserve TRS, namely, that there exist nontrivial time-reversal-invariant $4$D Chern insulators. In addition, the meaning of the second Chern number for electromagnetic response can be found in Refs.~\cite{Golterman1993,XLQi2008}.

Let us consider isolated gap-closing points in a $(2n+1)$D BZ, where the Berry connection is not well-defined. Although the Berry connection is singular at any gap-closing point, a $(2n)$D sphere $\mathbb{S}^{2n}$ can be chosen to enclose it, restricted on which the spectrum is gapped with the well-defined Berry connection. Accordingly the Chern number can be calculated on the $\mathbb{S}^{2n}$, and is referred to as the monopole charge of the singular point. For monopoles in $3$D space, the monopole charge can be calculated by the Abelian Berry connection $a^\mu=\mathrm{tr}\A^\mu$, and therefore are termed as Abelian monopoles. For instance the Weyl points described by the Hamiltonian $\H_{W}(\k)=\pm\k\cdot\sigma$ can be interpreted as unit Abelian monopoles in momentum space for the respective Abelian gauge field of valence band restricted on $\mathbb{S}^2$ surrounding the origin. The monopole charges defined in higher dimensions are introduced in the Appendix.

\subsection{Spin Chern number and $\Z_2$ topological invariants}\label{TIZ2}

We further consider particles with spin-1/2 (or pseudo-spin-1/2) in 2D momentum space.  
If the $U(1)$ spin-rotation symmetry to any specific direction (denoted as $z$-direction here) is preserved, the corresponding spin polarization $s =\uparrow,\downarrow$ is a good quantum number, and therefore the notation $|\mathbf{k},\alpha\rangle$ of valence bands used above should be refined as $|\mathbf{k},\alpha,s \rangle$. Then each spin $s$ can be individually assigned a Chern number $C_s$ as that of Eq.~\eqref{Chern-BZ}, which is the sum of the Chern numbers of all valence bands with the corresponding spin $s$ and naturally integer valued. As a topological insulator it is now characterized by two topological indices, the usual Chern number $C$ and the spin Chern number $C_s$ \cite{DNSheng2006}, respectively given by
\begin{equation}
C = C_\uparrow + C_\downarrow, \ \ \ \ C_s = (C_\uparrow - C_\downarrow)/2.
\end{equation}
Provided TRS is preserved (thus $C=0$) as well as the $U(1)$ spin-rotation symmetry, the spin Chern number $C_s$ is also integer valued. In this case $C_s$ can be used to characterized the quantum spin Hall effect~\cite{DNSheng2006}.


Notably, the $U(1)$ spin-rotation symmetry can usually be broken by generic spin-orbital couplings and therefore is not a good symmetry, but TRS is still preserved in the absence of magnetic field. In the general situation with only TRS, the spin Chern number $C_s$ is no longer well-defined and should be replaced by a $\Z_2$ topological invariant for characterizing the 2D topological insulators with TRS \cite{Kane2005b,Bernevig2006a,Bernevig2006b}, which was first proposed by Kane and Mele in Ref. \cite{Kane2005a}.  The $\Z_2$ topological invariants proposed there can be generalized to characterize 3D time-reversal-invariant topological insulators. These $\Z_2$ topological invariants are briefly introduced in the Appendix \ref{App}.

\subsection{The Hopf invariant}\label{HopfIndex}
There is a kind of topological insulator restricted in both two bands and three dimensions. For a two-band insulator, the Hamiltonian \eqref{M-2-b-model} at each $\k$ can be topologically regarded as a point $\hat{d}(\k)$ on a unit sphere $\mathbb{S}^2$, and thereby it
gives a mapping from the $3$D BZ to $\mathbb{S}^2$.
Because of the homotopy group $\pi_{3}(\mathbb{S}^2)\cong \Z$, there exist (strong) $3$D two-band topological insulators with $\Z$ classification, which is termed the Hopf insulators \cite{Moore2008}. The corresponding topological invariant is called the Hopf invariant~\cite{Wilczek1983,Moore2008}, and is given by
\begin{equation}
\nu_{H}=-\frac{1}{4\pi^2}\int_{\mathbb{T}^3}d^3k~\epsilon_{\mu\nu\lambda} \A^{\mu}\partial^{\nu}\A^{\lambda}, \label{Hopf-invariant}
\end{equation}
where $\A^\mu=\langle-,\k|\partial_{ k_\mu}|-,\k\rangle$ is the Berry connection of the valence band defined in Eq.~\eqref{gauge-potential}.


\section{Engineering the Hamiltonian of atoms}\label{SecIII}

For particles of mass $m$ and index $i$, charge $q$ and magnetic moment
$\mathbf{\mu}_B$, in an electromagnetic field described by the
vector potential $\mathbf{A} = (A_x, A_y, A_z)$ and scalar
potential $V(\mathbf{r})$, the Hamiltonian is given by
\begin{equation}
\label{HEA}
H=\sum_i\left[\frac{1}{2m}(\mathbf{p}_i-q\mathbf{A})^2+V(\mathbf{r}_i)-\mathbf{\mu}_B\cdot
\mathbf{B}(\mathbf{r}_i)\right]+U_{\text{int}},
\end{equation}
where $\mathbf{p}_i = -i\hbar \nabla_i$ is the momentum operator,
$\mathbf{B}$ is the magnetic field, and $U_{\text{int}}$ is the
Hamiltonian caused by the interaction between particles. One of
the great advantages of ultracold atomic systems is that, all
terms in the Hamiltonian (\ref{HEA}) are tunable in experiments, and
thus many exotic quantum phases, including various topological
phases addressed latter in this review, can be realized. In this
section, we first briefly review the methods to modify
the mean kinetic energy $\langle p_i^2\rangle/2m$ related to the temperature and  the interaction $U_{\text{int}}$, and
then address more detailed the approaches to engineer the so-called artificial gauge fields for neutral atoms (the vector potential $\mathbf{A}$, the scalar potential $V(\mathbf{r}_i)$, and the effective Zeeman field $\mathbf{B}(\mathbf{r}_i)$), which are fundamentally
important in creating various exotic topological phases.

\subsection{Laser cooling}

The mean kinetic energy of the atoms $\langle p_i^2\rangle/2m$ is mainly determined by
the temperature of the atomic cloud and can be controlled by
laser cooling, which refers to a number of techniques in
which atomic samples are cooled down to near absolute zero. Laser
cooling techniques rely on the fact that when an atom absorbs and
re-emits a photon its momentum changes. For an ensemble of
particles, their temperature is proportional to the variance in
their velocities. That is, more homogeneous velocities among
particles corresponds to a lower temperature. Laser cooling
techniques combine atomic spectroscopy with the mechanical effect
of light to compress the velocity distribution of an ensemble of
particles, thereby cooling the particles. A Nobel prize was
awarded to three physicists, S. Chu, C. N. Cohen-Tannoudji, and W. D. Phillips, for their achievements of laser cooling of atoms in 1997. 

The first proposal of laser cooling by H\"ansch and Schawlow in
1974 \cite{Hansch1975} was based upon Doppler cooling in a two-level
atom. It was suggested that the Doppler effect due to the thermal
motion of atoms could be exploited to make them absorb
laser light at a different rate depending on whether they moved
away from or toward the laser. Consider an atom irradiated by
counterpropagating laser beams that are tuned to the low frequency side of
atomic resonance. The beam counterpropagating with the atom will
be Doppler shifted towards resonance, thus increasing the
probability of photon absorption. The beam co-propagating with the
atom will be frequency-shifted away from resonance, so there will
be a net absorption of photons opposing the motion of the atom.
The net momentum kick felt by the atom could then be used to slow itself down.
By surrounding the atom with three pairs of
counter-propagating beams along the $x$, $y$ and $z$ axes, one can  generate a drag
force opposing the velocity of the atom. The term
"optical molasses" was coined to describe this situation.

When this simple principle was finally applied in the early 1980s, it
immediately led to low temperatures only a few hundreds of
micro-Kelvins above absolute zero. As an example, the mean velocity
of a $^{87}$Rb atomic gas in temperature of $100$ mK is about $0.17$
m/s, which is much slower than the velocity of several
hundred meters per second at room temperature. Ultracold atoms
also turned out to be an ideal raw material for the realization of
magnetic traps for neutral atoms. Held in place by magnetic dipole
forces, such atomic gases can then be evaporatively cooled by
successively lowering the trap depth, thus letting the most
energetic atoms escape and allowing the remaining ones to
rethermalize. In this way, the fundamental limitations of laser
cooling due to photon scattering can be overcome and the temperature as low as a few nano-Kelvins can be reached. The mean
velocity of a $^{87}$Rb atomic gas in temperature of $1$ nK is about
$5.3 \times 10^{-4} $ m/s.

\subsection{Effective interactions}

The term $U_{\text{int}}$ in Eq. (\ref{HEA}) is induced by interatomic interactions
and can be manipulated with a powerful method called the Feshbach resonance (for a review, see Ref. \cite{Chin2010}).
The fundamental result for the atom-atom scattering is that under appropriate conditions, the
effective interaction potential $U_{\text{int}}(\mathbf{r}=\mathbf{r}_i-\mathbf{r}_j)$ of two atoms (particle indices $i$ and $j$) of reduced mass $m_r$ can be replaced by a delta function of strength
$2\pi\hbar^2 a_s /m_r$, where $a_s$ is the low-energy $s$-wave
scattering length. As for two similar particles with mass $m$,
the commonly quoted form of the effective interaction is
\begin{equation}
\label{Uint} U_{\text{int}}(\mathbf{r})=\frac{4\pi a_s \hbar^2}{m}\delta
(\mathbf{r}).
\end{equation}
Alternative, it can be understood in the following
way: the mean interaction energy of the many-body
system is given by the expression
\begin{equation}
\label{E_int} \langle E_{\text{int}}\rangle=\frac{1}{2} \frac{4\pi a_s
\hbar^2}{m} \sum_{ij} |\Psi(r_{ij}\rightarrow 0)|^2,
\end{equation}
where $\Psi$ is the many-body wave function and the notation
$r_{ij}\rightarrow 0$ means that the separation $r_{ij}$ of the
two atoms, while large compared to $a_s$, is small compared to
any other characteristic length (e.g., thermal de Broglie wavelength,
interparticle spacing, etc).  The conditions necessary for the
validity of Eq. (\ref{E_int}) in the time-independent case are the
following: First, the orbital angular momentum $l \neq 0$ scattering must be negligible.
Second, the existence of the limit $r_{ij}\rightarrow 0$ implies the condition $k_c a_s \ll 1$, where $k_c$
is the characteristic wave-vector scale of the many-body wave
function $\Psi$ (for a very general argument, see Ref.
\cite{Leggett2001}).

The scattering length $a_s$ can be manipulated by a Feshbach
resonance \cite{Chin2010}. It occurs when the bound molecular state in the closed channel energetically
approaches the scattering state in the open channel. Then even
weak coupling can lead to strong mixing between the two channels.
The energy difference can be controlled via a magnetic field when
the corresponding magnetic moments are different. This leads to a
magnetically tuned Feshbach resonance. The magnetic tuning method
is the common way to achieve resonant coupling and it has found
numerous applications. A magnetically tuned Feshbach resonance without inelastic two-body
channels can be described by a simple expression, introduced by
Moerdijk et al. \cite{Moerdijk1995}, for the s-wave scattering
length $a_s$ as a function of the magnetic field strength $B$,
\begin{equation}
a_s(B) = a_0\left(1-\frac{\Delta_\text{rw}}{B-B_0}\right).
\end{equation}
The background scattering length $a_0$ represents the off resonant
value. The parameter $B_0$ denotes the resonance position, where
the scattering length diverges $(a_s\rightarrow\pm\infty)$, and
the parameter $\Delta_\text{rw}$ is the resonance width. Note that both
$a_0$ and $\Delta_\text{rw}$ can be positive or negative, thus the interaction energy $U_{\text{int}}$ can be
positive or negative and even infinity by just controlling the
magnetic field strength $B$. Alternatively, resonant coupling can be
achieved by optical methods, leading to optical Feshbach
resonances with many conceptual similarities to the magnetically
tuned case. Such resonances are promising for cases where
magnetically tunable resonances are absent.

\subsection{Dipole potentials and optical lattices}

\emph{The dipole potentials.} The potentials $V(\mathbf{r})$ in Eq. (\ref{HEA}) can be manipulated
with the laser beams. As for the topological band structures
reviewed in this paper, we are particularly interested in OLs formed by the light-atom interactions. OLs
and other optical traps work on the principle of the ac Stark
shift. In order to understand the origin of light-induced atomic
forces and their applications in laser cooling and trapping it is
instructive to consider an atom oscillating in an electric field.
When an atom is subjected to a laser field, the electric field
$\mathbf{E}$ induces a dipole moment $\mathbf{p}_d$ in the atom as
the protons and surrounding electrons are pulled in opposite
directions. The dipole moment is proportional to the applied
field, $\mathbf{p}_d = \alpha (\omega) \mathbf{E}$, where the
complex polarizability of the atoms $\alpha (\omega)$ is a
function of the laser light's angular frequency $\omega$. The
potential felt by the atoms is equivalent to the ac Stark shift
and is defined as
\begin{equation}
\label{V_dip}
V(\mathbf{r})=-\frac{1}{2}\langle\mathbf{p}_d\cdot\mathbf{E}\rangle=-\frac{1}{2}\alpha
(\omega) \langle\mathbf{E}^2(t)\rangle,
\end{equation}
where the angular brackets $\langle\cdot\rangle$ indicate a time
average in one cycle.

For a two-level atomic system, away from resonance and with
negligible excited state saturation, the dipole potential can be
derived semiclassically. To perform such a calculation, the
polarizability is obtained by using Lorentz¡¯s model of an
electron bound to an atom with an oscillation frequency equal to
the optical transition angular frequency $\omega_0$. The natural
line width has a Lorentzian profile as the Fourier transform of an
exponential decay is a Lorentzian. Then the dipole potential
calculated by the two-level model is given as
\begin{equation}
\label{V_dip2} V (\mathbf{r})=-\frac{3\pi
c^2\Gamma}{2\omega_0^3}\left(
\frac{1}{\omega_0-\omega}+\frac{1}{\omega_0+\omega}   \right) I
(\mathbf{r}),
\end{equation}
where $\Gamma$ is the natural line width of the excited state and
has a Lorentzian profile, and $I(\mathbf{r})=\epsilon_0
c|\mathbf{E}(\mathbf{r})|^2/2$ is the laser intensity at the
position $\mathbf{r}$. For small detuning $\Delta_d=\omega-\omega_0$
and $\omega/\omega_0 \approx 1$, the rotating wave approximation
can be made and the $1/(\omega_0 + \omega)$ term in
Eq. (\ref{V_dip2}) can be ignored. Under such an assumption, the
scale of the dipole potential $V(\mathbf{r}) \propto
I(\mathbf{r})/\Delta_d$. Therefore, a blue-detuned laser (i.e., the
frequency of the light field is larger than the atomic transition
frequency ($\Delta_d>0$)) will produce a positive AC-stark shift. The resulting dipole potential will be such that its gradient,
which results in a force on the atom, points in the direction of
decreasing field. On the other hand, an atom will be attracted to the
red-detuned ($\Delta_d<0$) regions of high intensity.

\emph{Optical lattices.} A stable optical trap can be realized by simply focusing a laser
beam along the $z$ direction to a waist of size $w$ under the
red-detuned condition. If the cross section of the laser beam is
a Gaussian form, with ${w}_0$ and $z_R={w}^2_0\pi/\lambda$ being the spot
(waist) and Rayleigh lengths, respectively, the resulting dipole
potential is given as
\begin{equation}
V (r,z)=V_0\exp \left(
-\frac{2r^2}{{w}^2_0\sqrt{1+(z/z_R)^2}}\right),
\end{equation}
where the trap depth $V_0=I_p/\Delta_d$  with $I_p$ being the peak
intensity of the beam. Expanding this expression at the waist
$z=0$ around $r=0$, we obtain that in the harmonic approximation
the radial trap frequency in such a potential is given by
$\omega_{\perp}=\sqrt{2V_0/m}/w_0$. Besides this radial trapping
force, there is also a longitudinal force acting on the atoms.
However, this force is much less than the radial one owing to the
much larger length scale given by the Rayleith length $z_R$. To
confine the atoms tightly in all spatial directions, one can use
several crossed dipole traps or superpose an additional magnetic
trap.

The possibility to create dipole potentials proportional to the
laser intensity allows for the creation of OL potentials from standing light waves \cite{Bloch2005}, as artificial crystals of light to trap ultracold atoms. As an example, we first address how to realize a 1D lattice created by two counterpropagating laser
beams with wave vectors $\mathbf{k}_L$ and $-\mathbf{k}_L$.  We
consider two identical laser beams of peak intensity $I_p$ and
make them counterpropagate in such a way that their cross sections
completely overlap. In addition, we also arrange their
polarizations to be parallel. In this case, the two beams can
create an interference pattern, with a distance $\lambda_L/2$
($\lambda_L=2\pi/k_L$ and $k_L=|\mathbf{k}_L|$) between two
maxima or minima of the resulting light intensity. Therefore, the
potential seen by the atoms is simply given by
\begin{equation}
\label{V_one} V_{\text{lat}}(x)=V_0\cos^2(\pi x/d),
\end{equation}
where the lattice spacing $d=\lambda_L/2$ and $V_0$ is the lattice depth.

Note that mimicking solid-state crystals with an OL
has the great advantage that, in general, the two obvious
parameters in Eq.(\ref{V_one}), the lattice depth $V_0$ and the
lattice spacing $d$ can be easily controlled by changing the
laser fields. Rather than directly calculating the lattice depth
$V_0$ from the atomic polarizability in Eq. (\ref{V_dip}), one
typically uses the saturation intensity $I_0$ of the transition
and obtains $V_0=\eta\hbar\Gamma\frac{\Gamma}{\Delta}\frac{I_p}{I_0}$, where
the prefactor $\eta$ of the order unit depends on the level structure
of the atom in question through the Clebsh-Gordan coefficients of
the various possible transitions between sublevels. Thus, the
lattice depth $V_0$ is proportional to the laser intensity $I_p$,
which can be easily controlled by using an acousto-optic
modulator. This device allows for a precise and fast (less than a
microsecond) control of the lattice beam intensity and introduces
a frequency shift of the laser light of tens of MHz. Typically,
the lattice depth is measured in units of the recoil energy
$E_R=\pi^2\hbar^/(2md^2)$, and often the dimensionless parameter
$s=V_0/E_R$ is used. It corresponds to the kinetic energy required
to localize a particle on the length of a lattice constant $d$.
Recoil energies are of the order of several kilohertz, roughly
corresponding to microkelvin or several picoelectron volts. The
lattice depth can take values of up to hundreds of recoil
energies. On the other hand, the lattice spacing $d=\lambda_L/2$
between two adjacent wells of a lattice can be enhanced by making
the two counterpropagating beams intersect at an angle $\theta
<\pi$. Assuming that the polarizations of the two beams are
perpendicular to the plane spanned by them, this will give rise to
a periodic potential with lattice constant
$d(\theta)=d/\cos(\theta/2)\ge d$.

In experiments, a 1D OL can be created in several ways. The simplest way is to take a linearly polarized
laser beam and retro-reflect it with a high-quality mirror. If the
retro-reflected beam is replaced by a second phase-coherent laser
beam, which can be obtained by dividing a laser beam in two with a
polarized beam splitter, we can introduce a frequency shift
$\delta\nu_L$ between the two lattice beams. The periodic lattice
potential will now no longer be stationary but move at a velocity
$v_{\text{lat}}=\delta\nu_L d$. If the frequency difference is varied at
a rate $\delta\dot{\nu}_L$, the lattice potential will be
accelerated with $a_{\text{lat}}=\delta\dot{\nu}_L d$. Therefore, there
will be a force $F=ma_{\text{lat}}=m\delta\dot{\nu}_L d$, acting on the
atoms in the rest frame of the lattice. We shall see latter that
this gives a powerful tool for manipulating the atoms in an
OL.

A superlattice or disordered lattice can be realized with two pairs of counterpropagating beams. We consider two counterpropagating beams, where the polarizations are perpendicular and the wave vectors are $k^L_1$ and $k^L_2$, respectively. In this case, each pair can form a lattice which is similar to that of Eq. (\ref{V_one}), and the resulting total potential is then given by
\begin{equation}
\label{V12} V_{\text{lat}}(x)=s_1E_{R1}\cos^2(\pi x/d_1)+s_2
E_{R2}\cos^2(\pi x/d_2),
\end{equation}
where $d_j=\pi/k_j^L$ $(j=1,2)$, and $s_1$ and $s_2$ measure the
height of the lattices in units of the recoil energies. A
superlattice with the period $pq$ is created when the ratio
$d_1/d_2=p/q$ (with $p, q$ being integers) is a rational number. For
instance, a dimerized lattice with two sites per unit cell is
realized when $d_1/d_2=1/2$, which is the famous Su-Schrieffer-Heeger model with a topological band structure (see Sec. \ref{SSHModel}). On
the other hand, a disordered lattice can be formed when the ratio
$d_1/d_2$ is an irrational number. Especially, when $s_2\ll s_1$ the disordering lattice has the
only effect to scramble the energies, which are nonperiodically
modulated at the length scale of the beating between the two
lattices $(2/\lambda_1^L-2/\lambda_2^L)^{-1}$ with
$\lambda_j^L=2\pi/k_j^L$. Theoretical and experimental works have
demonstrated that in finite-sized systems this quasi-periodic
potential can mimic a truly random potential and allow the
observation of a band gap \cite{Fallani2007,Roati2008}. Alternatively, for
a system of ultracold atoms in a lattice one can introduce
controllable disorders by using laser speckles \cite{Billy2008}.

By combining standing waves in different directions or by creating more complex interference patterns, one
can create various 2D and 3D lattice structures.
To create a 2D lattice potential for example, one can  use two
orthogonal sets of counter propagating laser beams. In
this case the lattice potential has the form
\begin{equation}
\label{V_two}
\begin{split}
V_{\text{lat}}(x,y) = & V_0 [\cos^2(k_L x) + \cos^2(k_L y)+ 2 \mathbf{\epsilon}_1\cdot\mathbf{\epsilon}_2\cos\phi
\cos(k_L
x) \cos(k_L y)],
\end{split}
\end{equation}
where $\mathbf{\epsilon}_1$ and $\mathbf{\epsilon}_2$ are
polarization vectors of the counter propagating set and $\phi$ is
the relative phase between them. In derivation of this equation, we have assumed
that the two pairs of laser beams have the same wave vector magnitude
$k_L$ and the same laser density $I_p$. A simple square lattice can be
created by choosing orthogonal polarizations between the standing
waves. In this case the interference term vanishes and the
resulting potential is just the sum of two superimposed 1D lattice
potentials. Even if the polarization of the two pair of beams is
the same, they can be made independent by detuning the common
frequency of one pair of beams from that the other. A more general
class of 2D lattices can be created from the interference of
three laser beams \cite{Blakie2004,SLZhu2007,Tarruell2012,Windpassinger2013}, which in
general yield non-separable lattices. Such lattices can provide
better control over the number of nearest-neighbor sites and allow for
the exploration of richer topological physics, such as the
honeycomb lattices for the Haldane model or Kane-Mele model.
Moreover, 3D lattices can be created with more laser
beams. For example, a simple cubic lattice $$V_{\text{lat}}(x,y,z) = V_0
[\cos^2(k_L x) + \cos^2(k_L y)+\cos^2(k_L z)]$$ can be formed with
three orthogonal sets of counter propagating laser beams when they
have the same wave vector magnitude $k_L$ and the same laser density
$I_p$, but have orthogonal polarizations.

\emph{The tight-binding Hamiltonian.} A useful tool to describe the particles in OLs is the
tight-binding approximation. It deals with cases in which the
overlap between localized Wannier functions at different sites is
enough to require corrections to the picture of isolated particles
but not too much as to render the picture of localized wave
functions completely irrelevant. In this regime, one can only take into account overlap between
Wannier functions in nearest neighbor sites as a very good
approximation. Wannier functions are
a set of orthonormalized wave functions that fully describes
particles in a band that are maximally localized at the
lattice sites. They can form a useful basis to describe the
dynamics of interacting atoms in a lattice. Furthermore, if
initially the atoms are prepared in the lowest band, the dynamics
can be restricted to remain in this band. In the absence of the
gauge potential and Zeeman field, the Hamiltonian in Eq. (\ref{HEA})
for the interacting particles in OLs is given by
\begin{equation}
\label{H_second}
H=-\frac{\hbar^2}{2m}\sum_i\nabla^2_i+V_{\text{lat}}(\mathbf{x})+V(\mathbf{x})+U_{\text{int}},
\end{equation}
where $V_{\text{lat}}(\mathbf{x})$ is the periodic lattice potential, and $V
(\mathbf{x})$ denotes any additional slowly-varying external
potential that might be present (such as a harmonic confinement
used to trap the atoms). In the grand canonical ensemble, the
second-quantized Hamiltonian reads
\begin{equation}
\label{H_2} H_{2} = \int  \Psi^\dagger (\mathbf{x}) \left[\frac{\hbar^2}{2m}\nabla^2+V_{\text{lat}}(\mathbf{x})+V(\mathbf{x})+U_{\text{int}}-\mu\right] \Psi
(\mathbf{x})d\mathbf{x},
\end{equation}
where $\Psi^\dagger (\mathbf{x})$ is the bosonic or fermionic field
operator that creates an atom at the position $\mathbf{x}$, and
$\mu$ is the chemical potential and acts as a Lagrange multiplier
to the mean number of atoms in the grand canonical ensemble.

We first consider the noninteracting situation. For sufficiently deep
lattice potentials, the atomic field operators can be expanded in
terms of localized Wannier functions. Assuming that the
vibrational energy splitting between bands is the largest energy
scale of the system, atoms can be loaded only in the lowest band,
where they will reside under controlled conditions. Then one can
restrict the basis to include only lowest band Wannier functions
$w_0(\mathbf{x})$, i.e., $\Psi(\mathbf{x})=\sum_j a_j w_0
(\mathbf{x}-\mathbf{x}_j)$, where $a_j$ is the annihilation
operator at site $j$ which obeys bosonic or fermonic canonical
commutation relations. The sum is taken over the total number of
lattice sites. If $\Psi(\mathbf{x})$ in this form is inserted in
Eq. (\ref{H_2}), and only the tunneling processes between nearest
neighbor sites are kept (Next-nearest-neighbor tunneling
amplitudes are typically two orders of magnitude smaller than
nearest-neighbor ones and they can be neglected.), one obtains the single-particle Hamiltonian
\begin{equation}
\label{H_single}
H=-\sum_{\langle i,j\rangle} J_{ij} a_i^\dagger a_j+\sum_j
(V_j-\mu) a_j^\dagger a_j,
\end{equation}
where $V_j=V(\mathbf{x}_j)$ and the notation $\langle i,j\rangle$
restricts the sum to nearest-neighbor sites.  $J_{ij}$ is the
tunneling matrix element between the nearest neighboring lattice sites
$i$ and $j$
\begin{equation}
J_{ij}=-\int dx w_0^{*}(x-x_i)\left[-\frac{\hbar^2}{2m}\nabla^2+V_{\text{lat}}(\mathbf{x})\right] w_0(x-x_{i+1}),
\end{equation}
Equation (\ref{H_single}) is a general noninteracting tight-binding Hamiltonian for atoms in
OLs.

\emph{The Hubbard models.} For interacting atoms in an OL, the Hubbard model can be
considered an improvement on the single-particle tight-binding model \cite{Jaksch2005,Bloch2008}. The Hubbard model was
originally proposed in 1963 to describe electrons in solids and
has since been the focus of particular interest as a model for
high-temperature superconductivity. The particles can either be
fermions, as in Hubbard's original work and named the (Fermi-)
Hubbard model, or bosons, which is referred to as the Bose-Hubbard
model. For strong interactions, it can give behaviors qualitatively different from those of the single-particle model
and correctly predict the existence of the so-called Mott insulators,
which are prevented from becoming conductive by the strong
repulsion between the particles. The Hubbard model is a good
approximation for particles in a periodic potential at
sufficiently low temperatures where all the particles are in the
lowest Bloch band, as long as any long-range interactions between
the particles can be ignored. If interactions between particles on
different sites of the lattice are included, the model is often
referred to as the ``extended Hubbard model".

The simplest nontrivial model that describes interacting bosons in
a periodic potential is the Bose-Hubbard Hamiltonian. It can be
derived from Eq. (\ref{H_single}) with the additional interacting
term $U_{\text{int}}$. In the grand canonical ensemble and
assuming the interactions are dominated by s-wave interactions,
i.e., $U_{\text{int}}=\frac{2\pi a_s\hbar^2}{m}|\Psi(\mathbf{x})|^2$, the
Bose-Hubbard Hamiltonian is given by \cite{Jaksch1998},
\begin{equation}
\label{H_BH}
H_{\text{BH}}=-J\sum_{\langle i,j\rangle} b_i^\dagger b_j+\sum_j
(V_j-\mu) b_j^\dagger b_j+\frac{U}{2}\sum_j b_j^\dagger
b_j^\dagger b_j b_j,
\end{equation}
where $U=(4\pi a_s\hbar^2/m)\int |w_0(\mathbf{x})|^4d\mathbf{x}$
accounts for interatomic interactions and measures the strength of
the repulsion of two atoms on the same lattice site. To express
that the atoms are bosons, the notation of the annihilation
operator in Eq. (\ref{H_BH}) is explicitly denoted as $b_j$.
While the parameter $J$ decreases exponentially with lattice depth $V_0$, $U$
increases as a power law of $V_0^{D/4}$, where $D$ is the
dimensionality of the lattice. The Bose-Hubbard model has been
used to describe many different systems in solid-state physics,
such as short correlation length superconductors, Josephson
arrays, critical behaviors of $^4$He and, recently, cold atoms in
OLs. The Bose-Hubbard Hamiltonian exhibits a quantum
phase transition from a superfluid to a Mott insulator state
\cite{Fisher1989}. Its phase diagram has been intensively studied via analytical
and numerical approaches with many different techniques and experimentally confirmed using ultracold atomic systems in 1D, 2D, and 3D lattice geometries \cite{Jaksch2005,Bloch2008}.

The ultracold atomic system also provides an almost ideal
experimental realization of the originally proposed Fermi-Hubbard
model with highly tunable parameters \cite{Esslinger2010}. To
simulate the spin-1/2 electrons in condensed matter physics, we may need
two-component 
Fermi gas trapped in OLs. The Fermi-Hubbard
Hamiltonian then takes the form
\begin{equation} H_{\text{FH}}=-J\sum _{\langle i,j\rangle ,\sigma }({
{c}}_{i,\sigma }^{\dagger }{ {c}}_{j,\sigma }+h.c.)+U\sum
_{i=1}^{N}{ {n}}_{i,\uparrow }{ {n}}_{i,\downarrow }+\sum_j
\epsilon_j n_j.
\end{equation}
%
Here the annihilation operator for
spin $ \sigma $ on $j$-th site is denoted as
$c_{j,\sigma}$, and $ n_{j,\sigma }=c_{j,\sigma }^{\dagger }
c_{j,\sigma}$ is the spin-density operator, with the total density operator $ n_j=n_{j,\uparrow }+
n_{j,\downarrow }$. The last term takes account of the additional
confinement $V(\mathbf{x})$ of the atom trap, which is usually
harmonic, with $\epsilon_j$ the corresponding energy offset on the $j$-th lattice site.

Experimentally, the tunnel amplitude in the Hubbard models is
controlled by the intensity of the standing laser waves. This
allows for a variation of the dimensionality of the system and
enables tuning of the kinetic energy. The energy width of the
lowest band is $W=4JD$. Due to the low kinetic energy of the atoms, two
atoms of different spins usually interact via s-wave scattering and
the coupling constant is given by $g=4\pi a_s/ m$. 
With this, the Hubbard interaction $U$ can be tuned
to negative or positive values by exploiting Feshbach resonances.
However, a single component Fermi gas is effectively
noninteracting because Pauli's principle does not allow $s$-wave
collisions of even parity.

\subsection{Artificial magnetic fields and spin-orbit couplings}\label{AMFSOC}

A magnetic field plays a crucial role in topological quantum
matter with broken TRS, whereas an SOC is a basic ingredient for those having TRS. Atoms are, however, electrically neutral; therefore, it is highly desirable to make them behave as charged particles in
an electromagnetic field. This capability has been explored and
demonstrated in a series of publications, including several nice
review papers \cite{Cooper2008,Dalibard2011,Galitski2013,Goldman2014,HZhai2015,XZhou2015}. In this section we describe three typical methods (geometric gauge potentials, laser-assisted tunneling and periodically driven OLs) to
generate artificial magnetic fields and SOCs for ultracold neutral atoms.

\subsubsection{Geometric gauge potentials}

When a quantum particle with internal structure moves
adiabatically in a closed path, Mead \cite{Mead1980} and Berry \cite{Berry1984} discovered
that a geometric phase, in addition to the usual dynamic phase, is accumulated on
the wave function of the particle. This geometric phase is a
generalization of Aharonov-Bohm phase \cite{Aharonov1959} that
a charged particle moving in a magnetic field acquires.
Therefore, an artificial magnetic field can
emerge in cold atom systems when the atomic center-of-mass motion
is coupled to its internal degrees of freedom through laser-atom
interaction. Based on this geometric
phase approach, Refs. \cite{SLZhu2006,Juzeliunas2006,Juzeliunas2004,Spielman2009,Gunter2009}
proposed setups for systematically engineering vector potentials associated with a non-zero artificial magnetic field for quantum
degenerate gases, and they have been experimentally realized for both bosonic
\cite{YJLin2009a,YJLin2009b} and fermionic atoms \cite{ZFu2011}.
When the local atomic internal states dressed by the laser fields
have degeneracies, effective non-Abelian gauge potentials can be
formed \cite{Ruseckas2005,Juzeliunas2008,Stanescu2007,Vaishnav2008,SLZhu2009},
manifesting as artificial SOCs in Bose-Einstein condensations \cite{CWu2008,Stanescu2008,CWang2010,Ho2011,HHu2012} or degenerate Fermi gases \cite{HHu2011,SLZhu2011}. The artificial SOCs have been
experimentally realized by several groups \cite{YJLin2011b,JYZhang2012,PWang2012,Cheuk2012,Beeler2013,Qu2013,Hamner2014,Olson2014,LHuang2016},
and they lead to an atomic spin Hall effect \cite{SLZhu2006,XJLiu2007}, which has been experimentally demonstrated \cite{Beeler2013}.

To understand these artificial gauge fields, we consider the
adiabatic motion of neutral atoms with $N$ internal levels in
stationary laser fields. The full Hamiltonian of the atoms reads
\begin{equation}
\label{H_original}
H=\frac{\mathbf{p}^{2}}{2m}+V(\mathbf{r})+H_{\text{AL}}
\end{equation}
where 
$H_{\text{AL}}$ represents the laser-atom interaction.
$H_{\text{AL}}$ depends on the position of the atoms and is a $N\times N$
matrix in the representation of the internal energy levels
$|j\rangle$. In addition, the potential $V(\mathbf{r})$ is assumed
to be diagonal in the internal states $|j\rangle$ with the form
$V(\mathbf{r})=\sum_{j=1}^N V_j(\mathbf{r})|j\rangle \langle j|$.
In this case, the full quantum state of the atoms (including
both the internal and the motional degrees of freedom) can then be expanded to $%
|\Phi (\mathbf{r})\rangle =\sum_{j=1}^{N}\phi
_{j}(\mathbf{r})|j\rangle $.

We may discuss the problem in the representation of the dressed
states $|\chi_n\rangle$ that are eigenvectors of the Hamiltonian
$H_{\text{AL}}$, that is, $H_{\text{AL}}$
$|\chi_n\rangle=\varepsilon_n|\chi_n\rangle$. Then the dressed
states $|\chi \rangle =(|\chi_1 \rangle,|\chi_2
\rangle,\cdots,|\chi_N \rangle)^{\top} $ (with $\top$ denoting the
transposition) are related to the original internal states
$|j\rangle$ with the relation $|\chi \rangle =U(|1\rangle
,|2\rangle ,\cdots,|N\rangle )^{\top}$, where the transform matrix
$U$ is a unitary operator. In the new basis $|\chi \rangle $, the
full quantum state of the atom $|\Phi (\mathbf{r})\rangle $ is
written as $|\Phi (\mathbf{r})\rangle =\sum_{j}\psi
_{j}(\mathbf{r})|\chi _{j}(\mathbf{r})\rangle $, where the wave functions $%
|\Psi\rangle =(|\psi _{1}\rangle,|\psi _{2}\rangle,\cdots,|\psi
_{N}\rangle)^{\top}$ obey the Schr\"{o}dinger equation $i\hbar
\frac{\partial}{\partial {t}}|\Psi\rangle =H_{\text{eff}}|\Psi\rangle $,
with the effective Hamiltonian $H_{\text{eff}}=UHU^{\dagger}$ taking the
following form:
\begin{equation}
H_{\text{eff}}=\frac{1}{2m}(-i\hbar \nabla -\mathbf{A})^{2}+\varepsilon I_N+\tilde{V}(%
\mathbf{r}).  \label{H_eff}
\end{equation}
Here $\mathbf{A}=i\hbar U\nabla U^{\dagger }$, $\tilde{V}(%
\mathbf{r})=UV(\mathbf{r})U^{\dagger }$,
$\varepsilon=(\varepsilon_1,\varepsilon_2,\cdots,\varepsilon_N)^{\top}$,
and $I_N$ is the $N\times N$ unit
matrix \cite{Ruseckas2005,Wilczek1984,SLZhu2006}. In the derivation
we have used the operator identity
$U\mathbf{P}^{2}U^\dagger=\left(-i\hbar\nabla-i\hbar U\nabla
U^\dagger \right)^2$ because of $\nabla (U^\dagger U)=0$. From
Eq. (\ref{H_eff}), one can see that in the
dressed basis the atoms can be considered as moving in an induced (artificial) vector potential $%
\mathbf{{A}}$ and a scalar potential $\tilde{V}(\mathbf{r})$, where
the potential $\mathbf{A}$ is usually called the Mead-Berry vector
potential \cite{Berry1984,Mead1980}. They come from the spatial
dependence of the atomic dressed states with the elements
\begin{equation}
\mathbf{A}_{mn} = i\hbar \langle \chi_m(\mathbf{r})|
\nabla\chi_n(\mathbf{r}) \rangle, \ \ \ \tilde{V}_{mn} = \langle
\chi_m(\mathbf{r})|V(\mathbf{r})|\chi_n(\mathbf{r})\rangle .
\label{eq:A}
\end{equation}

\emph{Abelian gauge potential.} An Abelian $U(1)$ gauge potential is induced for each dressed states provided that
the off-diagonal elements of the matrices $\mathbf{A}$ and $\tilde{V}$ are
much smaller than the energy difference between any pair of the
dressed states, which implies that the eigenstates must be
non-degenerate. In this case an adiabatic approximation can be
applied which is equivalent to neglecting the transitions between the
specific dressed state $|\chi_n\rangle$ and the remaining
$|\chi_l\rangle$ with $n\not= l$. Therefore, atoms in the dressed
state $|\chi_n\rangle$ evolve according to a separately effective
Hamiltonian $H_n$. We project the full Hamiltonian in
Eq. (\ref{H_eff}) to the specific state $|\chi_n\rangle$ and obtain
an effective Hamiltonian given by
\begin{equation}
H_n=\frac{1}{2m}(-i\hbar \nabla
-\mathbf{A}_{n})^{2}+\varepsilon_n+\tilde{V}_n+\tilde{V}^\prime_n,
\end{equation}
where $\mathbf{A}_{n}=\mathbf{A}_{mn}\delta_{nn}$, $\tilde{V}_n
=\tilde{V}_{mn}\delta_{nn}$ and $ \tilde{V}_n^\prime
=\frac{1}{2m}\sum_{l\not= n }^{N}\mathbf{A}_{n,l}\cdot
\mathbf{A}_{l,n}$. So an Abelian gauge potential $\mathbf{U}(1)$
is induced for the neutral atoms.

\emph{Non-Abelian gauge potential.} A non-Abelian gauge potential introduced by Wilczek and Zee
\cite{Wilczek1984} can also be induced in this way if
there are degenerate (or nearly degenerate) dressed
states \cite{Ruseckas2005}. In this case the adiabatic
approximation fails and then the off-diagonal couplings between
the degenerate dressed states can no longer be ignored. Assume
that the first $q$ atomic dressed states among the total $N$
states are degenerate, and these levels are well separated from
the remaining $N-q$ states, we neglect the transitions from
the first $q$ atomic dressed states to the remaining states. In this way,
we can project the full Hamiltonian onto this subspace.
Under this condition, the wave function in the subspace
$\tilde\Psi=\left(\psi_1,\dots,\psi_q\right)^\top$
is again governed by the Schr\"odinger equation
$i\hbar\frac{\partial}{\partial t}\tilde\Psi=\tilde{H}_{\text{eff}}\tilde\Psi$,
where the effective Hamiltonian reads
\begin{equation}
\tilde{H}_{\text{eff}}=\frac{1}{2m}
(-i\hbar\nabla -\mathbf{A})^2 +\varepsilon I_q +  \tilde{V} +
\tilde{V}^\prime. \label{H_non}
\end{equation}
Here the matrices $\mathbf{A}$, $\varepsilon I_q$, and $\tilde{V}$
are the truncated $q\times q$ matrices in
Eq. (\ref{H_eff}). The projection of the term $\mathbf{A}^2$ in
Eq. (\ref{H_eff}) to the $q$ dimensional subspace cannot entirely be expressed in terms of the truncated
matrix $\mathbf{A}$. This gives rise to an additional scalar
potential $\tilde{V}^\prime$ which is also a $q\times q$ matrix,
\begin{equation}
\tilde{V}^\prime _{n,j}
=\frac{1}{2m}\sum_{l=q+1}^{N}\mathbf{A}_{n,l}\cdot
\mathbf{A}_{l,j}
=\frac{\hbar ^2}{2m}\left( \langle\nabla\chi
_{n}|\nabla\chi_{j}\rangle +\sum_{k=1}^{q}\langle\chi
_{n}|\nabla\chi _{k}\rangle \langle\chi _{k}|\nabla\chi
_{j}\rangle \right)  \label{eq:fi1} 
\end{equation}
with $n,j \in(1,\dots,q)$.
Since the adiabatic states $|\chi_1\rangle \dots |\chi_q\rangle$
are degenerate, any basis generated by a local unitary
transformation $U(\mathbf{r})$ within the subspace is equivalent.
The corresponding local basis change as
%
%
$\tilde\Psi \rightarrow U(\mathbf{r})\tilde\Psi,$
%
%
which leads to a transformation of the potentials according to
\begin{equation}
\mathbf{A} \rightarrow  U(\mathbf{r}) \mathbf{A}U^{\dag}(\mathbf{r})
-i\hbar\left[\nabla U(\mathbf{r})\right] U^{\dag}(\mathbf{r}),~~
\tilde{V} \rightarrow  U(\mathbf{r}) \tilde{V} U^{\dag}(\mathbf{r}) .
\end{equation}
These transformation rules show the gauge character of the
potentials $\mathbf{A}$ and $\tilde{V}$. The vector potential
$\mathbf{A}$ is related to a curvature (an effective ``magnetic"
field) $\mathbf{B}$ as:
\begin{equation}
B_i = \frac{1}{2}\epsilon_{ikl} F_{kl},~~ F_{kl} =
\partial_k A_l-\partial_l A_k
-\frac{i}{\hbar}[A_k,A_l]. \label{eq:B}
\end{equation}
Note that the term $\frac{1}{2}\varepsilon_{ikl}[A_k,A_l] =
(\mathbf{A}\times \mathbf{A})_i$ does not vanish in general, since
the components of $\mathbf{A}$ do not necessarily commute. This term reflects the non-Abelian character of the gauge
potentials. The generalized ``magnetic" field transforms under
local rotations of the degenerate dressed basis as $ \mathbf{B} \rightarrow
U(\mathbf{r})\mathbf{B} U^{\dag}(\mathbf{r}).$ Thus, as expected,
$\mathbf{B}$ is a true gauge field. In the following, we
employ this general scheme to create laser-induced gauge
potentials for ultracold atoms using two typical laser-atom
interacting configurations.

\begin{figure}
\begin{centering}
\includegraphics[width=1.0\columnwidth]{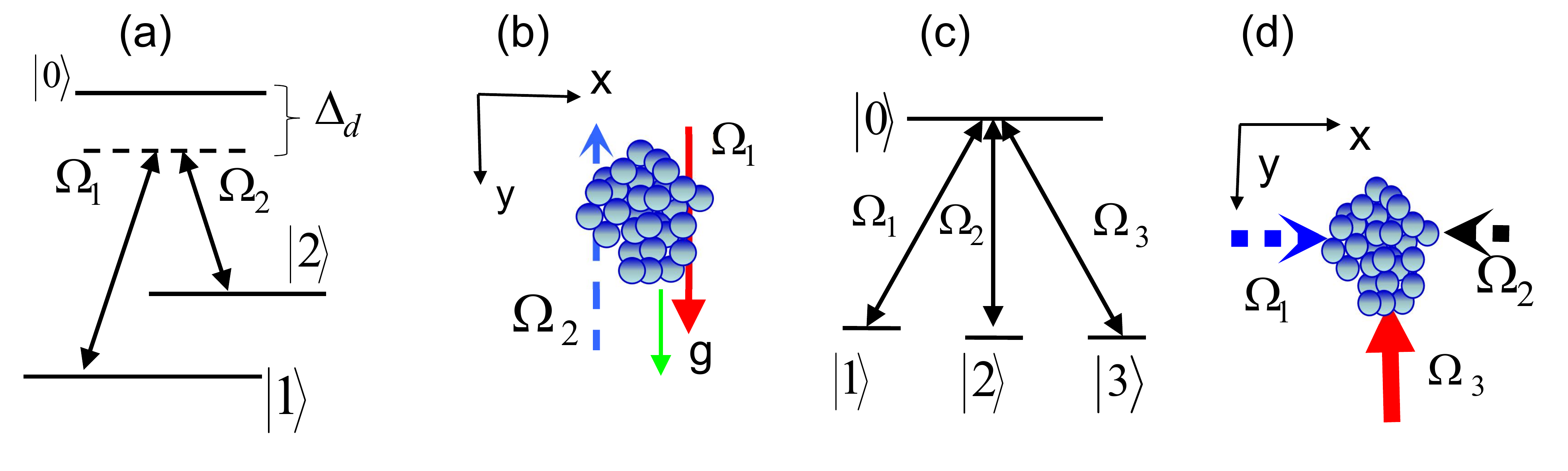} 
\par \end{centering}
\caption{(Color online) Schematic of atom-laser interactions for artificial gauge potentials. (a) Three-level
$\Lambda$-type atoms interacting with laser beams characterized by
the Rabi frequencies $\Omega_1$ and $\Omega_2$ through the Raman-type
coupling with a large single-photon detuning $\Delta_d$. (b) The
configuration of the Raman laser beams for a spin Hall effect.  (c)
and (d) Atoms with tripod-level configuration interacting with
three laser beams characterized by the Rabi frequencies
$\Omega_1$, $\Omega_2$, and $\Omega_3$.}\label{Lambda_tripod}
\end{figure}

\emph{Spin-dependent gauge potentials in three-level $\Lambda$-type atoms}. We
first take an atomic gas with each atom having a $\Lambda $-type level configuration as an example to illustrate the above idea \cite{Juzeliunas2004,SLZhu2006,Juzeliunas2006}. As shown in Fig. \ref{Lambda_tripod}(a), the ground states $|1\rangle $ and $%
|2\rangle $ are coupled to an excited state $|0\rangle $ through
spatially varying laser fields, with the corresponding Rabi
frequencies $\Omega _{1}$ and $\Omega _{2}$, respectively. We
assume off-resonant couplings for the single-photon
transitions with the same large detuning $\Delta_d$. In this case
the atom-laser interaction Hamiltonian $H_{\text{AL}}$  in the basis $\left\{ |1\rangle ,|2\rangle ,|0\rangle \right\} $ is given by
\begin{equation}
H_{\text{AL}}=\left(
\begin{array}{lll}
0 & 0 & \Omega _{1} \\
0 & 0 & \Omega _{2} \\
\Omega _{1}^{\ast } & \Omega _{2}^{\ast } & \Delta_d
\end{array}
\right).  \label{H_int}
\end{equation}

We may parameterize the Rabi frequencies through $\Omega _{1} =
\Omega \sin \theta e^{i\varphi }$ and $ \Omega _{2} = \Omega \cos
\theta$,
with $\Omega =\sqrt{%
|\Omega _{1}|^{2}+|\Omega _{2}|^{2}}$ ($\theta $ and $\varphi $
are in general spatially varying). We are interested in the
subspace spanned by the two lowest dressed states $\left\{ |\chi
_{1}\rangle ,|\chi _{2}\rangle \right\} $ (called respectively the
dark and the bright states). This gives an effective spin-1/2
system, and in the spin language we also denote $|\chi _{\uparrow
}\rangle \equiv |\chi _{1}\rangle $ and $|\chi _{\downarrow
}\rangle \equiv |\chi _{2}\rangle $. In the case of a large
detuning ($\Delta_d \gg\Omega $), both states $|\chi _{\uparrow
}\rangle $ and $|\chi _{\downarrow }\rangle $ have negligible
contribution from the initial excited state $|0\rangle $, so they
are stable under atomic spontaneous emission. Furthermore, we
assume the adiabatic condition, which requires that the off-diagonal
elements of the matrices $\mathbf{\tilde{A}}$ and $\tilde{V}$ are
much smaller than the eigenenergy differences $\left| \lambda _{i}-\lambda _{j}\right| $ ($%
i,j=1,2,0$) of the states $|\chi _{i}\rangle $. This gives the
quantitative condition $\Delta_D\ll \Omega ^{2}/\Delta_d $, where $\Delta_D=\cos ^{2}\theta |\mathbf{v}%
\cdot \nabla (\tan \theta e^{i\varphi })|$ ($\mathbf{v}$ is the
typical velocity of the atom) represents the two-photon Doppler
detuning \cite{Juzeliunas2006}. Under this adiabatic condition,
the effective Hamiltonian for the wave function $\Psi$ in the
subspace spanned by $\left\{ |\chi _{\uparrow }\rangle,|\chi _{\downarrow }\rangle \right\} $ is \cite{SLZhu2006}
\begin{equation}
H_{\text{eff}}=\left(
\begin{array}{cc}
H_{\uparrow } & 0 \\
0 & H_{\downarrow }
\end{array}
\right) ,  \label{H}
\end{equation}
where $H_{\sigma }=\frac{1}{2m}(-i\hbar \nabla -\mathbf{A}_{\sigma
})^{2}+V_{\sigma }(\mathbf{r})$ $\left( \sigma =\uparrow
,\downarrow \right)$. The gauge potentials $\mathbf{A}_{\sigma}$
can be obtained as
$\mathbf{A}_{\uparrow }=-\mathbf{A}_{\downarrow }=-\hbar \sin
^{2}\theta \nabla \varphi, $ and the related gauge field
\begin{equation}
\mathbf{B}_{\sigma }=\nabla \times \mathbf{A}_{\sigma }=-\eta
_{\sigma }\hbar \sin (2\theta )\nabla \theta \times \nabla
\varphi,
  \label{B}
\end{equation}
where $\eta _{\uparrow }=-\eta _{\downarrow }=1$. We obtain precisely a
spin-dependent gauge field that is critical for the spin Hall
effect. A typical scheme to generate atomic spin Hall
effect is shown in Fig \ref{Lambda_tripod}(b), which was
demonstrated experimentally in Ref. \cite{Beeler2013}.

\emph{Spin-orbit couplings in a tripod configuration}. The second example we address is an $SU(2)$ non-Abelian gauge field
created in a tripod-level configuration \cite{Juzeliunas2008,Stanescu2007,Vaishnav2008,SLZhu2009}. Consider
the adiabatic motion of atoms in $x$-$y$ plane with each having a
tripod-level structure in a laser field as shown in Fig.
\ref{Lambda_tripod}(c) and (d). The atoms in three lower levels
$|1\rangle$, $|2\rangle$ and $|3\rangle$ are coupled with an
excited level $|0\rangle$ through three laser beams characterized by the Rabi frequencies
$\Omega_1=\Omega\sin\theta\mathrm{e}^{-i\kappa x}/\sqrt{2}$,
$\Omega_2=\Omega\sin\theta\mathrm{e}^{i\kappa x}/\sqrt{2}$, and
$\Omega_3=\Omega\cos\theta\mathrm{e}^{-i\kappa y}$, respectively,
where $\Omega=\sqrt{|\Omega_1|^2+|\Omega_2|^2+|\Omega_3|^2}$ is
the total Rabi frequency and the mixing angle $\theta$ defines
the relative intensity. The atom-laser interaction Hamiltonian
$H_{\text{AL}}$ in the interaction representation reads
\begin{equation}
H_{\text{AL}}=-\hbar\sum_{j=1}^{3}\Omega_j|0\rangle\langle j
|+\mathrm{H.c.}.
\end{equation}
Diagonalizing this Hamiltonian yields two degenerate dark states with zero energy as well as two
bright states separated from the dark states by the energies $\pm
\hbar\Omega$. If $\Omega$ is sufficiently large compared to the
two-photon detuning due to the laser mismatch and/or Doppler
shift, the adiabatic approximation is justified and one can
safely study only the internal states of an atom evolving within the
dark state manifold. In this case, the non-Abelian gauge potential
$\mathbf{A}$ in the present configuration of the light field can
be obtained as
\begin{equation}
 \mathbf{A}  = \hbar\kappa\left(
\begin{array}{cc}
\mathbf{e}_y & -\mathbf{e}_x\cos\theta\\ -\mathbf{e}_x\cos\theta &
\mathbf{e}_y\cos^2\theta
\end{array}\right).
\end{equation}
Furthermore, let the mixing angle $\theta=\theta_0$ with $\cos\theta_0=\sqrt{2}-1$, such that
$\sin^2\theta_0=2\cos\theta_0$. Thus, the vector potential takes a symmetric form
$\mathbf{A}=\hbar\kappa^{\prime}(-\mathbf{e}_x\sigma_x+\mathbf{e}_y\sigma_z)
+\hbar\kappa_0\mathbf{e}_yI_2,$
where $\kappa^{\prime}=\kappa\cos\theta_0$ and
$\kappa_0=\kappa(1-\cos\theta_0)$. Using a unitary transformation
$\tilde{H}'=U^{\dag}\tilde{H}U$ with
$U=\exp(-i\kappa_0y)\exp\left(-i\frac{\pi}{4}\sigma_x\right)$, one
obtains the Hamiltonian for the atomic motion
\begin{equation}
H=\frac{1}{2m}[(p_x+\hbar\kappa'\sigma_x)^2+(p_y+\hbar\kappa'\sigma_y)^2]+V.
\end{equation}
This Hamiltonian provides a coupling between the atomic center-of-mass
motion and the internal pseudospin degrees of freedom, thus
giving rise to an effective SOC.

However, the two degenerate dark states in this tripod
configuration are not the lowest-energy states, so the atoms may quickly decay out of the dark states due
to collisions and other relaxation processes. This problem may be
solved by using the blue-detuned lasers \cite{YZhang2012} or a
closed loop Raman coupling configuration \cite{Campbell2011}.
Furthermore, the combination of an SOC and an
effective perpendicular Zeeman field is required for the emergence
of topological superfluid. To this end, five or two additional laser
beams superposing into the above tripod configuration were
proposed \cite{CZhang2010,SLZhu2011}. However, locking the phases of these laser
beams is challenging in experiments. It was thus proposed and then
experimentally demonstrated that controlling polarizations of the
Raman lasers is sufficient to generate simultaneously an effective
SOC and a perpendicular Zeeman field for atoms
\cite{LHuang2016,ZMeng2016}.

\subsubsection{Laser-assisted tunneling}\label{Laser-assisted-tunnelings}

\begin{figure}[htbp]\centering
\includegraphics[width=0.9\columnwidth]{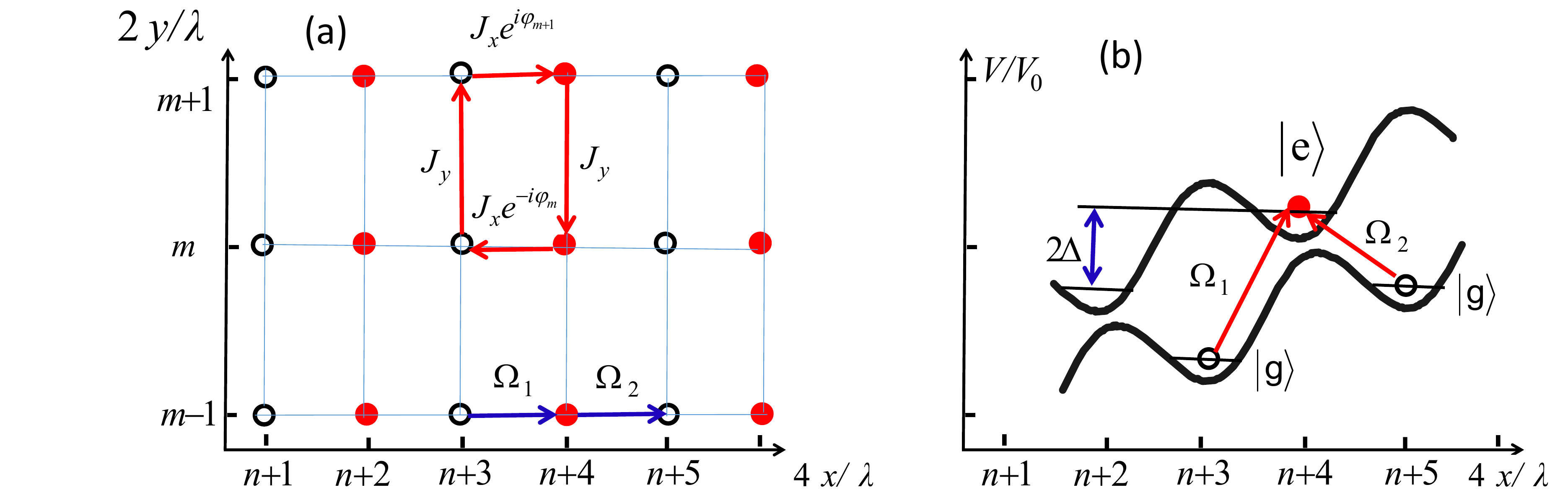}
\caption{(Color online) Scheme for realizing artificial magnetic fields based on
laser-assisted tunneling \cite{Jaksch2003}. Open (closed) circles denote atoms in
state $|g\rangle$ $(|e\rangle)$. (a) Hopping in the $y$-direction
$J_y$ is the same for particles in states $|e\rangle$ and
$|g\rangle$, while the $x$-direction hopping is laser-assisted. (b)
Laser-assisted tunneling along the $x$-direction. Adjacent sites
are set off by an energy $\Delta$. The laser $\Omega_1$ is
resonant for transitions $|g\rangle$ and $|e\rangle$ while
$\Omega_2$ is resonant for transitions $|e\rangle$ and $|g\rangle$
due to the offset of the lattice sites. The atoms hopping around
one plaquette get phase shifts of $2\pi\alpha$ due to the created
artificial magnetic fields.}\label{JZ}
\end{figure}

Laser-induced tunneling was the first method
proposed to generate artificial magnetic fields in OLs \cite{Jaksch2003}, and it was used to experimentally
realize the Hofstadter-Harper model \cite{Aidelsburger2014}.
Furthermore, this method was further proposed to simulate artificial SOCs in OLs \cite{Bermudez2010b,Goldman2010,XJLiu2014},
and was implemented very recently in an experiment realizing
artificial 2D SOC in a Raman OL
\cite{ZWu2016}. In this section we illustrate the basic idea of laser-induced tunneling,
following the original proposal in Ref. \cite{Jaksch2003}. We then introduce recent developments in Sec. \ref{SecIV}.

Consider a gas of atoms trapped in a 3D OL created by standing wave laser fields, which generates a
potential for the atoms $V(\mathbf{r}) =
V_{0x} \sin^2(kx) + V_{0y} \sin^2(ky) + V_{0z} \sin^2(kz)$ with $k
= 2\pi/\lambda$ being the wave vector of the light. We assume the
lattice to trap atoms in two different internal hyperfine states
$|e\rangle$ and $|g\rangle$ and the depth of the lattice in the
$x-$ and $z-$directions to be so large that hopping in these
directions due to kinetic energy is prohibited. Furthermore, we
assume that adjusting the polarization of the lasers that confine
the particles in the $x$-direction allows us to place the potential
well trapping atoms in the different internal states at distances
$\lambda/4$ with respect to each other, as shown in
Fig. \ref{JZ}(a). Therefore, the resulting 2D lattice has a lattice
constant (disregarding the internal state) in the $x$-direction of
$a_x = \lambda/4$ and in the $y$-direction of $a_y = \lambda/2$. We
focus on one layer of the OL in the $xy$-plane since in
the following, there will neither be hopping nor interactions
between different layers. The dynamics of atoms occupying the
lowest Bloch band of this OL can be described by the
Hamiltonian
 $$H_l=\sum_{n,m} J_y
(a^{\dagger}_{n,m} a_{n,m-1} + h.c.)+\sum_{n\in even,m}\omega_{eg}
a^{\dagger}_{n,m} a_{n,m}, $$
where $J_y$ is the hopping strength for particles to tunnel
between adjacent sites along the $y$-direction. The energy
difference between the two hyperfine states is $\omega_{eg} > 0$
and the operators $a_{n,m}$ ($a^\dagger_{n,m})$ are destruction
(creation) operators for atoms in the lowest motional band located
at the site  $x_{n,m} = (x_n, y_m)$, where $x_n = n\lambda/4$ and
$y_m = m\lambda/2$.

In addition, there is an energy offset of $\Delta$ between
two adjacent sites in the $x$-direction, as shown in Fig.
\ref{JZ}(b). This can be achieved by accelerating the OL
along the $x$-axis with a constant acceleration $a_{\text{acc}}$, which
induces an additional potential energy term $H_{\text{acc}} = Ma_{\text{acc}}x$ with $M$ being the mass of the atoms.
Alternatively, if both of the internal atomic states $|e\rangle$
and $|g\rangle$ have the same static polarizability $\mu$ an
inhomogeneous static electric field of the form $E(x) = \delta
Ex$, where $\delta E$ is the slope of the electric field in the
$x$-direction, can be applied to the OL, which leads to
a potential energy term $H_{\text{acc}} = \mu_p \delta Ex$. We keep this
additional potential energy small compared to the OL
potential and treat $H_{\text{acc}}$ as a perturbation. In second
quantization this yields $H_{\text{acc}} = \Delta\sum_{n,m} n
a^\dagger_{n,m}a_{n,m}$, where $\Delta =\mu_p\delta E\lambda/4$ in
the case of an inhomogeneous electric field and $\Delta =
Ma_{\text{acc}}\lambda/4$ when the lattice
is accelerated. The condition for this perturbation treatment to
be valid is $\delta<<\nu_x$ with $\nu_x = 4\sqrt{E_R V_{0x}}$ being the
trapping frequency of the OL in the $x$-direction. Here
$E_R = k^2/2M$ is the recoil energy.

Finally, the laser-induced tunneling can be activated along
the $x$-direction by coupling two internal states $|g\rangle$ and
$|e\rangle$ with two additional lasers forming Raman transitions.
The Raman beams consist of two running plane waves chosen to give
space-dependent Rabi frequencies of the form $\Omega_{1,2} =
\Omega e^{\pm iqy},$ where $\Omega$ denotes the magnitude of the
Rabi frequencies, and $\pm \Delta$ is the detuning. We assume the
lasers not to excite any transitions to higher-lying Bloch bands
with detuning of the order of $\Delta$, i.e.
$\Omega\ll\Delta\ll\nu_x$. Then the lasers $\Omega_{1(2)}$ will only
drive transitions $n-1 \leftrightarrow n$ if $n$ is even (odd) and
we can neglect any influence of the nonresonant transitions. Then
one can find the following Hamiltonian describing the effect of the Raman lasers
$$
H_{\text{AL}}=\sum_{n,m}(\gamma_{n,m}a^{\dagger}_{n,m} a_{n-1,m}+h.c.)
-\sum_{n,m}\Delta a^{\dagger}_{n,m} a_{n,m}.
$$
Here the matrix elements $\gamma_{nm}$ can be written as
$$\gamma_{n,m} = \frac{1}{2} e^{2i\pi \alpha m}\Omega\Gamma_y
(\alpha)\Gamma_x,$$
where $\alpha = q\lambda/4\pi$, and the matrix elements
$$\Gamma_x = \int dx w^\star
(x)w(x- \lambda/4), \ \ \
\Gamma_y(\alpha) = \int dy w^\star(y)
\cos(4\pi\alpha m)w(y),$$
with $w(\mathbf{r}) = w(x)w(y)w(z)$ being the Wannier function. To achieve a symmetric
Hamiltonian, we assume hopping amplitudes $J_x
=\Omega\Gamma_x\Gamma_y/2 = J_y=J$, and thus the total Hamiltonian
describing the configuration is given by
$$
H_\alpha=J\sum_{n,m}\left( e^{2i\pi\alpha m}a^{\dagger}_{n,m}
a_{n+1,m} + a^{\dagger}_{n,m} a_{n,m+1}+h.c.\right).
$$
This Hamiltonian $H_\alpha$ is equivalent to the Hamiltonian for
electrons with charge $e$ moving on a lattice in an external
magnetic field $B = 2\pi\alpha/A_{\text{cell}}e$, where $A_{\text{cell}}$ is the area of one
elementary cell.

\subsubsection{Periodically driven systems}

Driving cold-atom systems periodically in time is a powerful
method to engineer effective magnetic fields or SOCs, and thus can trigger topological quantum phases. For
instance, the OL shaking method has been used to
experimentally realize the Hofstadter model \cite{Aidelsburger2011,Aidelsburger2013,Miyake2013}.
Modulating a honeycomb OL also led to the
experimental realization  of the Haldane model \cite{Jotzu2014}.

We first describe two simple examples to illustrate the basic concept
of creating artificial gauge fields with the periodically driven method. In the
first example we consider ultracold atoms trapped in a 1D shaken
bichromatic OL \cite{FMei2014}. This lattice is
generated by the superposition of two shaken OLs. The
single-particle Hamiltonian of an atom in this 1D shaken lattice
system reads
\begin{equation}
H_s= \frac{p_x^2}{2m}+V_1\sin^2[k_1(x-x_1(t))]+V_2\sin^2[k_2(x-x_2(t)+\phi)],\label{HMei}
\end{equation}
where $V_i$, $k_i=2\pi/\lambda_i$, and $\lambda_i\ (i=1,2)$ are
the lattice depth, laser wave vector and wavelength, respectively; and $\phi$ is
the phase of the second laser, $x_i(t)=b\sin(\omega t)$ is the
periodic time-dependent lattice shaking. Here we assume that the
two lattices experience the same shaking amplitude $b$ and
frequency $\omega$. Experimentally, a shaking sinusoidal lattice
can be realized through a modulation of the driving frequency and
by changing the relative phase of the acousto-optic modulators.
The tunneling between neighboring sites decreases exponentially
with the intensity of the lasers creating the lattice, whereas the
shape of the wavepacket (the Wannier functions) has a much weaker
dependence. Therefore, by varying the laser intensity, one can rapidly vary the
tunneling. With a unitary
rotation, the Hamiltonian is transferred to a new frame
$x\rightarrow x+b\sin(\omega t)$
\begin{equation}
H_r= \frac{(p_x-A_x)^2}{2m}+V_1\sin^2(k_1 x)
+V_2\sin^2(k_2x+\phi),
\end{equation}
with a shaking-induced vector potential $A_x=m\omega\cos(\omega
t)$ \cite{FMei2014}.

The second example is the topological phases of a 2D
honeycomb lattice proposed by Haldane \cite{Haldane1988}, which can also be realized
with the method of shaking lattices \cite{Jotzu2014,WZheng2014}. We
consider the following time-dependent lattice potential
\begin{eqnarray}
\label{HaldaneF}
V(x,y,t)&&=-V_{\overline{X}}\cos^2[k_r(x+b\cos\omega t)+\theta/2]-V_{{X}}\cos^2[k_r(x+b\cos\omega
t)]\\
&&-V_{Y}\cos^2[k_r(y+b\sin\omega t)]
-2\alpha\sqrt{V_X V_Y}\cos[k_r(x+b\cos\omega
t)]\cos[k_r(y+b\sin\omega t)],\nonumber
\end{eqnarray}
which leads to a honeycomb lattice realized by the ETH group when
$b=0$ \cite{Tarruell2012}. Here $\theta$ controls the energy offset
between two sublattices $A$ and $B$ in the honeycomb lattice.
The $b\not=0$ case describes a shaking lattice in both $x$ and $y$
directions with a phase difference $\pi/2$. Similar to the
1D case, transferring into the moving frame
$x\rightarrow x+b\cos(\omega t)$ and $y\rightarrow y+b\sin(\omega
t) $, one obtains a Hamiltonian with time-dependent vector
potential term
\begin{equation}
H(t)=\frac{1}{2m}\left[\mathbf{p}-\mathbf{A}(t)\right]+V(x,y),
\end{equation}
where $A_x(t)=m\omega b\sin(\omega t)$ and $A_y(t)=-m\omega
b\cos(\omega t)$ \cite{WZheng2014}. It is equivalent to a
Hamiltonian that describes a particle in an ac electrical field
in the 2D plane $\mathbf{E}(t)=m\omega^2 b
(\cos(\omega t),\sin(\omega t))$. The phase diagram in this
Hamiltonian has been calculated in Ref. \cite{WZheng2014}, and it
shows a similar phase diagram with that of the Haldane model \cite{Haldane1988};
i.e., it contains topological trivial and nontrivial phases
characterized by a Chern number.
This shaking lattice method has been experimentally used to realize the Haldane model \cite{Jotzu2014},
as addressed in detail in Sec. \ref{HaldaneModel}.

After addressing the basic ideas, we now turn to some general frameworks that describe periodically driven quantum
systems. A general theoretical treatment of periodically driven
quantum systems is based on the Floquet theory. For a periodically
driven Hamiltonian ${H}(t)$ with period $\tau$, its Floquet
operator ${\hat{F}_o}$ is defined as
\begin{equation}
\label{Floquet} {\hat{F}_{o}}\equiv
U(\tau_i+\tau,\tau_i)=\mathcal{T}\exp\left[
-i\int_{\tau_i}^{\tau+\tau_i} H(t)dt\right],
\end{equation}
where $\tau_i$ is the initial time, and $\mathcal{T}$ denotes the
required time-ordered integral as the Hamiltonian at
different times do not necessarily commute. The eigenvalue and
eigenstates of the Floquet ${\hat{F}_o}$ are given by
\begin{equation}\label{Fo} {\hat{F}_o}|\varphi_n\rangle= e^{-i\epsilon_n
\tau}|\varphi_n\rangle,\end{equation}
where $\epsilon_n\in (-\pi/\tau,\pi/\tau)$ is the quasi-energy. A
general method to explore the topological phases, which is free
from any further approximation, is to numerically evaluate Floquet
operator $\hat{F}_o$ according to Eq. (\ref{Floquet}) and determine its
eigenvalues and eigenfunctions from Eq. (\ref{Fo}). If a
periodically driven system exhibits nontrivial topology, there
must be in-gap quasi-energies $\epsilon_n$ and their corresponding
wave functions $\varphi$ are spatially well localized at the edge
of the system \cite{WZheng2014}.

A physically more transparent method is introducing a
time-independent effective Hamiltonian $H_{\text{eff}}$ via the Floquet
operator \cite{WZheng2014,Goldman2014b}
\begin{equation}
\label{HF}
U(\tau_f,\tau_i)=e^{-iK(\tau_f)}e^{-i\tau H_{eff}}
e^{iK(\tau_i)},
\end{equation}
where we impose that (1) $H_{\text{eff}}$ is a time-independent operator,
(2) $K(t)$ is a time-periodic operator $K(t+\tau)=K(t)$ with zero
average over one period, and (3) $H_{\text{eff}}$ does not dependent on
the starting time $\tau_i$, which can be realized by transferring
all undesired terms into the kick operator $K(\tau_i)$. Similarly,
$H_{\text{eff}}$ does not depend on the final time $\tau_f$. Equation
(\ref{HF}) shows that the initial (final) phase of the Hamiltonian
at time $\tau_i$ ($\tau_f$) may have an important impact on the
dynamics. However, the topological phenomena in the periodically
driven systems can be connected to those in equilibrium systems described by the
effective Hamiltonian $H_{\text{eff}}$. Consider a static system
described by a Hamiltonian $H_0$ that is driven by a time-periodic
modulation $V(t)$, whose period $\tau=2\pi/\omega$ is assumed to be
much smaller compared to any characteristic time scale in the
problem. In this high-frequency regime, one can obtain the
effective Hamiltonian by using a perturbation expansion.

We consider a time-periodic Hamiltonian $H(t)=H_0+V(t)$ with
$$
V(t)=\sum_{j=1}^\infty [V^{(j)} e^{i j\omega t}+V^{(-j)}e^{-i
j\omega t}]
$$
between times $\tau_i$ and $\tau_f$, with the period of the driving
$\tau=2\pi/\omega$. The time-dependent potential has been
explicitly expanded with its Fourier's form. By using a
perturbation expansion in powers of $1/\omega$, one can obtain \cite{Goldman2014b}
\begin{eqnarray}
H_{\text{eff}}&=&H_0+\frac{1}{\omega}\sum_{j=1}^\infty \{
\frac{1}{j}[V^{(j)},V^{(-j)}]  \nonumber \\
 &+&\frac{1}{2\omega^2}([[V^{(j)},H_0],V^{(-j)}]+[[V^{(-j)},H_0],V^{(j)}] \}+\mathcal{O}(\tau^3), \label{Hshake}\\
K(t)&=&\int^{t} V(t^\prime)dt^\prime + \mathcal(\tau^2).
\label{HshakeK}
\end{eqnarray}
In Eq. (\ref{Hshake}), the second-order terms that mix different
harmonics have been omitted.

To understand the emergence of topological nontrivial phases, we
usually write the Hamiltonian into momentum space. As for two-band
systems, the general Hamiltonian in momentum space $\mathbf{k}$
can be rewritten as
$H_{\text{eff}}=\mathbf{B}(\mathbf{k})\cdot\mathbf{\sigma}$. By using
this kind of perturbation expansion, one can obtain the explicit expressions of
$\mathbf{B}(\mathbf{k})$ for the models described in
Eqs. (\ref{HMei}) and (\ref{HaldaneF}) \cite{FMei2014,WZheng2014}, which shows that the nontrivial
topological phases can be induced in the periodically driven OLs.

The perturbation expression in Eq. (\ref{Hshake}) can also be used
in the derivation of the effective Hamiltonian for the general
situation where a pulse sequence is characterized by the
repeated $N$-step sequence
\begin{equation}
\gamma_N=\{H_0+V_1,H_0+V_2,\cdots,H_0+V_N   \},
\end{equation}
where the $V_m$'s are arbitrary operators \cite{Goldman2014b}. For
simplicity, we assume that the duration of each step is $\tau/N$,
and we further impose that $\sum_{m=1}^N V_m=0$. The Hamiltonian
$\gamma_N$ can be expanded in terms of the harmonics
$H(t)=H_0+\sum_{j\not= 0} V_j e^{ij\omega t},$ where
$$
V_j=\frac{1}{2\pi i}\sum_{m=1}^N \frac{1}{j} e^{-i2\pi j
m/N}(e^{ij(2\pi/N)}-1)V_m.
$$
By applying Eqs. (\ref{Hshake}) and (\ref{HshakeK}), one can derive the effective
Hamiltonian and the initial-kick operator as
\begin{eqnarray}
H_{\text{eff}}&=&H_0 +\frac{2\pi i}{N^3\omega}\sum_{m<n=2}^N
\mathcal{C}_{m,n}[V_m,V_n]+\frac{\pi^2(N-1)^2}{6N^4\omega^2}\sum_{m=1}^N
[[V_m,H_0],V_m] \nonumber\\
&& \ \ \ + \frac{\pi^2}{6N^4\omega^2}\sum_{m<N=2}^N
\mathcal{D}_{m,n}
\left([[V_m,H_0],V_n]+[[V_n,H_0],V_m]\right)+\mathcal{O}(1/\omega^3),\\
K(0)&=&\frac{2\pi}{N^2\omega}\sum_{m=1}^N
mV_m+\mathcal{O}(1/\omega^2),
\end{eqnarray}
where $\mathcal{C}_{m,n}=N/2+m-n$ and
$\mathcal{D}_{m,n}=1+N^2-6N(n-m)+6(n-m)^2$. The above equations
show that the initial kick $K(0)$ depends on the way the pulse
sequence starts, whereas the effective Hamiltonian $H_{\text{eff}}$ is
independent of this choice: redefining the operators
$V_m\rightarrow V_{m+j}$ with an integer $j$ results in a change
in $K(0)$ but leaves $H_{\text{eff}}$ invariant. The expressions are
useful for engineering effective Hamiltonians with
artificial gauge fields. In view of
this fact, we illustrate the case $N=4$ in the following.

Consider the following four-step sequence:
\begin{equation}
\label{Ps} \gamma_4= \{ {H}_0+{L_1}, {H}_0+ {L_2}, {H}_0- {L_1},
{H}_0- {L_2}\},
\end{equation}
we obtain

\begin{eqnarray}
 {H}_{\text{eff}}&=& {H}_0+\frac{i\pi}{8\omega}[ {L_1}, {L_2}]
 +\frac{\pi^2}{48\omega^2}\left([[ {L_1},H_0], {L_1}]+[[
{L_2},H_0], {L_2}]\right)
+\mathcal{O}(1/\omega^3),\label{Hem}\\
  {K}(0)&=&-\frac{\pi}{4\omega}( {L_1}+ {L_2})+\mathcal{O}(1/\omega^2).
\end{eqnarray}
We first consider atoms moving in a two-dimensional free space, such
that $ {H}_0=( {p}_x^2+ {p}_y^2)/2m$. We drive the system with a
pulse sequence (\ref{Ps}) with the operator
$$ {L_1}=( {p}_x^2- {p}_y^2)/2m, \ \  {L_2}=\kappa
 {x} {y}.$$
The corresponding effective Hamiltonian is given by Eq.
(\ref{Hem}), which yields, up to the second order $(1/\omega^2)$
\begin{equation}
\nonumber
 {H}_{\text{eff}}=\frac{1}{2m}[( {p}_x-A_x)^2+( {p}_y-A_y)^2]+\frac{m\omega_h^2}{2}( {x}^2+ {y}^2),
\end{equation}
where $\mathbf{A}=(A_x,A_y)=(-m\Omega y,m\Omega x)$ with
$\Omega=\pi\kappa/(8m\omega)$ and $\omega_h=\sqrt{5/3}\Omega$. It
corresponds to the realization of a perpendicular and uniform
artificial magnetic field
$\mathbf{B}=2m\Omega\mathbf{e}_z=\pi\kappa/(4\omega)\mathbf{e}_z$.
An early version of this scheme for artificial magnetic fields was
proposed to realize fractional QHE with bosonic
atoms in OLs \cite{Sorensen2005}.

The similar four-step sequence can also be used to generate
SOCs. Considering the operators $ {H}_0=( {p}_x^2+
{p}_y^2)$ and
$$
 {L_1}=( {p}_x^2- {p}_y^2),\ \
 {L_2}=\kappa( {x}\sigma_x- {y}\sigma_y).
$$
The time evolution of the driven system is characterized by the
effective Hamiltonian
\begin{equation}
 {H}_{\text{eff}}=\frac{ {p}_x^2+ {p}_y^2}{2m}+\lambda_R(\sigma_x {p}_x+\sigma_y {p}_y)
+\Omega_{SO} {L}_z\sigma_z+\mathcal{O}(1/\omega^3),
\end{equation}
where $\lambda_R=\pi\kappa/(8m\omega)$ and
$\Omega_{SO}=-(8m/3)\lambda_R^2$. The term ${H}_R=\lambda_R
{\mathbf{p}}\cdot\mathbf{\sigma}$ is the Rashba SOC, and ${H}_{L\sigma}=\Omega_{SO} {L}_z\sigma_z$ is the
so-called "intrinsic" or "helical" SOC, which is
responsible for the quantum spin Hall effect in topological
insulators. The combination of these two terms appears in the Kane-Mele model (see Sec. \ref{KMModel}).

\section{Topological quantum matter in optical lattices}\label{SecIV}

In the previous section, we introduced the techniques for engineering the Hamiltonian of cold atoms, specially the techniques of creating artificial magnetic fields and SOCs. The use of these techniques in OLs has led to the realization and characterization of some topological states for cold atoms. Compared with conventional solid-state systems, cold atoms offer an ideal platform with great controllability to study topological models. For instance, the laser fields that couple hyperfine states of atoms can be used to synthesize effective physical fields, such as gauge fields, SOCs, and Zeeman fields. The forms and strengths of those synthetic fields are tunable as they are determined by the atom-laser coupling configurations. The structure of an OL can be designed via several counterpropagating lasers to realize various unconventional lattice potentials, which include the double-well superlattices, honeycomb lattices, spin-dependent lattices, and so on.

This section systematically discusses some important lattice models with topological quantities originally introduced in condensed matter theories and describes their proposed schemes as well as current implementation methods. The topological bands and phenomena in these models can be created and detected with cold atoms in OLs. These lattice models range from 1D to 3D and even higher-dimension geometries, which can be implemented with OLs of various geometric structures. These systems mainly focus on energy bands in the absence of interactions, and hence the topological phenomena addressed here correspond to the single-particle physics. Some advances in their extension to the interacting regime will also be briefly discussed.

This section is divided into five parts. In the first, we describe some basic topological models with nontrivial bands realized in 1D OLs, which include the famous Su-Schrieffer-Heeger (SSH) model and its implementation for topological pumping. In the second part we discuss the physics of Dirac fermions, the topological properties of the Hofstadter model, Haldane model and Kane-Mele model, and their experimental realization and detection in 2D OLs. Some typical 3D topological insulating states of $\mathbb{Z}_2$ or $\mathbb{Z}$ types and topological gapless (semimetal) states with emergent Dirac or Weyl fermions in 3D OLs are presented in the third part. The last two parts are respectively devoted to topological states in higher dimensions with the newly developed synthetic dimension technique and unconventional topological quasi-particles with higher pseudospins for cold atoms in OLs, both of which are currently absent or extremely challenging to realize in condensed matter systems.


\subsection{One-dimension}\label{1DModel}

\subsubsection{Su-Schrieffer-Heeger model and Rice-Mele model}\label{SSHModel}

\begin{figure}[http]
\centering
\includegraphics[width=0.95\textwidth]{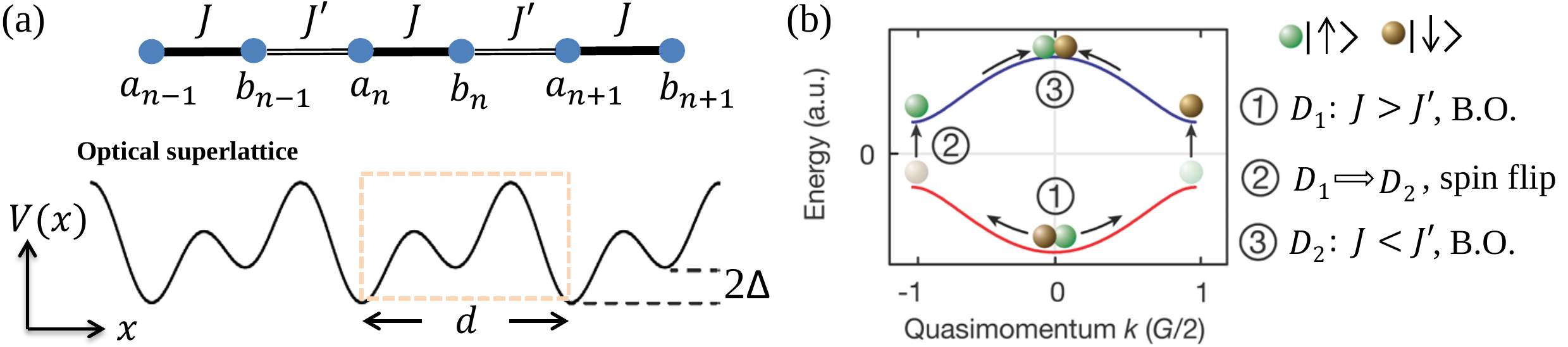}
\caption{(Color online) (a) Schematic illustration of the 1D optical
superlattice for realizing the SSH model ($\Delta=0$) and the Rice-Mele model ($\Delta\neq0$) in the experiment \cite{Atala2013}. The yellow box denotes the unit cell containing two sites with staggered hopping strengths ($J$ and $J'$) and tunable energy offset (2$\Delta$). (b) Schematic illustration of experimental three-step sequence of measuring the
Zak phase difference $\delta\varphi_{Zak}$, based on a combination of spin-dependent Bloch oscillation (B.O.) and Ramsey interferometry \cite{Atala2013}. The preparation and state evolution of the atomic gas in a
superposition of two spin-states with opposite magnetic moment
(brown and green balls) are described in the text.}\label{SSH-OL}
\end{figure}

The SSH model \cite{WPSu1979,Heeger1988}
for polyacetylene is the simplest 1D model of band topology in
condensed matter physics. Such a model describes the polyacetylene
with free fermions moving in a 1D chain with dimerized tunneling
amplitudes. The essence of the SSH model is manifested by two
topological characters. The first character is the nontrivial Zak
phase that describes distinct topological phases in 1D lattice
systems with zero-energy edge modes in a finite chain with open
boundaries. The second one is the topological solitons with
fractional particle numbers, which emerge on the domain walls in
the lattice potential to separate two dimerization structures. The
physics of such a dimerized lattice with two sites per unit cell
is captured by the SSH Hamiltonian
\begin{equation}\label{SSH}
H_{\text{SSH}}=-\sum_{n}(Ja_{n}^{\dag}b_{n}+J'a_{n}^{\dag}b_{n-1}+h.c.),
\end{equation}
where $J$ and $J'$ denote the modulated hopping amplitudes, and $a_{n}^{\dag}$($b_{n}^{\dag}$)
are the creation operators for a particle on the sublattice site $a_{n}$($b_{n}$) in the $n$th lattice cell, as shown in Fig. \ref{SSH-OL}(a). Written in momentum space, the Hamiltonian (\ref{SSH}) takes the form $H_{\text{SSH}}=\sum_k\Psi_k^{\dag}\mathcal{H}_{\text{SSH}}(k)\Psi_k$ with $\Psi_k^{\dag}=\left(a_{k}^{\dag},b_{k}^{\dag}\right)$ and
\begin{equation}\label{SSH-k}
\mathcal{H}_{\text{SSH}}(k)=-[J+J'\cos(ka)]\sigma_x-J'\sin(ka)\sigma_y,
\end{equation}
where $a$ denotes the lattice spacing. Consequently, there are two bands with the energy dispersion $E_{\pm}=\pm\sqrt{[J+J'\cos(ka)]^2+[J'\sin(ka)]^2 }$.

It can be found that $\mathcal{H}_{\text{SSH}}(k)$ possesses the
chiral symmetry
$\sigma_z\mathcal{H}_{\text{SSH}}(k)\sigma_z=-\mathcal{H}_{\text{SSH}}(k)$
and the TRS
$\hat{T}\mathcal{H}_{\text{SSH}}(k)\hat{T}^{-1}=\mathcal{H}_{\text{SSH}}(-k)$,
where $\hat{T}=\hat{\mathcal{K}}$ with $\hat{\mathcal{K}}$ being
the complex conjugate operator. Note that the chiral symmetry here is a sublattice symmetry and requires that hoppings only exist between two sublattices. The chiral symmetry gives rise to
an additional particle-hole (charge-conjugation) symmetry because
for any eigenstate $|u_E\rangle$ with energy $E$ there exists a
corresponding eigenstate $|u_{-E}\rangle=\sigma_z|u_E\rangle$ with
energy $-E$. Thus, the SSH model is classified in the BDI class of
topological insulators \cite{Schnyder2008}. It is known that the
SSH model has two topologically distinct phases with different
dimerization configurations, $D1$ for $J>J'$ and $D2$ for $J<J'$,
separated by a topological phase transition point at $J=J'$. The
topological features can be characterized by the Zak phase
\cite{Zak1989}
\begin{equation}\label{Zak}
\varphi_{Zak}=i\int^{G/2}_{-G/2}
\langle u_{\pm}(k)|\partial_k|u_{\pm}(k)\rangle dk,
\end{equation}
where $G=2\pi/d$ is the reciprocal lattice vector with $d=2a$ and $|u_{\pm}(k)\rangle$ denote the Bloch wave functions of the higher ($+$) and lower ($-$) bands.
The Zak phase in each lattice configuration is a gauge dependent quantity depending on the choice of origin of the unit cell. For our choice, the Zak phase $\varphi_{Zak}=0$ for $D1$ and there is no edge state, yielding a trivial insulating phase. For $D2$, $\varphi_{Zak}=\pi$ and the system hosts two degenerate zero-energy edge states, yielding a gapped topologically nontrivial phase. The difference between the Zak phases for the two dimerization configurations is well defined as $\delta\varphi_{Zak}=\pi$, which is gauge invariant and thus can be used to identify the different topological characters of the Bloch bands.  In the topologically nontrivial phase, there are two degenerate zero-energy modes respectively localized at two edges of the system under the open boundary condition.

Another topological feature in the SSH model is that a kink
(anti-kink) domain between the two dimerization configurations
gives rise to an undegenerate, isolated soliton (anti-soliton)
state on the domain, which is a zero-energy mid-gap state. Due to
the particle-hole ambiguity of the energy spectrum, the zero mode
takes the fractional fermion number $\mathcal{N}=\frac{1}{2}$ ($\mathcal{N}=-\frac{1}{2}$)
when this mode is occupied (unoccupied). The
soliton state is topologically protected in the sense that it is
impossible to remove it without closing the bulk energy gap,
which is due to the fact that it is on an interface between
topologically distinct phases. Historically, such topological
solitons with fractional particle numbers were first found in a 1D
modified Dirac equation in the context of the field theory by
Jackiw and Rebbi \cite{Jackiw1976} (see Sec. \ref{JRModel&TS}),
and the SSH model provides the first physical demonstration of
this remarkable phenomenon in lattice systems.

The SSH model was generalized to describe the linearly
conjugated diatomic polymers with the Rice-Mele Hamiltonian
\cite{Rice1982}
\begin{eqnarray}\label{RM}
H_{\text{RM}}&=&-\sum_{n}(Ja_{n}^{\dag}b_{n}+J'a_{n}^{\dag}b_{n-1}+h.c)+\Delta\sum_{n}(a_{n}^{\dag}a_{n}-b_{n}^{\dag}b_{n}),
\end{eqnarray}
where $\Delta$ is the energy offset between neighboring lattice
sites. For a heteropolar dimer configuration with $\Delta\neq0$,
the particle-hole symmetry (and chiral symmetry) in the original
SSH model is broken. Consequently, the Zak phase is fractional
in units of $\pi$ and depends on the energy offset $\Delta$.  Strictly, the Rice-Mele model is not topological in the theory of topological classification with symmetry \cite{Chiu2016}. However, one can investigate the existence of edge modes to determine which configuration has nontrivial properties in this case, noting that the two
edge modes in the topological phase are no longer degenerate when
$\Delta\neq0$. If there is a domain wall in the Rice-Mele model,
where a Dirac Hamiltonian emerges in the continuum limit
as the generalized Jackiw-Rebbi model (see Sec. \ref{JRModel&TS}),
the unpaired soliton state in general has non-zero energy and
carries an irrational particle number
\cite{Rice1982,Goldstone1981,Jackiw1983}. The Rice-Mele/SSH model
with band topology (geometry) provides a paradigmatic system for studying
topological quantum pumps \cite{Thouless1983,Niu1984,Xiao2010} (see
Sec. \ref{Topopuming}).

Compared with conventional solid-state materials, cold atom
systems offer a perfectly clean platform with high controllability
to study topological states of matter. The first scheme to
simulate the SSH/Rice-Mele model along this direction was proposed to
engineer the spatial profiles of the hopping amplitudes for cold
atoms in a 1D optical superlattice in such a way that an optically
induced defect as the domain wall carries fractional particle
numbers in the lattice
\cite{Ruostekoski2002,Javanainen2003,Ruostekoski2008}. In the
proposed system, a two-species gas of fermionic atoms is trapped
in a state-dependent OL, where the internal states
are denoted by $|\uparrow\rangle$ and $|\downarrow\rangle$. The
two atomic species experience different optical potentials that
are shifted relative to each other by $\lambda/4$, where $\lambda$
is the wavelength of light of the confining OL. Such a state-dependent OL
is achieved when the laser beam is blue detuned from the internal
transition of the atoms in $|\uparrow\rangle$ and red detuned by
the same amount from the internal transition of the atoms in
$|\downarrow\rangle$. When the lattice is sufficiently deep, each
site is assumed to support one mode function that is weakly
coupled to two nearest-neighbor sites, such that the hopping of
the atoms between adjacent lattice sites only occurs as a result
of driving by coherent electromagnetic fields. The coupling could
be a far-off-resonant optical Raman transition via an intermediate
atomic level. In this way, the required dimerized OL with the alternating hopping
for atoms can be realized by using coupling lasers of proper
two-photon Rabi frequency. Since the laser phase directly
modulates the hopping configuration, when there is a jump in one
particular lattice site in the laser phase, it will generate a
domain wall to separate the $D1$ and $D2$ configurations \cite{Ruostekoski2002}. The topological soliton
states in the domain wall in this OL system are
tunable via the atom-laser coupling parameters. A similar proposal
based on a 1D spin-dependent OL for synthesizing
SSH/Rice-Mele models with fully tunable parameters in the absence of
domains was also presented in Ref.\cite{Zheng2017}.

Creation and measurement of the fractionalized soliton modes in the domain walls are experimentally challenging. It was suggested that the fractional particle number could be detected via far-off-resonant light scattering in the atomic gases trapped in OLs \cite{Javanainen2003,Ruostekoski2008}. By measuring the intensity of the scattered light, one can detect the fractional expectation value of the atom number and its fluctuations. It was recently demonstrated that the fractional particle number in the SSH/Rice-Mele model can be simulated in the momentum-time parameter space in terms of Berry curvature without a spatial domain wall \cite{DWZhang2015}. In the simulation, a hopping modulation is adiabatically tuned to form a kink-type configuration, and the induced current plays the role of an analogous soliton distributing in the time domain. Thus the mimicked fractional particle number is expressed by the particle transport and can be detected from the center-of-mass motion of an atomic cloud. Two feasible experimental setups of OLs for realizing the required SSH Hamiltonian with tunable parameters and time-varying hopping modulation were presented in Ref. \cite{DWZhang2015}.

Other schemes for creating 1D topological bands and soliton/edge states have recently been proposed by using cold atoms trapped
in double-well OLs. The authors in Ref. \cite{Li2013} considered a single species of fermionic atoms occupying an $sp$ orbital ladder of the two wells, where the staggered hopping pattern for realizing the topological phase naturally arises. In the noninteracting limit, the $sp$ orbital ladder naturally reproduces the SSH model with a quantized Zak phase and fractionalized zero-energy edge states. The stability of the topological phase against atomic interactions and the emergence of topological flat bands of edge modes in the presence of inter-ladder coupling were also discussed \cite{Li2013}. It was shown that an atomic gas of attractively interacting fermions in a 1D periodically shaken OL can give rise to the emergent Rice-Mele model with controlled domain walls, which comes from the density-wave ground state \cite{Przysiezna2015}. By using cold atoms in a spin-dependent optical double-well lattice, one may realize a two-leg generalized SSH model with glide reflection symmetry \cite{SLZhang2017}, which is topologically characterized by Wilson lines and automatical fractionalization without producing domains in the lattice due to the interplay between the glide symmetry and atomic repulsive interactions.

\begin{figure}
\centering
\includegraphics[width=0.95\linewidth]{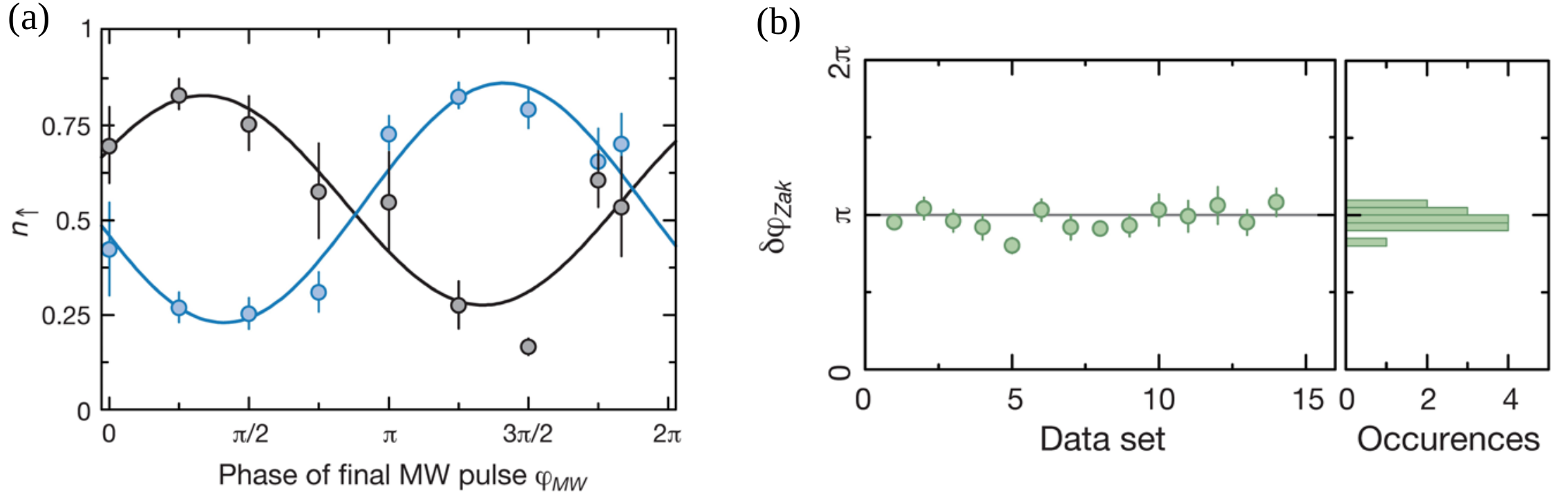}
\caption{(Color online) Determination of the Zak phase. (a) The atom number in the two spin states $N_{\uparrow,\downarrow}$ is measured following the sequence and the fraction of atoms in the $|\!\!\uparrow\rangle$ spin state $n_\uparrow=N_\uparrow/(N_\uparrow+N_\downarrow)$ is plotted as a function of the phase of the final microwave $\pi/2$-pulse. The difference in the phases of the two Ramsey fringes yields the Zak phase difference $\delta \varphi_\text{Zak}$. Blue (black) circles correspond to the fringe in which the dimerization was (not) swapped. (b) Measured relative phases for 14 identical experimental runs (left), which give an average value of $\delta\varphi_\text{Zak}=0.97(2)\pi$. The corresponding histogram is shown on the right with a binning of $0.05\pi$. Reprinted by permission from Macmillan Publishers Ltd: Atala {\it et al.}\cite{Atala2013}, copyright\copyright ~(2013). }\label{SSHZakData}
\end{figure}

The SSH/Rice-Mele model described by the Hamiltonian (\ref{RM})
with the tunable energy offset $\Delta$ has already been
experimentally realized with a Bose-Einstein
condensate (BEC) of $^{87}$Rb trapped in a 1D optical superlattice
\cite{Atala2013}. The superlattice potential shown in Fig. \ref{SSH-OL}(a)
is created by superimposing two standing optical waves of short-
and long-wavelengths differing by a factor of two
($\lambda_l=2\lambda_s=1534$nm), which leads to the total lattice
potential $V(x)=V_{l}\sin^{2}(k_Lx+\phi/2)+V_{s}\sin^{2}(2k_Lx+\pi/2)$,
where $k_L$ is the wave vector of the short wavelength trapping
lasers, and $V_{1}$ and $V_{s}$ are the corresponding strengths of the
two standing waves. The lattice potential can be controlled by
varying the laser intensity of long-wavelength standing-wave
lasers to make the system well described by the SSH/Rice-Mele Hamiltonian in
the tight-binding regime. Phase control between the two standing
wave fields enables one to fully control $\phi$ for tuning the
atomic hopping. This makes the OL into the $D1$ or $D2$
configuration with ease. In the experiment, switching between
$\phi=0$ and $\phi=\pi$ was used to rapidly access the two
different dimerized configurations with $\Delta=0$, whereas a
tunable energy offset $\Delta\neq0$ was also introduced by
tuning $\phi$ slightly away from these symmetry points.

Moreover, the Zak phase $\delta\varphi_{Zak}$ characterizing the topological Bloch
bands was detected, even though the atoms used in the experiment are bosons. A three-step
sequence shown in Fig. \ref{SSH-OL}(b) was employed, which is based on a combination
of spin-dependent Bloch oscillations and Ramsey interferometry
\cite{Atala2013}. The first step is to start with an atomic
condensate in the state $|\downarrow,k=0\rangle$ and bring it into
a coherent superposition state $1/\sqrt{2}(|
\uparrow,k=0\rangle+|\downarrow,k=0\rangle)$ using a microwave
$\pi/2$-pulse. Here $\sigma=\uparrow,\downarrow$ denotes two spin
states of the atoms with opposite magnetic moment and $k$ is the
central momentum of the condensate. Then a magnetic field gradient
is applied to create a constant force in opposite directions for
the two spin components, leading to spin-dependent Bloch
oscillations. In this process, the atomic wavepacket evolves into
the coherent superposition state
$1/\sqrt{2}(|\uparrow,k\rangle+e^{i \delta
\varphi}|\downarrow,-k\rangle)$. When the two states reach the band edge,
the differential phase between them is given by $\delta
\varphi = \varphi_{Zak}+\delta \varphi_\text{Zeeman}$, where
$\delta\varphi_\text{Zeeman}$ denotes the Zeeman phase difference
induced by the magnetic field. For the TRS
Hamiltonian here, the dynamical phase acquired during the
adiabatic evolution is equal for the two spin states and thus
cancels in the total phase difference. The second step is to
eliminate the Zeeman phase difference by applying a spin-echo
$\pi$-pulse and switching dimerization configurations following the
first step. For atoms located at the band edge $k=\pm G/2$, this
non-adiabatic dimerization switch induces a transition to the
upper band of the SSH/Rice-Mele model. The sequence is finally completed by
letting the spin components further evolve in the upper band until
they return to the band center $k=0$. At this point, a final
$\pi/2$-pulse with phase $\varphi_{MW}$ is applied in order to make
the two spin components interfere and read out their relative
phase $\delta \varphi_{Zak}$ through the resulting Ramsey fringe.
Experimental results for the two Ramsey fringes obtained with
and without dimerization swapping during the state evolution are
shown in Fig. \ref{SSHZakData}(a), and the obtained phase
differences are shown in Fig. \ref{SSHZakData}(b), together with
the corresponding histogram. Thus, the Zak phase difference
between the two dimerized configurations was determined to be
$\delta \varphi_{\text{Zak}} = 0.97(2)\pi$, which agrees well with
theoretical prediction of the topological Bloch bands in the SSH
model. This method was used to further study the dependence of the
Zak phase on the offset energy $\delta \varphi_{\text{Zak}}(\Delta)$, which
corresponds to the Rice-Mele model with the fractional Zak phase.
This work establishes a general approach for probing the
topological invariant in topological Bloch bands in OLs. The measurement technique can be extended to more
complicated topological models, such as detecting the Chern
numbers of the Hofstadter model and the Haldane model
\cite{Aidelsburger2014,Jotzu2014}, and the $\pi$ Berry flux
associated with a Dirac point in 2D OL systems
\cite{Duca2015}.

\begin{figure}
\begin{center}
\includegraphics[width=0.95\linewidth]{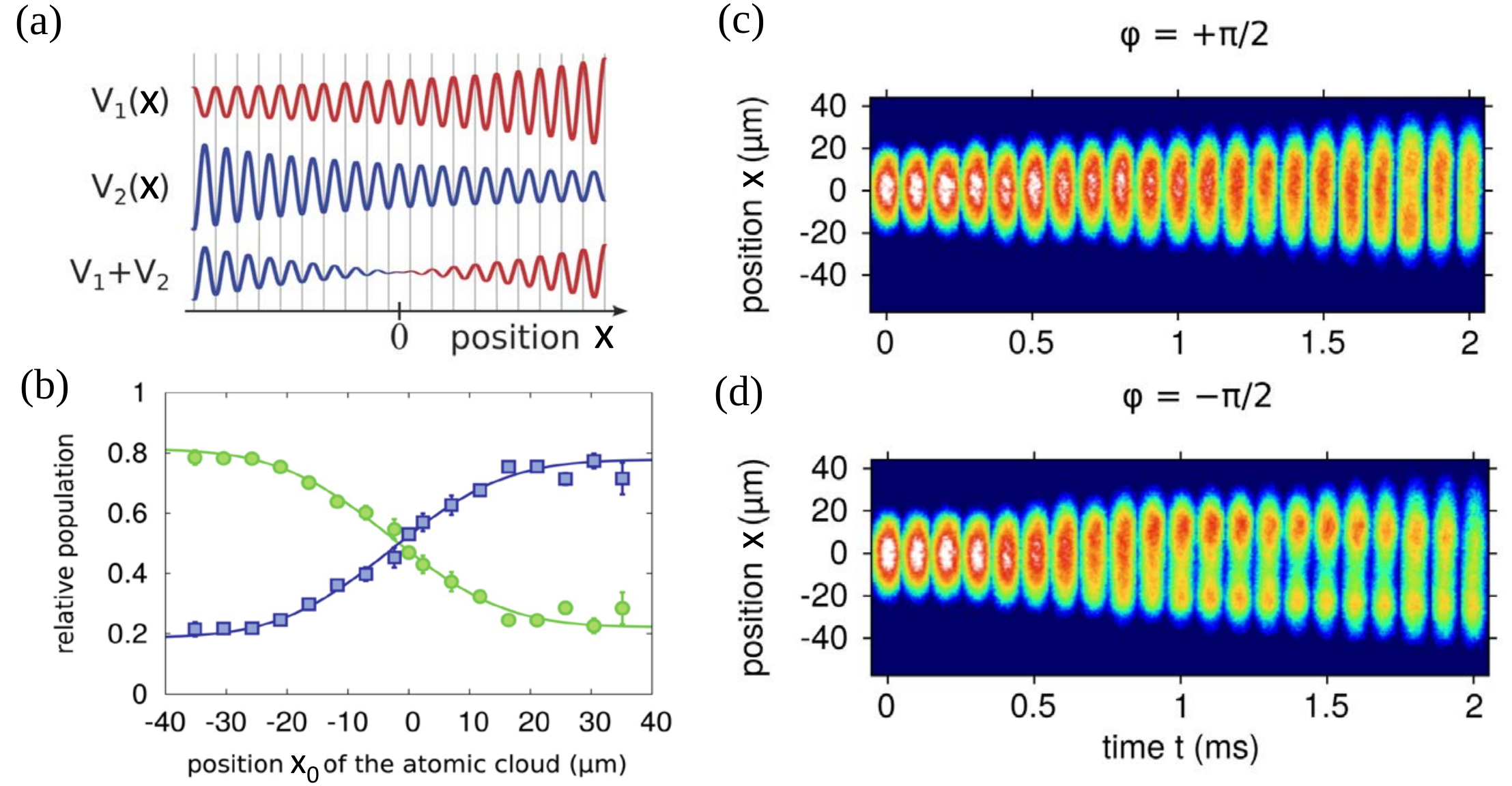}
\end{center}
\caption{(Color online) Realizing the interface and probing the topological bound state from temporal evolution of atomic clouds.
(a) The 1D OL with spatially varying lattice depth from the amplitude-chirped lattice potentials
$V_{1,2}(x)$, which increases (decreases) along $x$ for positive
(negative) values of the Raman detuning. (b) Measured spatial
variation of band ordering. Relative atomic population transferred
into the upper (green circles) and lower (blue squares) band on
loading from an initial state with an atomic cloud centered at
position $x_0$. Series of absorption images
for a relative phase of the initially prepared atomic wavepacket
of (c) $\varphi=+\pi/2$ and (d) $\varphi=-\pi/2$ for different holding
times in the lattice. For $\varphi=+\pi/2$, trapping of atoms in
the topological edge state was observed, while for $\varphi=-\pi/2$, the cloud
splits up. Reprinted with permission from Leder {\it et al.} \cite{Leder2016}.}
\label{SSH-Soliton-Lattice}
\end{figure}

After measuring the bulk topological index in the SSH model, cold
atom experiments have begun to probe topological boundary
states at the intersection of two different topological
phases. Recently, two different approaches for synthesizing and
observing topological soliton states in the SSH model using
BECs in 1D OLs were
experimentally developed \cite{Leder2016,Meier2016}. In the first experiment \cite{Leder2016}, the authors used a 1D
OL for cold rubidium atoms with a spatially chirped
amplitude to create a domain wall and then directly observed the
soliton state by optical real-space imaging of atoms confined at
the interface. The lattice potential is realized using a
rubidium atomic three level configuration with two ground states
of different spin projections and one excited state. To achieve a topological interface, two
four-photon potentials $V_1(x)$ and $V_2(x)$ with opposite spatial
variation of the Raman coupling are superimposed, as shown in Fig.
\ref{SSH-Soliton-Lattice}(a). This gives rise to a 1D lattice
potential $V(x)=V_1(x)+V_2(x)\approx2a\cdot x\cos(4k_Lx)$,
creating a zero crossing at $x=0$, where $k_L=2\pi/\lambda$ with $\lambda$ being the wavelength of the laser beams.
For $x>0$
($x<0$) the maxima (minima) of the potential are located at
integer multiples of $\lambda/4$, This phase change is reflected
in the inversion of ordering bands, which cannot be transformed
into each other by continuous deformation without closing the gap.
For such a situation a non-degenerate topologically protected
bound state localized around $x=0$, where the bands intersect, is
expected in the SSH model. The dynamics of
atoms in such a structure near the band crossing is described by
the Dirac Hamiltonian with a spatially dependent effective mass
(the Jackiw-Rebbi model) in the continuum limit with good
accuracy, as the width of the topological bound state is two
orders of magnitude larger than the lattice spacing
\cite{Leder2016}. To verify the band inversion on sign change of
$x$, the adiabatically expanded atomic cloud centered at different
lateral positions $x_0$ along the lattice beam axis was transferred
to the state $\phi_{\text
i}(x)=\frac{1}{\sqrt{2}}\phi_x(x-x_0)(e^{2ikx}+e^{-2ikx})$ via two
simultaneously performed Bragg pulses. The band populations
following activation of the lattice were determined. For $x<0$,
loading is enhanced in the lower band, while for $x>0$ most
atoms are transferred into the upper band, and near $x=0$ the
curves cross, as shown in Fig. \ref{SSH-Soliton-Lattice}(b). This
experimentally verifies the spatial variation of the band
structure and the exhibition of a topological interface at $x=0$.
Next in the experiments, the atoms were loaded
into the interface and the topological bound state was observed
via an initial state $\phi_{\text
i}(x)=\frac{1}{\sqrt{2}}\phi_x(x)(e^{2ikx}+e^{-i\varphi}e^{-2ikx})$
with $\varphi=\pi/2$. The atomic wavepacket was centered at $x=0$,
after which the lattice beams were activated and a series of atomic
absorption images were recorded after a variable holding time in
Fig. \ref{SSH-Soliton-Lattice}(c). This shows that the atomic cloud
remains trapped at the expected position of the topological bound
state. On the other hand, for a phase of $\varphi=-\pi/2$,
no such trapping in the bound state was observed, as shown in Fig.
\ref{SSH-Soliton-Lattice}(d). As expected, there is no overlap
with the topological bound state when the initially prepared
atomic wavepacket is $\pi$ out of phase, such that the wavepacket
is split into two spatially diverging paths. Further evidence for the successful population of the
topological bound state was obtained from the phase
($\varphi$) dependence and the dependence of the initial atomic
momentum width on loading efficiency.

\begin{figure}
\begin{center}
\includegraphics[width=0.95\linewidth]{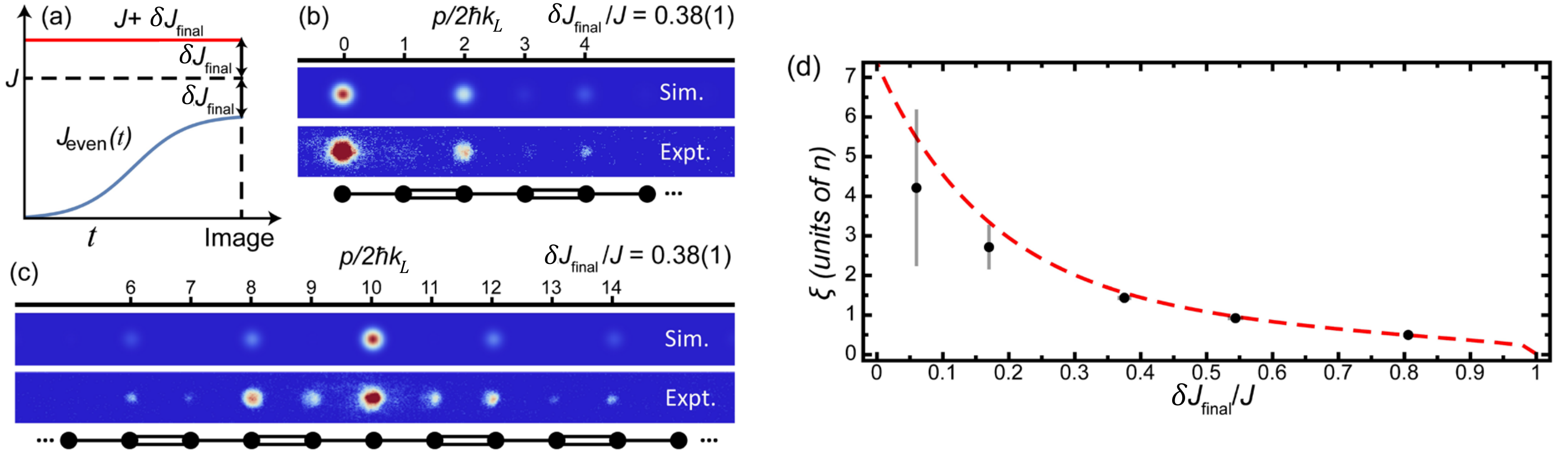}
\end{center}
\caption{(Color online) Adiabatic preparation of the topological bound state in the momentum-space lattice. (a) Time sequence of the smooth ramp of the weak tunneling links (blue), holding the strong (red) links fixed. (b) Simulated (top) and averaged experimental (bottom) absorption images for an adiabatically loaded edge-defect lattice. (c) Same as (b), but for an adiabatically loaded central-defect lattice. (d) Decay length of the atomic distribution on even sites of the edge-defect lattice vs. $\delta J_{\text{final}}/J$. The dashed line represents the results of a numerical simulation of the experimental sequence. Reprinted with permission from Meier {\it et al.}\cite{Meier2016}.} \label{SSH-Momentum}
\end{figure}

In another experiment \cite{Meier2016}, a momentum-space lattice was used to simulate the hard-wall boundaries in the SSH model and the simulated topological bound state was then observed by site-resolved detection of the populations in the lattice. The physics of the SSH lattice model was emulated by using the controlled evolution of momentum-space distributions of cold atomic gases \cite{Gadway2015,Meier2016a}. Controlled coupling between discrete free-particle momentum states is achieved through stimulated two-photon Bragg transitions, driven by counterpropagating laser fields detuned far from atomic resonance. The lasers coherently coupled 21 discrete atomic momentum states in the experiment \cite{Meier2016}, creating a "momentum-space lattice" of states in which atomic population may reside [see, e.g., Figs.~\ref{SSH-Momentum}(b,c)]. The momentum states are characterized by site indices $n$ and momenta $p_n = 2 n \hbar k_L$. The coupling between these states is fully controlled through 20 distinct two-photon Bragg diffraction processes, allowing simulation of 1D tight-binding models with local control of all site energies and tunneling terms. This enables one to create the hard-wall boundaries and lattice defects in the SSH model. The authors then directly probed the topological bound states in the SSH model simulated in such an OL through quench dynamics, phase-sensitive injection, and adiabatic preparation. The first detection method is to abruptly expose the condensate atoms initially localized at only a single lattice site and observe the ensuing quench dynamics with single-site resolution. When population was injected onto a defect site, a large overlap with the topological bound state was found, resulting in a relative lack of dynamics as compared to injection at any other lattice site. The second method is to probe the inherent sensitivity of the bound state to a controlled relative phase of initialized states, following a Hamiltonian quench. It was observed that the dynamics is nearly absent when the phase matches that of the bound state, while defect-site population is immediately reduced when the phase does not match. Lastly, the topological mid-gap bound state was directly probed through a quantum annealing procedure. The bound eigenstate was initially prepared in the fully dimerized limit of the time-dependent SSH Hamiltonian \cite{Meier2016}
\begin{equation}
H_{\text{SSH}}(t)=-\sum_{n\in \text{odd}}{(J+\delta J_{\text{final}}) (c^{\dagger}_{n+1}c_{n}+\text{h.c.})}-\sum_{n\in \text{even}}{J_{\text{even}}(t) (c^{\dagger}_{n+1}c_{n}+\text{h.c.})} \ ,
\end{equation}
with only the odd tunneling links present at a strength $J_{\text{odd}} = J+\delta J_{\text{final}}$. Atomic population was injected at the decoupled zeroth site, identically overlapping with the mid-gap state. Next, the tunneling on the even links was slowly increased from zero to $J-\delta J_{\text{final}}$, as depicted by the smooth ramp in Fig.~\ref{SSH-Momentum}(a). For adiabatic ramping, the atomic wavefunctions should follow the eigenstate of $H_{\text{SSH}}(t)$. This adiabatic preparation method was performed for a lattice with the defect on the left edge and at its center, as shown in Fig.~\ref{SSH-Momentum}(b) and \ref{SSH-Momentum}(c), respectively. As known in the SSH model, the amplitude of the mid-gap state wavefunction is largest at the defect site and decays exponentially into the bulk. The decay length $\xi$ (in units of the spacing between lattice sites) should scale roughly as the inverse of the energy gap. Thus, highly localized mid-gap states for $\delta J_{\text{final}}/J \sim1$ (dimerized limit) and full delocalization over all 21 sites for $\delta J_{\text{final}}/J \ll 1$ (uniform limit) are expected. By tuning the normalized tunneling imbalance in the experiment, one can achieve a direct exploration of the mid-gap state's localization decay length as a function of $\delta J_\text{final}/J$, as shown in Fig.~\ref{SSH-Momentum}(d). The technique of creating momentum-space lattices with direct and full control of tunneling configuration demonstrated in this work has been extended to create 2D artificial flux lattices for cold atoms \cite{FAn2017a}.

\subsubsection{Topological pumping}\label{Topopuming}

Quantum pumping as originally proposed by Thouless ~\cite{Thouless1983} entails the transport
of charge in a 1D periodic potential through an adiabatic cyclic evolution of the underlying
Hamiltonian. In contrast to classical pumping, the transported charge is quantized and purely determined by the topology of the pump cycle, which is related to the Chern number. Topological pumping can be generalized to interacting systems \cite{Niu1984}, and is robust against moderate perturbations and finite-size effects.

Topological Thouless pumping is closely related to the modern theory of polarization. Consider a lattice site at $x=R$ in a 1D periodic lattice, with the Bloch function of the lowest band $|\psi_k\rangle=e^{ikx}|u_k\rangle$, the corresponding Wannier function is given by
\begin{equation}
|R\rangle=\frac{1}{\sqrt{N}}\sum_{k=-\pi/d}^{\pi/d}e^{-ikR}|\psi_k\rangle
=\sqrt{\frac{d}{L}}\sum_{k=-\pi/d}^{\pi/d}e^{ik(x-R)}|u_k\rangle,
\end{equation}
where $N=L/d$ is the number of unit cells in the system with $L$ being the system length and $d$ being the lattice constant.
The expected shift of the Wannier center from the lattice site $R$ at time $t$ is denoted by the polarization
\begin{align}\label{eq:pol}
P(t)=\langle R(t)|x-R|R(t)\rangle=\frac{d}{L}\sum_{-\pi/d}^{\pi/d}\langle u_k(t)|i {\partial_{k}}|u_k(t)\rangle=d\int_{-\pi/d}^{\pi/d}\frac{dk}{2\pi}A_k(k,t).
\end{align}
Here $A_k(k, t)=i\langle u_k(t)|\partial_{k}|u_k(t)\rangle$ is the Berry connection, and its integration over the first BZ is the Zak phase in Eq. (\ref{Zak}). The spatial shift of the Wannier function after one pumping cycle at $t=T$ can be represented by the change of the polarization $\Delta P=\int_0^{T}dt\frac{\partial P}{\partial t}$, which can be obtained by using Stokes's formula as
\begin{equation}
\Delta P=\frac{d}{2\pi}\int_{-\pi/d}^{\pi/d} dk\int_{0}^{T} dt
\left[{\partial_{t}} A_k(k,t)-{\partial_{k}} A_t(k,t)\right].
\end{equation}
Such an integral over two parameters corresponds to a topological invariant, i.e., the first Chern number $\mathcal{C}$ in a $k$-$t$ BZ:
\begin{equation}
C=\frac{1}{2\pi}\int_{-\pi/d}^{\pi/d} dk\int_{0}^{T} dt \mathcal{F}(k,t),
\end{equation}
with $\mathcal{F}(k,t)={\partial_{t}} A_k(k,t)-{\partial_{k}}
A_t(k,t)$ as the Berry curvature. Thus the shift of the Wannier
center after one pumping cycle is quantized in units of the
lattice constant of a unit cell: $\Delta P = C d$. This
indicates that the transported particle in the adiabatic cyclic
evolution is quantized and related to topological Bloch bands. The
SSH/RM model can serve as a pictorial model for implementing the topological pumping \cite{Xiao2010}.

\begin{figure}\centering
\includegraphics[width=0.95\linewidth]{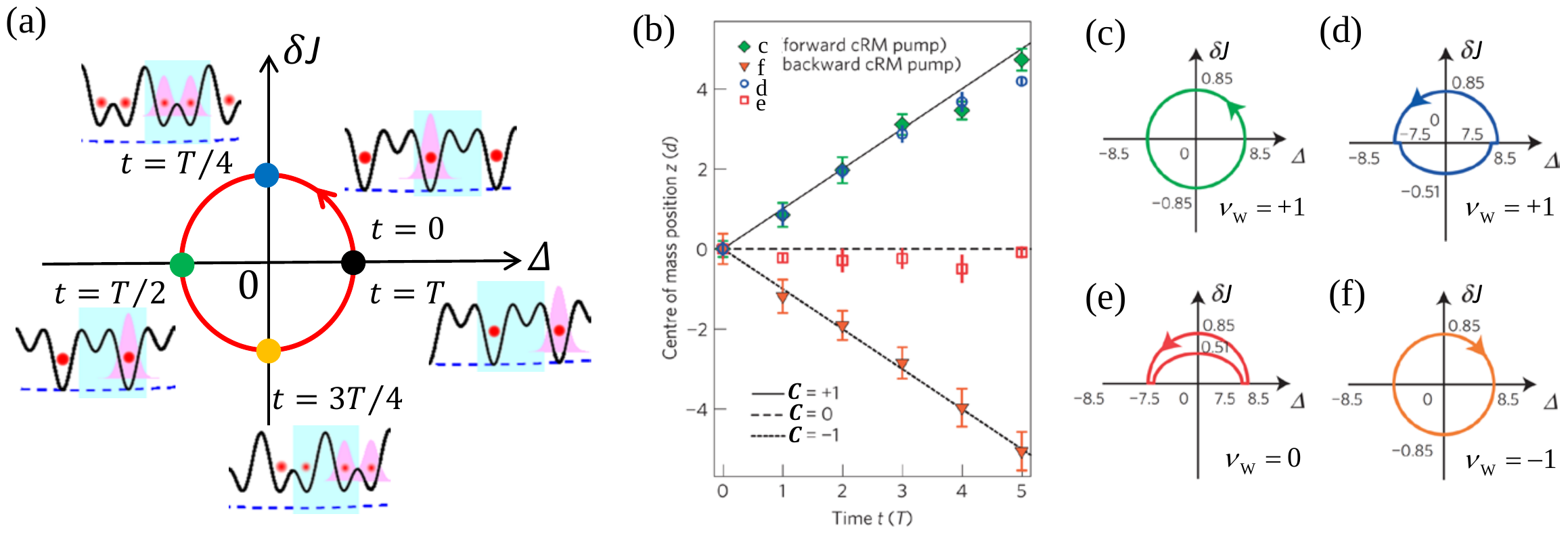}
\caption{ (Color online) Topological pumping in the Rice-Mele OL. (a)
A pump cycle sketched in $\delta J$-$\Delta$ space  with schematic of the
pumping sequence. The pink shaded packet indicates the wave
function of a particular atom initially localized at the unit cell. The wave function shifts to right as the pumping proceeds and
moves the atom to the next unit cell after one pump cycle. (b) Results of four typical topological/trivial pumping (characterized by the Chern number $C=\pm1,0$) with schematic pumping sequences in the $\delta J$-$\Delta$ plane shown in (c-f) and the winding number $\nu_{\text{w}}$ of each trajectory around the origin. (c) Charge pumped during a simple Rice-Mele pumping; (d) topologically nontrivial pumping; (e) topologically trivial pumping; and (f) negative sweep pumping. 
Reprinted by permission from Macmillan Publishers Ltd: Nakajima {\it et al.}\cite{Nakajima2016}, copyright\copyright~(2016).}\label{TopoPumpOL}
\end{figure}

Although topological charge pumping was first proposed more
than thirty years ago, it has not yet been directly realized in condensed
matter experiments. With ultracold atoms in tunable optical
superlattices, the implementation of topological pumping has
been extensively discussed
\cite{Qian2011,Wang2013,FMei2014,Grusdt2014,DWZhang2015,Marra2015,Zeng2015,Zeng2016,Taddia2017,Xu2017,Sun2017}.
In the context of the SSH/Rice-Mele model, it was theoretically
demonstrated that the quantized particle pumping characterized by
the non-zero Chern number can be realized in the cold atom systems
\cite{Wang2013,FMei2014,DWZhang2015}, which is robust with respect
to some perturbations in realistic experiments, such as nonzero
temperature and the effects of finite sizes, non-adiabatic
evolutions and trapping potentials. Moreover, as an extension of
Thouless pumping, the topological pumping of interacting bosoinc and fermionic atoms trapped in
OLs for specific models was also studied \cite{Qian2011,Grusdt2014,Zeng2015,Zeng2016,Taddia2017,YKe2017}. For
example, in the strongly interacting region, the bosonic atoms share the same transport
properties as non-interacting fermions with quantized transport \cite{Qian2011,YKe2017}. Due to the
degeneracy of the many-body ground states of the interacting
bosons or fermions, the so-called topological fractional pumping
(fractional values of the pumped particle) related to the
many-body Chern number can be realized at certain fractional
fillings \cite{Grusdt2014,Zeng2015,Zeng2016,Taddia2017}.

Recently, topological pumping has been realized in two
experiments \cite{Nakajima2016,Lohse2016} with ultracold fermionic
and bosonic atoms in 1D optical superlattices, respectively. In
the experiment in Ref. \cite{Nakajima2016}, an ultracold Fermi gas
of ytterbium atoms $^{171}$Yb was loaded into a dynamically
controlled optical superlattice, occupying the lowest energy band.
The superlattice is formed by superimposing a long lattice $V_L$
and a short lattice $V_S$ with periodicity difference by a factor of two,
and its time-dependent potential takes the form
\begin{equation}
V(x,t) = -V_S(t)\cos^2\left(\frac{2\pi x}{d}\right)
-V_L(t)\cos^2\left(\frac{\pi x}{d}-\phi(t)\right),
\end{equation}
where $\phi$ is the phase difference between the two lattices. By
slowly sweeping $\phi$ over time, the lattice potential returns to
its initial configuration whenever $\phi$ changes by $\pi$, thus
completing a pumping cycle. The ability to tune all parameters of
the lattice potential independently in a dynamic way offers the
opportunity to realize various pumping protocols. In the absence
of a static short lattice, $V(x,t)$ describes the simple sliding
lattice that Thouless originally proposed \cite{Thouless1983}.
In addition to this term, the double well lattice shown in Fig.
\ref{TopoPumpOL} is realized, which can be described by the
tight-binding Rice-Mele Hamiltonian in Eq. (\ref{RM}) for the alternating pumping.
Figure \ref{TopoPumpOL} shows the schematics of the continuous
pumping protocol as a closed trajectory in the
$\delta J$-$\Delta$ parameter plane, where $\delta J\equiv (J-J')/2$ is the modulation of the dimerization configurations. By sweeping the phase linearly in time $\phi(t)=\pi t/T$, one can periodically modulate the hopping amplitudes and on-site
energies. In the experiment,
topological pumping was detected as a shift of the center-of-mass of
the atomic cloud measured with {\sl in situ} imaging and the first Chern
number $C$ of the pumping procedure was extracted from the average
shift of the center-of-mass per pumping cycle, which is consistent
with the ideal value $C=0,\pm1$. The topological nature of the
pumping was revealed by the pumping trajectories' clear dependence on the topology
in parameter space (denoted by the winding number $\nu_{\text{w}}$) as to whether or not a
trajectory encloses the degenerate point. Pumping in the sliding lattice was
demonstrated to be topologically equivalent to the continuous
Rice-Mele pumping because of the same Chern numbers of the first
band, which can be connected by a smooth crossover without closing
the gap to the second band. It was also verified that the topological pumping indeed works in the quantum regime by varying the
pumping speed and the temperature in the experiment \cite{Nakajima2016}.

In another experiment \cite{Lohse2016}, the authors realized the topological pumping with ultracold bosonic atoms $^{87}$Rb forming
a Mott insulator in a similar dynamically controlled optical superlattice. Due to the large on-site interaction, each atom is
localized on an individual double well, resulting in homogeneous delocalization over the entire first BZ. By taking in
situ images of the atomic cloud in the lattice, they also observed a quantized deflection per pump cycle. The genuine quantum
nature of the pumps was revealed by a counterintuitive reversed deflection of particles in the first excited band and a controlled topological transition in higher bands when tuning the superlattice parameters.

Also with bosonic atoms, a quantum  geometric pump for a BEC in the lowest Bloch band of a tunable
bipartite magnetic lattice was realized \cite{HILu2016}. In contrast to the topological pumping yielding quantized pumping set
by the global topological properties of the filled bands, the geometric pumping for a BEC occupying just a
single crystal momentum state exhibits non-quantized particle pumping set by local geometrical properties of the band structure.
For each pump cycle, a non-quantized overall displacement and a temporal modulation of the atomic wave packet's position in each
unit cell (i.e., the polarization) were observed \cite{HILu2016}.

These cold atom systems can be extended to implement more complex quantum pumping schemes, including spin degrees of freedom and in higher dimensions \cite{Schweizer2016}. Analogous to Thouless pumping, topological pumping for spins without a net transport of charge may be constructed by imposing the TRS \cite{LFu2006}. A spin pump with spin conservation can be composed of two independent pumps, where up and down spins have inverted Berry curvature and are transported in opposite directions. Spin pumps could serve as spin current sources for spintronic applications.

\begin{figure}\centering
\includegraphics[width=0.95\linewidth]{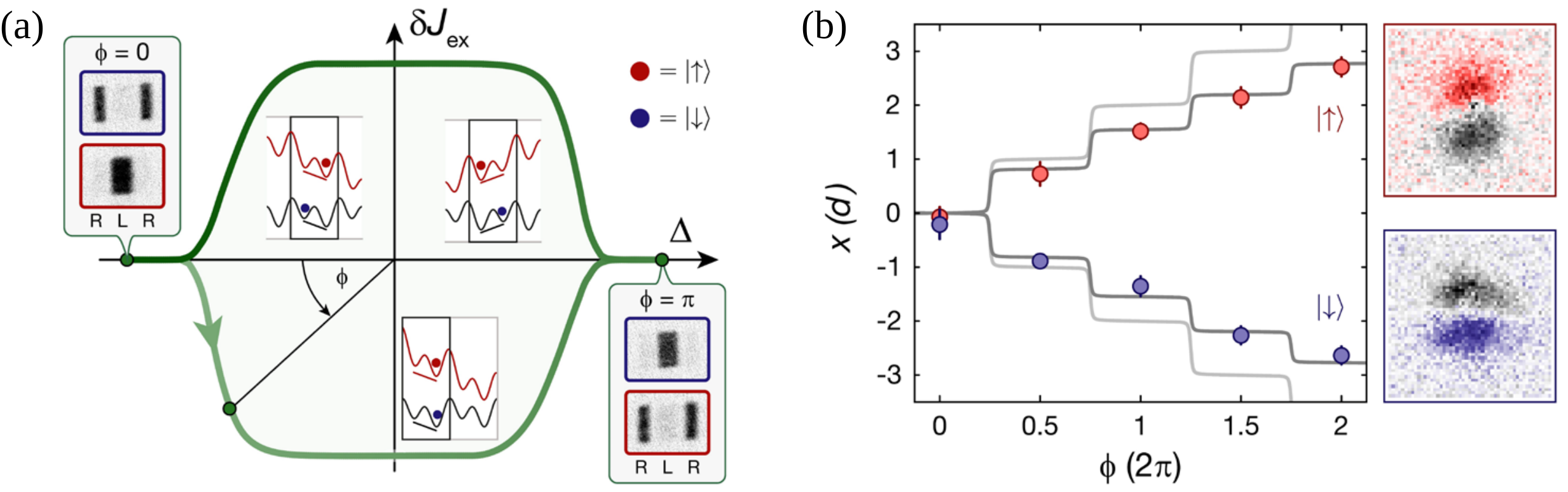}%
\caption{(Color online) Spin pumping in a spin chain of cold atoms. (a) A spin pump cycle in parameter space (green) of spin-dependent tilt $\Delta$ and exchange coupling dimerization $\delta J_\text{ex}$. The path can be parametrized by the angle $\phi$, acting as the pump parameter. The insets in the quadrants show the local mapping of globally tilted double wells to the corresponding local superlattice tilts with the black rectangles indicating the decoupled double wells.
Between $\phi=0$ and $\pi$, $\left|\uparrow\right>$ and $\left|\downarrow\right>$ spins exchange their position, which can be observed by site-resolved band mapping images detecting the spin occupation on the left (L) and right (R) sites.
(b) Center-of-mass position of up (red) and down (blue) spins as a function of $\phi$. Different absorption images of both sequences for $\left|\uparrow\right>$ and $\left|\downarrow\right>$ spins are shown on the right side. The solid lines depict the calculated motion of a localized spin for the ideal case (light gray) and for a reduced ground state occupation and a pump efficiency per half pump cycle that was determined independently through a band mapping sequence (gray). Reprinted with permission from Schweizer {\it et al.} \cite{Schweizer2016}. Copyright\copyright ~(2016) by the American Physical Society.}%
\label{SpinPump}%
\end{figure}

In a recent experiment \cite{Schweizer2016}, quantum spin pumping was implemented with ultracold bosonic atoms in two hyperfine states in a spin-dependent dynamically controlled optical superlattice, where each spin component is localized to a Mott insulator with negligible interspin interaction. In addition, the two spin components are coupled via spin-isotropic on-site interactions. For strong interactions $U\gg J$ (tunneling $J$ is suppressed) and unit filling, the system can be described by a 1D spin chain \cite{Schweizer2016}:
\begin{eqnarray}
    \begin{aligned}
        H_{\text{SP}} = -\frac{1}{4}\sum_{n}\left[J_\text{ex}+(-1)^{n}\delta J_\text{ex}\right]\left({S}^+_{n}{S}^-_{n+1}+\text{h.c.}\right)
        + \frac{\Delta}{2}\sum_{n} (-1)^{n} {S}^z_n
        \end{aligned}
\end{eqnarray}
with spin-dependent tilt $\Delta$ and alternating exchange coupling $\frac{1}{2}\left(J_\text{ex}\pm \delta J_\text{ex}\right)$.
For large tilts $\Delta\gg\frac{1}{2}\left(J_\text{ex} + \delta J_\text{ex}\right)$ the many-body ground state forms an antiferromagnetic ordered spins, while for strong exchange coupling $\frac{1}{2}\left(J_\text{ex} + \delta J_\text{ex}\right)\gg\Delta$ dimerized entangled pairs are favored. Varying $\delta J_\text{ex}$ and $\Delta$ during the pump cycle modulates ($\delta J_\text{ex},\Delta$) in the interacting 1D spin chain and encircles the degeneracy
point, as shown in Fig. \ref{SpinPump}(a). After a full cycle, the two spin components move by a lattice site in opposite directions. This leads to a quantized spin transport described by the $\Z_2$ invariant as in the topologically equivalent case of independent spins \cite{LFu2006}. Such a spin pump can be regarded as a dynamical version of the quantum spin Hall effect, where the parameter $\phi$ (the phase of the superlattice) is an additional dimension in a generalized 2D momentum space. The atomic spin current was measured and the net spin transport was further verified through {\sl in situ} measurements of the spin-dependent center-of-mass displacement, as shown in Fig. \ref{SpinPump}(b).

\subsubsection{1D AIII class topological insulators}

The topological or trivial character of a 1D insulator is completely determined by the presence or absence of chiral
symmetry \cite{Schnyder2008,Ryu2010,Chiu2016}. There are two distinct classes of 1D topological insulators, as chiral symmetry
is the composition of time-reversal and charge-conjugation (particle-hole) symmetries. The first class is invariant under
both the two symmetries and is called the BDI symmetry class represented by the SSH model. The second one is
the AIII class with broken time-reversal and charge-conjugation symmetries, which still lacks experimental realization in
condensed matter materials. The topology of the AIII class phase is quantified by an integer winding number. Recently, several
works have been presented to study the 1D AIII class topological insulator using cold fermionic atoms in OLs
\cite{Liu2013,Velasco2017,Zhou2017,BSong2018}.

A simple dimerized lattice model for realizing the AIII class topological
insulator was proposed in Ref. \cite{Velasco2017}. The Hamiltonian is given by
\begin{equation}\label{AIIIModel-1}
H_{\text{AIII}}=-\sum_{n}\left(Ja_{n}^{\dagger}b_{n}+J' e^{i\theta}a_{n}^{\dagger}b_{n-1}+\text{h.c.}\right),
\end{equation}
which is a generalization of the SSH model by introducing an
acquired complex phase factor $e^{i\theta}$  ($\theta$ is a phase difference between the inter-cell and extra-cell hoppings) when particles tunnel from
one unit cell to the next.
 It is worth noting that for open boundary conditions, the Hamiltonian $H_{\text{AIII}}$ in Eq. \ref{AIIIModel-1} is equivalent to $H_{\text{SSH}}$ in Eq. \ref{SSH} through the gauge transformations, $(a^\dagger_{n},b_n^\dagger)\rightarrow e^{-in\theta}(a^\dagger_{n},b_n^\dagger)$. However,  under the periodic boundary conditions, it is equivalent to the SSH model in a ring of $N$ lattice sites threaded by a magnetic flux $\phi=N\theta$.
Therefore, under the periodic boundary conditions,   this Hamiltonian still
corresponds to the SSH model  for $\theta=0$, while in the presence of this extra
phase $\theta\neq0,\pi$ (and $N\theta \neq 2m\pi$ with $m$ an integer), the model enters the AIII symmetry class.
This model can be realized with spinless cold atoms by combining a
1D optical superlattice with the Raman assisted tunneling
\cite{Velasco2017}. The Bloch Hamiltonian in this model is given by
$\mathcal{H}_{\text{AIII}}(k)=-\left[J+J'\cos(k-\theta)\right]\sigma_{x}-J'\sin(k-\theta)\sigma_{y}$,
which exhibits chiral symmetry for any value of $\theta$ since
$\sigma_{z}\mathcal{H}_{\text{AIII}}(k)\sigma_{z}=-\mathcal{H}_{\text{AIII}}(k)$.
For $\theta\ne 0, \pi$, it is not time reversal symmetric and
thereby not charge-conjugation symmetric because $ \mathcal{H}_{\text{AIII}}^{*}(-k)\ne
\mathcal{H}_{\text{AIII}}(k)$ and no $2 \times 2$ unitary transformation $U$ such that $U
\mathcal{H}_{\text{AIII}}^{*}(-k)U^{\dagger}=\mathcal{H}_{\text{AIII}}(k)$ in this case.



\begin{figure}\centering
\includegraphics[width=0.6\linewidth]{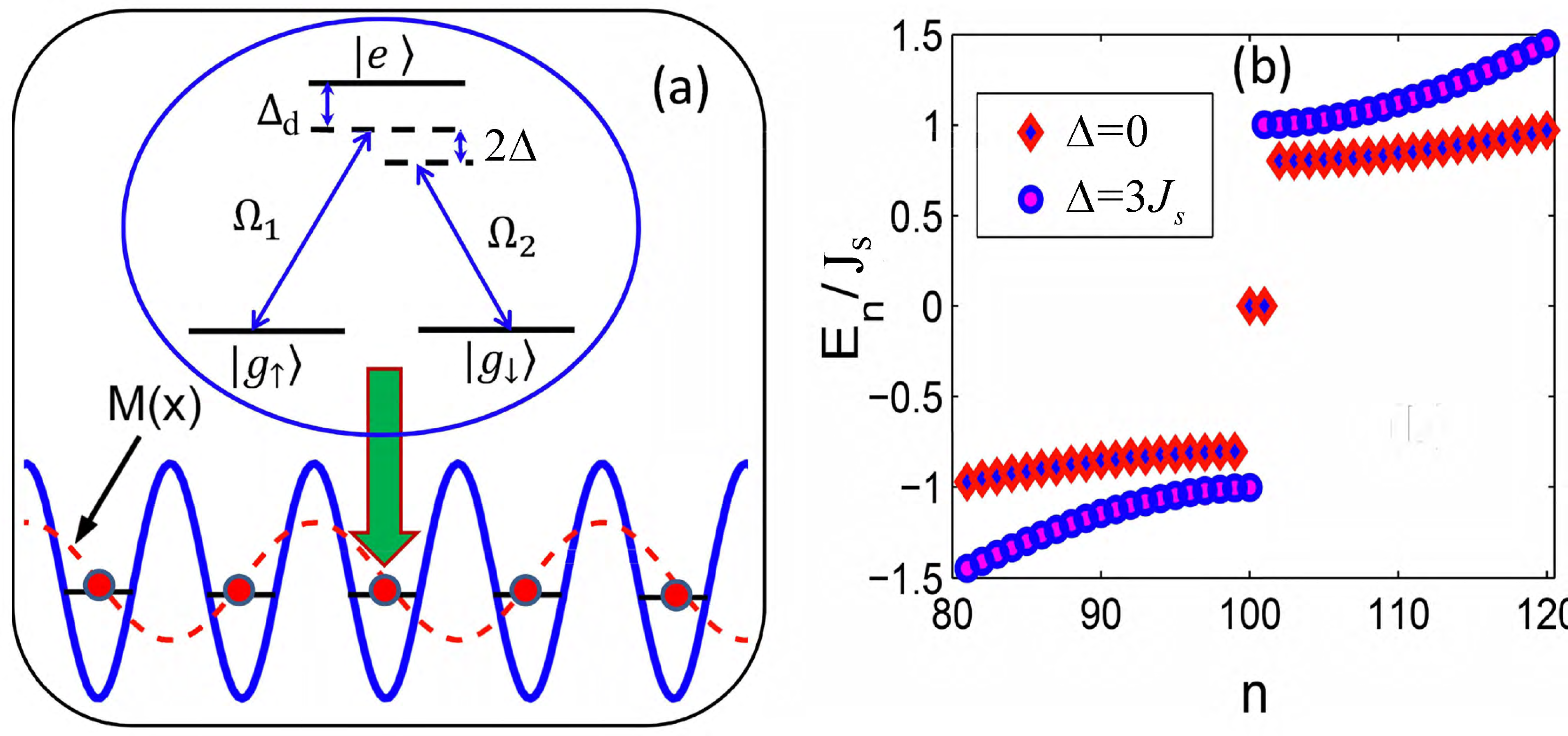}
\caption{(Color online) Proposed optical Raman lattice for realizing the Hamiltonian
(\ref{AIIIModel-2}). (a) Cold fermions trapped in 1D optical
lattice with internal three-level $\Lambda$-type configuration
coupled to radiation. (b) Energy spectra with open boundary
condition in the topological ($\Delta=0$) and trivial
($\Delta=3J_s$) phases. The SO coupled hopping $J_{\rm
so}=0.4J_s$. Reprinted with permission from Liu {\it et al.}\cite{Liu2013}. Copyright\copyright~(2013) by the American Physical Society.}
\label{AIII-RamanOL}
\end{figure}

Another proposed model of 1D AIII class topological insulator is
using spin-orbit-coupled fermionic atoms in an optical Raman
lattice \cite{Liu2013}. The atoms with the internal three-level
$\Lambda$-type configuration are coupled through the transitions
$|g_\uparrow\rangle,|g_\downarrow\rangle\rightarrow|e\rangle$
driven by the laser fields with Rabi-frequencies
$\Omega_1(x)=\Omega_{0}\sin(k_0x/2)$ and
$\Omega_2(x)=\Omega_{0}\cos(k_0x/2)$, as shown in
Fig.~\ref{AIII-RamanOL}(a). In the presence of a large one-photon
detuning $|\Delta_d|\gg\Omega_{0}$ and a small two-photon detuning
$2|\Delta|\ll\Omega_0$ for the transitions, the system Hamiltonian
reads $H=H_0+H_1$, with $H_0=\sum_{\sigma=\uparrow,\downarrow}\bigr[\frac{p_x^2}{2m}+V_\sigma(x)\bigr]|g_\sigma\rangle\langle
g_\sigma|+2\hbar\Delta|g_\downarrow\rangle\langle g_\downarrow|$, and $
H_1=\hbar\Delta_d|e\rangle\langle
e|-\hbar\bigr(\Omega_1|e\rangle\langle
g_\uparrow|+\Omega_2|e\rangle\langle g_\downarrow|+{\rm
H.c.}\bigr)$.
Here the potentials $V_{\uparrow,\downarrow}(x)=-V_0\sin^2(k_0x)$
form a 1D spin-independent OL. For $|\Delta_d|\gg\Omega_{0}$, the
lasers $\Omega_{1,2}$ induce a two-photon Raman transition between
$|g_{\uparrow}\rangle$ and $|g_\downarrow\rangle$. The effect of
the small two-photon detuning is equivalent to a tunable Zeeman field
along $z$ axis $\Gamma_z=\hbar\Delta$. Eliminating the excited state by
$|e\rangle\approx\frac{1}{\Delta}(\Omega_1^*|g_\uparrow\rangle+\Omega_2^*|g_\downarrow\rangle)$
yields the effective Hamiltonian
\begin{eqnarray}\label{AIIIModel-2}
H_{\rm eff}&=&\frac{p_x^2}{2m}+\sum_{\sigma=\uparrow,\downarrow}\bigr[V_\sigma(x)+\Gamma_z\sigma_z\bigr]|g_\sigma\rangle\langle g_\sigma|-\bigr[M(x)|g_\uparrow\rangle\langle g_\downarrow|+{\rm H.c.}\bigr],
\end{eqnarray}
where $M(x)=M_0\sin(k_0x)$ with $M_0=\hbar\Omega_0^2/2\Delta_d$ represents a transverse Zeeman field induced by the Raman process. In the tight-binding regime, the system Hamiltonian can be recast into \cite{Liu2013}
\begin{equation}
\tilde{H}_{\text{AIII}}=-J_s\sum_{<i,j>}(c_{i\uparrow}^{\dag}
c_{j\uparrow}-c_{i\downarrow}^{\dag}
c_{j\downarrow})
+\sum_{j}\Gamma_z(n_{j\uparrow}-n_{j\downarrow})
+\sum_{j}J_{{\rm so}}(c_{j\uparrow}^\dag c_{j+1\downarrow}-c_{j\uparrow}^\dag c_{j-1\downarrow})+ {\rm H.c.}.
\end{equation}
Here time reversal and charge conjugation operators are
respectively defined by ${\hat T}=i\hat K\sigma_y$ with $\hat K$ being the complex
conjugation, and ${\hat C}: (c_{\sigma},\
c^\dag_{\sigma})\longmapsto(\sigma_z)_{\sigma\sigma'}(
c^\dag_{\sigma'}, c_{\sigma'})$. The topological phase in this
free-fermion system belongs to the chiral AIII class because while
both ${\hat T}$ and ${\hat C}$ are broken in
$\tilde{H}_{\text{AIII}}$, the chiral symmetry, defined as their
product, is reserved since $({\hat
C \hat T})\tilde{H}_{\text{AIII}}({\hat
C \hat T})^{-1}=\tilde{H}_{\text{AIII}}$, with $({\hat C \hat T})^2=1$. In
particular, this Hamiltonian describes a 1D topological insulator
for $|\Gamma_z|<2J_s$ with two mid-gap zero edge modes and
otherwise a trivial insulator, with the bulk gap
$E_g=\mbox{min}\{|2J_s-|\Gamma_z||, 2|J_{\rm so}|\}$, as
shown in Fig.~\ref{AIII-RamanOL}(b). It was shown that the zero
edge modes in this 1D AIII topological insulator are spin
polarized, with left and right edge spins polarized to opposite
directions and forming a topological spin qubit of cold atoms
\cite{Liu2013}. A similar Raman lattice scheme was proposed to
simulate symmetry-protected topological states using
alkaline-earth-like atoms \cite{Zhou2017}. The interaction-driven
topological phase transition for interacting fermionic atoms and a
$\Z_4$ reduction of the 1D AIII class was also studied
\cite{Liu2013,Zhou2017}.

A recent experiment \cite{BSong2018} was reported to realize the
1D symmetry-protected topological state with cold fermionic atoms
of $^{173}$Yb in the optical Raman lattice. The experimental setup
is similar as that proposed in Ref. \cite{Liu2013}, except that
the potential forms a spin-dependent lattice with spin-dependent hopping strengths that explicitly break the
locally defined chiral symmetry. In
this case, the topological phase is protected by a magnetic group
symmetry (defined as the product of time-reversal and mirror
symmetries) and a nonlocal chiral symmetry. The topology of the
cold atom system was measured via Bloch states at symmetric
momenta and the spin dynamics after a quench between trivial and
topological phases \cite{BSong2018}. This work may open the way to
explore the symmetry-protected topological states with ultracold
atoms, including the chiral AIII class by considering
spin-independent rather than spin-dependent OLs.
Further generalization of this study to higher dimensional systems
or interacting regimes also offers the simulation of quantum
phases beyond natural conditions in solid-state materials.

\subsubsection{Creutz ladder model}

\begin{figure}\centering
\includegraphics[width=0.95\linewidth]{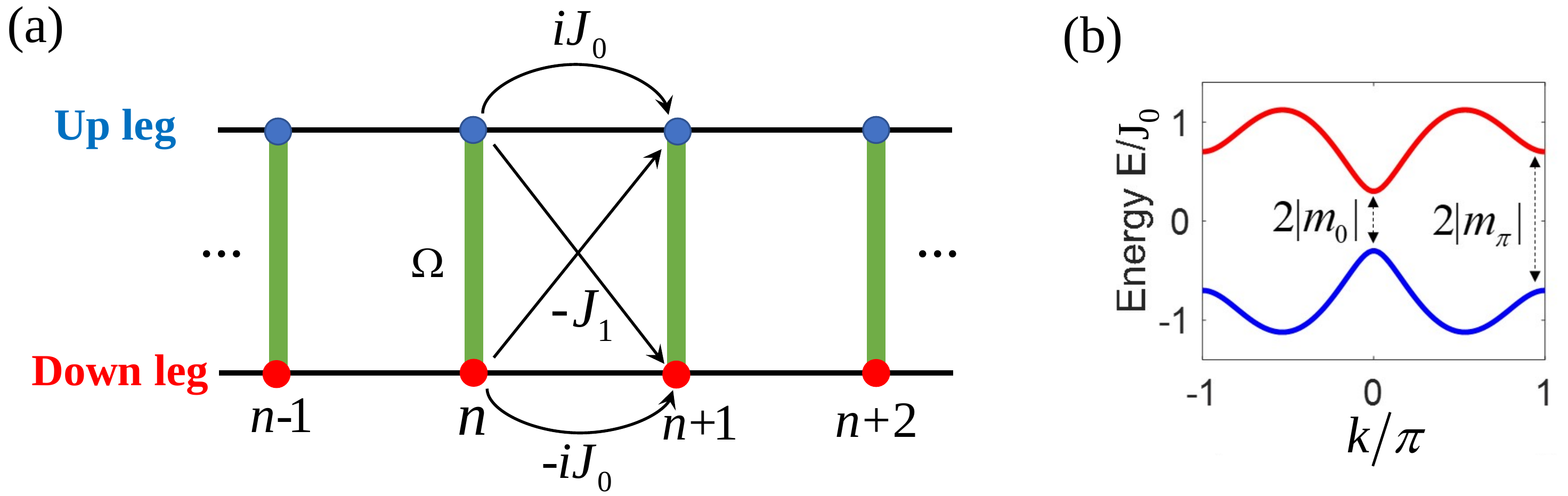}
\caption{(Color online) (a) The Creutz ladder with up and down legs. Each lattice site encompasses a vertical bond with two leg states, containing an on-site coupling $\Omega$ and a leg-conserving hopping $\pm iJ_0$ and a leg-flip hopping $-J_1$, respectively. (b) Energy dispersion of the Creutz ladder model with gapped Dirac cones at $k=0$ and $\pi$ for $\Omega\neq\pm J_1$. The 1D Dirac points exhibit when $\Omega=\pm J_1$.}
\label{CreutzModel}
\end{figure}

Initially introduced in the context of lattice gauge theory, the Creutz model \cite{Creutz1999} has gained a foothold in condensed matter physics as a versatile model to study fractionalization, Dirac fermions, and topological phases \cite{Sticlet2014}. The model describes free fermions hopping on a two-leg ladder pierced by a $\pi$ magnetic flux. The Creutz ladder shown in Fig. \ref{CreutzModel}(a) is described by the following Hamiltonian
\begin{equation}\label{HCreutz}
H_{\text{CL}}=\frac{1}{2}\sum_n\big[
\Omega c_n^\dag\sigma_x c_n+
c^\dag_{n+1}(iJ_0\sigma_z-J_1\sigma_x)c_{n}\big]+{\rm H.c.},
\end{equation}
where $n$ is the site index containing up and down legs as the spin basis for the Pauli matrices, with the particle annihilation operators $c_n=(c_{n u}, c_{n d})$. The fermions can jump from one site to the nearest-neighboring site within the same leg with a complex amplitude $\pm iJ_0$, i.e., gaining or losing a Peierls phase $\pi/2$. The fermions can hop between sites with amplitude $J_1$ while flipping between the legs, mimicking an SOC, and also horizontally along the legs with amplitude $\Omega$, mimicking a Zeeman field along $x$ axis. The corresponding Bloch Hamiltonian is given by $\mathcal{H}_{\text{CL}}(k)=J_0\sin k\sigma_z+(w-J_1\cos k)\sigma_x$. Consequently, there are two bands with the energy dispersion $E_\pm=\pm\sqrt{(J_0\sin k)^2+(\Omega-J_1\cos k)^2}$, as shown in Fig. \ref{CreutzModel}(b). For periodic boundary conditions, the bands display a pair of massive Dirac fermions with different Wilson masses $m_0=J_1-\Omega$ at $k=0$ and $m_{\pi}=J_1+\Omega$ at $k=\pi$. When $J_1=\Omega$ ($J_1=-\Omega$), the system is gapless with a Dirac cone at momentum $k=0$ ($k=\pi$). When both $J_1$ and $\Omega$ vanish, the system exhibits two Dirac cones.

The Creutz ladder model is classified in the BDI class of topological insulators since the Hamiltonian has the
particle-hole symmetry $\sigma_z
\mathcal{H}_{\text{CL}}^*(k)\sigma_z=-\mathcal{H}_{\text{CL}}(-k)$,
the TRS
$\sigma_x\mathcal{H}_{\text{CL}}^*(k)\sigma_x=
\mathcal{H}_{\text{CL}}(-k)$, and a chiral symmetry represented by
$\sigma_y$ with $\{\mathcal{H}_{\text{CL}}(k),\sigma_y\}=0$.
Therefore, the Creutz ladder is in the same symmetry class as
the SSH model, exhibiting nontrivial band topology and edge
states. In the SSH model, there are only two phases with different
dimerization configurations. In the Creutz ladder model of fully
gapped insulators, there are three phases distinguished by their
bulk band topology as characterized by the Zak phase for all
values of the parameters $(J_1,\Omega)$: a trivial insulator with
$\varphi_{Zak}=0$ when $|J_1/\Omega|<1$, two nontrivial insulators with
$\varphi_{Zak}=\pi$ when $|J_1/\Omega|<1$ and $J_1>0$, and with
$\varphi_{Zak}=-\pi$ when $|J_1/\Omega|<1$ and $J_1<0$. Equivalently, the
band topology can be characterized by the winding number
$\nu_{\text w}$ defined from the complex phase of the Bloch vectors: $\nu_{\text w}=\frac{1}{2}\big[\text{sgn}(\Omega+J_1)-\text{sgn}(\Omega-J_1)
\big]$. In the topologically nontrivial phase there are two zero-energy
bound states at the edges of the ladder with half fractional
particle numbers.

Several recent works proposed schemes to realize and study the topological properties of the Creutz ladder
model with cold atoms
\cite{Mazza2012,Huegel2014,Mazza2015,Mugel2016,Juenemann2017,Sun2017,Barbarino2018}.
The general scheme of using cold atoms with artificial SOCs in a ladder-like OL under a synthetic
magnetic field was suggested for simulation of the Creutz ladder
\cite{Mazza2012}. The experimental setups capable of implementing
the tunable Creutz ladder model were proposed in Refs.
\cite{Mazza2015,Juenemann2017,Sun2017}. Several protocols that can
be used to extract the topological properties in the Creutz model
from atomic density and momentum distribution measurements as well
as topological quantum pumping were presented
\cite{Huegel2014,Mazza2015,Sun2017}. By engineering a quantum walk
with cold atoms, one can observe the topological phases and the
bound states in the Creutz model \cite{Mugel2016}.
By adding an energy imbalance between the two legs of the ladder,
the symmetry class of the topological insulator changes from BDI
to AIII \cite{Juenemann2017}. Moreover, the interaction-induced
topological phase transition in the presence of interatomic
interactions in the optical Creutz ladder were investigated
\cite{Mazza2015,Juenemann2017}. A topological insulating phase protected by the inversion symmetry was found in a three-leg ladder model \cite{Barbarino2018}. Notably, the realization of an
optical two-leg ladder for ultracold bosonic atoms exposed to a uniform
artificial magnetic field created by laser-assisted tunnelling has
been experimentally achieved \cite{Atala2014}. In the experiment,
the atomic current on either leg of the ladder and the momentum
distribution were observed for demonstrating chiral Meissner-like
edge currents. Very recently, an experimental realization of a three-leg chiral ladder with ultracold fermionic atoms was reported \cite{JHKang2018}, where the legs were formed by the orbital states of a 1D optical lattice and the complex inter-leg links were generated by the orbital-changing Raman transitions.

\subsubsection{Aubry-Andr\'{e}-Harper model}\label{AAHModel}

Recently, the topological properties of 1D quasiperiodic lattices
have been theoretically and experimentally revealed in the context of
cold atoms and photonic quasicrystals \cite{Lang2012a,Kraus2012a}.
The system exhibits topological edge states and nontrivial Chern
numbers, equivalent to those of 2D quantum Hall systems on
periodic lattices. The tight-binding Hamiltonian of the 1D
quasiperiodic lattices takes the form:
\begin{equation}
H_{\text{AAH}}=-J\sum_n (c^\dagger_n c_{n+1}+
\mathrm{H.c.})+\sum_{n} V_n c^\dagger_n c_n \label{AAHam},
\end{equation}
where $V_n=V \cos(2\pi \alpha n+\phi)$ is the spatially modulated
on-site potential with $V$ being the strength, $\phi$ being the tunable modulation
phase, and $\alpha$ controlling the periodicity of the
modulation. When $\alpha$ is rational (irrational), the modulation
is commensurate (incommensurate). Alternatively, the system
Hamiltonian can be rewritten as
\begin{equation}
H_{\text{AAH}}\psi_n=-J(\psi_{n+1}+\psi_{n-1})+V \cos(2\pi \alpha
n+\phi)\psi_n,
\end{equation}
where $\psi_n$ is the wave function at site $n$. Historically,
this model is known as the Aubry-Andr\'{e}  model \cite{Aubry1980}
or Harper model \cite{Harper1955}.

\begin{figure}[tbp]\centering
\includegraphics[width=0.95\linewidth]{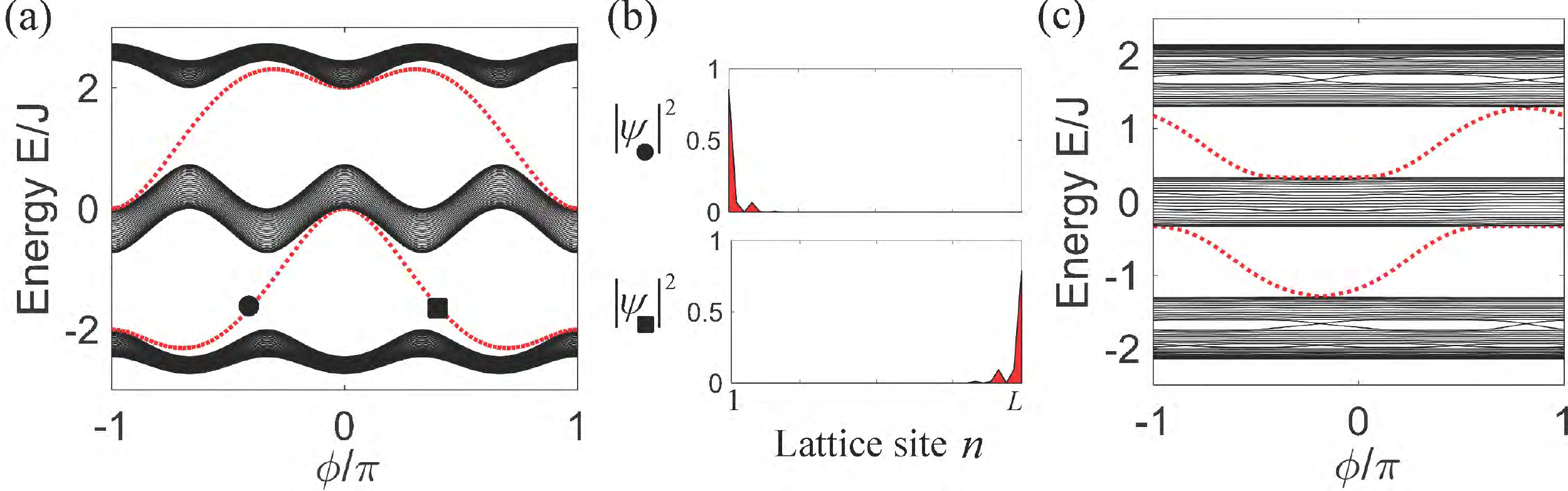}
\caption{(Color online) Energy spectrum of the Aubry-Andr\'{e}-Harper model as a
function of the phase $\phi$ for (a) $\alpha=1/3$, $V=J=1$; and (b) $\alpha=(\sqrt{5}+1)/2$, $V=0.5$, $J=1$
under the open boundary condition in a lattice of site $L=98$. The density distribution of two typical in-gap edge states (red dotted lines) is depicted.}
\label{AAH-Spectrum}
\end{figure}

Figure \ref{AAH-Spectrum} shows the energy spectrum of the
Aubry-Andr\'{e}-Harper model as a function of the modulation phase
$\phi$ for a finite lattice of the length $L$ under the open
boundary condition. In the commensurate potential ($\alpha=1/3$),
the spectrum changes periodically as $\phi$ varies from $0$ to
$2\pi$. The position of the edge states in the gaps also varies
continuously with the change of $\phi$, which is localized either
on the left or on the right boundary of the system. In the
incommensurate potential ($\alpha=(\sqrt{5}+1)/2$), the spectrum
is broken into a fractal set of bands; however, the in-gap edge
states also exhibit and sweep across the gaps.

Due to the bulk-edge correspondence, the in-gap edge states are
generally associated with the topologically nontrivial bulk bands.
It was revealed that the topological properties in the 1D
quasiperiodic lattices are assigned nontrivial Chern numbers in a
2D quantum Hall system \cite{Lang2012a,Kraus2012a}. Specifically,
the Aubry-Andr\'{e}-Harper model is connected to the Hofstadter
model (see Sec. \ref{HofModel}), which describes electrons hopping
on a 2D square lattice in a perpendicular magnetic field, with the
eigenvalue problem described by the Harper equation: $-J_x
(\psi_{n-1}+\psi_{n+1})-2J_y \cos( 2 \pi \alpha n- k_y ) \psi_n=
E(k_y) \psi_n$, where $J_x$ ($J_y$) is the hopping amplitude along
the $x$ ($y$) direction. By substitutions of $J \rightarrow J_x$,
$V \rightarrow -2J_y$, and $\phi \rightarrow -k_y$, the current 1D
problem can be mapped to the lattice version of the 2D integer
QHE problem. Adiabatically varying $\phi$ from $0$ to
$2\pi$ for each Bloch band forms an effective 2D manifold of
Hamiltonian $H(k,\phi)$ in the $k$-$\phi$ parameter space,
where the first Chern number can be defined.

For cold atoms, the Aubry-Andr\'{e}-Harper model
has been experimentally realized in 1D optical superlattices for studying
localization phenomena \cite{Roati2008,Schreiber2015}. In the experiments,
the quasiperiodic lattices were created using a primary 1D OL $V_1$ and an additional weak lattice $V_2$, with
the wave numbers $k_1$ and $k_2$, respectively. For deep
potentials, the atomic system is governed by the tight-binding
Hamiltonian $H=H_1+H_2$, with $H_1=-J\sum_{n} (c_{n+1}^{\dagger}
c_{n} +c_{n-1}^{\dagger} c_{n})$ from the primary lattice, and
$H_{\text{2}}= V_2 \sum_{n} c^{ \dagger}_{n} c_{n} \cos(2 \pi
\alpha n +\phi)$ from the interference of the perturbation
lattice, where $V_2 \sim J \ll V_1$, $\alpha=k_{2}/k_1$, and
$\phi$ is the tunable relative phase. It was suggested that the energy spectrum and the Chern numbers can be
revealed by observing the density profile of trapped fermionic
atoms \cite{Lang2012a}, which display plateaus with their positions uniquely
determined by varying the parameter $\alpha$ of the optical
superlattices. Another method to create and study quasiperiodic
OLs underlying all quasicrystals by the abstract
cut-and-project construction was recently proposed \cite{Singh2015}.

It was shown that the commensurate off-diagonal
Aubry-Andr\'{e}-Harper model is topologically nontrivial in the
gapless regime and supports zero-energy edge modes
\cite{Ganeshan2013}, which is attributed to the topological
properties of the 1D Majorana chain of $\Z_2$ class. By
generalizing the spinless Aubry-Andr\'{e}-Harper model to a
spinfull version, which can be realized with spin-1/2 atoms in a
spin-dependent quasiperiodic OL, one can realize
the $\Z_2$ topological insulators and the topological spin pumping \cite{Mei2012}. The phases of ultracold spin-1/2
bosons with SOCs in the quasiperiodic optical
lattice were also studied \cite{Ray2017}. In the presence of the pairings
or interactions, the topological superconducting phase with
Majorana end modes \cite{Lang2012b,Cai2013,DeGottardi2013}, the fractional
topological phases connecting to the 2D fractional QHE \cite{Xu2013}, and the topological Bose-Mott insulators
\cite{SLZhu2013} in the quasiperiodic lattices have been theoretically investigated.

\subsection{Two-dimension}

\subsubsection{Graphene-like physics and Dirac fermions}\label{Diracfermion}
\begin{figure}[htbp]\centering
\includegraphics[width=0.9\columnwidth]{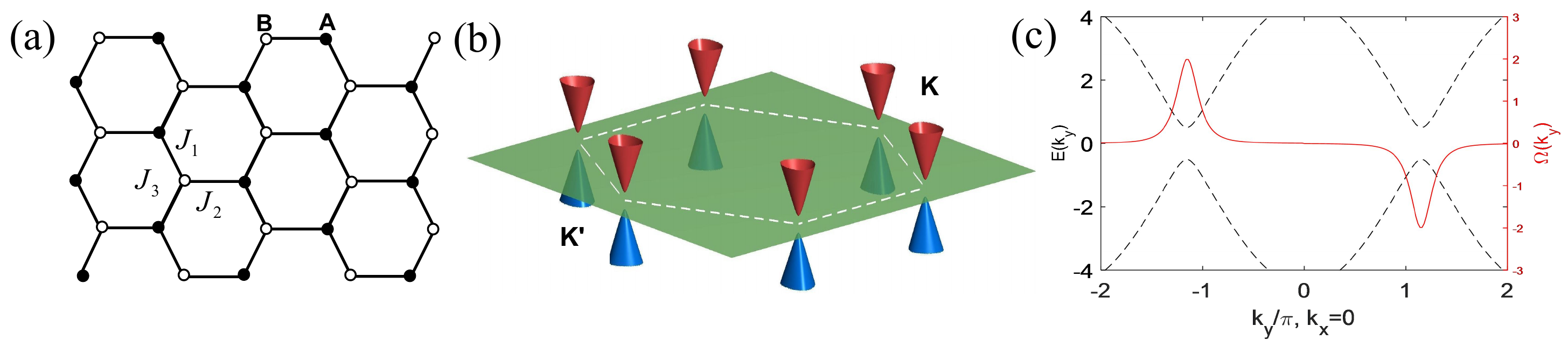}
 \caption{(Color online) (a) Crystal
structure of a honeycomb lattice consisting of sublattices A (solid)
and B (open). The nearest-neighbor hopping amplitudes denoted as
$J_1$, $J_2$, $J_3$ corresponding to the three different
directions. (b) Energy bands of the low-energy excitations with a
gap. The first BZ is outlined by the dashed line, and
two inequivalent valleys are labelled $\mathbf{K}$ and
$\mathbf{K}'$, respectively. Reprinted with permission from Xiao {\it et al.}\cite{DXiao2007}. Copyright\copyright ~(2007) by the American Physical Society. (c) The energy spectrum (dashed) and Berry curvature (solid) of the
conduction bands of a honeycomb lattice with broken inversion
symmetry.}
\label{Grapheneberryphase}
\end{figure}

The graphene material, formed with a single layer of carbon atoms
arranged in a honeycomb lattice
with its low-energy quasipaticles described by the massless Dirac
equation \cite{Novoselov2005}, has recently attracted strong interest in
condensed-matter physics. The crystal structure of graphene, as
shown in Fig. \ref{Grapheneberryphase}(a), consists of sublattices
A (solid) and B (open). Its energy spectrum is shown in Fig.
\ref{Grapheneberryphase} (b), where two inequivalent points
denoted as $\mathbf{K}$ and $\mathbf{K}^\prime$ are Dirac points.
At a Dirac point, two energy bands intersect linearly and the
quasiparticles in the vicinity of these points are described by
the Hamiltonian $H_D$ and are frequently called ``Dirac
fermions", where $H_D$ is the Dirac Hamiltonian in two spatial
dimensions given by
\begin{equation}\label{DiracHamiltonian}
H_D=v_x\sigma_xp_x+v_y\sigma_yp_y+\Delta_g\sigma_z.
\end{equation}
Here $\sigma_{x,y,z}$ are the three Pauli matrices. Compared with
the standard energy-momentum relation for the relativistic Dirac
particles, here $\Delta_g$ and $v_{x,y}$ denote rest energy and
the effective velocity of light respectively. For $\Delta_g\neq0$,
the energy spectrum with a gap  denotes the massive Dirac
fermions , as shown in Fig. \ref{Grapheneberryphase}(b). In
graphene, by contrast, the Dirac points appear at the corners
($\mathbf{K}$ and $\mathbf{K}'$ points) where the dispersion
relation of the honeycomb lattice shows the conical intersection
between the first and second bands. Here, the low-energy
fermionic excitations are the massless Dirac fermions described by
the Dirac Hamiltonian (\ref{DiracHamiltonian}) with $\Delta_g=0$,
$v_x=\pm v_y=\pm v$. Moreover, the associated Berry phases of the
Dirac points are $\pm\pi$ \cite{Novoselov2005,YZhang2005,Duca2015}
where the corresponding Berry curvatures have the form
\begin{equation}
\Omega_n=\pm\pi\delta\big(\mathbf{k}-\mathbf{K}_{n}\big).
\end{equation}
Here $\mathbf{K}_n$ denotes the Dirac points $\mathbf{K}$ or
$\mathbf{K}'$. One can see that the Berry curvatures tend to
infinity at Dirac points.  Here the Dirac points with the quantization of Berry phase (to $0$ and $\pi$) require symmetry protection, which can be the composition of inversion and time reversal symmetries. This is in contrast to Weyl points in 3D (see Sec. \ref{WeylSM}), which do not require any symmetry protection at all.

Since the relativistic Dirac fermions were found in graphene, a substantial
amount of effort has been devoted to the understanding of exotic
relativistic effects in solid-state systems and other artificial
quantum systems \cite{Hasan2010,Wehling2014,DWZhang2011}. Given these exciting results and the state-of-the-art
technologies in quantum control of atoms, one topic that naturally
arises is how to mimic the graphene and the relativistic
quasiparticles with cold atoms in a similar 2D hexagonal
lattice \cite{SLZhu2007,CWu2008b,Wunsch2008,KLLee2009,Poletti2011}. In cold atomic systems, it is easy
to realize the anisotropy Hamiltonian with mass term
$\Delta_g\neq 0$ and $v_x\neq v_y$ due to the highly controllable
experimental parameters \cite{SLZhu2007}. Furthermore, the
detection of the Berry curvature can spread over a finite range,
which provides a feasible way to measure the Berry phase over the
first BZ in reciprocal space \cite{Duca2015}. In addition to the honeycomb lattice, Dirac fermions can emerge in some lattices of other geometric
structures \cite{Bercioux2009,RShen2010,Satija2008,JMHou2009,Lim2008,Goldman2009,XJLiu2010,Kennett2011,Goldman2011}.

\begin{figure}[htbp]\centering
\includegraphics[width=0.75\columnwidth]{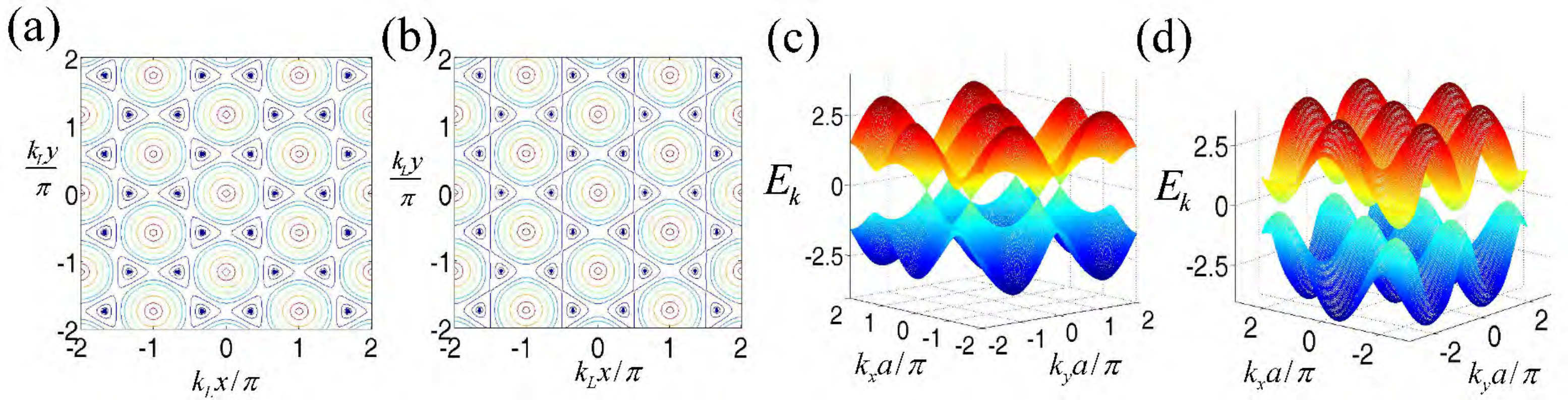}
 \caption{(Color online)
The honeycomb optical lattices. (a),(b) The contours with three
potentials described in Eq. (\ref{Graphenept}). The minima of the
potentials are denoted by the solid dots. All $V^0_j$ are the same
in (a), and $V^0_1=V^0_2=0.91V^0_3$ in (b). The dispersion
relations are shown in (c) for $\beta=1$ (gapless state) and (d)
for $\beta=2.5$ (gapped state). Reprinted with permission from Zhu {\it et al.}\cite{SLZhu2007}. Copyright\copyright ~(2007) by the American Physical Society.}\label{Graphenelike}
\end{figure}
Simulating Dirac equations with cold atoms loaded in a honeycomb
OL was initially proposed in Ref. \cite{SLZhu2007}. In
the proposal, single-component fermionic atoms (e.g.,
spin-polarized atoms $^{40}$K, $^{6}$Li, etc.) trapped in a
2D ($x$-$y$ plane) hexagonal OL are
considered. The hexagonal OL is constructed by three
standing-wave laser beams with the corresponding potential
\begin{equation}\label{Graphenept}
V(x,y)=\sum_{j=1,2,3}V_j^0\sin^2\Big[k_L(x
\cos\theta_j+y\sin\theta_j)+\frac{\pi}{2}\Big],
\end{equation}
where $\theta_1=\pi/3$, $\theta_2=2\pi/3$, $\theta_3=0$, and $k_L$
is the optical wave vector. It is easy to tune the potential
barriers $V^0_j$ by varying the laser intensities along different
directions to form a standard hexagonal lattice for
$V^0_1=V^0_2=V^0_3$, or a hexagonal lattice with a finite
anisotropy for different $V^0_j$ as depicted in Fig.
\ref{Graphenelike}(a) and \ref{Graphenelike}(b), respectively. The
tight-binding Hamiltonian of the system is then given by
\begin{equation}\label{Grephenetb}
H=-\sum_{\langle i,j\rangle}J_{ij}(a^\dagger_i b_j+{\rm H.c.}),
\end{equation}
where $\langle i,j \rangle$ represents the neighboring sites,
$a_i$ and $b_j$ denote the fermionic mode operators for the
sublattices A and B, respectively. The tunneling amplitudes
$J_{ij}$ depend on the tunneling directions in an anisotropic
hexagonal lattice, denoted as $J_1$, $J_2$, and $J_3$ corresponding to
the three different directions as illustrated in Fig.
\ref{Grapheneberryphase}(a). For simplicity, assume $J_1=J_2=J$
and $J_3=\beta J$ with $\beta$ being the anisotropy parameter. As
the atomic tunneling rate in an OL is exponentially
sensitive to the potential barrier, this control provides an
effective method to control the anisotropy of the atomic tunneling
by laser intensities. The first BZ of this system also has
a hexagonal shape in the momentum space with only two of the
six corners in Fig. \ref{Grapheneberryphase}(b) being inequivalent,
corresponding to two different sites $A$ and $B$ in each cell in
the real hexagonal lattice, usually denoted as $\bf K$ and ${\bf
K}'$. One can choose ${\mathbf K}=(2\pi/a)(1/\sqrt{3},1)$ and
${\mathbf K'}=-{\mathbf K}$, where $a=2\pi/(\sqrt{3}k_L)$ is the
lattice spacing. Taking a Fourier transform
$a^\dag_i=(1/\sqrt{N})\sum_{{\mathbf k}}{\rm exp}(i{\mathbf
k}\cdot{\mathbf A}_i)a_{\mathbf k}^\dag$ and
$b^\dag_j=(1/\sqrt{N})\sum_{{\mathbf k}}{\rm exp}(i{\mathbf
k}\cdot{\mathbf B}_j)b_{\mathbf k}^\dag$, where ${\mathbf A}_i$
(${\mathbf B}_j$) represents the position of the site in
sublattice $A$ ($B$) and $N$ is the number of sites of the
sublattice, the Hamiltonian (\ref{Grephenetb}) can be diagonalized
with the expression of energy spectrum\cite{SLZhu2007}
\begin{equation}\label{Graphenesprectrum}
E_{\mathbf k}=\pm J\sqrt{2+\beta^2+2\cos(k_y a)+4\beta
\cos(\sqrt{3}k_x a/2)\cos(k_ya/2)}.
\end{equation}
As plotted in Fig. \ref{Graphenelike}(c) and
\ref{Graphenelike}(d), there are two branches of the dispersion
relation, corresponding to the $\pm$ sign in Eq.
(\ref{Graphenesprectrum}). When $0<\beta<2$, the two branches
touch with each other, and a Dirac cone structure appears around the touching points. It has the same standard Dirac cones as the
graphene material with $\beta=1$
\cite{Novoselov2005,YZhang2005,Semenoff1984}. The Dirac cones squeeze in the $x$ or $y$ direction when $\beta$ deviates from $1$,
but they still touch each other. When $\beta>2$, a finite energy
gap $\Delta_g=|J|(\beta-2)$ appears between the two branches. So,
across the point $\beta=2$, the topology of the Fermi surface
changes, corresponding to a quantum phase transition without breaking any
local symmetry \cite{XGWen2004}. Such a topological phase
transition associated with the production or annihilation of a
pair of Dirac points has been investigated in
Ref.\cite{Bermudez2010a}. The evolution of the Dirac points in the
hexagonal lattice by varying the asymmetric hopping and the
resulting phase transition was also studied in
Ref.\cite{Wunsch2008}. With this phase transition, the system
changes its behavior from a semimetal to an insulator at the half
filling case (which means one atom per cell; note that each cell has two
sites). Around half filling, the Fermi surface is close to the
touching points, and one can expand the momentum ${\mathbf k}$
around one of the touching points ${\mathbf K}\equiv(k_x^0,k_y^0)$
as $(k_x,k_y)=(k_x^0+q_x,k_y^0+q_y)$. Up to the second order of
$q_x$ and $q_y$, the energy spectrum (\ref{Graphenesprectrum})
becomes
\begin{equation}
E_{\mathbf q}=\pm \sqrt{\Delta_g^2+v_x^2 q_x^2+v_y^2q_y^2},
\end{equation}
where $\Delta_g=0$, $v_x=\sqrt{3}\beta J a/2$, and $v_y=J a
\sqrt{1-\beta^2/4}$ for $0<\beta<2$; $\Delta_g=|J|(\beta-2)$,
$v_x= J a \sqrt{3\beta/2}$, and $v_y=J a \sqrt{\beta/2-1}$ for
$\beta>2$. This simplified energy sprectrum $E_{\mathbf q}$ is
actually a good approximation (named long wavelength
approximation) as long as $q_x$, $q_y\lesssim 1/2a$. The wave
function for the quasiparticles around half filling then
satisfies the Dirac equation $i\hbar\partial_t \Psi=H_D\Psi$,
where the relativistic Hamiltonian $H_D$ is the Dirac Hamiltonian
with the form
\begin{equation}\label{Diracpoint}
H_D=\tau_zv_x\sigma_x q_x+v_y\sigma_y q_y+\Delta_g\sigma_z,
\end{equation}
where $\tau_z=\pm1$ labels the two inequivalent valleys in Fig.
\ref{Graphenelike}(d). The corresponding Berry curvature is
concentrated in the valleys and has opposite signs in the two
inequivalent valleys. The symmetry property of the Berry curvature
$\Omega(k)$: it is an odd function in the presence of TRS and even in the presence of inversion symmetry,
as shown in Fig. \ref{Grapheneberryphase}(c). From Eq.
(\ref{Diracpoint}), we can obtain the Berry curvature near the
valleys for the conduction band\cite{DXiao2007}
\begin{equation}
\Omega(q)=\tau_z\frac{v_xv_y\Delta_g}{2(\Delta_g^2+v_xv_yq^2)^{3/2}}.
\end{equation}

Through an analogy to the graphene physics, one can realize both
massive and massless Dirac fermions and observe the phase
transition between them by controlling the lattice anisotropy \cite{SLZhu2007}. This proposal was demonstrated  to be
experimentally feasible in Ref. \cite{KLLee2009}, where the
temperature requirement and critical imperfections in the laser
configuration were considered in detail. Even in the presence of a
harmonic confining potential, the Dirac points are found to
survive \cite{Block2010}. In the presence of atomic interactions,
the many-body physics of Dirac particles in graphene-type
lattices, such as novel BCS-BEC crossover \cite{EZhao2006},
topological phase transition between gapless and gapped
superfluid \cite{Poletti2011} and even charge and bond ordered
states with the $p$-orbital band of lattices \cite{CWu2008b,CWu2007},
have been investigated. Notably, a honeycomb lattice has been realized and investigated using a BEC \cite{Soltan-Panahi2011,Soltan-Panahi2012}, but no signatures of Dirac points were observed. Even so,
these important theoretical works pave the way for mimicking
relativistic Dirac fermions and the aforementioned beyond-graphene
physics with controllable systems.

\begin{figure}[htbp]\centering
\includegraphics[width=0.9\columnwidth]{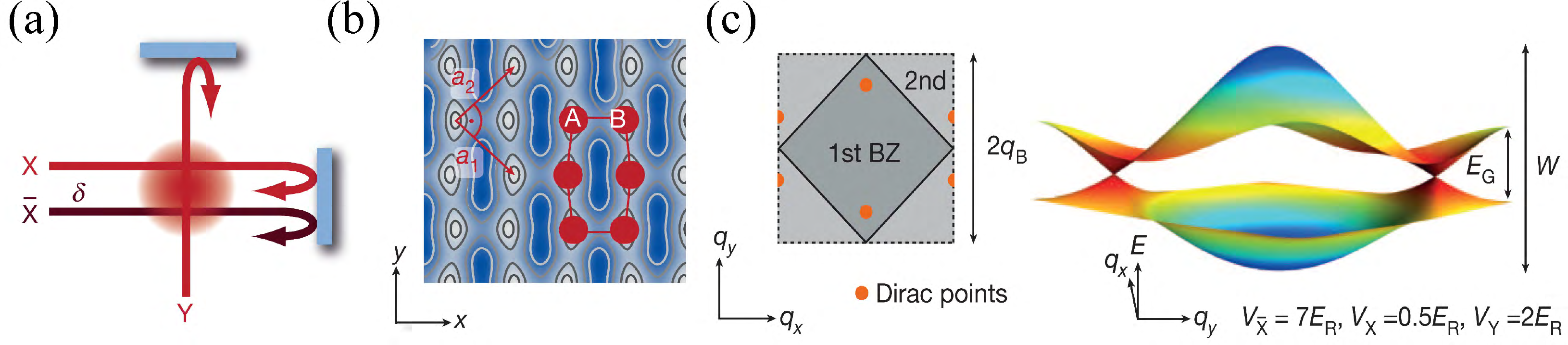}
 \caption{(Color online)
Optical lattice with adjustable geometry. (a) Three
retro-reflected laser beams create the
2D lattice potential of Eq.(\ref{Graphenepotential}). (b) The real-space potential of the
honeycomb lattice. (c) Left: sketch of the first and second
BZs of the honeycomb lattice, indicating the
positions of the Dirac points. Right: the energy spectrum of
honeycomb lattice in the first BZ showing the linear intersection of the
bands at the two Dirac points. The color scale illustrates lines
of constant energy. $W$ and $E_G$ denote the full bandwidth and
the minimum energy gap at the edges of the BZ,
respectively; $q_B=2\pi/\lambda$ is the Bloch wavevector. Reprinted by permission from Macmillan Publishers Ltd:  Tarruell {\it et al.}\cite{Tarruell2012}, copyright\copyright~(2012).
}\label{GrapheneDirac}
\end{figure}

Subsequently, an experiment to realize Dirac points with
adjustable properties using single-component ultracold fermionic atoms in a tunable honeycomb
OL was reported in Ref. \cite{Tarruell2012}. Furthermore, an artificial graphene consisting of a two-component ultracold atomic Fermi gas with tunable interactions was realized \cite{Uehlinger2013a}. To
create and manipulate Dirac points, the authors studied an
ultracold Fermi gas of $^{40}$K atoms in a 2D tunable
OL \cite{Tarruell2012}. In the experimental setup, three retro-reflected
laser beams of wavelength $\lambda=1,064$nm are arranged to form
the adjustable honeycomb OL, as shown in Fig.
\ref{GrapheneDirac}(a). The interference of two perpendicular
beams $X$ and $Y$ can form a chequerboard lattice of spacing
$\lambda/\sqrt{2}$. The third beam $\bar{X}$, collinear with $X$
but detuned by a frequency $\delta$, creates an additional
standing wave with a spacing of $\lambda/2$. The yielding
potential takes the form
\begin{equation}\label{Graphenepotential}
\begin{aligned}
V(x,y)&=-V_{\bar{X}}\cos^2(kx+\theta/2)-V_{X}\cos^2(kx)\\
&-V_{Y}\cos^2(ky)-2\alpha\sqrt{V_XV_Y}\cos(kx)\cos(ky)\cos(\varphi),
\end{aligned}
\end{equation}
where $V_{\bar{X}}$, $V_X$ and $V_Y$ denote the single-beam
lattice depths, $\alpha$ is the visibility of the interference
pattern and $k=2\pi/\lambda$. Varying the relative intensities of
the beams can realize various lattice structures, such as the
chequerboard, triangle, square and honeycomb lattices. We focus on
the honeycomb lattice with real-space potential as shown in
Fig. \ref{GrapheneDirac}(b). The primitive lattice vectors are
perpendicular, leading to a square BZ with two Dirac
points inside, as shown in Fig. \ref{GrapheneDirac}(c). This lattice is called
a brick-wall lattice, which is topologically equivalent to the
honeycomb lattice.

\begin{figure}[htbp]\centering
\includegraphics[width=0.9\columnwidth]{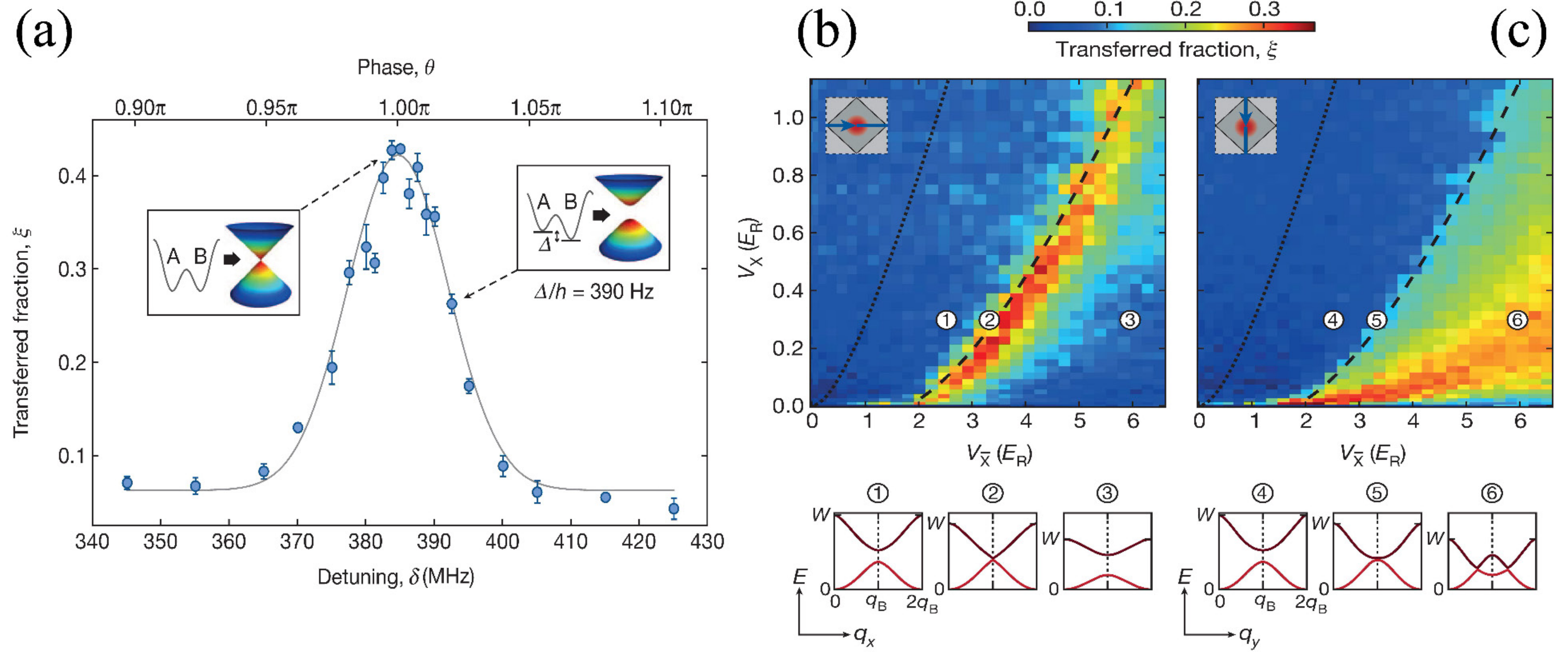}
\caption{(Color online) Energy offset and topological transition.
(a) The total fraction $\xi$ of atoms transferred to the second
band as a function of the detuning $\delta$, which controls the sublattice energy offset $\Delta$.
The maximum indicates the Dirac point. Insets: away from the peak, the atoms behave as Dirac fermions with a tunable mass. Solid line is a Gaussian fit to the data.The topological transition occurs along $q_x$ (b) and $q_y$ (c) directions in the quasi-momentum space.  The dashed line is the theoretical result for the transition line, and the dotted line indicates the transition from the triangular lattice to the dimer lattice. The bottom diagrams show cuts of the band structure along the $q_x$ axis ($q_y$;(b)) and $q_y$ axis ($q_x$; (c)) for the values of $V_X$ and $V_{\bar X}$ indicated. Reprinted by permission from Macmillan Publishers Ltd:  Tarruell {\it et al.}\cite{Tarruell2012}, copyright\copyright~(2012). }\label{Graphenephase}
\end{figure}

The Dirac points here were characterized by probing the energy
split between the two lowest-energy bands through inter-band
transitions. They are topological defects in the band structure
with the associated Berry phases $\pm\pi$, which guarantee their stability while a perturbation only moves the
positions of Dirac points. However, breaking the inversion symmetry
of the potential by introducing an energy offset $\Delta$ between
the sublattices opens an energy gap at the Dirac points, as shown
in the insets of Fig. \ref{Graphenephase}(a).
The band structure can be measured with the Bloch-Landau-Zener-oscillation technique \cite{Tarruell2012,Uehlinger2013,Lim2012}
(see Sec. \ref{Detectmethod}), and the results are plotted in
Fig. \ref{Graphenephase}(a), where the total fraction of atoms
transferred to the second band $\xi$ is plotted as a function of
the detuning $\delta$. The maximum indicates the point of
inversion symmetry, where $\Delta=0$ and the gap at the Dirac
point vanishes. Therefore, one can identify the points of maximum
transfer with the Dirac points. To investigate how breaking the
inversion symmetry of the lattice affects the Dirac points, the
authors varied the sublattice offset $\Delta$, which is controlled
by the frequency detuning $\delta$ between the lattice beams, and
measured the total fraction of atoms transferred $\xi$. The
population in the second band decreases symmetrically on both
sides of the peak as the gap increases, indicating the transition
from massless to massive Dirac fermions.

The position of the Dirac points inside the BZ and the slope of the associated linear dispersion relation are
determined by the relative strength of the tunnel couplings (i.e.,
$J_2/J_1,J_3/J_1$) \cite{SLZhu2007,Wunsch2008,Montambaux2009},
which can be adjusted simply by controlling the intensity of the
laser beams. Therefore, it was observed that the positions of the
Dirac points continuously approach the corners of the BZ when the tunnelling in the $x$ direction gradually increases
by decreasing the intensity of $\bar{\text{X}}$. When they reach
the corners of the BZ, the two Dirac points merge,
annihilating each other. Beyond this critical point, a finite
band gap appears for all quasimomenta of the BZ. This
situation signals the transition between band structures of two
different topologies, one containing two Dirac points and the
other containing none. This corresponds to a Lifshitz phase transition from a semimetallic phase to a band-insulating phase in 2D honeycomb lattices at half-filling  \cite{SLZhu2007,Wunsch2008}. The topological transition line was
experimentally mapped out by recording the fraction of atoms
transferred to the second band, $\xi$, as a function of the
lattice depths $V_{\bar{\text{X}}}$ and $V_X$, while keeping the
value of $V_Y/E_R$.  The results are shown in Fig.
\ref{Graphenephase}(b). There the onset of population transfer to
the second band signals the appearance of Dirac points in the band
structure of the lattice. For a given value of $V_X$, the
transferred fraction, $\xi$, decreases again for large values of
$V_{\bar{\text{X}}}$, as the Dirac points lie beyond the momentum
width of the cloud. The $\xi$ as a function of $q_y$ for a
1D lattice structure ($V_{\bar{\text{X}}}
\gg V_\text{X}$) was obtained in Fig. \ref{Graphenephase}(c), where the
transition line is clearly demonstrated. This work opens the way to realize and investigate other
topological models with cold atoms in OLs, such as the Haldane model \cite{Haldane1988} and Kane-Mele
model \cite{Kane2005a,Kane2005b} to be addressed in Sec. \ref{HaldaneModel} and
\ref{KMModel}, respectively.

\begin{figure}[htbp]\centering
\includegraphics[width=0.9\columnwidth]{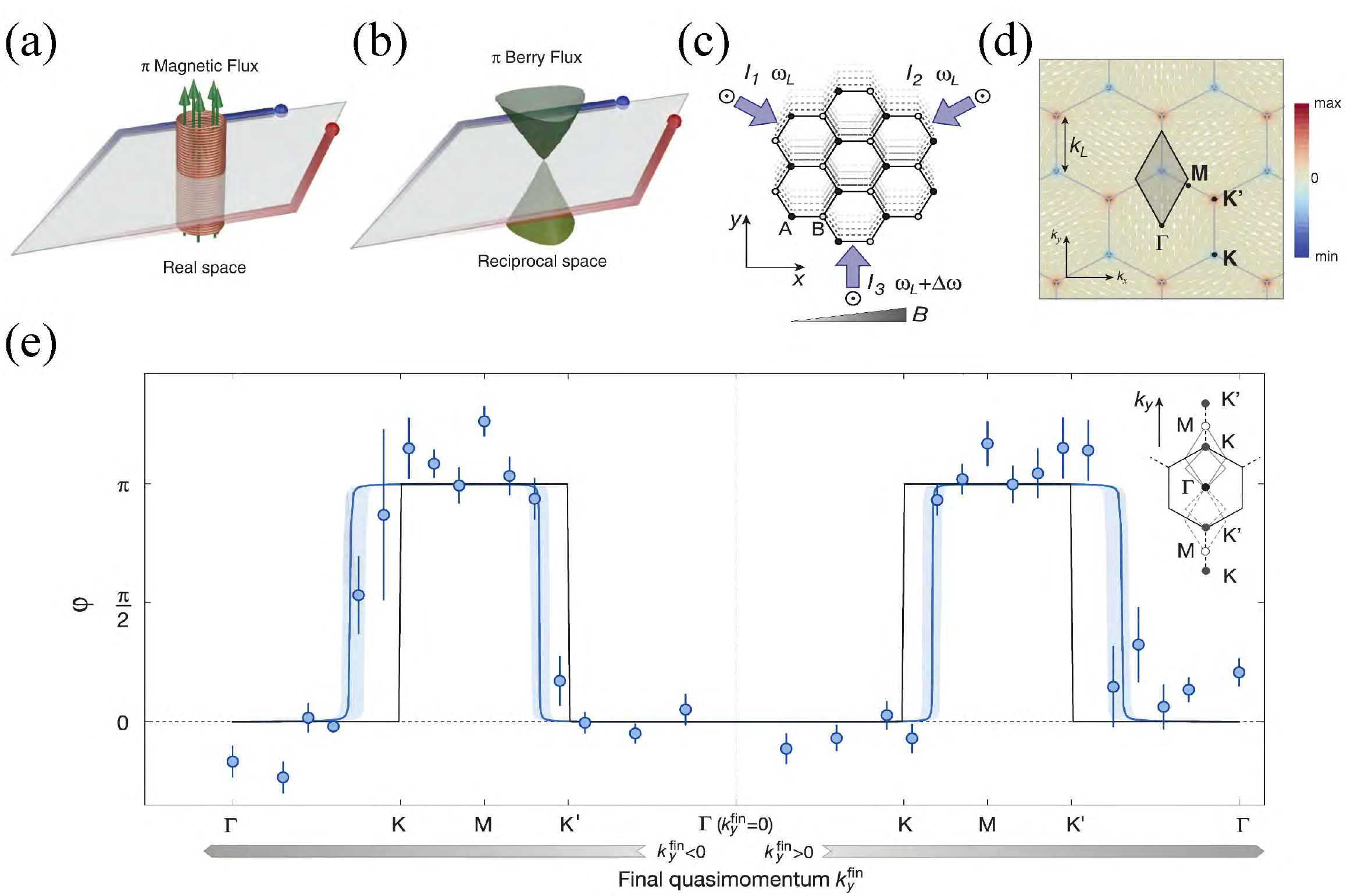}
 \caption{(Color online) Aharonov-Bohm analogy and geometric properties of the hexagonal lattice. (a) In the Aharonov-Bohm
effect, electrons encircle a magnetic flux in real space. (b) In
the interferometer, the particles encircle the $\pi$ Berry flux of
a Dirac point in reciprocal space. (c) Sketch of the hexagonal
lattice in real space, which is realized by interfering three laser beams
(arrows) of intensity $I_i$ and frequency
$w_L$, with linear out-of-plane polarizations. A linear frequency
sweep of the third lattice beam creates a uniform lattice
acceleration along the y direction. A magnetic field gradient along the $x$ axis creates an additional
spin-dependent force. (d) Dirac points are located at the corners
($\mathbf{K}$ and $\mathbf{K}'$ points) of the BZ (gray hexagons).
Black diamond is a typical interferometer path.
(e) Summary of phase shifts measured relative to the zero-area reference interferometer for different final
quasimomenta $k_y^{fin}$ (Inset shows the interferometer paths).  Lines follow ab initio theory using a full
band structure calculation with no momentum spread $\sigma_k=0$
and perfectly localized Berry curvature $\delta k_{\Omega} = 0$
(black) or $\sigma_k = 0.21k_L$ and $\delta k_{\Omega} \simeq
10^{-4}k_L$ (blue).
Reprinted from Duca {\it et al.}\cite{Duca2015}. Reprinted with permission from AAAS. }\label{GrapheneBerryphase}
\end{figure}

After the realization of the Dirac points in the honeycomb OL
using ultracold atoms, a further experiment to detect the $\pi$
Berry flux located at each Dirac point by realizing an atomic
interferometer was
reported \cite{Duca2015}. The idea of detecting the Berry flux is
analogous to using an Aharonov-Bohm interferometer to measure a
magnetic flux in real space. As we know, the Aharonov-Bohm effect
describes a charged particle wave packet being split
into two parts that encircle a given area in real space [Fig.
\ref{GrapheneBerryphase}(a)]. Any magnetic flux through the
enclosed area gives rise to a measurable phase difference between
the two components. In analogy to the magnetic field, the Berry
curvature $\Omega_n$ for a single Bloch band in the reciprocal
space can be probed by forming an interferometer on a closed path
in reciprocal space [Fig. \ref{GrapheneBerryphase}(b)]. The
geometric phase acquired along the path can be calculated from the
Berry connection $\mathbf{A}_n(\mathbf{k})$, which is given by
$\mathbf{A}_n(\mathbf{k})=\langle
u^n_{\mathbf{k}}|i\nabla_{\mathbf{k}}|u^n_{\mathbf{k}}\rangle$.
Here, $|u^n_{\mathbf{k}}(r)\rangle$ is the cell-periodic part
of the Bloch wave function
$|\psi^n_{\mathbf{k}}(\mathbf{r})\rangle=e^{i\mathbf{k}\cdot\mathbf{r}}|u^n_{\mathbf{k}}(r)\rangle$
with quasimomentum $\mathbf{k}$ in the $n$th band. Accordingly,
the phase along a closed loop in reciprocal space is
\begin{equation}
\varphi_
\text{Berry}=\oint_{\mathcal{L}}\mathbf{A}_n(\mathbf{k})d\mathbf{k}=\int_{S^2}\Omega_n(\mathbf{k})d^2\mathbf{k}.
\end{equation}
where $S$ is the area enclosed by the path $\mathcal{L}=\partial_S$, and
$\Omega_n(\mathbf{k})=\nabla\times\mathbf{A}_n(\mathbf{k})$.

In the experimental setup, the graphene-like hexagonal OL for ultracold $^{87}$Rb atoms is implemented by
superimposing three linearly polarized blue-detuned running waves
at 120(1)$^\circ$ angles (Fig. \ref{GrapheneBerryphase}(c)).
Fig. \ref{GrapheneBerryphase}(d) shows that the resulting
dispersion relation includes two inequivalent Dirac points with
opposite Berry fluxes $\pm\pi$ located at $\mathbf{K}$ and
$\mathbf{K}'$, respectively. The interferometer sequence begins
with an almost pure $^{87}$Rb BEC in the state
$|\uparrow\rangle=|F=2,m_F=1\rangle$ at quasimomentum $k=0$ in a
$V_0=1$ $E_r$ deep lattice, where $E_r=h^2/(2m\lambda^2_L)\approx
h\times4$ kHz is the recoil energy and $h$ is Planck's constant.
In fact, the method of detecting the Berry phase here is a
2D extension of the Zak phase's measurement addressed
in Sec. \ref{SSHModel}. The first step is to create a coherent
superposition of $|\uparrow\rangle$ and $|\downarrow\rangle=|F=1,
m_F=1\rangle$ states by using a resonant $\pi/2$-microwave pulse.
Then a magnetic field gradient along $x$ axis is applied to create
a constant force in opposite directions for the two spin
components. Meanwhile, an orthogonal, spin-independent force from
lattice acceleration is created by a linear frequency sweep of the
third lattice beam and can move atoms along the $y$
direction [Fig. \ref{GrapheneBerryphase}(c,d)]. As a consequence, the
two spin components move symmetrically along the interferometer
path in reciprocal space. After an evolution time $\tau$, a spin-echo $\pi$-microwave pulse is applied to swap the states
$|\uparrow\rangle$ and $\downarrow\rangle$. Subsequently, the two
atomic wave packets experience opposite magnetic forces in the $x$
direction, such that both spin components arrive at the same
quasimomentum $k^{fin}$ after an additional evolution time $\tau$.
At this point, the coherent superposition state is given by
$|\psi^{fin}\rangle\propto|\uparrow,k^{fin}\rangle+e^{i\varphi}|\downarrow,k^{fin}\rangle$
with relative phase $\varphi$. Finally, a second $\pi/2$-pulse
with a variable phase $\varphi_{MW}$ is applied in order to close
the interferometer and convert the phase information into spin
population fractions $n_{\uparrow,\downarrow}\propto 1\pm
\cos(\varphi+\varphi_{MW})$. Notably, the phase difference
$\varphi=\varphi_B+\varphi_d$ consists of the geometric phase
$\varphi_B$ and any difference in dynamical phases $\varphi_d$
between the two paths of the interferometer and is
equal to the Berry phase of the region enclosed by the
interferometer. This is because the dynamical contribution should
vanish due to the symmetry of the paths and the use of the
spin-echo sequence. To ascertain that the measured phase is truly
of geometric origin, the authors additionally employed a
¡°zero-area¡± reference interferometer, which comprises a V-shaped path
produced by reversing the lattice acceleration after the
$\pi$-microwave pulse.

The experimental results for detecting the Berry phase (flux) of
the varying region enclosed by the interferometer are shown in
Fig. \ref{GrapheneBerryphase}(e). The results show that
a phase difference $\varphi\simeq\pi$ when a Dirac point is
enclosed in the measurement loop, which agrees well with the
theoretical prediction of the Berry phase for a single Dirac
point. In contrast, the phase difference vanishes when enclosing
zero or two Dirac points.
The shift in the phase jump results from the momentum spread
$\sigma_k$, the broadening of the edges is caused by $\delta
k_{\Omega}$, and the shaded area accounts for a variation in
$\sigma_k=0.14-0.28 k_L$. The contrast is limited by inhomogeneous
broadening of the microwave transition, the finite momentum spread
of the condensate, and, for large final quasimomenta, the
dynamical instability of the Gross-Pitaevskii equation.

\subsubsection{Hofstadter model}\label{HofModel}

The celebrated Hofstadter model (also named as the
Hofstadter-Harper model) describes charged particles moving in a
2D periodic lattice under a uniform magnetic flux per
unit cell \cite{Hofstadter1976}. In the tight-binding regime, the
single-particle energy spectrum depends sensitively on the number
of flux quanta per unit cell and a band splits into narrow
magnetic bands. At high magnetic fields, the self-similar energy
spectrum was predicted to emerge, known as the Hofstadter
butterfly. Moreover, for filled bands of non-interacting fermions
when the Fermi energy lies in one of the band gaps, the Hall
conductance of the system is quantized \cite{Thouless1982}. In
this case, the Hofstadter model realizes the paradigmatic example
of a topological insulator that breaks TRS and
can be characterized by the first Chern numbers.

\begin{figure}[tbp]\centering
\includegraphics[width=0.45\columnwidth]{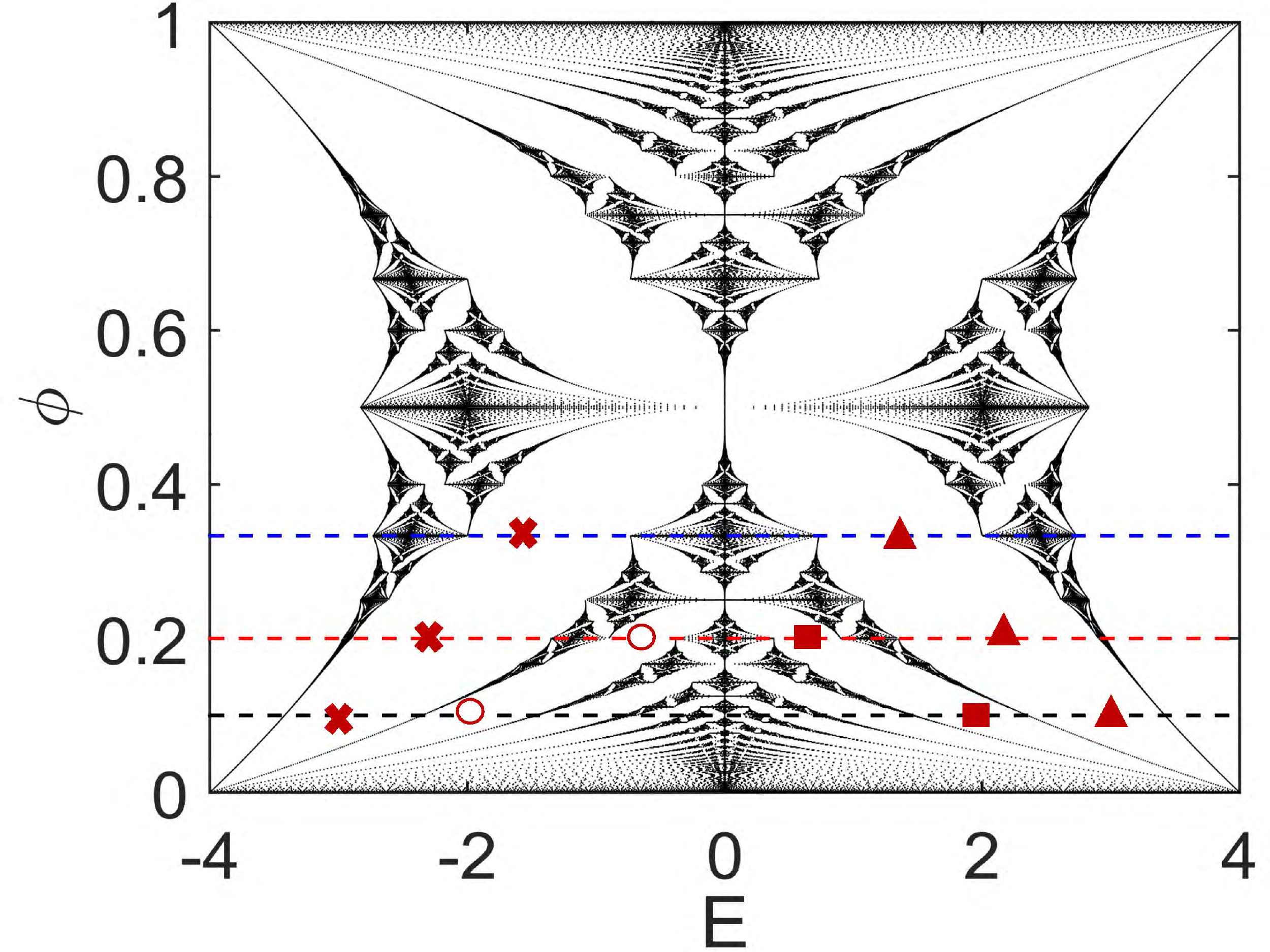}
\caption{(Color online) Hofstadter-butterfly energy spectrum. Dashed lines
represent the Fermi energy for different values of $\phi$, namely
$\phi = 1/3$, $1/5$, and $1/10$. Regions marked by $\times$ and
$\blacktriangle$ have the band Chern numbers $C=\pm1$; those marked by $\circ$ and
{\tiny $\blacksquare$} have $C=\pm2$. }
\label{Hofs-Butterlfy}
\end{figure}

Consider the non-interacting spinless particles moving in a 2D
square lattice in the presence of an artificial magnetic field,
which are described by the Hofstadter Hamiltonian
\cite{Hofstadter1976}
\begin{equation} \label{HofstadterHam}
H_{H}=-J\sum_{m,n} \big( a^{\dagger}_{m+1,n} a_{m,n} +
e^{i\varphi_{m,n}} a^{\dagger}_{m,n+1} a_{m,n}
+\mathrm{h.c.} \big),
\end{equation}
where  $ a^{\dagger}_{m,n}$ ($a_{m,n}$) is the creation (annihilation) operator of a particle at
lattice site $(m,n)$, and $\varphi_{m,n}$ denotes the
spatially-varying hopping phase induced by a magnetic flux
$2\pi\phi$. Taking the Landau gauge, the Hofstadter Hamiltonian
can be rewritten as
\begin{equation}
H_{H}=-J\sum_{m,n}\left(a^{\dagger}_{m+1,n} a_{m,n}
                     +\mathrm{e}^{ i 2\pi m\phi} a^{\dagger}_{m,n+1}
                     a_{m,n}+\mathrm{h.c.}\right).
\end{equation}
With $y=na$ as the periodic coordinates on the system, this
Hamiltonian can be diagonalized as a block Hamiltonian
$H_{H}=\bigoplus {H}_x(k_y)$, where $k_y$ is the quasimomenta
along the periodic directions. The decoupled block Hamiltonian
takes the form
\begin{equation}
{H}_x(k_y)=-J\sum_m ( a^{\dagger}_{m+1} a_{m} +
\text{h.c.}) - \sum_m V_m a^{\dagger}_{m} a_{m},
\end{equation}
where $V_m=2J\cos(2\pi\phi m+k_ya)$. The single-particle wave
function is written as $\Psi_{mn}=e^{ik_yy}\psi_m$, and then the
Schr\"{o}dinger equation $H_x(k_y)\Psi_{mn}=E\Psi_{mn}$
reduces to the Harper equation \cite{Harper1955}
\begin{equation}\label{Eigenfun1}
-J(\psi_{m-1}+\psi_{m+1})-V_m\psi_m = E \psi_m.
\end{equation}
For rational fluxes $\phi=p/q$ with $p$ and $q$ being relatively
prime integers, and under the periodic boundary condition along
$x$ axis, the wave function $\psi_m$ satisfies
$\psi_m=e^{ik_xx}u_m(\mathbf{k})$ with
$u_m(\mathbf{k})=u_{m+q}(\mathbf{k})$. In this case, the spectrum
of this system consists of $q$ energy bands and each band has a
reduced (magnetic) BZ: $-\pi/qa\leq k_x\leq\pi/qa$,
$-\pi/a\leq k_y\leq\pi/a$. In term of the reduced Bloch wave
function $u_m(\mathbf{k})$, Eq. (\ref{Eigenfun1}) becomes
\begin{equation} \label{Eigenfun2}
-J(e^{ik_x}u_{m-1}+e^{-ik_x}u_{m+1})-V_m u_m = E(\mathbf{k}) u_m.
\end{equation}
Since $u_m(\mathbf{k})=u_{m+q}(\mathbf{k})$, the problem of
solving the equation (\ref{Eigenfun2}) reduces to solving the
eigenvalue equation, $M\Upsilon=E\Upsilon$, where
$\Upsilon=(u_1,...,u_{q})$ is the Bloch wave function for the $q$
bands and $M$ is the $q\times q$ matrix. The Hofstadter energy
spectrum is displayed in Fig. \ref{Hofs-Butterlfy}, where the band
gaps form continuous regions in the $\phi-E$ plane. When the Fermi
energy lies in a gap, the system is an insulator, and the
topological nature and the Hall conductance of the insulator do
not change as long as the Fermi level remains within the same gap
\cite{Thouless1982}. When the Fermi energy is in the gap between two
bands $N$ and $N+1$, the quantized Hall conductance is
$\sigma_{xy}=Ce^2/h$ with the topological Chern number
\begin{equation}
C=\frac{1}{2\pi}\sum_{n\leqslant
N}\int_{-\pi/qa}^{\pi/qa}dk_x\int_{-\pi/a}^{\pi/a}dk_{y}
F_{xy}^{(n)}(\textbf{k}),
\end{equation}
where $F_{xy}^{(n)}$ is the Berry curvature of the $n$-th subband. As
marked in Fig. \ref{Hofs-Butterlfy}, the largest two gaps
correspond to topological insulators with the Chern number
$C=\pm1$, and the second largest ones have $C=\pm2$.

Despite its mathematical elegance, the Hofstadter butterfly and
the topological Hofstadter insulator can hardly be realized in
traditional solid-state systems because the magnetic field needs
to be thousands of Tesla for electrons in order to create a
magnetic flux comparable to one flux quantum per unit
cell. However, some quantum Hall features associated with the
fractal Hofstadter spectrum for low-energy Dirac fermions were
recently observed in graphene superlattices and van der Waals
heterostructures. \cite{Dean2013,Hunt2013}.

Recent theoretical and experimental advances in the creation of synthetic
gauge fields for neutral atoms provide an excellent platform to
simulate the physics of charged particles in magnetic fields. The
concept of coupling two or several internal states to realize
artificial magnetic fields was suggested in 2D
OLs \cite{Jaksch2003,Mueller2004,Osterloh2005}, which
can be used to realize the Hofstadter model. In the proposals, the
crucial element is the laser-assisted hopping between neighboring
sites by Raman transitions. Because of the spatial variation of
the Raman coupling, the wave function of an atom tunneling from
one lattice site to another acquires an effective
spatially-varying Berry phase. This method can also create
non-Abelian U(2) gauge potentials acting on cold atoms in the
OLs, leading to generalized Hofstadter butterfly
spectra with new fractal structures and topological properties
\cite{Osterloh2005,Goldman2009b}.

The laser-assisted
hopping scheme was further proposed by using a long-lived
metastable excited state for alkaline-earth or ytterbium atoms in
an optical superlattice to produce uniform magnetic fields for
realizing the Hofstadter Hamiltonian \cite{Gerbier2010}. It was
also suggested to realize the uniform synthetic magnetic fields
for neutral atoms by periodically shaking square OLs,
and thus provided the Floquet realization of the Hofstadter
Hamiltonian
\cite{Bilitewski2015,Bukov2014,Goldman2014b,Zhou2014b,Creffield2016}.
The atomic gas of noninteracting spinless fermions in a rotating
OL was considered to study the Hofstadter butterfly
and to measure the quantized Hall conductance of the Hofstadter
insulator from density profiles using the St\v{r}eda formula
\cite{Umucalilar2008}. The evolution of the Hofstadter butterfly
in a tunable OL among the square, checkerboard, and
honeycomb structures was studied \cite{Yilmaz2015,Yilmaz2017}. A
method for detecting topological Chern numbers in the Hofstadter
bands by simply counting the number of local maxima in the
momentum distribution from time-of-flight images of ultracold
atoms was presented \cite{Zhao2011}. The detection of the fractal
energy spectrum of the Hofstadter model from the density
distributions of ultracold fermions in an external trap and under
finite temperatures was analyzed \cite{Wang2014b}. The chiral edge
states in the Hofstadter insulator may be created by using a steep
confining potential in the OLs, and then they can be
detected from the atomic Bragg spectroscopy and from their
dynamics after the potential is suddenly removed
\cite{Goldman2012,Goldman2013}.

\begin{figure}
\centering
\includegraphics[width=0.9\columnwidth]{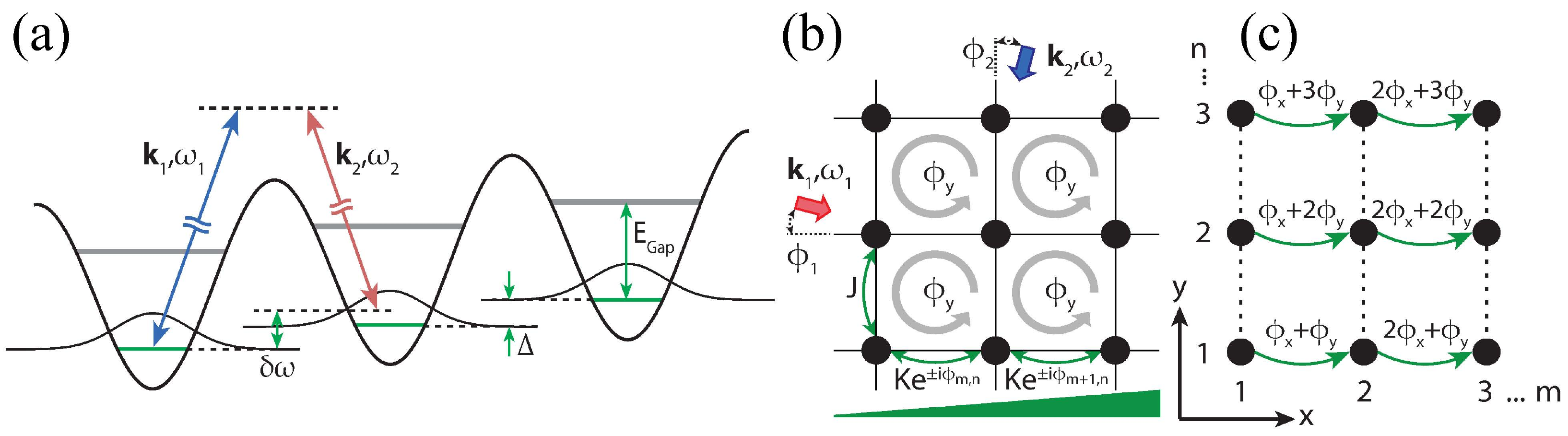}
\caption{(Color online) (a) Raman-assisted tunneling in the lowest band of a
tilted lattice with an energy offset $\Delta$ between neighboring
sites.
(b) Experimental geometry to generate
uniform magnetic fields using a pair of far-detuned laser beams
and a uniform potential energy gradient.
(c) A schematic depicting
the position-dependent phases of the tunneling process with
resulting magnetic flux quanta per unit cell $\alpha=\phi_y/2\pi$. Reprinted with permission from Miyake {\it et al.}\cite{Miyake2013}. Copyright\copyright~(2013) by the American Physical Society.
} \label{Hofs-OL}
\end{figure}

Experimentally, the laser-assisted technique was used to
generate large staggered magnetic fields for ultracold bosonic
atoms \cite{Aidelsburger2011}, where the two internal states in
the proposals \cite{Jaksch2003,Gerbier2010} were replaced by
doubling the unit cell of the OL using superlattices.
Two experiments were subsequently implemented for realizing the
Hosftadter Hamiltonian based on the generation of homogeneous and
tunable artificial magnetic fields with ultracold atoms in tilted
OLs \cite{Miyake2013,Aidelsburger2013}. In the
experiments, the bosonic atoms were loaded in a square OL with a tilt potential along the $y$ direction, as shown in
Fig. \ref{Hofs-OL}. The atomic tunneling in this direction was then
suppressed by the linear tilt of energy per lattice site
$\Delta\gg J$, which
can be created with magnetic field gradients, gravity, or an ac
Stark shift gradient. The tunneling is resonantly restored by the
laser-assisted hopping method with two far-detuned Raman beams of
two-photon Rabi frequency $\Omega$, frequency detuning
$\delta\omega = \omega_1 - \omega_2=\Delta/\hbar$, and momentum
transfer $\delta \mathbf{k} = \mathbf{k_1} -
\mathbf{k_2}\equiv(\delta k_x,\delta k_y)$, as shown in Fig.
\ref{Hofs-OL}(a). Here the two Raman beams couple different sites,
but do not change the internal state of the atoms, similar
to the scheme proposed in Ref. \cite{Kolovsky2011}. In the dressed
atom picture (for resonant tunneling along the $x$ axis) and
high-frequency limit ($\delta\omega\gg J/\hbar$), the tilt
disappears and time averaging over rapidly oscillating Raman beams
yields an effective time-independent Hamiltonian for the lattice
system, which takes the single-band form ($\Delta<E_{\text{Gap}}$)
of the Hofstadter Hamiltonian:
\begin{equation} \label{Hofs-ExpHam}
\tilde{H}_{H}=-\sum_{m,n} \big( Ke^{-i\phi_{m,n}}
a^{\dagger}_{m+1,n} a_{m,n} + J a^{\dagger}_{m,n+1}
a_{m,n} +h.c. \big).
\end{equation}
The induced hopping strength $K$ along the $x$ axis and the
spatially-varying phase $\phi_{m,n}  = \delta
\mathbf{k}\cdot\mathbf{R}_{m,n}= m\phi_x + n\phi_y$ correspond
to the vector potential $\mathbf{A} =  \hbar(\delta k_x x +\delta
k_y y)/a\: \hat{\mathbf{x}}$ instead of the simple Landau gauge.
Adding up the accumulated phases around a closed path leads to an
enclosed phase $\phi_{y}=\delta k_y a$ per lattice unit cell of
area $a^2$, thus realizing the Hofstadter Hamiltonian with the
magnetic flux $\alpha=\phi_{y}/2 \pi$. When the frequencies of the
Raman beams are similar to those used for the OL, one
can tune the magnetic flux over the full range between zero and
one by adjusting the angle between the Raman beams.

In the experiments, the laser-assisted tunneling processes was
characterized by studying the expansion of the atoms in the
lattice \cite{Miyake2013}, and the local distribution of fluxes
were determined through the observation of cyclotron orbits of the
atoms on lattice plaquettes \cite{Aidelsburger2013}. Since the
laser-assisted hopping used does not require near-resonant light
for connecting hyperfine states, this method can be implemented
for any atoms, including fermionic atoms. Moreover, for two atomic
spin states $|\uparrow,\downarrow\rangle$ with opposite magnetic
moments and the titled potential created by a magnetic field
gradient, two different spin components experience opposite
directions of the magnetic field
\cite{Aidelsburger2013,Kennedy2013}, and the system naturally
realizes the time reversal symmetric spinfull Hofstadter
Hamiltonian:
\begin{equation}
H_{\uparrow,\downarrow}=-\sum_{m,n} \big( Ke^{\pm i\phi_{m,n}}
a^{\dagger}_{m+1,n} a_{m,n} + J a^{\dagger}_{m,n+1}
a_{m,n} +h.c. \big),
\end{equation}
which gives rise to the quantum spin Hall effect, topologically
characterized by a $\Z_2$ spin Chern number. In a recent experiment
\cite{Kennedy2015}, the weakly interacting ground state of the
Hofstadter Hamiltonian (\ref{Hofs-ExpHam}) was studied, which for
bosonic atoms is a superfluid BEC.

\begin{figure}
\centering
\includegraphics[width=1.0\linewidth]{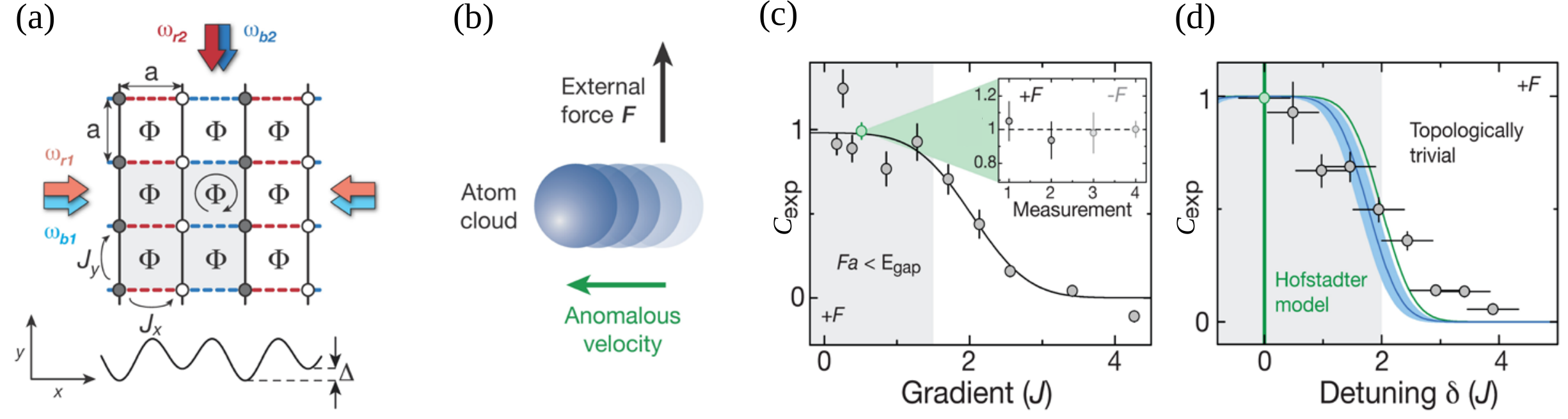}
\caption{(Color online) Measurements of the Chern number of Hofstadter bands with
ultracold bosonic atoms. (a) The setup consists of a
2D OL with a staggered potential.
The magnetic unit cell (gray shaded area) is four times larger than
the usual lattice unit cell. (b) The Chern number is extracted
from the transverse displacement of the atomic cloud, in response
to an external force generated by an optical gradient.
%
Measured Chern number $C_{\mathrm{exp}}$ as a function of (c) gradient strength $Fa$; and (d) staggered detuning $\delta$.
Reprinted by permission from Macmillan Publishers Ltd: Aidelsburger {\it et al.}\cite{Aidelsburger2014}, copyright \copyright~(2014).}
\label{Hofs-ChN}
\end{figure}

The Chern numbers of the Hofstadter bands have been measured with
ultracold bosonic atoms from the transverse deflection of an
atomic cloud as a Hall response \cite{Aidelsburger2014}. The
experimental setup consisted of an ultracold gas of $^{87}$Rb
atoms loaded into a two-dimensional lattice created by two
orthogonal standing waves with wavelength
$\lambda_s=767\,\mathrm{nm}$. An additional standing wave with
twice the wavelength $\lambda_L=2\lambda_s$ was superimposed along
$x$ to create the staggered potential as shown in Fig.
\ref{Hofs-ChN}(a), with an energy offset $\Delta$ much larger than
the bare tunneling $J_x$. The modulation restoring resonant
tunneling was created by two additional pairs of far-detuned laser
beams with wave number $k_L=2\pi/\lambda_L$ and frequency $\omega
= \Delta / \hbar$. This system realized an effective
time-independent Hofstadter Hamiltonian with the magnetic flux
$\alpha=\pi/2$. In contrast to previous experiments generating
uniform flux in tilted OLs
\cite{Miyake2013,Aidelsburger2013}, this scheme does not rely on
magnetic field gradients, providing a higher degree of
experimental control. The incoherent distribution of bosonic atoms
(where the population within each band is homogeneous in momentum
space) was then loaded into the lowest Hofstadter band via an
experimental sequence using an auxiliary superlattice potential,
which introduces a staggered detuning $\delta$ along both
directions \cite{Aidelsburger2014}. For $\delta > 2J$ the topology
of the bands is trivial and all Chern numbers are zero. When
crossing the topological phase transition at $\delta=2J$ (the
spectral gaps close), the system enters the topologically
non-trivial regime, where the lowest band has a Chern number
$C=1$. The Chern numbers were finally extracted from the
transverse Hall drift by exploiting Bloch oscillations. Under a
constant force $\mathbf{F}=F \hat{\mathbf{e}}_y$, atoms on a
lattice undergo Bloch oscillations along the $y$ direction. The cloud
also experiences a net perpendicular (Hall) drift shown in Fig.
\ref{Hofs-ChN}(b) when the energy bands have a nonzero Berry
curvature, which leads to an anomalous velocity that can be
isolated by uniformly populating the bands. In the absence of
inter-band transitions, the contribution of the $n$-th band to the
center-of-mass motion along the $x$ direction can be written in
terms of its band Chern number $C_{n}$ \cite{Aidelsburger2014}:
\begin{equation}
x_{n}(t)=-\frac{4a^2 F}{h}\,C_{n} \, t = - 4a \,C_{n} \,
\frac{t}{\tau_B},
\end{equation}
where the factor $4a^2$ corresponds to the extended unit cell and
$\tau_B = h/(Fa)$ is the characteristic time scale for Bloch
oscillations. The center-of-mass evolution of the atomic cloud was
measured in-situ with opposite directions of the flux $\alpha$ for
subtracting the differential shift $x(t,\alpha)-x(t,-\alpha)=2
x(t)$, as shown in Fig. \ref{Hofs-ChN}(c), where the deflection is
symmetric with respect to the direction of the applied force and
gives an experimental Chern number $C_{\mathrm{exp}}\approx1$.
The measured drifts for $\alpha=0$ and for a staggered-flux
distribution do not show any significant displacement,
corresponding to zero Chern number. The dependence of the
Chern-number measurement with respect to the force was studied, as
shown in Fig. \ref{Hofs-ChN}(d).

The chiral edge states of the Hofstadter lattice were
experimentally observed in a ribbon geometry with an ultracold gas
of neutral fermions \cite{Mancini2015} and bosons \cite{Stuhl2015}
subjected to an artificial gauge field and a synthetic dimension (see Sec. \ref{sythdimention} for synthetic dimensions). Very
recently, the following interacting Hofstadter model of bosons in
the two-body limit was realized in OLs
\cite{Tai2017}:
\begin{equation}
H=-\sum_{i,j} \big( Ke^{-i\phi_{i,j}} a^{\dagger}_{i+1,j} a_{i,j} + J a^{\dagger}_{i,j+1} a_{i,j} +h.c. \big)
+\frac{U}{2}\sum_{i,j} n_{i,j}(n_{i,j}-1),
\end{equation}
where $U$ is an on-site repulsive interaction energy. Through microscopic
atomic control and detection \cite{Tai2017}, it was shown that the
inter-particle interactions affect the populating of chiral bands,
giving rise to chiral dynamics whose multi-particle correlations
indicate both bound and free-particle characteristics. The novel form of
interaction-induced chirality observed in these experiments
provides the key piece for future investigations
of highly entangled topological phases of many-body systems. The superfluid pairing
and vortex lattices for interacting fermions in OLs under a uniform magnetic field was studied \cite{Zhai2010}. The
Hofstadter-Hubbard model on a cylinder geometry with fermionic
cold atoms in OLs was shown to allow one to probe the Hall response as a
realization of Laughlin's charge pump \cite{Lacki2016}.


\subsubsection{Haldane model}\label{HaldaneModel}

\begin{figure}[htbp]\centering
\includegraphics[width=0.95\columnwidth]{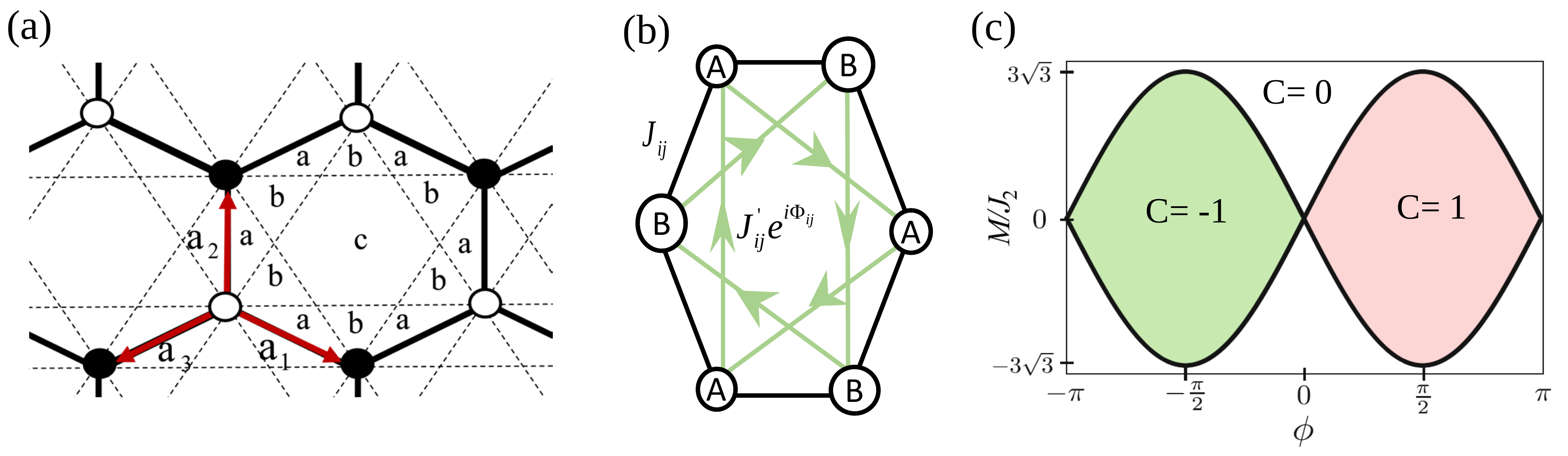}
 \caption{(Color online)  (a) The Haldane honeycomb model showing nearest-neighbor bonds (solid lines) and next-nearest-neighbor bonds (dashed lines).
 The white and black dots represents the two sublattice sites $A$ and $B$ with different on-site energy $M$ and $-M$.
 The areas $a$ and $b$ are threaded by the magnetic flux $\phi_a$ and $\phi_b=-\phi_a$, respectively.
 The area $c$ has no flux. (b) A distorted honeycomb lattice realized in the experiment \cite{Jotzu2014}.
(c) Phases of the Haldane model.} \label{Haldane}
\end{figure}

The well-known QHE in 2D electron systems is usually associated with
the presence of a uniform externally generated magnetic field,
which splits the electron energy spectrum into discrete Landau
levels. In order to realize the integer QHE  seen
in the Landau-level problem while keeping the translational
symmetry of the lattice, in 1988, Haldane proposed a spinless
fermion model for the integer QHE without Landau
levels\cite{Haldane1988}. He proposed that the QHE
may result from the broken TRS without any net
magnetic flux through the unit cell of a 2D hexagonal lattice, as
illustrated in Fig. \ref{Haldane}(a). The Haldane model based on
breaking both time reversal and inversion symmetries is the
first example of a topological Chern insulator, and the
Hamiltonian is as follows
\begin{equation}
\begin{aligned}
H=J_1\sum_{\langle{i,j}\rangle}c_i^\dag{c_j}+J_2\sum_{\langle\langle{i,j}\rangle\rangle}e^{-iv_{ij}\phi}{c_i^\dag{c_j}}+\sum_{i}{\epsilon_ic_i^\dag{c_i}}.
\end{aligned}
\end{equation}
Here the on-site energy $\epsilon_i$ is $\pm{M}$, depending on
whether $i$ is on the $A$ or $B$ sublattice, $J_1$ is the
nearest-neighbor-hopping energy, $J_2$ is the
next-nearest-neighbor energy, and
\begin{equation}\label{phasesign}
v_{ij}=\text{sgn}(\hat{\mathbf{d}}_i\times\hat{\mathbf{d}}_j)_z=\pm1,
\end{equation}
where $\hat{\mathbf{d}}_{i,j}$ are the unit vectors along the two
bonds constituting the next-nearest neighbors the particle
traverses going from site $j$ to $i$. As depicted in Fig.
\ref{Haldane}(a), a periodic magnetic flux density $\mathbf{B}(r)$
is added normal to the plane with the full symmetry of the lattice
and with zero net flux through the unit cell. Thus, the flux
$\phi_a$ and the flux $\phi_b$ in the regions $a$ and $b$
respectively has the relation $\phi_b=-\phi_a$: since the net
flux is zero and the next-neighbor hoppings form closed loops in the
hexagonal cell, the hopping terms $J_1$ are not affected but the
hopping terms $J_2$ acquire a phase
$\phi=2\pi(2\phi_a+\phi_b)/\phi_0$ where $\phi_0$ is the flux
quanta.

Under the periodic condition, we can diagonalize the Haldane
Hamiltonian by using the basis of a two-component spinor
$c^\dagger_k=(c^\dagger_{k,A},c^\dagger_{k,B})$ of Bloch states
constructed on the two sublattices. Let $\mathbf{a}_1$,
$\mathbf{a}_2$, $\mathbf{a}_3$ be the displacements from a $B$
site to its three nearest-neighbor $A$ sites, as shown in Fig.
\ref{Haldane}(a), then
$\mathbf{a}_1=(\frac{\sqrt{3}}{2}a,-\frac{1}{2}a),$
$\mathbf{a}_2=(0,a),$ $
\mathbf{a}_3=(-\frac{\sqrt{3}}{2}a,-\frac{1}{2}a),$
where $a$ is the bond length. Taking a Fourier transform
$c^\dagger_i=(1/\sqrt{N})\sum_k{e^{i\mathbf{k}\cdot\mathbf{r}_i}}c^\dagger_k$,
where $\mathbf{r}_i$ represents the position of the site in
sublattice $A$ or $B$ and $N$ is the number of sites of the
sublattice, the Haldane Hamiltonian can be expressed as
\begin{equation}\label{HaldaneHam}
\mathcal{H}(\mathbf{k})=\epsilon(\mathbf{k})+\mathbf{d}(\mathbf{k})\cdot\sigma,
\end{equation}
where
\begin{equation}
\begin{aligned}
\epsilon(\mathbf{k})&=2J_2\cos{\phi}\sum^3_{i=1}{\cos(\mathbf{k}\cdot\mathbf{b}_i)},
~~d_1(\mathbf{k})=J_1 \sum^3_{i=1}\cos(\mathbf{k}\cdot\mathbf{a}_i),\\
d_2(\mathbf{k})&=J_1 \sum^3_{i=1}\sin(\mathbf{k}\cdot\mathbf{a}_i),
~d_3(\mathbf{k})=M-2J_2\sin\phi\sum^3_{i=1}\sin(\mathbf{k}\cdot\mathbf{b}_i),
\end{aligned}
\end{equation}
with $\mathbf{b}_1=\mathbf{a}_1-\mathbf{a}_2$,
$\mathbf{b}_2=\mathbf{a}_3-\mathbf{a}_2$, and
$\mathbf{b}_3=\mathbf{a}_1-\mathbf{a}_3$. The BZ is a
hexagon rotated $\frac{\pi}{2}$ with respect to the Wigner-Seitz
unit cell: At its six corners
$(\mathbf{k}\cdot\mathbf{a}_1,\mathbf{k}\cdot\mathbf{a}_2,
\mathbf{k}\cdot\mathbf{a}_3)$ is a permutation of
$(0,\frac{2\pi}{3}, -\frac{2\pi}{3})$. The two distinct corners
$\mathbf{k}_\alpha^0$, are defined so that
$\mathbf{k}_\alpha^0\cdot\mathbf{b}_i =\alpha\frac{2\pi}{3}$,
$\alpha=\pm 1$. The energy spectrum of this system can be easily
obtained by diagonalizing the Hamiltonian (\ref{HaldaneHam}).
There are two bands that touch only if all three Pauli matrix
terms have vanishing coefficients, and only occur at the zone
corner $\mathbf{k}_\alpha^0$ while
$M=\alpha3\sqrt{3}J_2\sin{\phi}$.

To guarantee that the two bands never overlap and are separated
by a finite gap unless they touch, in the following, we consider
the case for $|J_2/J_1|<1/3$. One can choose the corner point
$\mathbf{K}=\frac{2\pi}{3a}(1/\sqrt{3},1)$, then
$(\mathbf{K}\cdot\mathbf{a}_1,\mathbf{K}\cdot\mathbf{a}_2,
\mathbf{K}\cdot\mathbf{a}_3)=(0,\frac{2\pi}{3}, -\frac{2\pi}{3})$.
We expand the Haldane Hamiltonian around the point $\mathbf{K}$ to
linear order in $\mathbf{q}=\mathbf{k}-\mathbf{K}$:
\begin{equation}
\mathcal{H}_+=v(q_x\sigma_x-q_y\sigma_y)+m_+\sigma_z
\end{equation}
where $v=\frac{3}{2}J_1{a}$ and $m_+=M-3\sqrt{3}J_2\sin{\phi}$.
Hereafter, we ignore the $\mathbf{k}$-independent term $-3J_2\cos{\phi}$, which plays no role in topology.
At the other point $\mathbf{K}'=-\frac{2\pi}{3a}(1/\sqrt{3},1)$,
$(\mathbf{K}'\cdot\mathbf{a}_1,\mathbf{K}'\cdot\mathbf{a}_2,
\mathbf{K}'\cdot\mathbf{a}_3)=(0,-\frac{2\pi}{3},
\frac{2\pi}{3})$, around $\mathbf{K}'$ we have
\begin{equation}
\mathcal{H}_-=v(-q_x\sigma_x-q_y\sigma_y)+m_-\sigma_z
\end{equation}
where $v=\frac{3}{2}J_1{a}$ and $m_-=M+3\sqrt{3}J_2\sin{\phi}$.
The Chern number of the whole system is determined by
\begin{equation}
C=\frac{1}{2}[\text{sgn}(m_-)-\text{sgn}(m_+)].
\end{equation}
The phase diagram of the Haldane model as a function of
$M/J_2$ and $\phi$ is shown in Fig. \ref{Haldane}(c). For
$\phi=0,\pi$, the model (\ref{HaldaneHam}) is under time reversal, and
the two mass $m_+$ and $m_-$ are equal, the system is trivial with
$C=0$. Moreover, the system has the inversion symmetry when
$M=0$. If  $M$ and (or) $J_2\sin{\phi}$ vanish, the two bands
touch with gapless Dirac fermions.
The model can have the nontrivial phases with $C=\pm1$ only if
$|M|<3\sqrt{3}J_2\sin{\phi}$ and $\phi\neq0,\pi$. Note that along
the critical lines in the phase diagram where either $m_+$ or
$m_-$ vanishes, the system experiences a topological
phase transition, and has a low-lying massless spectrum around
$\mathbf{K}$ or $\mathbf{K}'$ simulating nondegenerate
relativistic chiral Dirac fermions.

\begin{figure}[htbp]\centering
\includegraphics[width=0.9\columnwidth]{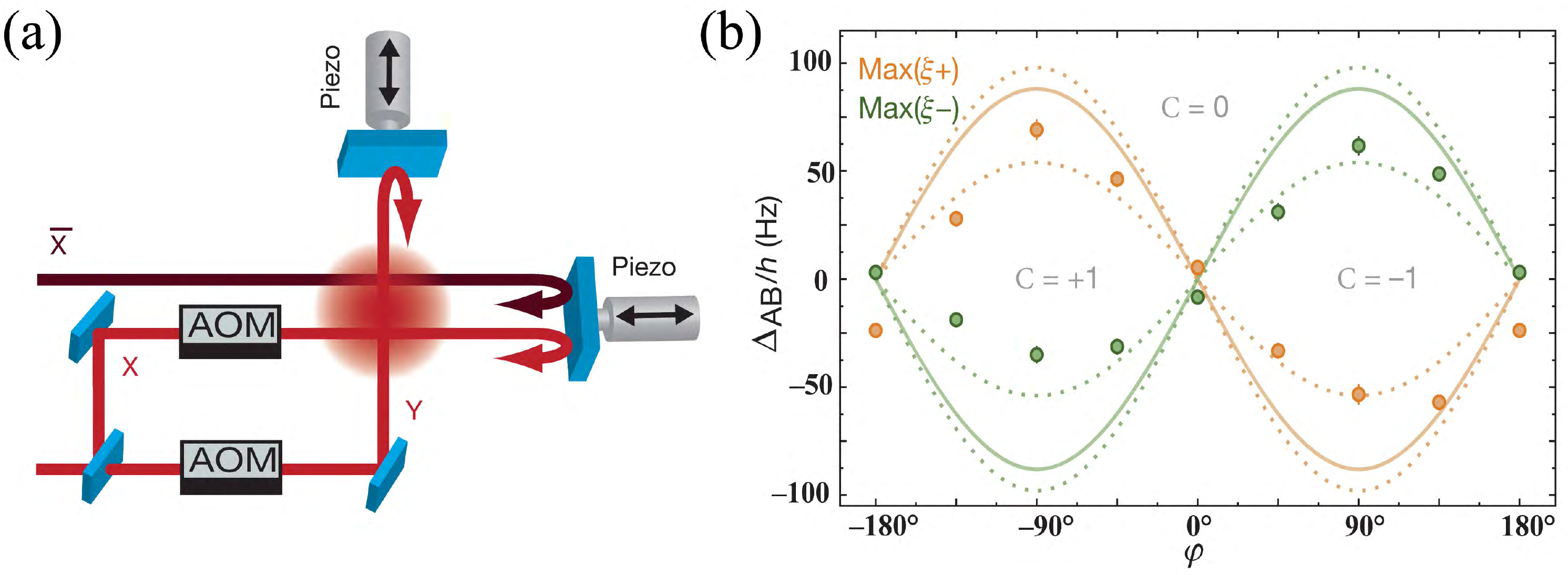}
 \caption{(Color online)
(a) Laser beam set-up for forming the OL. The laser
$\bar{X}$ is frequency-detuned from the other beams.
Piezo-electric actuators sinusoidally modulate the
retro-reflecting mirrors, with a controllable phase difference
$\varphi$ that induces the  complex tunneling phase $\Phi$.
Acousto-optic modulators ensure the stability of the
lattice geometry.
(b) Topological phase diagram measured in the experiments. Solid
lines show the theoretically computed topological transitions. 
Dotted lines represent the uncertainity
of the maximum gap originating from the uncertainty of the lattice
parameters. Data are the points of maximum transfer for each Dirac
point. Reprinted by permission from Macmillan Publishers Ltd: Jotzu {\it et al.}\cite{Jotzu2014}, copyright\copyright~(2014). } \label{Haldaneep}
\end{figure}

Although the Haldane model has been proposed for nearly $30$
years, it has not been realized in any condensed
matter systems since it is extremely hard to realize the required
staggered magnetic flux assumed in the model. The technology of ultracold atoms in an OL provides an
approach to realize and explore this model originally proposed in condensed matter physics. The idea of realizing and detecting the QHE of the Haldane model in an OL was first proposed in Ref.
\cite{LBShao2008}. In the proposal, three standing-wave
laser beams are used to construct a honeycomb OL where
different on site energies in two sublattices required in the
model can be implemented through tuning the phase of one laser
beam. The other three standing-wave laser beams are used to create
the staggered magnetic field. Firstly, to generate the honeycomb
lattice with different on site energies in sublattices $A$ and
$B$, the three laser beams with the same wave length but different
polarizations are applied along three different directions:
$\mathbf{e}_y$ and
$\frac{\sqrt{3}}{2}\mathbf{e}_x\pm\frac{1}{2}\mathbf{e}_y$,
respectively. Thus, the corresponding potential is given by
\begin{equation}
V=V_0[\sin^2(\alpha_++\frac{\pi}{2})+\sin^2(yk^L_0+\frac{\pi}{3})\sin^2(\alpha_--\frac{\chi}{2})],
\end{equation}
where $\alpha_\pm=\sqrt{3}xk^L_0/2\pm{yk^L_0/2}$, $V_0$ is the
potential amplitude and $k^L_0$ is the wave vector of the laser.
The ingenuity of the design is that the different site
energies of sublattices $A$ and $B$ are controllable by the phase
of the laser beam $\chi$. An interesting method to realize the
staggered magnetic field in the Haldane model is to use two
opposite-travelling standing-wave laser beams to induce Berry
phase \cite{SLZhu2006}, which can create effective staggered
magnetic fields with zero net flux per unit cell.  For the two
laser beams with Rabi frequencies
$\Omega_1=\Omega_0\sin(yk^L_2+\frac{\pi}{4})e^{ixk^L_1}$ and
$\Omega_2=\Omega_0\sin(yk^L_2+\frac{\pi}{4})e^{-ixk^L_1}$, the
effective gauge potential is generated as
$\mathbf{A}_1(\mathbf{r})=\hbar{k^L_1}\sin(2yk^L_2)\mathbf{e}_x.$
Here, $k^L_1=k^L\cos\theta$ and $k^L_2=k^L\sin\theta$ with $k^L$ being
the wave vector of the laser and $\theta$ being the angle between the
wave vector and the $\mathbf{e}_x$ axis. The choice of wave vector
$k^L_2$ of the laser beams must be a multiple of
$\frac{2\sqrt{3}\pi}{3a}$ in order to be commensurate with the
OL, such as $k^L_2=\frac{2\sqrt{3}\pi}{3a}$.
Since the lattice has the symmetry of point group $C_{3v}$, the
other two vector potentials can be rotated by $\pm\frac{2\pi}{3}$
from the vector potential $\mathbf{A}_1$. Finally, the total
accumulated phases along the nearest-neighbor directions are found
to cancel each other out because of the symmetry of the honeycomb lattice.
However, the total accumulated phases for the
next-nearest-neighbor hopping are preserved as the hopping
phase $\varphi=k^L_1a\sin{\frac{ak^L_2}{\sqrt{3}}}$. 
Consequently, the total Hamiltonian of this cold atomic system is
described by the Haldane model. However, in the proposal, the
lasers for the honeycomb lattice and artificial magnetic fields
are different and thus the required lasers are extremely
complicated and hard to realize in practice.

Another scheme to realize the Haldane-like model was proposed
in Ref. \cite{CWu2008a}, with an orbital analogue of the
anomalous QHE arising from orbital angular momentum polarization without Landau levels.
This effect arises from the energy-level splitting between the on
site $p_x-ip_y$ and $p_x+ip_y$ orbitals by rotating each OL site around its own center. At large rotation angular
velocities, this model naturally reduces to two copies of
Haldane's quantum Hall model. An improved experimental proposal
to realize the generalized Haldane model using OLs
loaded with fermionic atoms in two internal states was proposed in
Ref. \cite{Alba2011}. In this simulation, the original phase
factors in the next-nearest-neighbor hopping in Haldane's paper
are replaced by that in the nearest-neighbor, whose phase depends
on the momentum imparted by the Raman lasers. An experimental scheme to realize the quantum anomalous Hall effect in an anisotropic square OL was proposed \cite{XJLiu2010}.

Instead of using the extra laser
beams, one can create effective magnetic fields in the honeycomb OL by shaking the lattice \cite{WZheng2014,Jotzu2014}. 
In $2014$, the first experimental realization of the Haldane model and the characterization of its topological band
structure were reported \cite{Jotzu2014}, which used ultracold fermionic atoms in a periodically
modulated honeycomb OL. In the experiment, the spin-polarized non-interacting ultracold
Fermi gas of $^{40}$K atoms was prepared in the OL created by several laser beams at wavelength
$\lambda=1064$ nm. The lattice potential is given by
\begin{equation}
\begin{aligned}
V(x,y,z)=-V_{\bar{X}}\cos^2(k_{\text{lat}}x+\theta/2)-V_X\cos^2(k_{\text{lat}}x)-V_Y\cos^2(k_{\text{lat}}y)
\\-2\alpha\sqrt{V_XV_Y}\cos(k_{\text{lat}}x)\cos(k_{\text{lat}}y)\cos(\varphi_{\text{lat}})-V_{\tilde{Z}}\cos^2(k_{\text{lat}}z),
\end{aligned}
\end{equation}
where $V_{\bar{X},X,Y,\tilde{Z}}$ are the single-beam lattice
depths and $k_{\text{lat}}=2\pi/\lambda$. The energy offset
$\Delta_{AB}$ can be controlled by varying $\theta$ around $\pi$
and changing the frequency detuning $\delta$ between the $\bar{X}$
and the $X$ (which has the same frequency as $Y$) beams using an
acousto-optic modulator \cite{Tarruell2012}, as depicted in Fig. \ref{Haldaneep}(a). $\varphi_{\text{lat}}$ is
the relative phase of the two orthogonal retro-reflected beams $X$
and $Y$, the geometry of the lattice  is actively stabilized at
$\varphi_{\text{lat}}=0$, and the visibility of the interference
pattern is $\alpha = 0.81(1)$.

The crucial point for realizing the Haldane Hamiltonian in
this experiment is the creation of the next-nearest neighbor
tunneling, which is the complex tunneling with phase $\Phi_{ij}$.
The complex tunneling $e^{i\Phi_{ij}}J'_{ij}$ (see Fig.
\ref{Haldane}(b)) can be induced by circular modulation of
the lattice position. The modulation applied in this experiment
consists of moving the lattice along a periodic trajectory
$\mathbf{r}_\text{lat}(t)$. Here, the time-dependence of the
lattice position
\begin{equation}
\mathbf{r}_{\text{lat}}(t)=-A(\cos[\omega{t})\mathbf{e}_x+\cos(\omega{t}-\varphi)\mathbf{e}_y],
\end{equation}
where $A$ is the amplitude of
the motion, and $\omega/{2\pi}$ denotes the modulation frequency. Thus, after
the atoms are loaded into the honeycomb lattice, a phase-modulated
honeycomb lattice will be realized by ramping up the sinusoidal
modulation of the lattice position $\mathbf{r}_{\text{lat}}$ along
the $x$ and $y$ directions with a final amplitude of
$0.087(1)\lambda$, frequency of $4.0$ kHz, and phase difference
$\varphi$. This gives access to linear ($\varphi=0^\circ$ or
$180^\circ$), circular ($\varphi=\pm90^\circ$) and elliptical
trajectories.

At this point, the effective Hamiltonian of this phase-modulated
honeycomb lattice can be well described by the Haldane model,
where the energy offset
$\Delta_{AB}\gtrless{0}$ between sites of the $A$ and $B$
sublattices breaks inversion symmetry and opens a gap
$|\Delta_{AB}|$. TRS can be broken by changing
$\varphi$. This controls the imaginary part of the
next-nearest-neighbor tunneling, whereas its real part, as well
as the nearest-neighbor tunneling $J_{ij}$ and $\Delta_{AB}$, are mostly unaffected.
To explore the topological properties of this system, the authors measured the band structure and probed the Berry curvature for the lowest band with different parameter $\Delta_{AB}$ and
$\varphi$ by applying a constant force to the atoms, and it was found
that orthogonal drifts are analogous to a Hall current. Meanwhile, one can map out the
transition lines in the topological phase diagram of the Haldane model, as shown
in Fig. \ref{Haldaneep}(c), by identifying the
vanishing gap at a single Dirac point.

\subsubsection{Kane-Mele model}\label{KMModel}
In $2005$, Kane and Mele \cite{Kane2005a,Kane2005b} generalized the Haldane model into the
time reversal symmetric electron systems with spin. They introduced
the spin-orbit interaction between electron spin and momentum
to replace the periodic magnetic flux and predicted a new quantum
phenomenon -- the quantum spin Hall effect. Unlike the QHE where the magnetic field breaks TRS, the spin orbit
interaction preserves TRS.

The Kane-Mele model takes the tight-binding Hamiltonian
\begin{equation}\label{KMHamiltonian}
\begin{aligned}
H_{\text {KM}}&=&J\sum_{\langle{i,j}\rangle}c_i^\dag{c_j}+i\lambda_{SO}\sum_{\langle\langle{i,j}\rangle\rangle}v_{ij}{c_i^\dag{s_z}{c_j}}+i\lambda_R\sum_{\langle{i,j}\rangle}c_i^\dag(\mathbf{s}\times{\mathbf{\hat{d}}_{ij}})_z{c_j}+\lambda_v\sum_{i}{\xi_ic_i^\dag{c_i}}.
\end{aligned}
\end{equation}
The first term is the usual nearest neighbor hopping term on a
honeycomb lattice [Fig. \ref{Haldane}(a)], where
$c^\dagger_i=(c^\dagger_{i,\uparrow},c^\dagger_{i,\downarrow})$.
The second term connecting next-nearest neighbors with a spin
dependent amplitude is a mirror symmetric spin-orbit interaction.
Here $v_{ij}=\pm1$ is still the sign of hopping phases determined
by the same form as Eq. (\ref{phasesign}). $s_i$ are the Pauli
matrices describing the electron's spin. The third term is a
nearest neighbor Rashba term, which explicitly violates the
$z\rightarrow{-z}$ mirror symmetry. The last term is a staggered
sublattice potential with $\xi_i=\pm1$ depending on whether $i$ is
the $A$ or $B$ site, which will describe the transition between
the quantum spin Hall phase and the simple insulator. If the
Rashba term vanishes, the Kane-Mele model then reduces to
independent copies for each spin of a Haldane model. In this case with $s_z$ being conserved, the distinction between graphene and
a simple insulator is easily understood. Each spin has an
independent Chern number $C_{\uparrow}$ and
$C_{\downarrow}$. The TRS gives rise to
$C_{\uparrow}+C_{\downarrow}=0$, but the difference
$C_{\uparrow}-C_{\downarrow}$ is nonzero and defines a
quantized spin Hall conductivity. This characterization breaks
down when $s_z$ non-conserving terms are present
($\lambda_R\neq0$), which makes the system more complicated. The
electrons with spin-up and spin-down are coupled, and thus
the spin Hall conductance is not quantized.

\begin{figure}[htbp]\centering
\includegraphics[width=8cm]{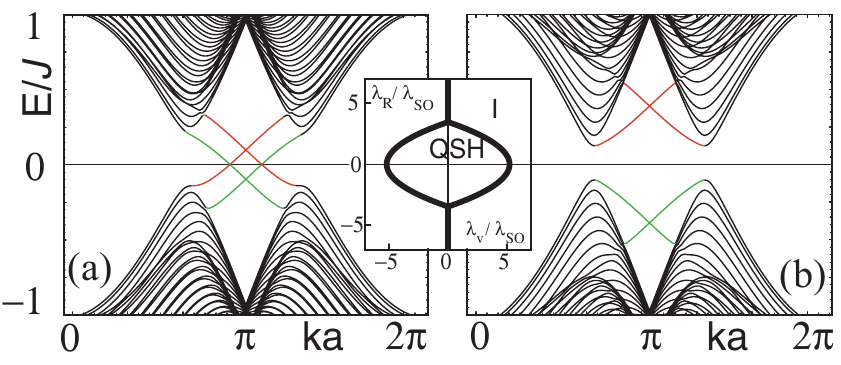}
 \caption{(Color online) Energy spectrum for a 1D zigzag strip in the (a) Quantum spin Hall phase $\lambda_v=0.1J$ and (b) the
insulating phase $\lambda_v=0.4J$. In both cases
$\lambda_{SO}=0.06J$ and $\lambda_R=0.05J$. The inset shows the
phase diagram as a function of $\lambda_v$ and $\lambda_R$ for
$0<\lambda_{SO}\ll{J}$. Reprinted with permission from Kane {\it et al.}\cite{Kane2005a}. Copyright\copyright~(2005) by the American Physical Society.}
\label{KaneMele}
\end{figure}

Following the method introduced in the Haldane model, we diagonalize the
Hamiltonian by using a basis of the four-component spinor
$c^\dagger_k=(c^\dagger_{k,A\uparrow},c^\dagger_{k,A\downarrow},c^\dagger_{k,B\uparrow},c^\dagger_{k,B\downarrow})$
of Bloch states constructed on the two sublattices and two spins.
The generic $4\times{4}$ Hamiltonian can be written in the
terms of Dirac matrices
\begin{equation}
\mathcal{H}_{\text{KM}}(\mathbf{k})=\sum^{5}_{a=1}d_a(\mathbf{k})\Gamma^a+\sum^{5}_{a<b=1}d_{ab}(\mathbf{k})\Gamma^{ab}.
\end{equation}
Here, the five Dirac matrices are defined as
\begin{equation}
\Gamma^a=(\sigma_x\otimes{s_0},\sigma_z\otimes{s_0},\sigma_y\otimes{s_x},\sigma_y\otimes{s_y},\sigma_y\otimes{s_z})
\end{equation}
(a=1,2,3,4,5); their $10$ commutators
$\Gamma^{ab}=\frac{1}{2i}[\Gamma^a,\Gamma^b]$, where the Pauli
matrices $\sigma_i$ represent the sublattice indices; $\sigma_0$
and $s_0$ are the identity matrices. In this representation, the
time reversal operator is given by
$\hat{T}=i(\sigma_0\otimes{s_y})\K$, where $\K$ is the complex
conjugation operator. The five Dirac matrices are even under the
time reversal operator $\hat{T}$,
$\hat{T}\Gamma^a\hat{T}^{-1}=\Gamma^a$, while the $10$ commutators
are odd, $\hat{T}\Gamma^{ab}\hat{T}^{-1}=-\Gamma^{ab}$. To obtain
a Hamiltonian preserving the TRS:
$\T\H(\k)\T^{-1}=\H(-\k)$, the coefficients must satisfy the
relations,
\begin{equation}
d_a(\mathbf{k})=d_a(-\mathbf{k}),~~
d_{ab}(\mathbf{k})=-d_{ab}(\mathbf{-k}).
\end{equation}
Therefore, the nonzero coefficients of Kane-Mele model are given by

\begin{equation}
\begin{aligned}
d_1(\mathbf{k})&=J(1+2\cos{x}\cos{y}), d_2(\mathbf{k})=\lambda_v,~~ d_3(\mathbf{k})=\lambda_R(1-2\cos{x}\cos{y}), \\ d_4(\mathbf{k}) &=-\sqrt{3}\lambda_R\sin{x}\sin{y}, d_{12}(\mathbf{k})=-2J\cos{x}\sin{y}, d_{15}(\mathbf{k})=2\lambda_{SO}(\sin{2x}-2\sin{x}\cos{y}), \\ d_{23}(\mathbf{k})&=-\lambda_R\cos{x}\sin{y},
~~~~d_{24}(\mathbf{k})=\sqrt{3}\lambda_R\sin{x}\cos{y},
\end{aligned}
\end{equation}
with $x=\frac{\sqrt{3}}{2}k_x$, $y=-\frac{3}{2}k_y$, which are
obtained through the variable transformations
$\mathbf{k}\cdot\mathbf{b}_1=y+x$ and $\mathbf{k}\cdot\mathbf{b}_2=y-x$.
Here $\mathbf{b}_1=(\frac{\sqrt{3}}{2}a,-\frac{3}{2}a)$ and
$\mathbf{b}_2=(-\frac{\sqrt{3}}{2}a,-\frac{3}{2}a)$ are the
lattice translation vectors in the honeycomb lattice shown  in
Fig. \ref{Haldane}(a).

The four-band system becomes insulating when the lower two bands
are fully occupied and an energy gap exists between the middle two
bands. The phase diagram of the Kane-Mele model is
shown in the inset of Fig. \ref{KaneMele}. For $\lambda_R=0$, the
Hamiltonian reduces into two independent copies of Haldane
Hamiltonian with different spins;
we can define a spin-dependent Chern number $C_s$. For
$\lambda_v>3\sqrt{3}\lambda_{SO}$, the gap is dominated by
$\lambda_v$, and the system is a normal insulator since both Chern
numbers $C_\uparrow$ and $C_\downarrow$ are zero. 
In contrast, for $\lambda_v<3\sqrt{3}\lambda_{SO}$, the
corresponding Chern number becomes nonzero,
$C_{\uparrow}=-C_{\downarrow}=\text{sgn}(\lambda_{SO})$.
Although the total Chern number $C=C_{\uparrow}+C_{\downarrow}=0$,
their difference $C_{\uparrow}-C_{\downarrow}=\pm2$, which
describes the quantum spin Hall phase with a pair of edge
states crossing the bulk gap, as depicted in Fig.
\ref{KaneMele}(a). For $\lambda_R\neq0$, the $s_z$ symmetry is
broken, and electrons with spin-up and spin-down mix together.
Thus, we cannot introduce the spin-dependent Chern number to
describe this system. Instead, Kane and Mele introduced the $\Z_2$
invariant (see Sec. \ref{TIZ2}) to describe it.

Although the Kane-Mele model has been proposed for more than a
decade, physicists still have not realized it or found such materials
in condensed matter physics. To directly implement the Kane-Mele model
is difficult, but the quantum spin Hall effect predicted
by Kane and Mele was first experimentally realized in HgTe quantum
wells \cite{Konig2007}. Compared with conventional solid-state
systems, cold atomic systems provide a perfectly clean platform
with high controllability to construct and investigate the
Kane-Mele-like model. There are several works in recent
years proposing schemes to realize and study the topological
properties of the quantum spin Hall insulators with cold
atoms \cite{GCLiu2010,Goldman2010,Cocks2012,Kennedy2013}. An
experimental scheme to simulate and detect the 2D quantum spin
Hall insulator in a kagome OL was
proposed in Ref. \cite{GCLiu2010}. In this proposal, a kagome OL with the trimer and SOC terms can host the 2D
quantum spin Hall insulator phase with only the nearest-neighbor
hopping instead of the next-nearest-neighbor hopping in the Kane-Mele
model. Moreover, the nearest-neighbor intrinsic SOC generated by the laser-induced-gauge-field method can be directly implemented in cold atomic experiments. Based on the
investigation of the Hofstadter model on a 2D square OL \cite{Aidelsburger2013,Miyake2013}, one can construct the
Kane-Mele-like model from two time-reversal copies of the spinless
Hofstadter model \cite{Goldman2010,Cocks2012,Kennedy2013}. Here, we
briefly introduce a recent proposal on realizing time-reversal
invariant topological insulators in alkali atomic
gases \cite{Goldman2010}, where quantum spin Hall states will
emerge. In this model, particles with spin
$\sigma=\uparrow,\downarrow$ experience a uniform magnetic flux
per plaquette, but opposite in sign for the two spin components.
The corresponding Hamiltonian is given by\cite{Goldman2010}
\begin{equation}
H=-J\sum_{m,n}c^\dagger_{m+1,n}e^{i\hat{\theta}_x}c_{m,n}+c^\dagger_{m,n+1}e^{i\hat{\theta}_y}c_{m,n}+\text{H.c.}
+\lambda_{\text{s}}\sum_{m,n}(-1)^mc^\dagger_{m,n}c_{m,n},
\end{equation}
where $c_{m,n}^\dagger$ is a two-component creation operator for
fermonic atoms defined on a lattice site $(ma,na)$. The last term describes an on-site
staggered potential with amplitude $\lambda_{\text{s}}$, along
the $x$ direction, which has been introduced to drive transitions
between different topological phases. The Peierls phases
$\hat{\theta}_x=2\pi\gamma\sigma_x$ and $\hat{\theta}_y=2\pi
m\Phi\sigma_z$ resulted from an artificial gauge field, are engineered within
this tight-binding model to simulate the analog of SOCs. The effect of the $SU(2)$ link variable
$\hat{U}_y(m)=e^{i\hat{\theta}_y}\propto \sigma_z$ is therefore
analogous to the intrinsic SOC in Eq.(\ref{KMHamiltonian}). It corresponds to opposite ¡°magnetic¡±
fluxes $\pm\Phi$ for each spin component and generates quantum
spin Hall phases. The link variable
$\hat{U}_x=e^{i\hat{\theta}_x}\propto \sin(2\pi \gamma)\sigma_x$
plays a role similar to the Rashba SOC in Eq.
(\ref{KMHamiltonian}). For $\gamma=0$, this model corresponds to
two decoupled copies of the spinless Hofstadter model. Besides,
$\hat{\theta}_x$ mix the two spin components as they tunnel from
one site to its nearest-neighbor sites. This model therefore captures
the essential effects of the Kane-Mele model in a multi-band framework and offers the practical
advantage of only involving nearest-neighbor hopping on a square
lattice.

\subsection{Three-dimension}

\subsubsection{3D Dirac fermions}

As introduced in the previous sections, the 2D Dirac fermions have
been extensively studied in graphene and honeycomb OLs. In recent years, it is of great interest to search for
relativistic quasiparticles in 3D materials or artificial systems
with stable band touching points, such as 3D Dirac(-like)
fermions, which can exhibit transport properties different from
those of 2D Dirac fermions. The Dirac equation in the Weyl
representation is written as $i\hbar{\partial\Psi}/{\partial t}= H_{D}\Psi,$
where $\Psi$ denotes the four-component bispinor for 3D Dirac
fermions and the Dirac Hamiltonian is given by
\begin{eqnarray}
H_D  = \left( \begin{array}{cc}  v_F{\boldsymbol \sigma} \cdot {\mathbf p} &  m \\
  m  & -v_F{\boldsymbol \sigma} \cdot {\mathbf p}
\end{array} \right), \label{3DDirac}
\end{eqnarray}
with the linear dispersion $E^{\pm}_{\rm D} = \pm\sqrt{v_F^2 p^2 +
m^2}$. Here ${\boldsymbol \sigma}=(\sigma_x,\sigma_y,\sigma_z)$
are Pauli matrices, ${\mathbf p}=(p_x,p_y,p_z)$ is the 3D
(quasi-)momentum, and the Fermi velocity $v_F$ and the mass term
$m$ represent the effective speed of light and rest energy,
respectively. Notably, the off-diagonal term $m$ in Eq.
(\ref{3DDirac}) mixes the two Weyl fermions (see the following
section) of opposite chirality.

The Hamiltonian (\ref{3DDirac}) can describe
the transition between a 3D topological insulator and a trivial
insulator in the critical case $m=0$. Recently, it was revealed that the Dirac points with fourfold degeneracies can be protected by certain symmetries \cite{Young2012,ZWang2012,ZWang2013} such as rotation or nonsymmorphic symmetries, which are not accidental band crossings at the transition between topological and trivial insulators. Thus, one has a topological Dirac semimetal with four band degeneracy, which can be viewed as 3D graphene, possessing 3D Dirac fermions in the
bulk with linear dispersions along all momentum directions. These
massless ($m=0$) 3D Dirac fermions have recently been observed in
some compounds (for a comprehensive review of Dirac semimetals in 3D solids, see Ref. \cite{Armitage2018}).

Before the Dirac semimetals were discovered in materials, several
schemes for simulating massless and massive 3D
Dirac fermions with cold atoms in 3D OLs have been
proposed \cite{MYang2010,Lepori2010,Bermudez2010b,Mazza2012}. With
cold fermions in an edge-centered cubic OL for
proper parameters, the linear dispersion characterizing 3D
Dirac-like particles with tunable mass can exhibit
\cite{MYang2010}. The system was proposed to realize 3D massless Dirac
fermions in a cubic OL subjected to a synthetic
frustrating magnetic field \cite{Lepori2010}, and the mass term  may be induced by coupling the ultracold atoms
to Bragg pulses in the system. It was suggested that the massless
and massive 3D relativistic fermions can also be simulated with
ultracold fermionic atoms in 3D optical superlattices with
Raman-assisted hopping \cite{Bermudez2010b,Mazza2012}. Moreover,
by tuning the Raman laser intensities, the system may allow the
decoupling of fermion doublers from a single Dirac fermion through
inverting their effective mass \cite{Bermudez2010b}, providing a
quantum simulation of Wilson fermions \cite{Wilson1977}.

\subsubsection{Weyl semimetals and Weyl fermions}\label{WeylSM}

The Dirac equation (\ref{3DDirac}) for massless
particles can be rewritten in a simpler form:
\begin{eqnarray}
i\hbar\frac{\partial\psi_{\pm}}{\partial t}= H_{W\pm}
\psi_{\pm},~~ H_{W\pm}=\pm v_F{\boldsymbol \sigma} \cdot {\mathbf
p},
\end{eqnarray}
where $\psi_{\pm}$ are effectively two-component vectors acting as two
chiral modes. This is the Weyl equation and $\psi_{\pm}$ are
referred to as Weyl fermions, which propagate parallel (or
antiparallel) to their spin and thus defines their chirality.
There are no fundamental particles currently found to be massless
Weyl fermions.
In some 3D lattice systems, Weyl fermions can
emerge as low-energy excitations near band crossings, and they
always arise in pairs with opposite chirality and separated
momenta. These systems are the so-called Weyl semimetals
\cite{XWan2011,Burkov2011,Burkov2011b}, which have
been intensively investigated in the last couple of years. For more details, see the
comprehensive review of vast theoretical and experimental studies
of Weyl semimetals in 3D solids \cite{Armitage2018} and
the references therein.

\begin{figure}[htbp]\centering
\includegraphics[width=0.4\columnwidth]{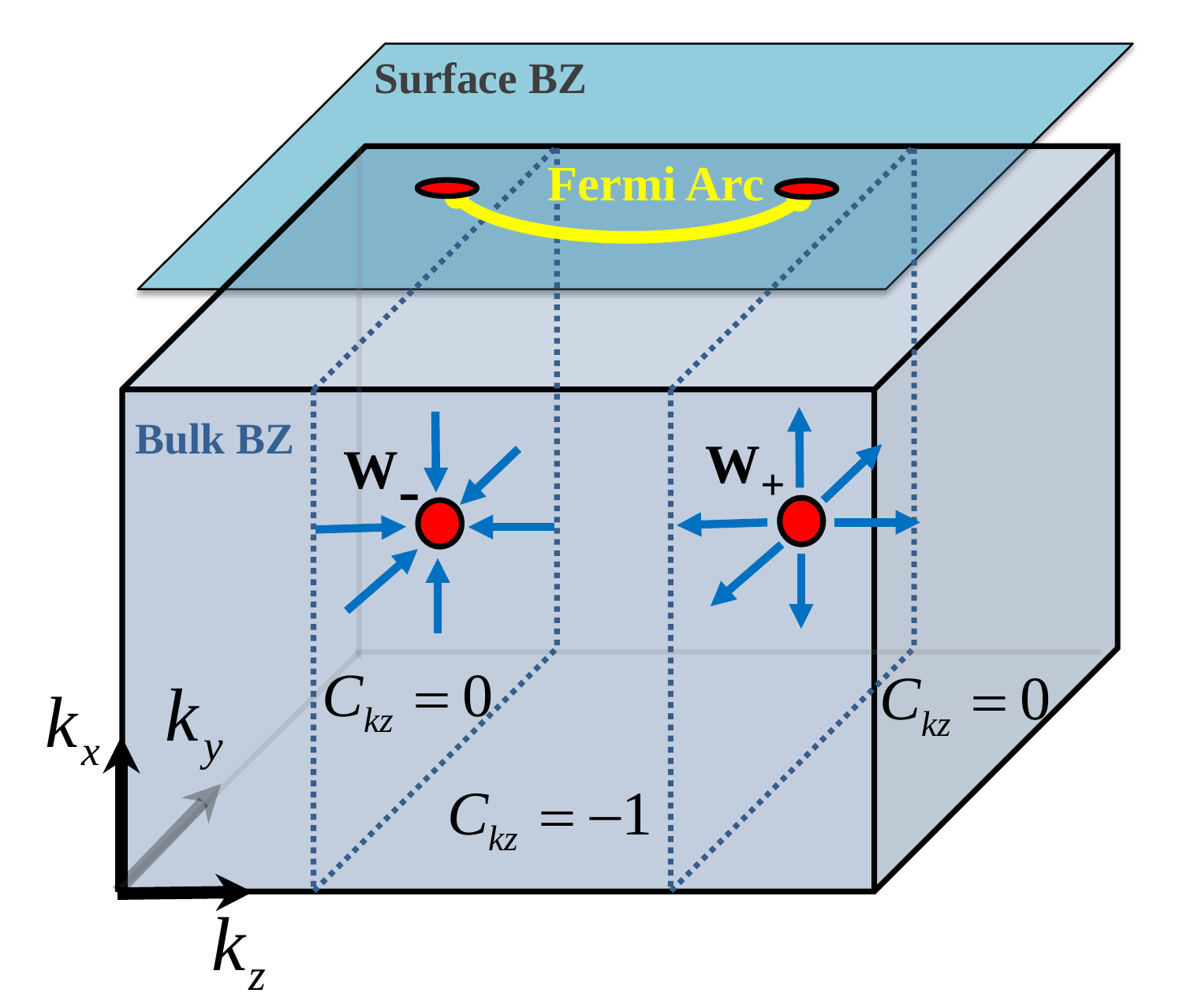}
\caption{(Color online) A pair of Weyl points as the Berry flux
monopole and anti-monopole in the bulk BZ. They are
connected by surface Fermi arcs. } \label{WeylPoint}
\end{figure}

Let us use the following minimal two-band model for
discussing the properties of Weyl semimetals in a simple cubic
lattice. The Bloch Hamiltonian is given by

\begin{eqnarray}
\mathcal{H}(\mathbf{k})=d_x(\mathbf{k})\sigma_x+d_y(\mathbf{k})\sigma_y+d_z(\mathbf{k})\sigma_z
\label{WeylBloch1}
\end{eqnarray}
with the Pauli matrices acting on two (pseudo-)spin states and the
Bloch vectors
\begin{equation}\label{WeylBloch2}
d_x=2J_s\sin k_x,\ \ d_y=2J_s\sin k_y,\ \ d_z=m_z-2 J_0(\cos k_x+\cos k_y+\cos k_z).
\end{equation}
Here $J_s$ and $J_0$ denote the hopping amplitudes of
spin-flipping hopping in the $xy$ plane and spin-dependent hopping
along all the three dimensions, respectively, and $m_z$ represents
the Zeeman field. For $2J_0<m_z<6J_0$, the bands
$E_{\pm}(\mathbf{k})=\pm \sqrt{d_x^2+d_y^2+d_z^2}$ have two
crossings in the first BZ located at
$\mathbf{W}_{\pm}=(0,0,\pm\arccos k_w)$ with $k_w=(m-4J_0)/2J_0$.
They are actually a pair of Weyl points (nodes) since one can
obtain the effective Weyl Hamiltonian for low-energy excitations
by approximating $H_{W\pm}(\mathbf{k})\approx
\mathcal{H}(\mathbf{W}_{\pm}+\mathbf{q})$:
\begin{eqnarray}
H_{W\pm} = v_xq_x\sigma_y+v_yq_y\sigma_x\pm v_zq_z\sigma_z,
\end{eqnarray}
where $v_x=v_y=2J_s$ and $v_z=2J_0$. When the Fermi level is at
$E_F=0$, the Fermi surface consists solely of two Weyl points and
the system is the Weyl semimetal with emergent Weyl fermions. The
Hamiltonian $H_{W\pm}$ can be written as $H_{\mathrm{W},\pm} =
\sum_{i,j}q_i\alpha_{ij}\sigma_j$, where $[\alpha_{ij}]$ is a
$3\times3$ matrix with elements $\alpha_{xy}=\alpha_{yx}=2J_s$,
$\alpha_{zz}=\pm2J_0$ and zero otherwise. Thus the chirality of
the two Weyl points $\mathbf{W}_{\pm}$ can be defined as
$\chi_{\pm}=\mathrm{sgn}(\det[\alpha_{ij}])=\pm1$, respectively.
One can see that there is no matrix that anticommutes with
$H_{W\pm}$ and opens a band gap since all three Pauli matrices are
used up in $H_{W\pm}$. To further realize the topological
stability of Weyl points, one can obtain the Berry flux
\begin{equation}
\mathbf{F}_{\pm}(\mathbf{k}) = \pm \frac{\mathbf{k}}{2
|\mathbf{k}|^3},
\end{equation}
which is the source and sink of the Berry curvature, forming vector
fields in momentum space that wraps around the Weyl points
$\mathbf{W}_{\pm}$. As shown in Fig. \ref{WeylPoint}, the pair of
Weyl points act as the monopole and anti-monopole in the bulk
BZ, which are characterized by the topological charges
(Chern numbers)
\begin{equation}
\label{eq:2} C_{W\pm} = \frac{1}{2\pi}\oint_S
\mathbf{F}_{\pm}(\mathbf{k})\cdot d\mathbf{S}=\pm1=\chi_{\pm},
\end{equation}
through any surface $S$ enclosing the points. This implies that
the Weyl points always exhibit in pairs of opposite chirality,
because the field lines of the Berry curvature must begin and end
somewhere within the BZ. The only way to eliminate the
Weyl points is to annihilate them pairwise by moving them at the
same point in momentum space. Therefore, the stability of the Weyl
points comes from their intrinsical topology.  It is worth noting that unlike Dirac points, Weyl points necessitate the breaking of either (or both) time-reversal or space-inversion symmetry in lattice systems.

To further discuss the topological properties of Weyl semimetals,
we consider the Bloch Hamiltonian
(\ref{WeylBloch1},\ref{WeylBloch2}) using the dimension reduction
method. Treating $k_z$ as an effective parameter ($k_z$ being
a good quantum number), we can reduce the original system to a
($k_z$-modified) collection of effective 2D subsystems described
by $\mathcal{H}_{k_z}(k_x,k_y)$. If $k_z\neq \pm k_c$, the 2D bulk
bands of $\mathcal{H}_{k_z}(k_x,k_y)$ are fully gapped and thus
can be topologically characterized by the first Chern number
\begin{equation} \label{ChN}
\nonumber C_{k_z} = \frac{1}{4\pi} \int_{-\pi}^{\pi}dk_x\int_{-\pi}^{\pi}d k_y~\hat{d}\cdot\left(\partial_{k_x}\hat{d}\times\partial_{k_y}\hat{d}\right)=\left\{
                                  \begin{array}{ll}
                                    -1, ~~& -k_c<k_z<k_c; \\
                                    0, ~~& |k_z|>k_c,
                                  \end{array}
                                \right.
\end{equation}
where $\hat{d}\equiv \vec{d}/|\vec{d}|$. For any plane
$-k_c<k_z<k_c$, one has $C_{k_z}=-1$ characterizing a Chern
insulator, while elsewhere $C_{k_z}=0$ signals a trivial
insulator. The value of $C_{k_z}$ changes for a
topological phase transition only when the bulk gap closes at
$\mathbf{W}_{\pm}=(0,0,\pm k_c)$. From this point of view, the two
Weyl points appear to be the critical points for the topological
phase transitions. An important consequence is that the Fermi arc
surface states arise in Weyl semimetals and terminate at the pair
of Weyl points. When each of the 2D Hamiltonians
$\mathcal{H}_{k_z}(k_x,k_y)$ represents a 2D Chern insulator, if
one considers a surface perpendicular to the $x$ direction (still
labeled by $k_y$ and $k_z$), each of the 2D Chern insulators
will have a gapless chiral edge mode near the Fermi energy
$E_F=0$. The Fermi energy will cross these states at $k_y=0$ for
all $-k_c<k_z<k_c$, which leads to a Fermi arc that ends at the
Weyl point projections on the surface BZ ($k_y$-$k_z$
plane), as shown in Fig. \ref{WeylPoint}. In this particular
model, the Fermi arc is a straight line. The Weyl fermions near
the Weyl points and the Fermi arcs in a Weyl semimetals are
fundamentally interesting and can give rise to exotic phenomena
absent in fully gapped topological phases, such as anomalous
(topological) electromagnetic responses
\cite{XWan2011,Burkov2011,Burkov2011b,Armitage2018}.

To realize this model of Weyl semimetals, a scheme has been proposed
by using ultracold fermionic atoms in a 2D square OL
subjected to experimentally realizable SOC and an
artificial dimension from an external parameter space (acting as
$k_z$) \cite{DWZhang2015}. It was further shown that
in the cold atom system, the simulated Weyl points can be
experimentally detected by measuring the atomic transfer fractions
in a Bloch-Landau-Zener oscillation, and the topological
invariants of the Weyl semimetals can be measured with the
particle pumping approach \cite{DWZhang2015}. Another proposal
to construct a Weyl semimetal was to stack 1D topological phases in
double-well OLs with two artificial dimensions
\cite{Ganeshan2015}. Similar 3D lattice models for realizing Weyl
semimetals with cold atoms from stacking 2D layers of Chern
insulators in checkerboard or honeycomb OLs with
synthetic staggered fluxes were suggested in Refs.
\cite{JHJiang2012,WYHe2016,JMHou2016,BZWang2018}. The realization of chiral
anomaly by using a magnetic-field gradient in the system was also
discussed \cite{WYHe2016}. In these schemes, the spin degree of
freedom can be encoded by two atomic internal states or
sublattices, and then the required hopping terms can be realized
by synthetic SOC or magnetic fields. These
ingredients are well within current experimental reach of
ultracold gases. It was illustrated that Weyl excitations can also
emerge in 3D OL of Rydberg-dressed atomic fermions or
dipolar particles \cite{XLi2015,Syzranov2016}. The Weyl points may
automatically arise in the Floquet band structure during the shaking of
a 3D face-centered-cubic OL without requiring
sophisticated design of the tunneling \cite{LJLang2017}.

\begin{figure}[htbp]\centering
\includegraphics[width=0.95\columnwidth]{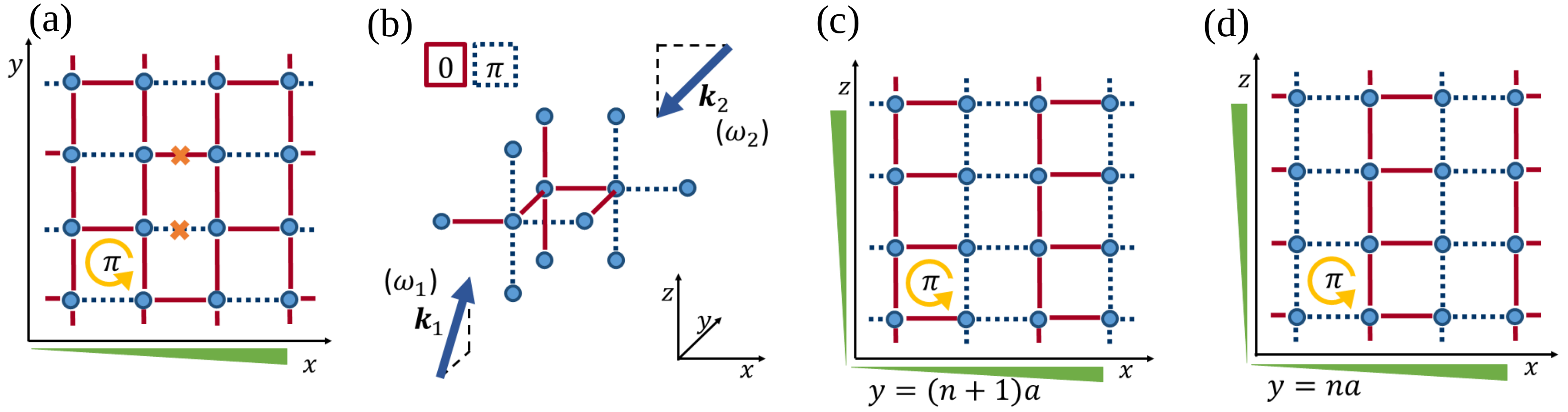}
\caption{(Color online) Sketch of the 3D cubic lattice with
engineered hopping along $x$ and $z$ directions, which possesses
Weyl points in momentum space. Dashed and solid lines depict hopping
with acquired phase $\pi$ and $0$, respectively. (a) The $xy$ planes
of the lattice are equivalent to the lattice of the
Hofstadter-Harper Hamiltonian with $\alpha=1/2$ flux. Green
triangles along the axes denote the tilted directions. (b) A pair
of Raman lasers enabling laser-assisted tunneling is sketched with
arrows. The 3D lattice alternates stacks of 2D lattices
parallel to the $xz$ plane, which is shown in (c) and (d); the
hopping between these planes (along $y$) is regular. The hopping
along $z$ is alternating with phase $0$ or $\pi$, depending on the
position in the $xy$ plane with broken inversion symmetry. Reprinted with permission from Dubcek {\it et al.} \cite{Dubcek2015}. Copyright\copyright~(2015) by the American Physical Society.
} \label{WeylLattice}
\end{figure}

An alternative scheme to realize the Weyl semimetal phase by stacking
2D Hofstadter-Harper systems in cubic OLs was
proposed in Refs.\cite{Dubcek2015,DWZhang2017}. A sketch of the 3D
lattice with laser-assisted tunneling along both $x$ and $z$
directions is shown in Fig. \ref{WeylLattice}. To realize this
hopping configuration, natural tunneling ($J_x,J_z$) along these
directions is first suppressed by introducing a large linear tilt
of energy $\Delta$ per lattice site ($J_x,J_z\ll\Delta\ll
E_{gap}$), which can be obtained by a linear gradient potential
(e.g., gravity or magnetic field gradient) along the $\hat x +
\hat z$ direction. The tunneling is then resonantly restored by
two far-detuned Raman beams of frequency detuning
$\delta\omega=\omega_1-\omega_2=\Delta/\hbar$, and momentum
difference $\delta{\bf k}={\bf k}_1-{\bf k}_2$
\cite{Aidelsburger2013,Miyake2013}. It yields an effective 3D
Hamiltonian for the system \cite{Dubcek2015}
\begin{align}
\label{Weyl H} & H_{3D} = -\sum_{m,n,l} \left(\right.
K_x e^{-i\Phi_{m,n,l}} a_{m+1,n,l}^{\dagger} a_{m,n,l} + \\
& J_y a_{m,n+1,l}^{\dagger} a_{m,n,l}+ K_z e^{-i\Phi_{m,n,l}}
a_{m,n,l+1}^{\dagger} a_{m,n,l}+h.c. \left.\right). \nonumber
\end{align}
Here, $a_{m,n,l}^{\dagger}$ ($a_{m,n,l}$) is the creation
(annihilation) operator on the site $(m,n,l)$, and $\Phi_{m,n,l} =
\delta{\bf k} \cdot {\bf R}_{m,n,l} = m \Phi_x+n \Phi_y+l \Phi_z$
are the nontrivial hopping phases, dependent on the positions
${\bf R}_{m,n,l}$. Next, the directions of the Raman lasers are
chosen such that $(\Phi_x,\Phi_y,\Phi_z) = \pi (1,1,2)$, i.e.
$\Phi_{m,n,l} =(m+n)\pi$ (modulo $2\pi$), as schematically shown
in Fig.~\ref{WeylLattice}(b). The 3D system can be viewed as an
alternating stack of two types of 2D lattices, parallel to the
$xz$ plane, as illustrated in Figs. \ref{WeylLattice}(c) and
\ref{WeylLattice}(d); hopping between these planes is regular
(along $y$). Another view is stacking of 2D lattices described by
the Hofstadter-Harper Hamiltonian with $\alpha=1/2$ flux in Fig.
\ref{WeylLattice}(a), such that the hopping along $z$ has phases
$0$ or $\pi$, for $m+n$ even or odd, respectively. The 3D lattice
has two sublattices (A-B) with broken inversion symmetry and has
the Bloch Hamiltonian
\begin{equation}
\mathcal{H}({\bf k})= -2 ( J_y \cos k_y\sigma_x + K_x \sin
k_x\sigma_y - K_z \cos k_z\sigma_z ). \label{HW}
\end{equation}
The energy spectrum of the Hamiltonian has two bands,
\begin{equation}
E({\bf k})=\pm 2 \sqrt{K_x^2\sin^2k_x
+J_y^2\cos^2k_y+K_z^2\cos^2k_z},
\end{equation}
which touch at four Weyl points within the first BZ at
$(k_x,k_y,k_z)=(0, \pm\pi/2, \pm \pi/2)$. The schemes for the
experimental detection of the Weyl points with both ultracold
bosons and fermions in the lattice were suggested
\cite{Dubcek2015}. The dynamics of Weyl quasiparticles in this
model was investigated \cite{ZLi2016}.

In 3D crystals, the generic energy dispersion near the Weyl points
takes the form of the following low-energy Hamiltonian
\begin{equation}
\tilde{H}_{W\pm}= v_0 q_j\sigma_0 \pm v_F {\mathbf
q}\cdot{\boldsymbol \sigma},
\end{equation}
where an additional term with the Fermi velocity $v_0$ introduces an
overall tilt of the Weyl cones. This term is forbidden by Lorentz
symmetry for the Weyl Hamiltonian in vacuum but it can generically
appear when expanding the Bloch Hamiltonian at the Weyl points.
The energy spectrum is given by $E_{\pm}=v_0 q_j\pm
v_F|{\mathbf q}|$ for particle and hole bands. When $|v_0|<|v_F|$,
the energy of the particle (hole) band is still positive
(negative) and the Weyl point is called a type-I Weyl point as
discussed previously. When $|v_0|>|v_F|$, the Weyl point is still
there, but the two bands now overlap in energy in certain regions,
forming particle and hole pockets. In this case, the Weyl point
becomes a point at which a particle and a hole pocket touch and
is dubbed a type-II Weyl point. These Weyl semimetals with broken
Lorentz symmetry that have no analog in quantum field theory are
called type-II Weyl semimetals \cite{Soluyanov2015}. Several
schemes have been proposed to realize and detect type-II Weyl
semimetals (points) with cold atoms in 3D OLs
\cite{YXu2016b,Shastri2017,XKong2017}.

There is an additional subclass of Weyl semimetals called the multi-Weyl semimetals
\cite{GXu2011,CFang2012,SHuang2016}. In these systems, the Weyl
points carry topological charges of higher magnitude, such as
$C_W=\pm2$ for double-Weyl points and $C_W=\pm3$ for triple-Weyl
points, which may be stabilized by certain point-group crystal
symmetries. Several materials have been predicted to be candidates
for double- and triple-Weyl semimetals, but they are yet to be
experimentally realized. The double-Weyl semimetals may be
simulated with ultracold atoms in 3D OLs in the
presence of synthetic non-Abelian SU(2) gauge potentials
\cite{Lepori2016}. The topological properties of double-Weyl semimetals in OLs were also explored \cite{XYMai2017}.

The Weyl excitations may exist in a superfluid of Fermi atoms or an atomic Bose-Hubbard system. For
instance, the manifestation of anisotropic Weyl fermions in sound
speeds of Fulde-Ferrell fermionic superfluids was studied
\cite{YXu2014}, an anisotropic Weyl superfluid state was shown to
be stable in a 3D dipolar Fermi gas \cite{BLiu2015}, and the
structured Weyl points may exhibit in the superfluid quasiparticle
spectrum of a 3D Fermi gas subject to synthetic SOCs and Zeeman fields \cite{YXu2015}. It was found that
the energy dispersion of Bogoliubov excitations has Weyl points in both the superfluid and Mott-insulator phases in a Bose-Hubbard extension of a Weyl semimetal \cite{YJWu2017}, which can be realized with ultracold bosonic atoms in 3D OLs.

\subsubsection{Topological nodal-line semimetals}

In 3D lattice systems, there is another kind of topological
semimetals in addition to Weyl and Dirac semimetals. In contract
to Weyl and Dirac semimetals that have band touching at isolated
points, they have the band touching along lines in the 3D BZ, termed nodal-line semimetals
\cite{Burkov2011b,CFang2016,RYu2017}. The nodal lines can be
topologically stable under certain discrete symmetry, and each
carries a quantized $\pi$ Berry phase (a $\mathbb{Z}_2$
topological invariant). The topological nodal-line semimetal state
has been predicted and recently confirmed to exist in some
materials \cite{CFang2016,RYu2017,GBian2016,JHu2016,Pezzini2017}.

A simple two-band model of nodal-line semimetals in the continuum
can be written as
\begin{equation}
h(\mathbf{k})=[k_0^2-(k_x^2+k_y^2+k_z^2)]\sigma_z+ k_z\sigma_y,
\end{equation}
which has both inversion symmetry $\hat P=\sigma_z$ and time-reversal
symmetry $\hat T=\hat{K}$ with the complex-conjugate operator
$\hat{K}$, and thus the combined $\hat P \hat T$ symmetry. It is found
that the gapless points form a closed nodal line on the
$k_x$-$k_y$ plane with $k_z=0$, which may be enclosed by a loop
from the gapped region, such as a tiny circle on the $k_y$-$k_z$
plane. The circle is parametrized as
$(0,k_0+\rho\cos\phi,\rho\sin\phi)$, with $\rho$ being the radius
and $\phi$ the angle. If $\rho$ is sufficiently small, the
Hamiltonian restricted on the circle is expanded as $h(\phi)=
-2k_0\rho\cos\phi\sigma_z+\rho\sin\phi\sigma_y+\mathcal{O}(\rho^2)$.
The Berry phase of the occupied state wave function of such a
Hamiltonian is quantized in units of $\pi$, which is equal to one
modulo $2$, namely
\begin{equation}
\gamma=\frac{1}{\pi}\int \langle \psi( \phi)|i\partial_\phi|\psi(\phi)\rangle
d\phi= 1\mod 2,
\end{equation}
with $|\psi(\phi)\rangle$ being the occupied state of $h(\phi)$. In a
lattice system, the periodicity of the momentum coordinates allows
every large circle going inside the nodal loop to have nontrivial
topological charge $\gamma=1$. For straight lines inside, each may be regarded as corresponding to a 1D gapped system that
is of topological band structure, leading to gapless boundary
modes. If particle-hole symmetry is additionally present, these
modes form the drumhead-shape surface states as a flat band over
the surface BZ enclosed by the projection of the nodal
line. There are still nearly flat surface bands in the
absence of this additional symmetry.

Notably, the quantized $\pi$ Berry phase means that the
nodal line is in the topological class of the $\mathbb{Z}_2$
classification \cite{DWZhang2016a,YXZhao2016}. According to the
classification theory, the topological protection of the stability
of the nodal line requires only the combined $\hat P \hat T$ symmetry, rather
than both $\hat P$ and $\hat T$ symmetries. In other words, the nodal line
still exists for topological reasons when perturbations break
both $\hat P$ and $\hat T$ symmetries but preserve $\hat P \hat T$ symmetry. In
general, a $\sigma_y$ term with even functions of $\mathbf{k}$ and
a $\sigma_z$ term with odd functions of $\mathbf{k}$ break both
$\hat P$ and $\hat T$, which just change the shape and (or) position of the
nodal line.

Several schemes were proposed to realize topological
nodal-line semimetals in cold atom systems. Ref. \cite{DWZhang2016a} proposed to realize tunable
$\hat P \hat T$-invariant topological nodal-loop states with ultracold atoms
in a 3D OL, which is described by the Bloch
Hamiltonian
\begin{equation}
\mathcal{H}(\mathbf{k}) = f_z(\mathbf{k})\sigma_z-2J_s\sin
k_z\sigma_y-f_0(\mathbf{k})\sigma_0.
\end{equation}
Here $f_z(\mathbf{k})=m_z-\alpha_{-}(\cos k_x+\cos
k_y)-\alpha_{+}\cos k_z$ and $f_0(\mathbf{k})=\alpha_{+}(\cos
k_x+\cos k_y)+\alpha_{-}\cos k_z$, where $m_z$ and
$\alpha_{\pm}\equiv J_{\uparrow}\pm J_{\downarrow}$ are tunable
parameters for adjusting the nodal rings, and
$J_{\uparrow,\downarrow}$ are the natural hopping strengths for
two spins. To realize this Hamiltonian, atoms with two hyperfine
spin states are loaded in a spin-dependent 3D OL and
two pairs of Raman lasers are used to create spin-flip hopping
with a site-dependent phase along the $z$ direction. It was also
demonstrated that the characteristic nodal ring can be detected
from Bloch-Landau-Zener oscillations, the topological invariant
may be measured based on the time-of-flight imaging, and the
surface states may be probed via Bragg spectroscopy.

A four-band model allowing Dirac or Weyl rings was also suggested
to realize with cold atoms in 3D OLs, where the superfluidity of attractive Fermi gases
in the model exhibits Dirac and Weyl rings in the
quasiparticle spectrum \cite{YXu2016}. The atomic topological superfluid with
ring nodal degeneracies in the bulk was proposed in Ref.
\cite{WYHe2018}. An alternative  model of 3D topological semimetals
whose energy spectrum exhibits a nodal line acting as a vortex
ring was proposed in Ref. \cite{Lim2017}, which may be realized
with cold atoms. Even in a dissipative system with particle gain
and loss, a novel type of topological ring was theoretically
discovered \cite{YXu2017b}, which is dubbed a Weyl exceptional
ring consisting of exceptional points at which two eigenstates
coalesce. Such a Weyl exceptional ring is characterized by both a
Chern number and a quantized Berry phase, and may be
realized and measured in a dissipative cold atomic gas trapped in
an OL. Recently, it was theoretically found that there are other possible
configurations for 1D nodal lines of band touching, such as a
nodal chain \cite{Bzdusek2016} containing connected loops and a
nodal link \cite{WChen2017,ZYan2017,Ezawa2017} hosting linked
nodal rings in the BZ. A scheme to realize the
topological semimetal with double-helix nodal links using cold
atoms in an OL was also presented \cite{WChen2017}.

In a very recent experiment \cite{BSong2018b}, a 3D topological nodal-line semimetal phase for ultracold fermions with synthetic SOCs in an optical Raman lattice was realized. The nodal lines embedded in the semimetal bands were observed by measuring the atomic spin-texture. Moreover, the realized topological band structure was confirmed by observing the band inversion lines from the dynamics of the quench from a deep trivial regime to topological semimetal phases. This work demonstrated a promising approach to explore 3D band topology for ultracold atoms in OLs.

\subsubsection{3D $\Z_2$ topological insulators}\label{3DZ2TI}

Inspired by the study of 2D $\Z_2$ topological insulators \cite{Kane2005a,Bernevig2006b,Konig2007}, three groups of theorists independently proposed 3D generalizations of the quantum spin Hall insulators \cite{LFu2007a,Moore2007,Roy2009}. A single $\Z_2$ invariant $\nu$ characterizes the topology of a 2D topological insulator; in contrast, four $\Z_2$ topological invariants $\nu_0;(\nu_1\nu_2\nu_3)$ are needed to fully characterize a 3D $\Z_2$ topological insulator. As introduced in Sec. \ref{TIZ2}, the mathematical formulations of the four $\Z_2$ invariants can be obtained from the 2D case, which involves the quantities $\delta_i$ of 8 distinct inversion invariant points $\Gamma_i$ in the 3D BZ. Points $\mathbf{k}=\Gamma_i$ in the BZ are inversion invariant since $-\Gamma_i =\Gamma_i+\mathbf{G}=\Gamma_i$ for a reciprocal lattice vector $\mathbf{G}$. Thus, these points are also time reversal invariant $\hat T\H(\Gamma_i)\hat T^{-1}=\H(\Gamma_i)$ and are called time reversal invariant momenta. The eight $\Gamma_i$ are expressed in terms of primitive reciprocal lattice vectors as
\begin{equation}\label{8TRIM}
\Gamma_{i=(n_1n_2n_3)}=(n_1\mathbf{b}_1+n_2\mathbf{b}_2+n_3\mathbf{b}_3)/2,
\end{equation}
where $n_j=0,1$ and $\mathbf{b}_l$ are primitive reciprocal lattice vectors. They can be regarded as the vertices of a cube.

In the 2D case with $\mathbf{b}_3=0$, the $\Z_2$ invariant can be determined by the quantities
\begin{equation}
\delta_a=\frac{\sqrt{\text{det}[\mathcal{U}(\Gamma_a})]}{\text{Pf}[\mathcal{U}(\Gamma_a)]}=\pm1,
\end{equation}
where $\Gamma_a$ are the four time reversal invariant momenta with the form (\ref{8TRIM}) in the 2D BZ, and $\mathcal{U}$ is the so-called \emph{sewing matrix} defined by
\begin{equation}
\mathcal{U}_{mn}(\mathbf{k})=\langle u_m(-\mathbf{k})|U_{T}|u_n(\mathbf{k})\rangle^*,
\end{equation}
which builds from the occupied Bloch functions $|u_m(\mathbf{k})\rangle$  \cite{LFu2006}. At $\mathbf{k}=\Gamma_a$, $\mathcal{U}_{mn}=-\mathcal{U}_{nm}$, so the Pfaffian $\text{Pf}[\mathcal{U}]$ satisfying $\text{det}[\mathcal{U}]=\text{Pf}[\mathcal{U}]^2$ is well defined. At this time, the single $\Z_2$ invariant $\nu$ is given by $(-1)^\nu=\prod_{a=1}^4\delta_a$. Similarly, the four $\Z_2$
indices $\nu_0;(\nu_1\nu_2\nu_3)$ in the 3D BZ can be defined in term of $\delta_{n_1n_2n_3}$ as
\begin{equation}\label{Z2index}
(-1)^{\nu_0}=\prod_{n_j=0,1}\delta_{n_1n_2n_3},\ \ \ \ \
(-1)^{\nu_{i=1,2,3}}=\prod_{n_{j\neq i}=0,1;n_i=1}\delta_{n_1n_2n_3}.
\end{equation}
One can see that $\nu_0$ can be expressed as the product over all eight points, while the other three invariants $\nu_i$ are given by products of four $\delta_i$, with which $\Gamma_i$ reside in the same plane. If the lattice system has inversion symmetry, the problem of evaluating the $\Z_2$ invariants can be greatly simplified \cite{LFu2007b}. At the
special points $\Gamma_i$, the Bloch states $|u_{2m}(\Gamma_i)\rangle$ are also the parity eigenstates of the $2m$-th
occupied energy band with eigenvalue $\xi_{2m}(\Gamma_i)=\pm1$, which shares the same eigenvalue
$\xi_{2m}=\xi_{2m-1}$ with its Kramers degenerate part. The product involves the $2N$ occupied bands that can be divided into $N$ Kramers pairs. In this case, the $\Z_2$ invariants are still determined by Eq. (\ref{Z2index}) with
\begin{equation}\label{Z2inversion}
\delta_{n_1n_2n_3}=\prod_m^N\xi_{2m}(\Gamma_{i}).
\end{equation}

According to the parity of $\nu_0$, the system can be divided into two classes of phases.
For $\nu_0=0$, the system is referred to as a ``weak" topological insulator with an even number of Dirac cones
at the surfaces, which can be interpreted as stacked layers of the 2D quantum spin Hall insulators. The TRS does not protect their surface states and the system is not robust against disorder. For $\nu_0=1$, the crystal is called a ``strong" topological insulator with an odd number of Dirac cones on all surfaces of the BZ. The connection between the bulk topological indices and the presence of unique metallic surface states is established. The 3D $\Z_2$ topological insulators have been theoretically predicted and then experimentally discovered in several materials, such as $\text{Bi}_{1-x}\text{Sb}_x$ \cite{LFu2007b,Hsieh2008} as well as
the ``second generation" topological insulators in $\text{Bi}_2\text{Te}_3$, $\text{Sb}_2\text{Tb}_3$ \cite{HJZhang2009}, and $\text{Bi}_2\text{Se}_3$ \cite{HJZhang2009,YXia2009}.

The realization of the $\Z_2$ topological insulators in ultracold atomic systems will allow investigation of interesting properties that cannot readily be explored in solid-state materials in a controlled way, such as the strong correlations and other perturbations in $\Z_2$ topological insulators. Based on the optical flux lattice \cite{Cooper2011a} with synthetic SOC, a generic scheme was proposed to realize 2D and 3D $\Z_2$ topological insulators with cold atoms \cite{Beri2011}. Interestingly, the proposed lattice system work in the nearly free particle regime, which allows for large gaps with the size set by the recoil energy. For an atom of $N$ internal states in an optical potential with position $\mathbf{r}$ and momentum $\mathbf{p}$, the generic Hamiltonian of the atom-laser system can be written as
\begin{equation}
H=\frac{\mathbf{p}^2}{2m}\mathds{1}_N+V\hat{M},
\end{equation}
where $V$ has dimensions of energy, $\mathds{1}_N$ is the identity, and $\hat{M}(\mathbf{r})$
is a position-dependent $N\times N$ matrix acting on the internal states of the atom describing the interaction between the atom and the laser field. To realize a $\Z_2$ topological insulator, one requires $N$ to be even and the Hamiltonian invariant under time
reversal: $\hat{T}\hat{M}\hat{T}^{-1}=\hat{M}$ with $\hat{T}=i(\mathds{1}_2\otimes\hat{\sigma}_y)\K$. The smallest nontrivial case has $N=4$ with  \cite{Beri2011}
\begin{equation}\label{Mmatrix}
\begin{aligned}
\hat{M}=\begin{pmatrix}(A+B)\mathds{1}_2&C\mathds{1}_2-i\hat{\sigma}\cdot\mathbf{D}\\
C\mathds{1}_2+i\hat{\sigma}\cdot\mathbf{D}&(A-B)\mathds{1}_2\end{pmatrix},
\end{aligned}
\end{equation}
where $A$, $B$ and $C$ are real parameters, and $\mathbf{D}=(D_x,D_y,D_z)$
is a 3D vector.

The optical potential in Eq. (\ref{Mmatrix}) can be implemented by using four internal states of $^{171}$Yb atom (with nuclear spin $1/2$) \cite{Beri2011}. Both the ground state ($^1S_0=g$) and the long-lived excited state ($^3P_0=e$) have two internal states. The magnetic
field is considered to be sufficiently small that the Zeeman splitting is negligible, and all four $e$-$g$ transitions involve the same
frequency $\omega_0=(E_e-E_g)/\hbar$. Under this single photon coupling with the state-dependent potential $V_{\text{am}}$, the optical
potential in the rotating wave approximation \cite{Tannoudji1992} is given by
\begin{equation}\label{opticalpotential}
\begin{aligned}
V\hat{M}=\begin{pmatrix} (\frac{\hbar}{2}\Delta_d+V_{\text{am}})\mathds{1}_2 & -i\hat{\sigma}\cdot\varepsilon d_r\\
               i\hat{\sigma}\cdot\varepsilon d_r &  -(\frac{\hbar}{2}\Delta_d+V_{\text{am}})\mathds{1}_2\end{pmatrix},
\end{aligned}
\end{equation}
where $\Delta_d=\omega-\omega_0$ is the detuning, $d_r$ is the reduced dipole moment, and $\varepsilon$ represents the electric amplitude vector.
For the 3D case, this optical potential can be achieved by three standing waves of linearly polarized light at the coupling
frequency $\omega$: two of equal amplitude with wave vectors in the 2D plane ($\mathbf{K}_1$ for $y$ polarization and $\mathbf{K}_2$ for $z$ polarization)
and one with a wave vector $\mathbf{K}_3$ normal to the 2D plane
for $x$ polarization with an amplitude smaller by a factor of $\delta$. The corresponding
electric field, detuning, and state-dependent potential in Eq. (\ref{opticalpotential}) are given by
\begin{equation}\label{Z2OC}
d_r\varepsilon=V\big[\delta',\cos(\mathbf{r}\cdot\mathbf{K}_1),\cos(\mathbf{r}\cdot\mathbf{K}_2)\big],\ \
\hbar\Delta/2+V_{\text{am}}=V\big[c_{12}+\delta'(\mu+c_{13}+c_{23})\big],
\end{equation}
where $\delta'=\delta \cos(\mathbf{r}\cdot\mathbf{K}_3)$ and $c_{ij}=\cos\big[\mathbf{r}\cdot(\mathbf{K}_i+\mathbf{K}_j)\big]$, with $\mathbf{K}_1=(1,0,0)\kappa$, $\mathbf{K}_2=(\cos\theta,\sin\theta,0)\kappa$, and $\mathbf{K}_3=(0,0,1)\kappa$. The amplitudes
are chosen to have a common energy scale $V$, which can be interpreted as a measure of the Rabi coupling. Since $\omega\simeq\omega_0$, the magnitude of the wave vectors is $\kappa\simeq2\pi/\lambda_0$ with $\lambda_0=578$ nm being the wavelength
of the $e$-$g$ transition. The space-dependent $V_{\text{am}}(\mathbf{r})$ is set
by a standing wave at the anti-magic wavelength $\lambda_{\text{am}}$ \cite{Gerbier2010},
which fixes the angle $\theta=2\text{arccos}(\pm\lambda_0/\lambda_{\text{am}})$.
For Yb atoms, $\lambda_0/\lambda_{\text{am}}\simeq1/2$, so $\theta\simeq\pm\frac{2\pi}{3}$ (let $\theta=\frac{2\pi}{3}$). The optical coupling $\hat{M}$ preserves the symmetry of a monoclinic lattice, where the lattice vectors $\mathbf{a}_1=(\sqrt{3}/2,-1/2,0)a$, $\mathbf{a}_2=(0,1,0)a$, and $\mathbf{a}_3=(0,0,\sqrt{3}/2)a$, with $a\equiv4\pi/(\sqrt{3}\kappa)$. Thus, the eight time reversal invariant momenta of the 3D topological insulator in this lattice are given by $\Gamma_{mnl}=(m\mathbf{K}_1+n\mathbf{K}_2+l\mathbf{K}_3)/2$, with $m,n,l=0,1$ and the
reciprocal lattice vectors $\mathbf{K}_i$. In addition, the system has inversion symmetry $\hat{P}:\mathbf{r}\rightarrow-\mathbf{r}$ as the optical coupling is even under the spatial inversion. Thus the $\Z_2$ topological invariants in this system take a simple form based on Eqs. (\ref{Z2index}) and (\ref{Z2inversion}): The product $\prod_{m,n,l=0,1}\prod_{\alpha\in\text{filled}}\xi_{\alpha}(\Gamma_{mnl})=-1$ for $\nu_0=1$, where $\xi_{\alpha}(\Gamma_{mnl})$ are the parity eigenvalues of the $\alpha$th Kramers pair of bands at the momenta $\Gamma_{mnl}$.

\subsubsection{3D Chiral topological insulators}\label{3DCIScheme}

As we discussed in Sec. \ref{1DModel}, 1D chiral topological insulators classified in the AIII class have been extensively studied in condensed matter systems and OLs with cold atoms. Similarly, according to the ten-fold classification of topological insulators \cite{Schnyder2008,Ryu2010}, there are two distinct classes of 3D topological insulators protected by the chiral symmetry $\hat{S}$, which is the combination of time-reversal $\hat T$ and charge-conjugation $\hat C$ symmetries.
The first class is the class AIII in the 3D cases, and the second one is the class DIII which is invariant under both $\hat T$ and $\hat C$ symmetries. The realization of 3D chiral topological insulators in condensed matter materials has been studied \cite{BLi2017,Hosur2010}. In addition, the experimental schemes for implementing the class AIII and DIII chiral topological insulators using cold atomic gases in 3D OLs have been proposed \cite{Essin2012,STWang2014}.

In the proposal in Ref. \cite{Essin2012}, an optical potential coupling noninteracting atoms with two spin states was constructed, which is described by the model Hamiltonian
 \begin{equation}\label{AFTI}
  H(\mathbf{p},\mathbf{r})=\frac{p^2}{2m}+V\big[\text{cos}kx+\text{cos}ky+\text{cos}kz\big]+\mathbf{B}_Z(\mathbf{r})\cdot\sigma,
 \end{equation}
with $ \mathbf{B}_Z(\mathbf{r})=B_Z\sum_{i=1}^4\mathbf{b}_i\text{cos}(k\mathbf{b}_i\cdot\mathbf{r}).$
Here $k=2\pi/a$ denotes the length scale of the wave vector; $\mathbf{p}$ and $\mathbf{r}$ are
the single-particle momentum and position; $\mathbf{\hat{x}}$, $\mathbf{\hat{y}}$, and $\mathbf{\hat{z}}$
are orthogonal unit vectors; and $\sigma$ represents the Pauli matrices in spin space.
Moreover, the tetrahedral vectors $\mathbf{b}_i$ are represented as $\mathbf{b}_1=(-\mathbf{\hat{x}}+\mathbf{\hat{y}}+\mathbf{\hat{z}})/2$, $\mathbf{b}_2=(\mathbf{\hat{x}}-\mathbf{\hat{y}}+\mathbf{\hat{z}})/2$,
$\mathbf{b}_3 =(\mathbf{\hat{x}}+\mathbf{\hat{y}}-\mathbf{\hat{z}})/2$, and $\mathbf{b}_4=-(\mathbf{\hat{x}}+\mathbf{\hat{y}}+\mathbf{\hat{z}})/2$.
The potential $V(\mathbf{r})$ creates a spin-independent cubic lattice,
while the effective Zeeman term $\mathbf{B}_Z(\mathbf{r})\cdot\sigma$ creates an alternating magnetic hedgehog
texture around the wells of the lattice \cite{Essin2012}, leading to the lattice structure with the translation symmetry of a
face-centered-cubic lattice. Although the Zeeman field $\mathbf{B}_Z(\mathbf{r})$ breaks $\T$ since $\mathbf{\sigma}=-\sigma_y\mathbf{\sigma}^*\sigma_y$,
the symmetry is restored by a translation $T_{1/2}$ through $a$ along any of the cubic axes. This Hamiltonian then has the
combined symmetry $\Sigma=\hat{T}T_{1/2}$, which satisfies $\Sigma^2=-1$, and therefore keeps the necessary topological characteristic of a nontrivial topological insulator phase.

In the deep-well limit, the Hamiltonian (\ref{AFTI}) reduces to the
following tight-binding model on the fcc cubic lattice
\begin{equation}
\begin{aligned}
H_{\text{tb}}&=\sum_{\mathbf{r}\in A}\sum_{\mathbf{e}}c_{\mathbf{r}}^\dagger(J+J_M\mathbf{e}\cdot\sigma)c_{\mathbf{r}+\mathbf{e}}+\text{H.c.},\\
\end{aligned}
\end{equation}
where $J$ and $J_M$ (both are real) are the nearest-neighbor and spin-dependent hopping, respectively.
$\hat{c}_{\mathbf{r}}=(\hat{c}_{\mathbf{r},\uparrow},\hat{c}_{\mathbf{r},\downarrow}) $ is the fermionic annihilation operator at site $\mathbf{r}$,
and $\mathbf{e}\in{\pm\mathbf{\hat{x}}, \pm\mathbf{\hat{y}},\pm\mathbf{\hat{z}}}$; $A$ labels one of the two sublattices of
the fcc cubic lattice. The Bolch Hamiltonian in momentum space is given by
\begin{equation}
\mathcal{H}(k)=2\begin{pmatrix} 0&g(k)\\g^\dagger(k)&0\end{pmatrix},\ \ \ \ g(k)=\sum_{j\in x,y,z}(J\text{cos}k_j-iJ_M\text{sin}k_j\sigma_j).
\end{equation}
The energy spectrum of this Hamiltonian has two bands (with twofold degeneracy)
 \begin{equation}
 \epsilon(k)=\pm2\sqrt{J^2f^2(k)+J_M^2f_{M}(k)},
 \end{equation}
where $ f(k)=\sum_{j\in{x,y,z}}\text{cos}k_j$, and $f_{M}(k)=\sum_{j\in{x,y,z}}\text{sin}^2k_j.$
In this tight-binding regime, the system are protected by an extra chiral symmetry: $\hat{S}\mathcal{H}(k)\hat{S}^{-1}=-\mathcal{H}(k)$ with the operator $\hat{S}=\tau_z\otimes\sigma_0$, where $\tau_j$ are the Pauli matrices in the sublattice space and $\sigma_0$ is the identity matrix. Thus, this model belongs to the symmetry class DIII. The associated topological invariant of this system can be characterized by the 3D winding number \cite{Ryu2010,Silaev2010,Essin2011}
\begin{equation}
\nu_{\text w}=\pi\int\frac{d^3k}{(2\pi)^3}\frac{1}{3!}\epsilon_{abc}\text{Tr} \big(\hat{S}D_a D_b D_c\big)=1,
\end{equation}
where $D_a=\mathcal{H}^{-1}(k)\partial_{k_a}\mathcal{H}(k),$ and the integral is over the whole BZ. The difficulty in the proposed scheme is to realize the optical potential coupling on the two atomic internal states $\mathbf{B}_Z(\mathbf{r})\cdot\sigma$, which may be achieved with the optical flux lattice method similar as to the one used for the 3D DIII chiral topological insulators \cite{Beri2011}.

\begin{figure}[htbp]\centering
\includegraphics[width=0.9\columnwidth]{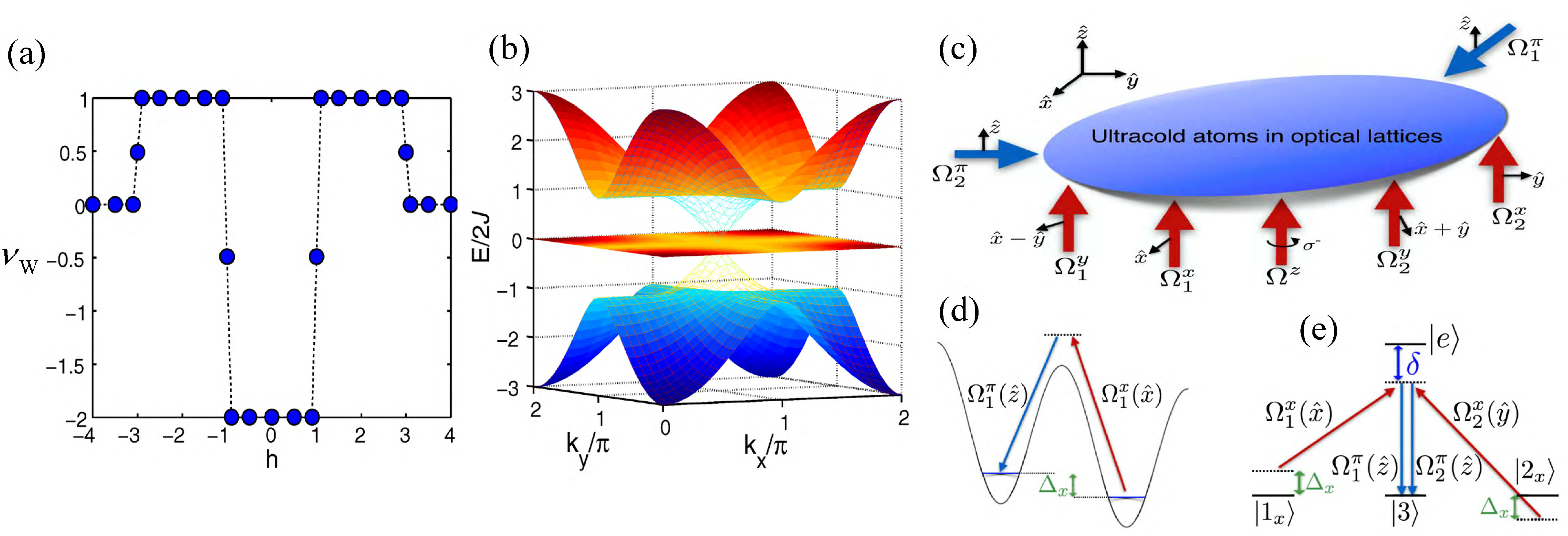}
 \caption{(Color online).  (a) The topological invariant $\nu_{\text w}$ as a function
of $h$. (b) Energy spectrum for three bulk bands
(surface plot) and surface states (mesh plot) at the boundary along
the $z$ direction for $h=2$.
(c) Proposed scheme to realize the Hamiltonian (\ref{AIIIHam}).  A linear tilt $\Delta_{x,y,z}$ per site in the lattice along each
direction. The detuning in $x$ direction matches the frequency
offset of the corresponding Raman beams, which are shown in
panel (d). Polarizations of each beam are
shown in brackets. Rabi frequencies for each beam are:
$\Omega_1^\pi=\Omega_0e^{ikx}$, $\Omega_2^\pi=\Omega_0e^{iky}$, $\Omega_1^x=i\sqrt{2}\Omega_0e^{ikz}$, $\Omega_2^x=-i\sqrt{2}\Omega_0e^{ikz}$,
$\Omega_1^y=-\sqrt{2}\Omega_0e^{ikz}$, $\Omega_1^y=\sqrt{2}\Omega_0e^{ikz}$, and $\Omega_z=2i\Omega_0e^{ikz}$.
Reprinted with permission from Wang {\it et al.}\cite{STWang2014}. Copyright\copyright~ (2014) by the American Physical Society.} \label{AIIICI}
\end{figure}

Another experimental scheme to realize a 3D AIII chiral topological insulator with cold fermionic atoms in an OL was proposed in Ref.\cite{STWang2014}. The proposed model Hamiltonian in momentum space is given by
\begin{equation}\label{AIIIHam}
\mathcal{H}(k)=\begin{pmatrix} 0& 0 & q_1-iq_2\\0 & 0& q_3-iq_0\\q_1+iq_2 & q_3+iq_0 & 0\end{pmatrix},
\end{equation}
where $q_0=2J(h+\text{cos}{k_xa}+\text{cos}{k_ya}+\text{cos}{k_za})$, $q_1=2J\text{sin}{k_xa}$, $q_2=2J\text{sin}{k_ya}$, and $q_3=2J\text{sin}{k_za}$, with $h$ being a tunable parameter. Here $\mathcal{H}(k)$ anticommutes with $\hat{S}$ and thus has a chiral symmetry with the operator $\hat{S}=\text{diag}(1,1,-1)$. Additionally, $\mathcal{H}(k)$ breaks TRS, and thus this model belongs to symmetry class AIII.  $\mathcal{H}(k)$ has three energy bands, with a zero-energy flat band protected by the chiral symmetry and the other two bands having energy dispersion $E_{\pm}=\pm\sqrt{q_0^2+q_1^2+q_2^2+q_3^2}$. This model is characterized by the $\Z$ topological invariant (winding number) \cite{Neupert2012,DLDeng2014}
\begin{equation}
\nu_{\text w}=\frac{1}{12\pi^2}\int_{\text{BZ}}d^3k\epsilon^{\alpha\beta\gamma\rho}\epsilon^{\mu\nu\tau}\frac{1}{E_+^4}q_{\alpha}\partial_{\mu}q_{\beta}
\partial_{\nu}q_{\gamma}\partial_{\tau}q_{\rho},
\end{equation}
where $\epsilon$ is the Levi-Civita symbol with $\{\alpha,\beta,\gamma,\rho\}$ and
$(\mu,\nu,\tau)$ labeling respectively the four components of the vector field $q$ and the three coordinates of the momentum
$\mathbf{k}$. The topological invariant $\nu_{\text w}$ as a function of $h$ is given by
\begin{equation}
\begin{aligned}
\nu_{\text w}(h)=\left\{\begin{matrix}
-2, &|h|<1\\
+1, &1<|h|<3\\
0, &|h|>3.
\end{matrix}
\right.
\end{aligned}
\end{equation}
As shown in Fig. \ref{AIIICI}(a), the system is gapped when $|h|\neq1,3$ and corresponds to the
topological nontrivial phases for $|h|<3$. The band gap closes for $|h|=1,3$, indicating topological quantum phase transitions.
Figure \ref{AIIICI}(b) shows the numerical results of the energy spectrum for the system, which keeps $x$ and $y$ directions in
momentum space with periodic boundaries and $z$ direction in real space with open
boundaries, revealing the macroscopic flat band as well as the surface states with Dirac cones.

To realize this model Hamiltonian, a three-species gas of noninteracting fermionic atoms (denoted by $|1\rangle$, $|2\rangle$ and $|3\rangle$) trapped in a 3D cubic OL is considered. The tight-binding Hamiltonian (\ref{AIIIHam}) in the real space has the following form
\begin{equation}
\begin{aligned}
H&=J\sum_{\mathbf{r}}[(2ihc^{\dagger}_{3,\mathbf{r}}c_{2,\mathbf{r}}+\text{H.c.})+H_{\mathbf{r\hat{x}}}+H_{\mathbf{r\hat{y}}}+H_{\mathbf{r\hat{z}}}],\\
H_{\mathbf{r\hat{x}}}&=ic_{3,\mathbf{r-\hat{x}}}^\dagger(c_{1,\mathbf{r}}+c_{2,\mathbf{r}})-ic_{3,\mathbf{r+\hat{x}}}^\dagger(c_{1,\mathbf{r}}-c_{2,\mathbf{r}})+\text{H.c.},\\
H_{\mathbf{r\hat{y}}}&=-c_{3,\mathbf{r-\hat{y}}}^\dagger(c_{1,\mathbf{r}}-ic_{2,\mathbf{r}})+c_{3,\mathbf{r+\hat{y}}}^\dagger(c_{1,\mathbf{r}}+ic_{2,\mathbf{r}})+\text{H.c.},\\
H_{\mathbf{r\hat{z}}}&=2ic^{\dagger}_{3,\mathbf{r-\hat{z}}}c_{2,\mathbf{r}}+\text{H.c.}.
\end{aligned}
\end{equation}
Here  $c_{j,\mathbf{r}}(j=1,2,3)$ denotes the
annihilation operator at the lattice site $\mathbf{r}$
with the spin state $|j\rangle$. The major difficulty for implementing this Hamiltonian is realization of the spin-transferring hopping
terms $H_{\mathbf{r\hat{x}}}$,$H_{\mathbf{r\hat{y}}}$,$H_{\mathbf{r\hat{z}}}$.
In principle, the required hopping can be realized by using the
Raman-assisted tunneling with proper laser-frequency
and polarization selections \cite{Dalibard2011,Galitski2013,Jaksch2003,Miyake2013,Aidelsburger2013,Kennedy2015}.

A scheme to realize this Hamiltonian was proposed in Ref. \cite{STWang2014}. As illustrated in Fig. \ref{AIIICI}(c), the atom-laser coupling configuration was suggested to realize the hopping terms, of which two beams $\Omega_1^{\pi}=\Omega_0e^{ikx}$ and $\Omega_2^{\pi}=\Omega_0e^{iky}$ constitute the $\pi$-polarized lights, propagating respectively along the $x$ and $y$ directions, where $k=2\pi/a$ is the magnitude of the laser wave vector. The other five beams $\Omega_{1,2}^{x,y,z}$ are all propagating along the $z$ direction with the polarizations shown in Fig. \ref{AIIICI}(c).
 Note that the required
broken parity (left-right) symmetry is achieved by titling
the lattice with a homogeneous energy gradient along the
$x$-, $y$-, $z$-directions. A different linear energy shift per site $\Delta_{x,y,z}$ along
different directions is required, such as $\Delta_z\approx1.5\Delta_y\approx3\Delta_x$.
Then the natural hopping is suppressed by the large tilt potential, and
the hopping terms are restored and engineered by applying two-photon Raman
coupling with laser beams of proper configurations.

\subsubsection{Hopf topological insulators}
In general, 3D topological insulators should be protected by certain kinds of symmetries \cite{Schnyder2008,Ryu2010}, such as time-reversal, particle-hole, or chiral symmetries. However, a special class of 3D topological insulators without any symmetry other than the prerequisite $U(1)$ charge conservation was theoretically proposed, called Hopf insulators \cite{Moore2008,DLDeng2013,CYWang2015,CXLiu2017,Ezawa2017}. The Hopf insulator is topologically characterized by a topological invariant termed Hopf index (also known
as Hopf charge or Hopf invariant) as discussed in Sec. \ref{HopfIndex}, and has zero Chern numbers.  A two-band  tight-binding model was first constructed on a cubic lattice to realize a special Hopf insulator with the Hopf index $\chi=1$ \cite{Moore2008}. Subsequently, a class of tight-binding Hamiltonians that realize arbitrary Hopf insulator phases
with any integer Hopf index $\chi$ were suggested \cite{DLDeng2013}. Recently, an experimental scheme to implement
a model Hamiltonian for Hopf insulators and  to measure the Hopf topology in ultracold atomic systems has been proposed in Ref. \cite{DLDeng2018}. The observation of topological links and Hopf fibration associated with Hopf insulators in a quantum simulator has been reported in Ref. \cite{XXYuan2017}.

The model Hamiltonian in momentum space discussed in Ref. \cite{DLDeng2018} is given by
\begin{equation}\label{HopfHam}
\mathcal{H}(\k)=\bf S(\k)\cdot\sigma,~~S(\k)=\eta^{\dagger}\sigma\eta,
\end{equation}
where $\bf S(\k)$ is the pseudospin field. It is defined in terms of the two
complex fields as $\eta=(\bar{\eta}_{\uparrow}^{p},\bar{\eta}_{\downarrow}^{q})^t$ with $p$ and $q$ being coprime integers, $\bar{\eta}_{\uparrow,\downarrow}=\eta_{\uparrow,\downarrow}^*$, where $\eta_{\uparrow}$ and $\eta_{\downarrow}$ are complex numbers given by
\begin{equation}\label{T3-S3}
\eta_{\uparrow}(\k)=\text{sin}{k_x}+i\text{sin}{k_y},\ \ \
\eta_{\downarrow}(\k)=\text{sin}{k_z}+i(\text{cos}{k_x}+\text{cos}{k_y}+\text{cos}{k_z}+h),
\end{equation}
with $h$ being a constant parameter. We introduce the standard $\text{CP}^1$ field $z(\k)=\eta/\sqrt{|\bf S(\k)|}=(z_{\uparrow}(\k),z_{\downarrow}(\k))^t$ and the normalized pseudospin ${\bf\hat{S}}(\k)={\bf S(\k)/|S(\k)|}=z^{\dagger}\sigma{z}$ with $|{\bf S(\k)}|=|\eta_{\uparrow}|^{2p}+|\eta_{\downarrow}|^{2q}$. It is easy to obtain the expression of $\bf\hat{S}(\k)$,
\begin{equation}\label{S3-S2}
\hat{S}_x+i\hat{S}_y={2\eta_{\uparrow}^p\bar{\eta}_{\downarrow}^{q}}/\eta_{+},
\ \ \ \hat{S}_z={\eta_{-}}/{\eta_{+}},
\end{equation}
where $\eta_{\pm}=|\eta_{\uparrow}|^{2p} \pm |\eta_{\downarrow}|^{2q}$. As we can see, the $\text{CP}^1$ field constructed by a four-component vector $N(\k)$ with the configuration $N_1=\text{Re}[z_{\uparrow}(\k)]$, $N_2=\text{Im}[z_{\uparrow}(\k)]$, $N_3=\text{Re}[z_{\downarrow}(\k)]$, and $N_4=\text{Im}[z_{\downarrow}(\k)]$, takes values on the 3D sphere $\mathbb{S}^3$, together with the normalization
condition $\sum_i N_i^2=1$. Therefore, Eq. (\ref{T3-S3}) forms a map $g$: $\mathbb{T}^3\rightarrow \mathbb{S}^3$,
where $\mathbb{T}^3$ is a 3D torus (describing the first BZ). On the other hand,
the normalized pseudospin $\bf\hat{S}(\k)$ expressed as $z^{\dagger}\sigma{z}$ defines a
mapping $f$: $\mathbb{S}^3\rightarrow \mathbb{S}^2$, where the $\mathbb{S}^3$ coordinates $N(\k)=(N_1,N_2,N_3,N_4)$ are mapped to $\mathbb{S}^2$
coordinates $(\hat{S}_x,\hat{S}_y,\hat{S}_z)$. Consequently, the underlying structure of the Hamiltonian (\ref{HopfHam})
constructs a composite map ${\bf\hat{S}}(\k)=f\circ g(\k)$: $\mathbb{T}^3\rightarrow\mathbb{S}^2$ from the BZ to the target
pseudospin space $\mathbb{S}^2$.

The topological properties of the Hamiltonian in Eq. (\ref{HopfHam}) are characterized by the Hopf index (see Sec. \ref{HopfIndex}), which has a simple integral expression \cite{Whitehead1947,Wilczek1983}
\begin{equation}
\nu_H({\hat{\bf S}})=-\int_{\text{BZ}}\mathbf{F}\cdot\mathbf{A}d^3{\bf k},
\end{equation}
where $\bf F$ is the Berry curvature defined as ${\bf F}_{\mu}=\frac{1}{8\pi}\epsilon_{\mu\nu\tau}\hat{\bf S}\cdot(\partial_{\nu}\hat{\bf S}\times\partial_{\tau}\hat{\bf{S}})$ with $\epsilon_{\mu\nu\tau}$ being the Levi-Civita symbol and $\partial_{\nu,\tau}\equiv\partial_{k_{\nu,\tau}}(\mu,\nu,\tau\in\{x,y,z\})$, and $\bf A$ is the Berry connection that satisfies $\nabla\times{\bf A}=\bf F$. One can prove \cite{DLDeng2013} that the Chern number $C_{\mu}=0$ in all three directions, and the Hopf index takes all integer values $\mathbb{Z}$ and has an analytic expression with $\nu_H(\hat{\bf S})=\pm pq$ when $1<|h|<3$, $\nu_H (\hat{\bf S})=\pm 2pq$ when $|h|<1$, $\nu_H(\hat{\bf S})=0$ otherwise. As we discussed before, $\hat{\bf S}(\k)$ is a composition
of two maps ${\bf\hat{S}}(\k)=f\circ g(\k)$. The generalized Hopf map $f$
from $\mathbb{S}^3\rightarrow \mathbb{S}^2$ has a known Hopf index $\chi(f)=\pm pq$ \cite{Whitehead1947}. Thus, we can decompose the composition map, $\nu_H(\hat{\bf S})=\nu_H(f)\Lambda(g)$, where $\Lambda(g)$ is the topological invariant classifying the maps $g$ from $\mathbb{T}^3\rightarrow\mathbb{S}^3$
\begin{equation}
\Lambda(g)=\frac{1}{12\pi^2}\int_{\text{BZ}}d{\bf k}\epsilon_{\mu\nu\rho\tau}\frac{{\epsilon_{\alpha\beta\gamma}}}{|\bs\eta|^4}\eta_{\mu}\partial_{\alpha} \eta_{\nu}\partial_{\beta}\eta_{\rho}\partial_{\gamma}\eta_{\tau}.
\end{equation}
Here $\bs{\eta}=(\text{Re}[\eta_{\uparrow}(\k)],\text{Im}[\eta_{\uparrow}(\k)],\text{Re}[\eta_{\downarrow}(\k)],\text{Im}[\eta_{\downarrow}(\k)])$. $\Lambda(g)=1$ when $1<|h|<3$, $\Lambda(g)=-2$ when $|h|<1$, and $\Lambda(g)=0$ otherwise. A geometric interpretation of such composition is as follows: $\Lambda(g)$ counts how many times $\mathbb{T}^3$ wraps around $\mathbb{S}^3$ nontrivially under the map $g$ and $\nu_H(f)$ describes how many times $\mathbb{S}^3$ wraps around $\mathbb{S}^2$ under $f$. This composite process ultimately gives the Hopf index $\nu_H(\hat{\bf S})$\cite{DLDeng2013}. Numerical results show that the topologically protected surface states and zero-energy modes
in these exotic nontrivial phases are robust against random perturbations \cite{DLDeng2013,Moore2008}. 

\begin{figure}[htbp]\centering
\includegraphics[width=0.9\columnwidth]{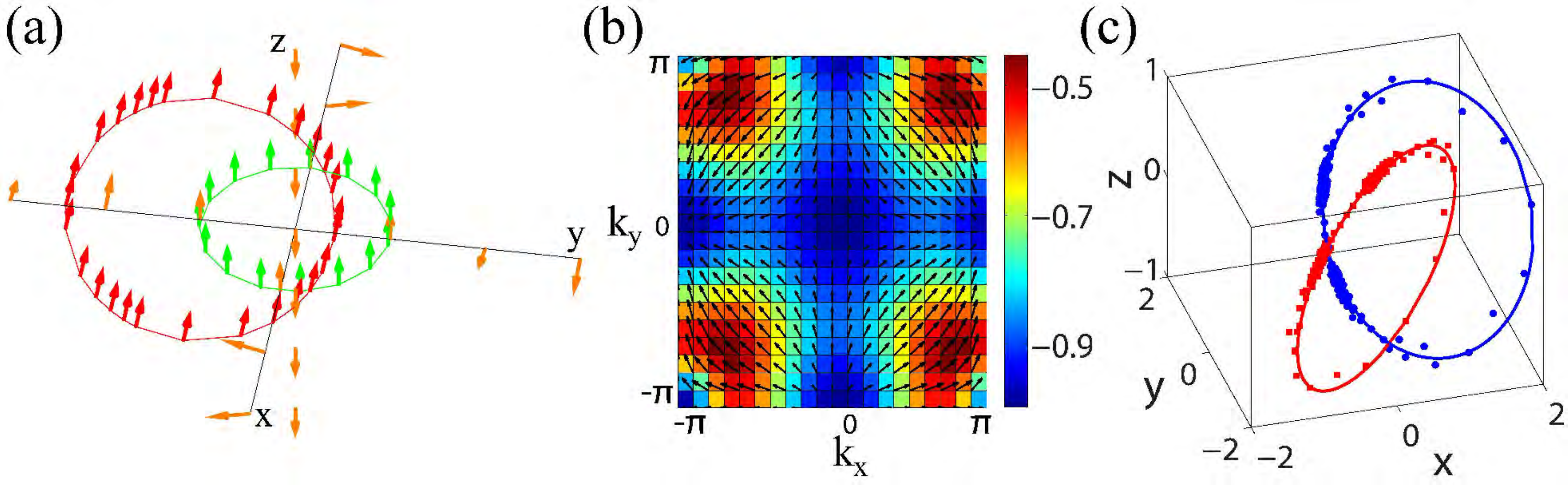}
\caption{(Color online). (a) Hopf links and spin texture in stereographic coordinates.
Spins residing on the red (green) circle point to
the $x$ ($z$) direction and those on the $z$ axis all point to the
south (negative $z$ direction). The red (green) circle represents the preimage
of $\hat{\bf S}_1=(1,0,0)$ ($\hat{\bf S}_2=(0,0,1)$), and the $z$ axis represent the preimage of $(0,0,-1)$.
(b)A cross-section of the measured spin
texture along the $k_z=0$ layer. The background color
scale labels the magnitude of the out-of-plane component $\hat{S}_z$,
and the arrows label the magnitude and direction of spins in
the $k_x$-$k_y$ plane. (c) Topological links between the preimages from two spin states on the Bloch sphere $\hat{\bf S}_1$ and $\hat{\bf S}_2$. The red (blue) circle denotes the theoretical preimage of $\hat{\bf S}_1$ ($\hat{\bf S}_2$) and the scattered red squares (blue dots) are numerically
simulated preimage of the $\epsilon$-neighborhood of $\hat{\bf S}_1$ ($\hat{\bf S}_2$), which can be observed from time-of-flight images.
The parameters are chosen as $p=q=1$ and $h=2$.
Reprinted by permission from Deng {\it et al.}\cite{DLDeng2018}.} \label{Hopflink}
\end{figure}

A geometrical image of the Hopf invariant can be obtained by noting that each point on $\mathbb{S}^2$ has a preimage that is a circle in $\mathbb{T}^3$, and that the linking number of two such circles taken from different points of $\mathbb{S}^2$ is the Hopf invariant $\nu_H(\hat{\bf S})$. To visualize such circles and knots more easily, one can work with $\mathbb{S}^3$ rather than $\mathbb{T}^3$ and probe the Hopf index $\nu_H(f)$. Similarly, the linking number of two preimage contours of distinct spin orientations is equal to the Hopf invariant $\nu_H(f)$. Nevertheless, $\mathbb{S}^3$ is a hypersphere in 4D space $\mathbb{R}^4$ where is difficult to visualize the circles in $\mathbb{S}^3$. So one can visualize the Hopf links by using a stereographic projection of $\mathbb{S}^3$ to $\mathbb{R}^3$, where the topological structure is retained \cite{Lyons2003,Hatcher2002}. The stereographic projection used in Ref. \cite{DLDeng2018} is defined as
\begin{equation}\label{sprojection}
(x,y,z)=\frac{1}{1+\eta_4}(\eta_1,\eta_2,\eta_3),
\end{equation}
where $(x,y,z)$ and $(\eta_1,\eta_2,\eta_3,\eta_4)$ are points of $\mathbb{R}^3$ and $\mathbb{S}^3$, respectively. Stereographic projection preserves circles and maps of Hopf fibers as geometrically perfect circles in $\mathbb{R}^3$, but there is one exception: the Hopf circle containing the projection point $(0,0,0,-1)$ maps to a straight line in $\mathbb{R}^3$ as a ``circle through infinity". Moreover, the preimage map $f^{-1}(\hat{\bf S})$ in $\mathbb{S}^3$ must be determined in order to obtain the stereographic projection of the point on $\mathbb{S}^3$ in $\mathbb{R}^3$ by using Eq. (\ref{sprojection}). A direct parametrization of the 3D-sphere employing the Hopf map is as follows \cite{Whitehead1947}
\begin{equation}\label{pS3}
\eta_{\uparrow}=|\eta_{\uparrow}|e^{iq\theta_1},\ \ \ \eta_{\downarrow}=|\eta_{\downarrow}|e^{ip\theta_2},
\end{equation}
or as follows in Euclidean $\mathbb{R}^4$
\begin{equation}\label{pS31}
\eta_1=|\eta_{\uparrow}|\text{cos}(q\theta_1),\ \ \eta_2=|\eta_{\uparrow}|\text{sin}(q\theta_1),\ \
\eta_3=|\eta_{\downarrow}|\text{cos}(p\theta_2),\ \ \eta_4=|\eta_{\downarrow}|\text{sin}(p\theta_2),
\end{equation}
where $\theta_{1,2}$ runs over the range $0$ to $2\pi$, with $|\eta_{\uparrow}|^2+|\eta_{\downarrow}|^2=1$.
A mapping of the above parametrization to the 2D sphere (according to Eq. (\ref{S3-S2})) is given by
\begin{equation}
\label{pS2}
\hat{S}_x=\frac{2|\eta_{\uparrow}|^p|\eta_{\downarrow}|^q}{\eta_{+}}\text{cos}[pq(\theta_1-\theta_2)],\
\hat{S}_y=\frac{2|\eta_{\uparrow}|^p|\eta_{\downarrow}|^q}{\eta_{+}}\text{sin}[pq(\theta_1-\theta_2)],\
\hat{S}_z=\frac{\eta_{-}}{\eta_{+}}
\end{equation}
 For $\hat{\bf S}_1=(1,0,0)\big(\hat{\bf S}_2=(0,1,0)\big)$, we have $\theta_1=\theta_2\big(\theta_1=\theta_2+\frac{\pi}{2pq}\big)$, $|\eta_{\uparrow}|^{2p}=|\eta_{\downarrow}|^{2q}$. By combining it with the
normalization condition $|\eta_{\uparrow}|^2+|\eta_{\downarrow}|^2=1$, we can obtain the values of $|\eta_{\uparrow}|$ and $|\eta_{\downarrow}|$.
One can easily verify that $|\eta_{\uparrow}|^2=1$, $|\eta_{\downarrow}|^2=0$ $\big(|\eta_{\uparrow}|^2=0,|\eta_{\downarrow}|^2=1\big)$ for $\hat{\bf S}=(0,0,1)$ $\big(\hat{\bf S}=(0,0,-1)\big)$.
With these preparations, we can check that a set of points ${\bf\eta}=(\text{cos}q\theta_1,\text{sin}q\theta_1,0,0)$ $\big({\bf\eta}=(0,0,\text{cos}p\theta_2,\text{sin}p\theta_2)\big)$ forming a ring in $\mathbb{R}^4$ is the preimage of the point $\hat{\bf S}=(0,0,1)\big(\hat{\bf S}=(0,0,-1)\big)$ on $\mathbb{S}^2$. If we denote the stereographic projection $s$: $\mathbb{S}^3\backslash (0,0,0,-1)\rightarrow \mathbb{R}^3$ given in Eq. (\ref{sprojection}), then $s\circ f^{-1}((0,0,1))$ is the unit circle in the $x$-$y$ plane, $s\circ f^{-1}((0,0,-1))$ is the $z$ axis, and for any other point on $\mathbb{S}^2$ not equal to $(0,0,1)$ or $(0,0,-1)$, $s\circ f^{-1}(\hat{\bf S})$
is a circle in $\mathbb{R}^3$ when we choose $p=q=1$. In Fig. \ref{Hopflink}(a), the simplest nontrivial
spin texture corresponding to $\nu_H(f)=1$ is sketched, where the parameters are chosen as $p=q=1$ and $h=2$. This spin texture twisted with $\nu_H(f)=1$ is nontrivial and cannot be continuously untwined unless a topological phase transition occurs. Following the above ideas, one can find more complex knots and links for larger $p$ and $q$, such as the well-known trefoil knot $(p=3,q=2)$ and the Solomon seal knot $(p=5,q=2)$ plotted in Ref. \cite{DLDeng2018} with nonunit knot polynomials \cite{Kauffman2013}.

In the previous discussion, a nonvanishing value of $\nu_H(\hat{\bf S})$ indicates that the pseudospin field $\hat{\bf S}$
has a nontrivial texture that cannot be continuously deformed into a trivial one. Since the spin textures for the model can be interpreted as $\hat{\bf S}=\langle\sigma\rangle$, they can be observed in cold-atom experiments through time-of-flight imaging \cite{Alba2011,DLDeng2014}. Fig. \ref{Hopflink}(b) shows a slice of the observed $\hat{\bf S}$ with $k_z=0$ for the simplest case of $p=q=1$, which provides a glimpse of the 3D twisting of the Hopfion \cite{Faddeev1997}.  With the obtained spin texture, one can reconstruct the topological links and knots by mapping out the preimages of two different orientations of $\hat{\bf S}(\k)$. However, the various kinds of noises involved in a real experiment may lead to inaccurate measurement results of $\hat{\bf S}(\k)$. To simulate real experiments,  the authors discussed the Hamiltonian (\ref{HopfHam}) in real space and considered a finite-size lattice with open boundaries, so the spin orientation $\hat{\bf S}(\k)$ in real experiment is always pixelized with a finite resolution, which means the observed $\hat{\bf S}(\k)$ can only be approximately rather than exactly equal to the specific orientations (such as $\hat{\bf S}_{1,2}$) at any momentum point $\k$. To circumvent these difficulties, one need to consider a small $\epsilon$-neigborhood of a specific orientation (e.g., $\hat{\bf S}_{1}$):
\begin{equation}
N_{\epsilon}(\hat{\bf S}_{1})=\{\hat{\bf S}:|\hat{\bf S}-\hat{\bf S}_{1}|\leq\epsilon\},
\end{equation}
where $|\hat{\bf S}-\hat{\bf S}_{1}|=[(\hat{S}_x-\hat{S}_{1x})^2+(\hat{S}_y-\hat{S}_{1y})^2+(\hat{S}_z-\hat{S}_{1z})^2]^{1/2}$ represents the distance between $\hat{\bf S}$ and $\hat{\bf S}_1$. The preimages of all orientations in $N_{\epsilon}(\hat{\bf S}_{1})$ are then denoted as a set of
$P_{\epsilon}(\hat{\bf S}_1) = (f\circ g)^{-1}[N_{\epsilon}(\hat{\bf S}_{1})]$ describing the points in $\mathbb{T}^3$. Due to the finite resolution and the discrete BZ, $P_{\epsilon}(\hat{\bf S}_1)$ contains finite momentum points. Therefore, one should choose an
appropriate value of $\epsilon$ to ensure that $P_{\epsilon}(\hat{\bf S}_1)$ contains a proper amount of momentum points that could depict the closed loop structure of $(f\circ g)^{-1}(\hat{\bf S}_{1})$. Fig. \ref{Hopflink}(c) shows the simulated Hopf link with linking number one of real experiments. In order to obtain such images, one should first examine the discrete $\hat{\bf S}(\k)$ at each momentum point $\k$ and then append $\k$
to the set of $P_{\epsilon}(\hat{\bf S}_1)$ ($P_{\epsilon}(\hat{\bf S}_2)$) while $\hat{\bf S}(\k)$ is in an $\epsilon$-neighborhood of $\hat{\bf S}_1$ ($\hat{\bf S}_2$). By plotting $g(P_{\epsilon}(\hat{\bf S}_1))$ and $g(P_{\epsilon}(\hat{\bf S}_2))$
in the stereographic coordinate system defined in Eq. (\ref{sprojection}), one can obtain Fig. \ref{Hopflink}(c) in $\mathbb{R}^3$.

On the other hand, with the observed $\hat{\bf S}(\k)$, one can directly extract the Hopf invariant. Since ${\bf F}_\mu=\frac{1}{8\pi}\epsilon_{\mu\nu\tau}\hat{\bf{S}}\cdot{\partial_{\nu}}\hat{\bf S}\times\partial_{\tau}\hat{\bf{S}}$, one can obtain the discrete Berry curvature $\bf F$ at each pixel of the BZ. Berry connection $\bf A(k)$ can be extracted from $\bf F$ by solving a discretized version of
the electrostatics equation $\nabla\times\bf A=\bf F$ in momentum space with the
Coulomb gauge $\nabla\cdot\bf A=0$. Finally, one can attain the value of the Hopf index $\nu_H(\hat{\bf S})$ by a discrete sum over all momentum points. It was also numerically demonstrated that a finite-size lattice of $10\times10\times10$ is already capable of producing highly accurate estimation of the quantized Hopf index and the detection method remains robust to experimental imperfections and the global harmonic trap \cite{DLDeng2018}.

The physical realization of the Hopf insulators is of great interest but also especially challenging. In principle, the model Hamiltonian of the Hopf insulators in Eq. (\ref{HopfHam}) (with $p=q=1$ as the simplest case) could be realized using the Raman-assisted hopping technique with ultracold atoms in OLs, which will involve a number of laser beams \cite{DLDeng2018}.

\subsubsection{Integer quantum Hall effect in 3D}\label{3DIQHE}

After the discovery of the QHE in 2D systems
\cite{Klitzing1980,Tsui1982}, it was shown that if
there is a band gap in a 3D periodic lattice, the integer QHE can also exhibit when the Fermi energy lies inside the
gap \cite{Halperin1987,Montambaux1990,Kohmoto1992,Koshino2001}. In
the 3D QHE, the Hall conductance in each crystal
plane can have a quantized Hall value defined on a torus spanned by the two quasi-momenta for
the crystal plane. It is hard to obtain the energy spectrum with
band gaps for the emergence of quantized Hall conductivities in 3D
periodic lattices since a motion along the third direction may
wash out the gaps of the perpendicular 2D plane. Therefore up to
now, the 3D QHE has been predicted or observed
only in systems with extreme anisotropy or unconventional toroidal
magnetic fields
\cite{Balicas1995,McKernan1995,Koshino2001,Bernevig2007,Mullen2015}.

\begin{figure}\centering
\includegraphics[width=0.9\columnwidth]{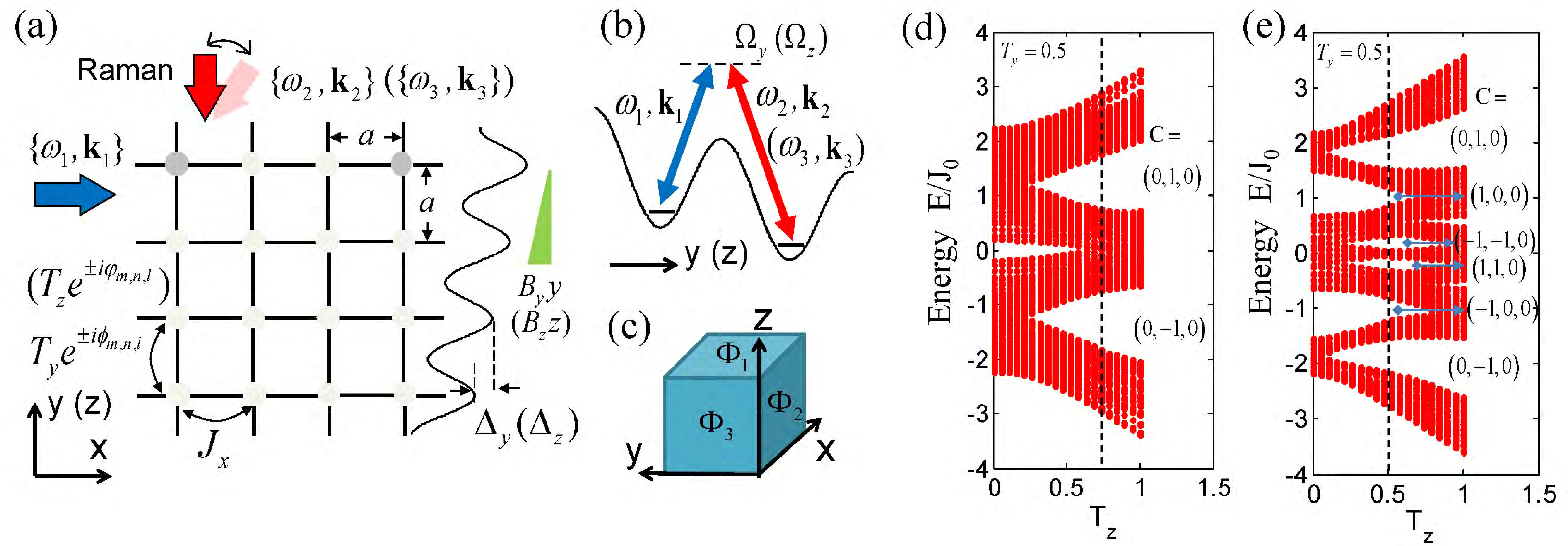}
\caption{(Color online) A cold-atom setup for realizing a
3D generalized Hofstadter model \cite{DWZhang2017}. (a) The
OL and hopping configuration.  Along the $s$ ($s=y,z$) axis, the
tilted lattice with large tilt potentials $\Delta_{s}$ can be
created by magnetic field gradients $B_ss$.
The natural hopping along the $s$ axis is suppressed and then be
restored by using three far-detuned Raman lasers denoted by
$\{\omega_j,\mathbf{k}_j\}$ ($j=1,2,3$), which give rise to
complex hopping amplitudes $T_ye^{\pm i\phi_{m,n,l}}$ and
$T_ze^{\pm i\varphi_{m,n,l}}$ with site indices ($m,n,l$). (b)
Laser-assisted tunneling between nearest neighboring sites along
the $s$ axis with the frequency differences
$\omega_2-\omega_1=\Delta_y/\hbar$ and
$\omega_3-\omega_1=\Delta_z/\hbar$ and the effective two-photon
Rabi frequency $\Omega_s$. (c) The effective magnetic fluxes
$\{\Phi_1,\Phi_2,\Phi_3\}$ in the three elementary plaquettes in
the $\{xy,xz,yz\}$ planes, respectively.  (d) (e) The energy spectra $E$ as a function of
the hopping strength $T_z$  for (d) $\Phi_1=1/2$, $\Phi_2=1/3$, and
$T_y=0.5$; (e) $\Phi_1=1/3$, $\Phi_2=1/5$, and $T_y=0.5$.  The dashed
lines are shown in (d) with $T_z=0.7$ and in (e) with $T_z=0.5$. The Chern numbers $\mathbf{C}=(C_{xy},C_{xz},C_{yz})$ when the Fermi level lies in each energy gap are shown. Reprinted with permission from Zhang {\it et al.}\cite{DWZhang2017}. Copyright\copyright~(2017) by the American Physical Society.}\label{3DQHE}
\end{figure}

A scheme was recently proposed to realize the 3D QHE in a tunable generalized 3D Hofstadter system that can be
simulated by engineering the Raman-assisted hopping of ultracold
atoms in a cubic OL \cite{DWZhang2017}. The optical
lattice is tilted along the $y$ and $z$ axis, as shown in Fig. \ref{3DQHE}(a). The atoms are
prepared in a hyperfine state of the ground state manifold, and
the tilt potentials with linear energy shift per lattice site
$\Delta_s$ ($s=y,z$) can be generated by the gravity or real
magnetic field gradients $B_ss$. For the case
$\Delta_s\gg J_s$ we considered, where $J_s$ denotes the bare
hopping amplitude along the $s$ axis, the atomic hopping between
neighboring sites in these two directions is suppressed. To
restore and engineer the hopping terms with tunable effective
phases, we can use the Raman-assisted tunneling technique, which
has been used to realize the original Hofstadter model in 2D
OLs
\cite{Miyake2013,Aidelsburger2013,Aidelsburger2014}. In order to
fully and independently engineer the atomic hopping along the $y$
and $z$ axes, one can use three
far-detuned Raman beams denoted by their frequencies and wave
vectors $\{\omega_j,\mathbf{k}_j\}$ ($j=1,2,3$), as shown in Fig.
\ref{3DQHE}(b). For resonant tunneling along different
directions, the frequency differences
$\omega_2-\omega_1=\Delta_y/\hbar$ and
$\omega_3-\omega_1=\Delta_z/\hbar$ with $\Delta_y\neq\Delta_z$ are required.
The momentum transfers $\mathbf{Q} = \mathbf{k_1} -
\mathbf{k_2}\equiv(Q_x,Q_y,Q_z)$ and $\mathbf{P} = \mathbf{k_1} -
\mathbf{k_3}\equiv(P_x,P_y,P_z)$ can be independently tunable, for
instance, through independently adjusting the angles of the second
and third Raman lasers with the first Raman laser being fixed, as
shown in Fig. \ref{3DQHE}(a). Therefore, the Raman lasers
induce atomic hopping along the $y$ and $z$ axes with tunable,
spatially dependent phases
$\phi_{m,n,l}=\mathbf{Q}\cdot\mathbf{R}=m\phi_x + n\phi_y+l\phi_z$
and $\varphi_{m,n,l}=
\mathbf{P}\cdot\mathbf{R}=m\varphi_x+n\varphi_y+l\varphi_z$,
respectively, where $\mathbf{R}=(ma,na,la)$ denotes the position
vector for the lattice site $(m,n,l)$, $\phi_{x,y,z}=aQ_{x,y,z}$
and $\varphi_{x,y,z}=aP_{x,y,z}$. This system realizes a
generalized 3D Hofstadter Hamiltonian with fully tunable hopping
parameters \cite{DWZhang2017}
\begin{equation} \label{3DHofstadter}
H=-\sum_{m,n,l}[J_x a^{\dagger}_{m+1,n,l} a_{m,n,l} + e^{i\phi_{m,n,l}}(T_y a^{\dagger}_{m,n+1,l} a_{m,n,l}
+T_z a^{\dagger}_{m,n,l+1} a_{m,n,l})
+ \text{H.c.}],
\end{equation}
where $J_x$ is the natural hopping along the $x$ axis,
$T_ye^{i\phi_{m,n}}$ ($T_ze^{i\varphi_{m,l}}$) denotes the
Raman-induced hopping along the $y$ ($z$) axis with the
spatially-varying phase $\phi_{m,n}$ ($\varphi_{m,l}$) imprinted
by the Raman lasers. The hopping strengths $T_s =
\Omega_s\lambda_s$ can also be tuned via the laser intensities,
with $\lambda_s$ denoting the overlap integral of Wannier-Stark
functions between neighbor sites along the $s$ axis. One can
introduce three effective magnetic fluxes
$\{\Phi_1,\Phi_2,\Phi_3\}$ through the three elementary plaquettes
in the $\{xy,xz,yz\}$ planes with the area $S=a^2$, as shown in
Fig. \ref{3DQHE}(c). The effective fluxes, in units of the
magnetic flux quantum, are determined by the phases picked up
anticlockwise around the plaquettes. They are obtained as
$\Phi_1=\frac{\phi_x}{2\pi}$, $\Phi_2=\frac{\varphi_x}{2\pi}$, and
$\Phi_3=\frac{\phi_z-\varphi_y}{2\pi}$, which can be independently
tuned. For certain hopping configurations, the bulk bands of the
system can respectively have Weyl points and nodal loops
\cite{DWZhang2017}, similar as the one proposed in
Ref. \cite{Dubcek2015}. This allows the study of both
nodal semimetal states within this cold atom system. Furthermore,
the system can exhibit the 3D QHE when the Fermi
level lies in the band gaps, which is topologically characterized
by one or two nonzero Chern numbers.

For simplicity, we consider $\Phi_3=0$ and rational fluxes
$\Phi_1=p_1/q_1$ and $\Phi_2=p_2/q_2$, with mutually prime
integers $p_{1,2}$ and $q_{1,2}$. In this case, the Hamiltonian
(\ref{3DHofstadter}) can be block diagonalized as
${H}=\bigoplus {H}_x(k_y,k_z)$, where $k_y$ and $k_z$ are
the quasimomenta along the periodic directions and the decoupled
block Hamiltonian is given by
\begin{equation}
{H}_x(k_y,k_z)=-\sum_m (J_x a^{\dagger}_{m+1} a_{m} +
\text{H.c.}) - \sum_m V_m  a^{\dagger}_{m} a_{m},
\end{equation}
where $V_m=2T_y\cos(2\pi\Phi_1m+k_ya)+2T_z\cos(2\pi\Phi_2m+k_za)$.
The corresponding single-particle wave function $\Psi_{mnl}$ is
written as $\Psi_{mnl}=e^{ik_yy+ik_zz}\psi_m$, and then the
Schr\"{o}dinger equation
$\hat{H}_x(k_y,k_z)\Psi_{mnl}=E\Psi_{mnl}$ reduces to a
generalized Harper equation:
$$-J_x(\psi_{m-1}+\psi_{m+1})-V_m\psi_m = E \psi_m.$$
This 1D reduced tight-binding system with two commensurabilities
$\Phi_1$ and $\Phi_2$ has a period of the least common multiple of
integers $q_1$ and $q_2$ denoted by $\tilde{q}=[q_1,q_2]$. Under
the periodic boundary condition along the $x$ axis, the wave
function $\psi_m$ satisfies $\psi_m=e^{ik_xx}u_m(\mathbf{k})$ with
$u_m(\mathbf{k})=u_{m+\tilde{q}}(\mathbf{k})$. Therefore in a
general case, the spectrum of the three-dimensional system in the
presence of the effective magnetic fluxes consists of $\tilde{q}$
energy bands and each band has a reduced (magnetic) BZ: $-\pi/\tilde{q}a\leq k_x\leq\pi/\tilde{q}a$, $-\pi/a\leq
k_y\leq\pi/a$, and $-\pi/a\leq k_z\leq\pi/a$. In term of the reduced
Bloch wave function $u_m(\mathbf{k})$, one has
$-J_x(e^{ik_x}u_{m-1}+e^{-ik_x}u_{m+1})-V_m u_m = E(\mathbf{k})
u_m$.

It was proven that every quantized invariant on a $d$-dimensional
torus $\mathbb{T}^d$ is a function of the $d(d-1)/2$ sets of Chern numbers
obtained by slicing $\mathbb{T}^d$ by the $d(d-1)/2$ distinct $\mathbb{T}^2$
\cite{Avron1983}. In this 3D Hofstadter system, the topological
invariants for the QHE are given by three Chern
numbers $\mathbf{C}=(C_{xy},C_{xz},C_{yz})$ for three 2D planes,
with $C_{yz}=0$ for the trivial $yz$ plane since $\Phi_3=0$.
As with the approach in Refs. \cite{Montambaux1990,Kohmoto1992},
when the Fermi energy lies in an energy gap between two bands $N$
and $N+1$ in this system, the other two Chern numbers $C_{xs}$
with $s=y,z$ are given by
\begin{equation}
C_{xs}=\frac{1}{2\pi}\sum_{n\leqslant N}\int_{-\pi}^{\pi}dk_{s'}
c_{xs}^{(n)}(k_{s'}),
\end{equation}
where $s'$ denotes replacing $s$ between $y$ and $z$, and the
Chern number $c_{xs}^{(n)}(k_{s'})$ for the $n$-th filling band
(or $n$-th occupied Bloch state) is defined on the torus $T^2$
spanned by $k_x$ and $k_s$:
$c_{xs}^{(n)}(k_{s'})=\frac{1}{2\pi}\int_{-\pi/\tilde{q}}^{\pi/\tilde{q}}dk_x\int_{-\pi}^{\pi}dk_{s}
F_{xs}^{(n)}(\textbf{k})$, where $F_{xs}^{(n)}(\textbf{k})$ is the
corresponding Berry curvature as a topological expression as a
generalization of the results in 2D \cite{Thouless1982}. In Figs.
\ref{3DQHE}(d) and (e), the three Chern numbers
$\mathbf{C}=(C_{xy},C_{xz},C_{yz})$ when the Fermi level lies in
each energy gap and  the spectra are plotted. The results demonstrate that the
QHE in this 3D Hofstadter system is topologically
characterized by one or two nonzero Chern numbers.

\subsection{Higher and synthetic dimensions}\label{sythdimention}

An important development in exploration of topological states with
ultracold atomic gases is the concept of ``synthetic dimensions".
As discussed in Sec. \ref{AAHModel}, the topological properties of
1D quasiperiodic OLs described by the
Aubry-Andr\'{e}-Harper model can be mapped to the 2D QHE. A key in the mapping is that a cyclical parameter
adiabatically varying from $0$ to $2\pi$ in the 1D lattice (such
as the phase of the lattice potential) plays the role of the
quasimomentum of the second dimension. In Thouless pumping,
the 1D system constitutes a Fourier component of a 2D quantum Hall
system at each point of the cycle, where an adiabatic periodic
pump parameter also acts as the quasimomentum. In these cases, the
cyclical parameter can effectively be considered a synthetic
dimension under the periodic boundary condition along this
dimension. This approach can be extended to studying topological
systems in higher dimensions $D=d_r+d_s\geqslant3$ in OLs of real spatial dimensions $d_r=\{1,2,3\}$ and synthetic
dimensions $d_s$. For example, it was proposed to simulate 3D Weyl
semimetal physics with cold atoms in $d_r=\{2,1\}$D OLs that are subjected to $d_s=\{1,2\}$D synthetic dimensions
from external cyclical parameters \cite{Ganeshan2015,DWZhang2015}.
More interestingly, the intriguing 4D quantum Hall physics
\cite{ZhangSC2001,XLQi2008} can be explored in a 2D topological
charge pump in 2D OLs with two cyclical parameters,
which together give an effective 4D BZ
\cite{Kraus2013b,Zilberberg2018,Lohse2018}.

\begin{figure}\centering
\includegraphics[width=0.85\columnwidth]{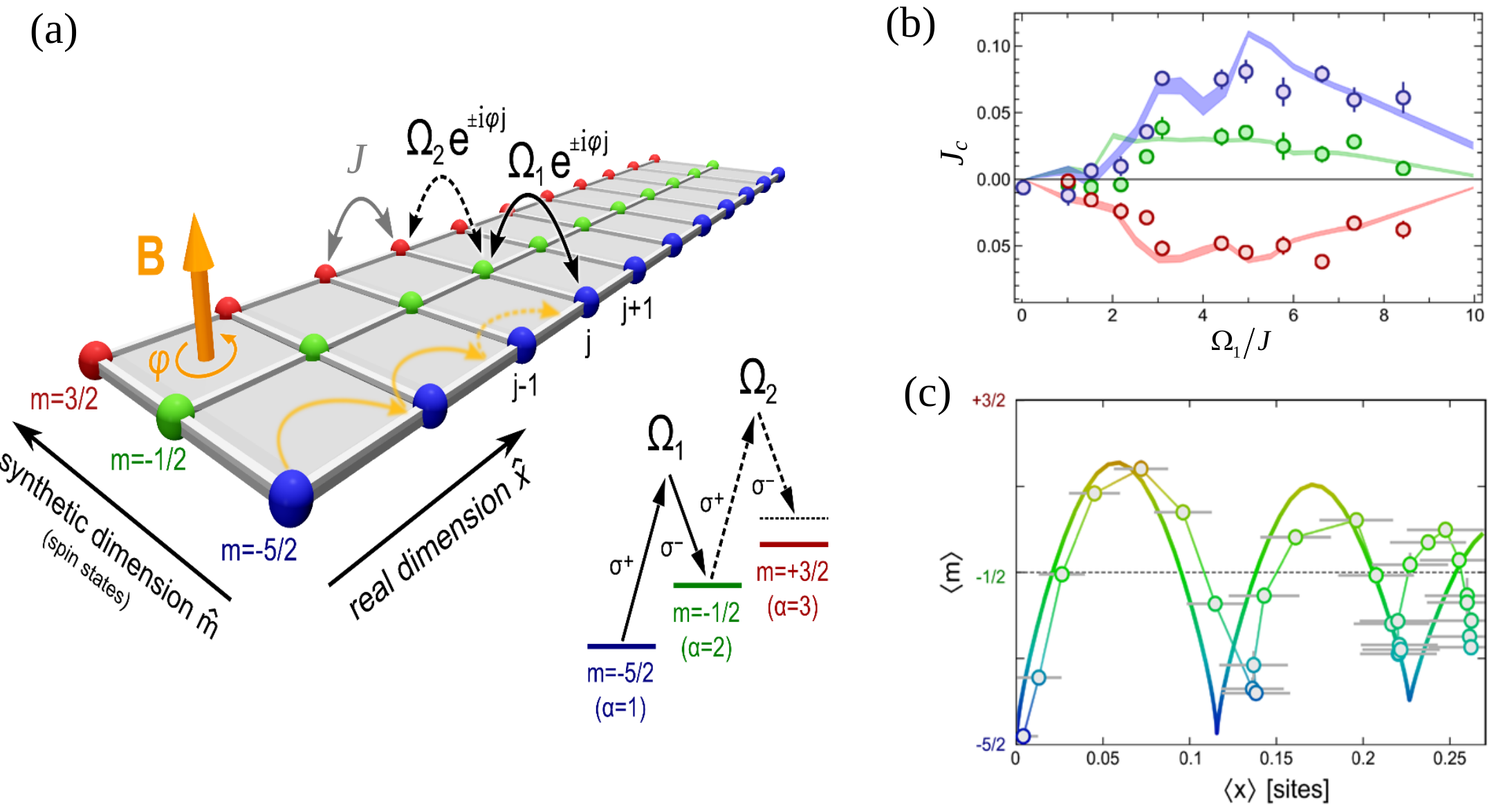}
\caption{(Color online) (a) A synthetic gauge field in a synthetic dimension.
$^{173}$Yb fermionic atoms are loaded in a hybrid lattice,
generated by an OL along a real direction $\hat{x}$
with tunneling $J$, and by a Raman-induced hopping between nuclear
spin states along a synthetic direction $\hat{m}$ with a complex
tunneling $\Omega_{1,2} e^{i\varphi j}$. (b) Experimental observation of chiral edge currents $J_c$. (c) Experimental observation of
edge-skipping orbits in the ladder. Reprinted from Mancini{\it et al.}\cite{Mancini2015}. Reprinted with permission from AAAS.} \label{SyntheticDim}
\end{figure}

Another kind of synthetic dimension is engineered by a set of
discrete internal atomic (spin) states as fictitious lattice sites
\cite{Boada2012,Celi2014}. In this approach, the atoms loaded into
a $d_r$-dimensional OL can potentially simulate
systems of $D=d_r+1$ spatial dimensions. The hopping processes
along the synthetic dimension can be induced by driving
transitions between different internal states with Raman lasers.
The laser-coupling between two internal atomic states has complex
coupling element, which represents the tunneling matrix in the
synthetic dimension picture. Hence, similar to the
Raman-assisted-tunneling scheme, this fictitious tunneling
contains a complex phase-factor, which can then be used to
simulate synthetic gauge fields in the synthetic dimensions and a
finite strip of the Hofstadter model \cite{Celi2014}. For the
atoms that have the hyperfine spin $F$, the Raman lasers couple
spin state $|m_F\rangle$ to $|m_F\pm1\rangle$, where $m_F$ takes
any value between $-F$ and $F$ with a total of $W=2F+1$
components. This provides the naturally sharp boundaries in the
extra dimension, while it is also possible to create periodic boundary
conditions in the synthetic dimension by using an additional
coupling to connect the extremal internal states. This
differs from the cyclical synthetic dimension previously
introduced. Therefore, the proposed 1D OL
\cite{Celi2014} that combines real and synthetic spaces offers a
key advantage to work with a finite-sized system with sharp and
addressable edges, such as the detection of chiral edge states
resulting from the synthetic magnetic flux.

\begin{figure}\centering
\includegraphics[width=0.9\columnwidth]{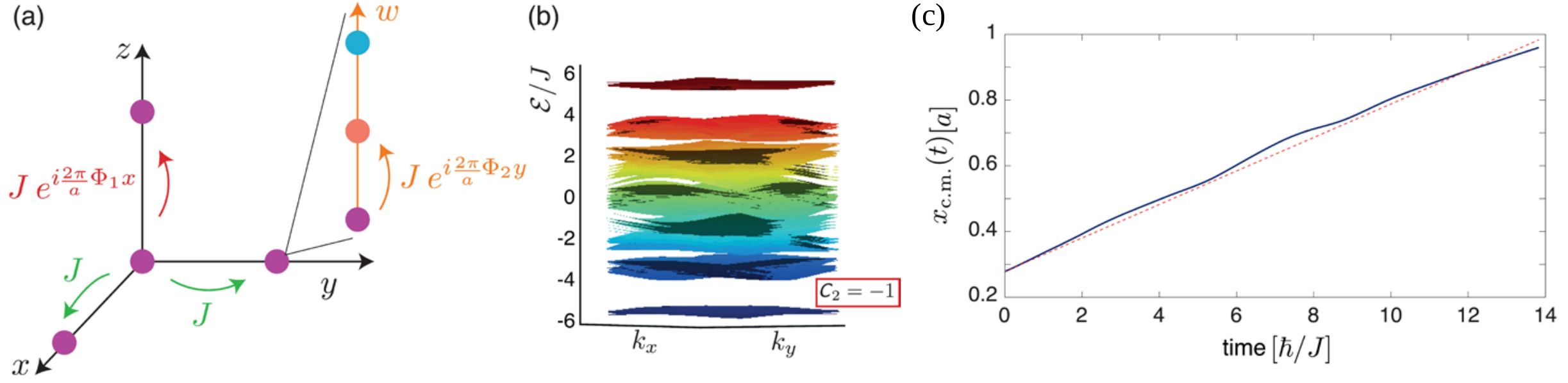}
\caption{(Color online) (a) The synthetic 4D lattice in the absence of perturbing
fields, with atoms in a 3D OL with hopping amplitude
$J$ and flux $2 \pi \Phi_1$ from $x$-dependent Peierls
phase-factors in $z$ hopping. The fourth dimension $w$ is a
synthetic dimension with flux $2 \pi \Phi_2$ in the $y\!-\!w$
plane by adding a $y$-dependent phase-factor to the Raman-induced
internal-state transitions. (b) Energy spectrum
$\mathcal{E}(k_x,k_y)$ for $\Phi_{1,2}\!=\!1/4$ and for many
values $k_{z,w}$, with the second Chern number of the lowest band
$C_2\!=\!-1$. (c) Simulation of the center-of-mass trajectory
$x_{\text{c.m.}} (t)$ after ramping up a perturbing ``electric"
and ``magnetic" field (the blue sold curve), with the predicted
drift for $C_2\!=\!-1$ (the red dotted curve). Reprinted with permission from Price {\it et al.}\cite{Price2015}. Copyright\copyright~(2015) by the American Physical Society.} \label{4D-QHE-Theo}
\end{figure}

The proposed synthetic-dimension scheme \cite{Celi2014} was
recently realized in two independent experiments with $^{173}$Yb
fermionic atoms \cite{Mancini2015} and $^{87}$Rb bosonic atoms
\cite{Stuhl2015}. As shown in Fig. \ref{SyntheticDim}(a), a system
of $^{173}$Yb fermionic atoms in an atomic Hall ribbon of tunable
width pierced by an effective gauge field was experimentally
synthesized \cite{Mancini2015}. One real dimension is realized by
an OL with the tunneling $J$ between different sites
along direction $\hat{x}$. The synthetic dimension is encoded in
the different internal spin states (the sites of the synthetic
dimension for the $F=5/2$ nuclear spin manifold are up to six),
which are coupled by a two-photon Raman transition with a coherent
controllable coupling $\Omega e^{i \varphi x}$ between different
spin components. The phase amounts to the synthesis of an
effective magnetic field with tunable flux $\varphi/2\pi$ (in
units of the magnetic flux quantum) per plaquette, mimicking the
2D Hofstadter model on a three-leg ladder. The Hamiltonian of the
system is given by
\begin{equation}
H= \sum_{j}\sum_{\alpha}\left[-J(c^{\dagger}_{j, \alpha}c_{j+1, \alpha}-\frac{\Omega_\alpha}{2}e^{i\varphi
j}c^\dagger_{j,\alpha}c_{j,\alpha+1}+\textrm{h.c.}) + \mu_j n_{j, \alpha}
 + \xi_\alpha
n_{j,\alpha}\right],
\end{equation}
where $c^\dagger_{j,\alpha}(c_{j,\alpha})$ are fermionic creation
(annihilation) operators on the site $(j,\alpha)$ in the real
($j$) and synthetic ($\alpha = 1, 2, 3$) dimension, and
$n=c^\dagger_{j,\alpha}c_{j,\alpha}$. The hopping parameters
$\Omega_{\alpha}$ are typically inhomogeneous due to the
Clebsch-Gordan coefficients associated with the atomic
transitions. Besides the tunneling terms, $\mu_j$ describes a weak
trapping potential along $\hat{x}$, while $\xi_\alpha$ accounts
for a state-dependent light shift, giving an energy offset along
$\hat{m}$. In the experiment, the chiral currents $J_c$ in the upper
and lower edge chains with opposite sign were observed, while the
central leg showed a suppressed net current in the bulk, as shown
in Fig. \ref{SyntheticDim}(b). This directly signals the existence
of chiral states propagating along the edges of the system,
reminiscent of the edge states in the QHE in the
Hofstadter model. In addition, the edge-skipping orbits with the
cyclotron-type dynamics in the ladder due to the presence of the
synthetic magnetic field were further detected through
state-resolved images of the atomic cloud, as shown in Fig.
\ref{SyntheticDim}(c). In the ladders, the finite-size effect
(such as the overlap of chiral edge modes in different edges)
is significant because the lattice size along the synthetic
dimension is small. To limit undesired finite-size effects, one
may use other atomic species with more addressable internal states
in the synthetic-dimension approach.

The synthetic three-leg ladder with the magnetic flux was also
realized for $^{87}$Rb bosonic atoms, and the chiral edge currents and
skipping orbits were both observed in the quantum Hall regime
\cite{Stuhl2015}. In addition, this synthetic-dimension approach
was demonstrated in a more recent experiment without two-photon
Raman transitions but instead, based on a single-photon optical clock
transition coupling two long-lived electronic states of
two-electron $^{173}$Yb atoms \cite{Livi2016}. These two systems
involve less heating, which would be important for further
studies, such as spectroscopic measurements of the Hofstadter
butterfly and realizations of Laughlin's charge pump.

There is an important difference between ordinary lattice systems
and systems involving a synthetic dimension. In the synthetic
dimension, interactions are generically long-ranged, in
contrast to the on-site interactions along the physical dimension.
With the effectively nonlocal interactions, the cold atom systems
realize extended-Hubbard models, which can display novel phases
due to the intriguing many-body effects unattainable in
conventional condensed matter setups
\cite{Barbarino2015,Ghosh2015,ZYan2015,Barbarino2016}. Such an
interacting fermion gas with multi-spin components coupled through
Raman beams in a 1D OL provides an ideal system to
realize the topological fractional pumping reflected by the
quantization to fractional values of the pumped charge and to
measure the many-body Chern number in a cold-atom experiment
\cite{TSZeng2015,Taddia2017}.


The synthetic-dimension technique offers a novel platform for
exploring topological states in higher dimensions \cite{ZhangSC2001,YLi2013a}, such as
detecting the 4D QHE \cite{Price2015}. In the 2D
QHE, the quantized Hall response induced by an
external electric field is topologically characterized by the
first Chern number. As one of the first predictions of the time-reversal symmetric
topological insulators \cite{ZhangSC2001,XLQi2008}, the QHE can
be generalized to 4D systems. In the 4D QHE, an
additional quantized Hall response appears, which is nonlinear and
described by a 4D topological index, the second Chern number. The
intriguing 4D QHE with the second Chern number was
also theoretically studied in other
models~\cite{XLQi2008,YLi2013,Edge2012,Kraus2013b}. In the
proposal \cite{Price2015}, as shown in Fig. \ref{4D-QHE-Theo}(a),
a synthetic 4D lattice contains atoms hopping in a 3D OL with Raman-coupling internal states as the fourth
dimension $w$, in which the $x\!-\!z$ and $y\!-\!w$ planes are
penetrated by synthetic uniform magnetic fluxes $\Phi_{1,2}$,
respectively. This corresponds to two copies of the Hofstadter
model defined in disconnected planes, described by the
tight-binding Hamiltonian
\begin{equation}
H_{4\text D}=-J \sum_{\bs r} \big( c^{\dagger}_{\bs r + a\bs e_x}c_{\bs r}+c^{\dagger}_{\bs r + a\bs e_y}c_{\bs r} \label{4D_ham}
 + e^{i 2 \pi \Phi_1 x/a} c^{\dagger}_{\bs r + a\bs e_z}c_{\bs
r}+ e^{i 2 \pi \Phi_2 y/a} c^{\dagger}_{\bs r + a\bs e_w}c_{\bs r}\big)
+ \text{h.c.} ,
\end{equation}
where $c^{\dagger}_{\bs r}$ creates a fermion at lattice site $\bs
r\!=\!(x,y,z,w)$. To realize this Hamiltonian, one requires $x$ ($y$)
dependent Peierls phases for tunneling along the $z$ ($w$)
direction, generating a uniform flux $\Phi_1$ ($\Phi_2$) in the
$x\!-\!z$ ($y\!-\!w$) plane. This can be created by combining the
laser-assisted hopping along the $z$ direction and the synthetic
gauge field in the synthetic dimension. As shown in Fig.
\ref{4D-QHE-Theo}(b), the bulk energy spectrum $\mathcal{E}(\bs
k)$ of the Hamiltonian is reminiscent of the two underlying 2D
Hofstadter models defined in the $x\!-\!z$ and $y\!-\!w$ planes,
where the lowest band can be non-degenerate and well isolated from
higher-energy bands for suitable fluxes $\Phi_{1,2}$. Moreover,
the lowest band $\mathcal{E}_1(\bs k)$ is characterized by a
non-zero second Chern number $C_2=-1$ from \cite{Price2015}
\begin{align}
C_2\!=\!\frac{1}{4 \pi^2} \int_{\mathbb{T}^4}
(\Omega^{xy}\Omega^{zw}\!+\! \Omega^{wx}\Omega^{zy}\!+\!
\Omega^{zx}\Omega^{yw}) \text{d}^4k,
\end{align}
where $\Omega^{\mu \nu}$ is the 4D generalized Berry curvature. If
there are additional perturbing ``electric" field $\!E_y$ and
``magnetic"  field $B_{zw}\!= - \!2 \pi \tilde{\Phi}/a^2$, the
current density along the $x$ dimension as a response for a filled
band is given by
\begin{align}
j^x=  \frac{C_2}{4 \pi^2} E_{y} B_{zw} = - \frac{C_2}{2 \pi
a^2} E_{y} \tilde \Phi,
\end{align}
which reveals a genuine non-linear 4D quantum Hall response and is
directly related to $C_2$. Hence, as shown in Fig.
\ref{4D-QHE-Theo}(c), the second Chern number in this system can
be measured from the center-of-mass drift along the $x$ direction
\cite{Price2015,Price2016}:
$x_{\text{c.m.}}(t)=x_{\text{c.m.}}(0)+v_{\text{c.m.}}t=x_{\text{c.m.}}(0)+j^xV_{\text{cell}}t$,
with $V_{\text{cell}}$ as the (magnetic) unit cell volume. For the
neutral atoms, the perturbing ``electric" field corresponds to a
linear gradient that can created either magnetically or optically, and the perturbing ``magnetic" field can be generated by
engineering additional Peierls phases.

\begin{figure}
\includegraphics[width=0.9\columnwidth]{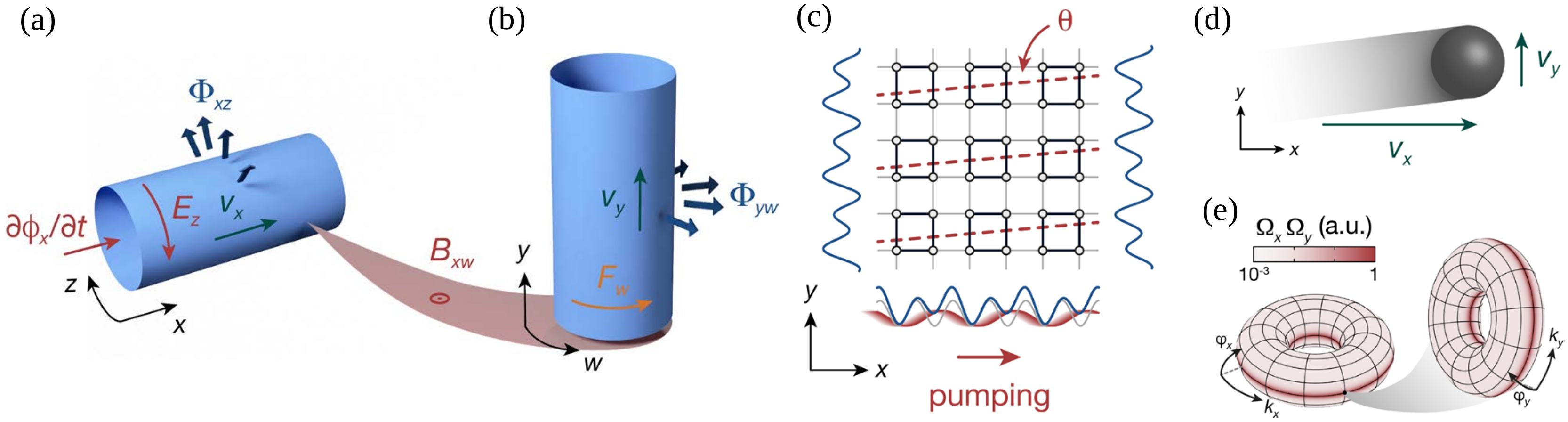}
 \caption{(Color online) 4D quantum Hall system and corresponding 2D topological charge pump.
(a) A 2D quantum Hall system on a cylinder pierced by a uniform
magnetic flux $\Phi_{xz}$, and an electric field $E_z$ on the
surface resulting in a linear Hall response along $x$ with
velocity $v_x$. (b) A 4D quantum Hall system can be composed of
two 2D quantum Hall systems in the $xz$- and $yw$-planes. (c) A
dynamical version of the 4D quantum Hall system can be realized
with a topological charge pump in a 2D superlattice (blue
potentials).
(d) The pumping gives rise to a motion of the atom cloud in the
$x$-direction, corresponding to the quantized linear response of a
2D quantum Hall system.
(e) The velocity of the non-linear response is determined by the
product of the Berry curvatures $\Omega^x \Omega^y$. The left
(right) torus shows a cut at $k_y = 0$, $\varphi_y = \pi/2$ ($k_x
= \pi/(2d_l)$, $\varphi_x = \pi/2$) through the generalized 4D
BZ spanned by $k_{x,y}$ and $\varphi_{x,y}$. Reprinted by permission from Macmillan Publishers Ltd: Lohse {\it et al.} \cite{Lohse2018}, copyright\copyright~(2018).}\label{2DPump-4DQHE}
\end{figure}

Recently, based on a 2D topological charge pump as proposed in
Ref. \cite{Kraus2013b}, a dynamical version of the 4D integer
QHE was realized by using ultracold bosonic atoms
in an angled optical superlattice, and the bulk quantized response
associated with the second Chern number was observed
\cite{Lohse2018}. The dynamical 4D quantum Hall system was also
experimentally realized with tunable 2D arrays of photonic
waveguides \cite{Zilberberg2018}. For the geometry in Figs.
\ref{2DPump-4DQHE}(a,b), the 2D subsystem as a Fourier component
of a 4D quantum Hall system is a square superlattice in Fig.
\ref{2DPump-4DQHE}(c). It consists of two 1D superlattices along
$x$ and $y$, each formed by superimposing two lattices $V_{s,\mu}
\sin^2 \left( \pi \mu / d_{s} \right) + V_{l,\mu} \sin^2 \left(
\pi \mu/d_{l} - \varphi_\mu/2 \right)$, where $\mu \in \{x,y\}$,
$d_{s,\mu}$, and $d_{l,\mu}=2d_{s,\mu}$ denote the lattice periods,
and $V_{s,\mu}$ ($V_{l,\mu}$) is the depth of the short (long)
lattice potential. The superlattice phases
$\varphi_\mu$ determined the position of the long lattices relative
to the short ones, and the phase $\varphi_x$ was chosen as the pump
parameter in the experiment \cite{Lohse2018}. This is equivalent
to threading $\varphi_x$ in the 4D model, which leads to a
quantized motion along $x$ as the linear response shown in Figs.
\ref{2DPump-4DQHE}(c,d). The magnetic perturbation $B_{xw}$
corresponds to a transverse superlattice phase $\varphi_y$ that
depends linearly on $x$, which is realized by tilting the long
$y$-lattice relative to the short one by a small angle $\theta$
in the $xy$-plane, with $\varphi_y (x) = \varphi_y^{(0)} + 2 \pi
\theta \, x/d_{l,y}$ to first order in $ \theta$. As the two
orthogonal axes are coupled, by varying $\varphi_x$, the motion
along $x$ changes $\varphi_y$. This is analogous to the Lorentz
force in 4D and induces a quantized non-linear response along $y$
\cite{Kraus2013b}. For a uniformly populated band in an infinite
system, the change in the center-of-mass position $\Delta{\bs r}$
during one cycle $\varphi_x = 0 \rightarrow 2\pi$ is given by
\cite{Lohse2018}
\begin{equation}
\Delta{\bs r}=C_1^x \, d_{l,x} \, \mathbf{e}_x + C_2 \, \theta
\, d_{l,x} \, \mathbf{e}_y.
\end{equation}
The first term proportional to the pump's first Chern number
$C_1^x$ describes the quantized linear response in the
$x$-direction. The second term is the non-linear response in the
$y$-direction, which is quantified by a 4D integer topological
invariant, the pump's second Chern number
\begin{equation}
C_2 = \frac{1}{4 \pi^2}\oint_{BZ} \Omega^{x} \Omega^{y}
\mathrm{d}k_x \mathrm{d}k_y \mathrm{d}\varphi_x
\mathrm{d}\varphi_y.
\end{equation}
The generalized 4D BZ is shown in Fig.
\ref{2DPump-4DQHE}(e), and the Berry curvature $\Omega^{\mu}
(k_{\mu}, \varphi_{\mu}) = i \left(
\langle{\partial_{\varphi_{\mu}} u}|{\partial_{k_{\mu}} u}\rangle
- \langle{\partial_{k_{\mu}} u}|{\partial_{\varphi_{\mu}}
u}\rangle\right)$, with $|{u (k_{\mu}, \varphi_x)}\rangle$ as the
eigenstate of a given non-degenerate band at $k_{\mu}$ and
$\varphi_{\mu}$. In the experiment \cite{Lohse2018}, the 2D
topological charge pump was realized with bosonic $^{87}$Rb atoms
forming a Mott insulator in the superlattice. The 4D-like
nonlinear response of the lowest subband with $C_2=+1$ was
experimentally observed from the atomic center-of-mass shift after
a pump circle. Furthermore, using a small cloud of atoms as a
local probe, the 4D geometric properties and the quantization of
the response were fully characterized via {\sl in situ} imaging
and site-resolved band mapping. This work paves the way for exploring
higher-dimensional quantum Hall systems with additional strongly
correlated topological phases, exotic collective excitations and
boundary phenomena such as isolated Weyl fermions
\cite{ZhangSC2001,XLQi2008}.

Finally we note that the concept of synthetic dimension has been
greatly extended, which can involve other kinds of degrees of
freedom. For instance, the synthetic dimensions may also be
engineered by a set of orbital angular momenta for light
\cite{XWLuo2015} or harmonic oscillator eigenstates for cold atoms
in a harmonic trap \cite{Price2017} as fictitious lattice sites.
Moreover, it was proposed to create an effective synthetic lattice
of sites in momentum space based on discrete momentum states of
neutral atoms, which can be parametrically coupled with
interfering Bragg laser fields \cite{Gadway2015,Meier2016a}. The
synthetic momentum-space lattice has been realized with cold atoms
and opened up new prospects in the experimental study of
disordered and topological systems
\cite{Meier2016,FAn2017a,FAn2017b,FAn2018,Meier2018}.

\subsection{Higher-spin topological quasiparticles}

In the previous sections, we focus on the spin-1/2 systems, such as the Dirac and Weyl fermions, which have rich topological features. Quasiparticles with higher spin numbers are also fundamentally important but rarely studied in condensed-matter physics or artificial systems \cite{Bradlyn2016,ZLan2011,LLiang2016}. These systems can potentially provide a quantum family to find relativistic quasiparticles that have no high-energy analogs, such as integer-(speudo)spin fermionic excitations. Recently, a series of work theoretically
predicted that unconventional fermions beyond the Dirac-Weyl-Majorana classification (also termed ``new fermions", which means no elementary particle analogs) can emerge in some band structures \cite{Bradlyn2016,HMWeng2016}. These works have set off a boom in investigating and realizing ``new fermions" in condensed matter and artificial systems \cite{BQLv2017,YQZhu2017,YXu2017a,XSTan2018,HPHu2017,Fulga2017,Fulga2017b,HPHu2018}.

A recent work to implement the pseudospin-1 fermions in cold-atom systems was  proposed in Ref. \cite{YQZhu2017}. In this paper, the authors constructed 2D and 3D tight-binding models realizable with cold fermionic atoms in OLs, where the low-energy excitations are effectively described by the spin-1 Maxwell equations in the Hamiltonian form. The Hamiltonian of these low-energy excitations is given by
\begin{equation}
H_M=v_xk_x\hat{S}_x+v_yk_y\hat{S}_y+v_zk_z\hat{S}_z,
\end{equation}
where $\hat{S}_\beta=(\hat{S}_{\alpha\gamma})^\beta=i\epsilon_{\alpha\beta\gamma}$,
and $\epsilon_{\alpha\beta\gamma}$ $(\alpha,\beta,\gamma=x,y,z)$ is the Levi-Civita symbol. This so-called Maxwell Hamiltonian $H_M$ originally describes a massless relativistic boson (photon) with spin one. Because quasiparticles in a lattice system are constrained only by certain subgroups (space groups) of the Poincar\'{e} symmetry rather than by Poincar\'{e} symmetry in high-energy physics \cite{Bradlyn2016}, there is the potential to find
free fermionic excitations described by $H_M$ in lattice systems. Such relativistic (linear dispersion) excitations with unconventional integer pseudospins are termed Maxwell fermions. For specificity and without loss of generality, we describe the topological features of Maxwell fermions in 2D and 3D OLs in some detail.

\begin{figure}\centering
\includegraphics[width=0.9\columnwidth]{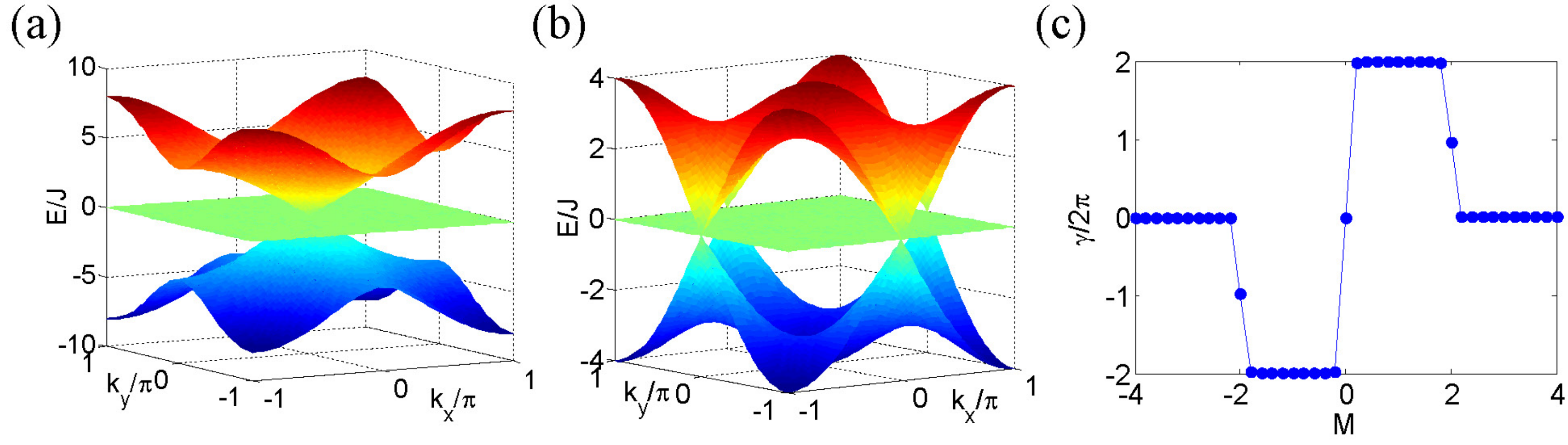}
 \caption{(Color online) The energy spectra and topological features of
the 2D Maxwell lattices \cite{YQZhu2017}. (a) The energy spectrum for $M=2$;
(b) The energy spectrum for $M=0$; (c) The Berry phase $\gamma$ as
a function of the parameter $M$, which corresponds to the
Chern number $\mathcal{C}_1=\gamma/2\pi$ when the 2D system is in the insulating phase with $M\neq0,\pm2$. Reprinted with permission from Zhu {\it et al.}\cite{YQZhu2017}. Copyright\copyright~(2017) by the American Physical Society.
}\label{Maxpd}
\end{figure}

The proposed Bloch Hamiltonian for a 2D case in momentum space is as follows \cite{YQZhu2017}
\begin{equation}
\mathcal{H}(\k)=\mathbf{R}(\k)\cdot\hat{\bf S}
\end{equation}
where the Bolch vector ${\bf R}(\k)=(R_x,R_y,R_z)$ is given by
\begin{equation}\label{MaxBv}
  R_x=2J~\sin k_x,\ \
  R_y=2J~\sin k_y,\ \
  R_z=2J(M-\cos k_x-\cos k_y),
\end{equation}
with $M$ being a tunable parameter. The
energy spectrum of this system is given by $E(\mathbf{k})=0,\pm|\mathbf{R}(\mathbf{k})|$, which has a
zero-energy flat band in the middle of the three bands. The three bands touch at a single point $\mathbf{K}_{\pm}=(0,0)/(\pi,\pi)$ when $M=\pm2$, as shown in Fig. \ref{Maxpd}(a), and touch at two points when $M=0$. The low-energy effective Hamiltonian near $\mathbf{K}_{\pm}$ can be expanded to linear order as
\begin{equation}
\mathcal{H}_{\pm}(q)=\pm v(q_x\hat{S}_x+q_y\hat{S}_y),
\end{equation}
where $v=2J$ is the effective speed of light and
$\mathbf{q}=\mathbf{k}-\mathbf{K}_{\pm}$. These fermionic excitations are described by the Maxwell Hamiltonian
$H_M$ in 2D. In this sense, these low-energy excitations can be named
Maxwell fermions, and the threefold degenerate point as Maxwell point. When the Fermi level lies near the Maxwell point, this system is also named Maxwell metal with a zero-energy flat band. To study the topological properties of Maxwell metal phase, one can evaluate the Berry phase circling around the Maxwell point
$\gamma=\oint_cd\mathbf{k} \cdot \mathbf{\mathbf{F}(k)},$
where the Berry curvature $\mathbf{F(k)}$ for the lower band in the
$k_x$-$k_y$ space has an expression of
\begin{equation}
F_{xy}=-\frac{1}{R^{3}}\mathbf{R}\cdot(\partial_{k_x}{\mathbf{R}}\times\partial_{k_y}{\mathbf{R}}).
\end{equation}
For $M=\pm2$, the gapless points contribute to non-trivial Berry phase $\gamma=\pm2\pi$.
In other words, each pseudospin-1 Maxwell point contributes to an integer Hall conductance when an external synthetic magnetic field along the $z$ axis is applied \cite{ZLan2011}. When $M=0$ and with the spectrum depicted in Fig. \ref{Maxpd}(b),
two Maxwell points touch at $(0,\pi)$ and $(\pi,0)$ with the
effective Hamiltonian
\begin{equation}
\mathcal{H}_0(\mathbf{q})=\pm v(q_x\hat{S}_x-q_y\hat{S}_y).
\end{equation}
In this case, the Berry phase for both Maxwell points is $\gamma=0$, which corresponds to a trivial metallic state.

The system is in an insulating state when $M\neq0,\pm2$ since there is a gap
between any two of the three bands. Then one can calculate the corresponding Chern number $\mathcal{C}_n$ for the
three bands:
\begin{equation}
C_n=\frac{1}{2\pi}\int_{BZ}{dk_xdk_y}F_{xy}(k_x,k_y)=\gamma/2\pi.
\end{equation}
Here $n=-1,0,1$ label the band index corresponding to the lowest, middle, and highest band, respectively. Direct calculation indicates that nonzero Chern numbers $C_{-1}=-C_1=2\text{sgn}(M)$ for $|M|<2$ and
$C_{-1}=C_1=0$ for $|M|>2$, and
thus the zero Chern number $C_0(M)=0$ for the flat band. Figure \ref{Maxpd}(d) shows the phase diagrams of this model characterized by $\gamma/{2\pi}$ of the lowest band as a function of the parameter $M$, which implies topological phase transition with band closing in this system when $M=-2,0,2$.
In addition, a correspondence between the helicity of these edge
states and the polarization of photons was found by investigating the edge modes between the first gap. For the case $0<M<2$, the system is a non-trivial insulator with $C=2$. Under the open boundary condition along the $x$ direction, there are two edge states for each edge and the corresponding effective Hamiltonian is given by
$\H_{\text{edge}}=v_yk_{y}\hat{S}_{y}.$
This edge Hamiltonian is the 1D Hamiltonian of circularly-polarized photons \cite{YQZhu2017}. The helicity operator defined as
\begin{equation}
h=\hat{\mathbf{S}}\cdot{\mathbf{k}}/{|\mathbf{k}|}=\text{sgn}(k_y)\hat{S}_y
\end{equation}
is the projection of the spin along the direction of the linear momentum \cite{ZLan2011}. Hence, the edge
quasiparticle-streams in this Maxwell topological insulator can be
treated as Maxwell fermion-streams with the same helicities
$h\equiv\langle \hat{h}\rangle=+1$ for opposite momenta.

The model Hamiltonian determined by the Bloch vector (\ref{MaxBv}) can be generalized to the 3D model by adding an external term $-2J\cos k_z$ to the $z$-component of ${\bf R}(\k)$. Thus the Bloch Hamiltonian preserves inversion symmetry ($\hat{P}$) represented as
$\hat{P}\mathcal{H}(\mathbf{k})\hat{P}^{-1}=\mathcal{H}(\mathbf{-k})$
and breaks the TRS ($\hat{T}$) since
$\hat{T}\mathcal{H}(\mathbf{k})\hat{T}^{-1}\neq\mathcal{H}(-\mathbf{k}),$
where $\hat{P}=\text{diag}(1,1,-1)$ and $\hat{T}=\hat{I}\K$, with
$\hat{I}=\text{diag}(1,1,1)$ and $\K$ being the complex
conjugate operator. The system is a Maxwell metal for $|M|<3$ and a normal insulator for $|M|>3$.
For simplicity, we consider the typical case of $M=2$, where the band spectrum hosts two Maxwell points in the first BZ at
$\mathbf{M}_{\pm}=\begin{pmatrix}0,0,\pm\frac{\pi}{2}\end{pmatrix}$. The corresponding low-energy
effective Hamiltonian now becomes
\begin{equation} \label{MWHam}
\mathcal{H}_{M_{\pm}}(\mathbf{q})=vq_{x}\hat{S}_{x}+vq_{y}\hat{S}_{y}\pm{vq_{z}\hat{S}_{z}}.
\end{equation}
The two 3D Maxwell points have topological monopoles $\mathcal{C}_{M_{\pm}}=\pm2$, which is defined in terms of a Chern number (defined by the lowest band) on a sphere enclosing the band touching point. There are two Fermi arcs connecting the two points under open boundary condition, which are similar to those in double-Weyl semimetals. The difference between them is that the 3D Maxwell
points in Maxwell metals have linear momenta along all three directions, while the dispersion near double-Weyl
points takes the quadratic form. Besides, the topological stability of Maxwell points is weaker than that of Weyl points. In this proposed model, the band gaps will be opened and Maxwell points will disappear when the inversion symmetry is broken by introducing a perturbation term with one of the other five $SU(3)$ Gell-Mann matrices.

Two different schemes were proposed to realize the spin-1 Maxwell fermions in OLs \cite{YQZhu2017}.
The first scheme is to use non-interacting fermionic atoms in a square or cubic OL and choose
three atomic internal states in the ground state manifold to
encode the three spin states. Using three atomic internal states to form the pseudospin-1 basis leads to the realization of Maxwell fermions in a
lattice of simplest geometry, i.e., a primitive square or cubic
lattice. Moreover, Maxwell fermions can be alternatively realized by using
single-component fermionic atoms in OLs with three
sublattices, where the pseudospin-1 basis is represented by the three
sublattices in a unit cell. Both schemes involve Raman-assisted hopping with proper laser-frequency and polarization selections \cite{YQZhu2017}, which is similar to the method we discussed in Sec. \ref{3DCIScheme}.

Inspired by the investigation of type-II Weyl semimetals, a recent work studied the topological triply-degenerate points \cite{Fulga2017b} in OLs induced by spin-tensor-momentum coupling \cite{HPHu2017}. These triply-degenerate points possess fermionic excitations with effective integer spins. As we mentioned above, a 3D Maxwell point with threefold degeneracy will be destroyed by a small spin-tensor perturbation $\thicksim N_{ij}$. Here the spin-1 matrices $\hat{\bf S}$ are also termed spin-vector-momentum-coupling, and the spin-tensor-coupling matrices $N_{ij}=(\hat{S}_i\hat{S}_j+\hat{S}_j\hat{S}_i)/2$ $(i,j=x,y,z)$ are equivalent to the so-called Gell-Mann matrices, which form a basis of the $SU(3)$ algebra. A simple Hamiltonian of a stable triply-degenerate point induced by a momentum-dependent term $k_i N_{ij}$ is as follows \cite{HPHu2017}
\begin{equation}\label{TDPHam}
\mathcal{H}(\k)=k_x\hat{S}_x+k_y\hat{S}_y+k_z(\alpha \hat{S}_z+\beta N_{ij}),
\end{equation}
where the spin-tensor $N_{ij}$ is coupled to the $k_z$ direction. At $k=0$, $\mathcal{H}(k)$ exhibits a triply-degenerate point with a topological charge $C$. The model Hamiltonian (\ref{TDPHam}) has a symmetry $\mathcal{H}(\k) =
-\mathcal{H}(-\k)$, indicating that the Chern number $C_{1}=-C_{-1}$ for the upper and lower bands and $C_0=0$ for the middle one. For convenience, we use the lower-band Chern number as an topological invariant for labeling triply-degenerate points, i.e., $C=C_{-1}$. For $\beta=0$, such a triply-degenerate point is the 3D Maxwell point we discussed before, which carries the topological monopole
\begin{equation}
C=\frac{1}{2\pi}\oint_{\mathcal{S}}\mathbf{F}_-(\k)\cdot d\mathcal{S}=2\text{sgn}(\alpha),
\end{equation}
where $\mathcal{S}$ is a surface enclosing a triply-degenerate point, and $\mathbf{F}_-(\k)=\nabla\times\langle{u_-(\k)|i\partial_{\k}|u_-(\k)\rangle}=\text{sgn}(\alpha)\k/|\k|^3$. Hereafter, the simplest triply-degenerate point with topological monopoles ${C}=\pm2$ is named type-I triply-degenerate point. In addition, a nonzero $\beta N_{ij}$ term will induce three types of triply-degenerate points \cite{HPHu2017} in Eq. (\ref{TDPHam}). The monopole charge of a type-I triply-degenerate point will not be changed by the three spin-tensors $N_{xx}$, $N_{yy}$, and $N_{xy}$. The tensor $N_{zz}$ induces a type-II triply-degenerate point with $C=\pm1$ for
$|\beta|>|\alpha|\neq0$. A type-III triply-degenerate point with $C=0$ can be induced by the tensor $N_{xz}$ or $N_{yz}$ for $|\beta|>2|\alpha|\neq0$.

It was proposed to realize type-II and type-III triply-degenerate points by coupling three atomic hyperfine states, based on the realization of spin-tensor-momentum coupling and spin-vector-momentum-coupling with spin-1 cold atoms \cite{Anderson2012,Campbell2016,BZWang2018,XWLuo2017}.
The realization of topological monopoles of three different types of triply-degenerate points and investigation of their geometric properties using the parameter space formed by three hyperfine states of ultracold atoms coupled by radio-frequency fields was proposed \cite{HPHu2018}. The Maxwell points has recently been experimentally realized in the parameter space of a superconducting qutrit \cite{XSTan2018}, where the other two types of triply-degenerate points may also be realized in this artificial atom system. Furthermore, there is potential to study even higher spin qusiparticles with untracold atoms. For instance, the spin-3/2 qusiparticles satisfy the so-called Rarita-Schwinger equation in Rarita-Schwinger-Weyl semimetal \cite{LLiang2016}, and Dirac-Weyl fermions with arbitrary spin in 2D optical superlattices \cite{ZLan2011}.

\section{Probing methods}\label{SecV}

Since the atoms are neutral, the traditional transport
measurements in solids that can be used to determine the
topological bands, such as measuring the Chern numbers via the
QHE, are very challenging in cold atomic
systems. Therefore, new methods of probing the topological
states of matter in cold atom systems are needed. On the other
hand, some topological invariants, such as the underlying Berry
curvature as the central measure of topology, could be directly
measured in OL systems, which is not easily
accessible in traditional condensed matter materials. In this
section, we review the methods developed to reveal intrinsic
properties of topological states and phenomena in cold atom systems. We
focus our discussion on the noninteracting atomic gases in OLs, highlighting those experimental probes of topological
invariants that are specific to the single-particle topological
Bloch bands, and will mention extensions to the topological atomic
states with interactions.

\subsection{Detection of Dirac points and topological transition}\label{Detectmethod}

\begin{figure}\centering
\includegraphics[width=0.8\columnwidth]{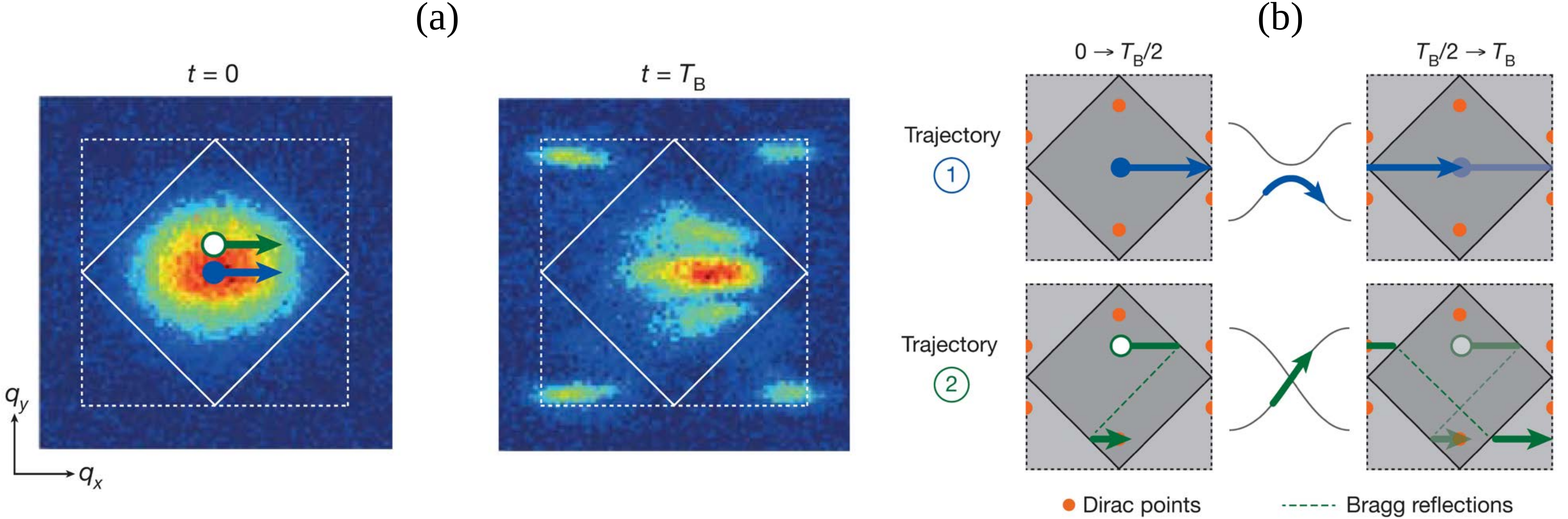}
 \caption{(Color online) Detecting Dirac points from Bloch-Landau-Zener transition.
  (a) The quasi-momentum distribution of the atoms before and after one Bloch oscillation,
  and (b) the corresponding trajectories.
  Reprinted by permission from Macmillan Publishers Ltd: Tarruell {\it et al.}\cite{Tarruell2012}, copyright\copyright~(2012).}
\label{BlochZener}
\end{figure}

It has been demonstrated that the Dirac points (massless Dirac
fermions) in a honeycomb OL can be probed from
measuring the atomic fraction tunnelling to the upper band in
Bloch oscillations \cite{Tarruell2012,Lim2012,Uehlinger2013}, which is
the Landau-Zener transitions between the two energy bands. The
starting point of the experiment is a ultracold gas of fermionic
$^{40}$K prepared in the lowest-energy band of a honeycomb OL. The atomic cloud is then subjected to a constant force
$F$ along the $x$ direction by application of a weak magnetic
field gradient, as the effect is equivalent to that produced by an
electric field in solid-state systems. As shown in Fig.
\ref{BlochZener}, the atoms are accelerated such that their
quasi-momentum $q_x$ increases linearly up to the edge of the
BZ, where a Bragg reflection occurs. The cloud
eventually returns to the centre of the band, performing one full
Bloch oscillation, and the quasi-momentum distribution of the
atoms in the different bands is measured. For a trajectory far
from the Dirac points, the atoms remain in the lowest-energy band.
In contrast, when passing through a Dirac point, the atoms are
transferred from the first band to the second because of the
vanishing energy splitting at the linear band crossing. Thus, the
points (position) of maximum transfer factions can be used to
identify the Dirac points as the transition probability in a
single Landau-Zener event increases exponentially as the energy
gap decreases. Moreover, the topological transition from gapless
(massless Dirac fermions) to gapped bands (massive Dirac fermions)
can also be mapped out by recording the fraction of atoms
transferred to the second band. This
Bloch-Landau-Zener-oscillation technique can be extended to detect
other band-touching points, such as the Weyl points and nodal
lines in 3D OLs
\cite{Zhang2015,WYHe2016,XYMai2017,DWZhang2016a}.

\begin{figure}\centering
\includegraphics[width=0.6\columnwidth]{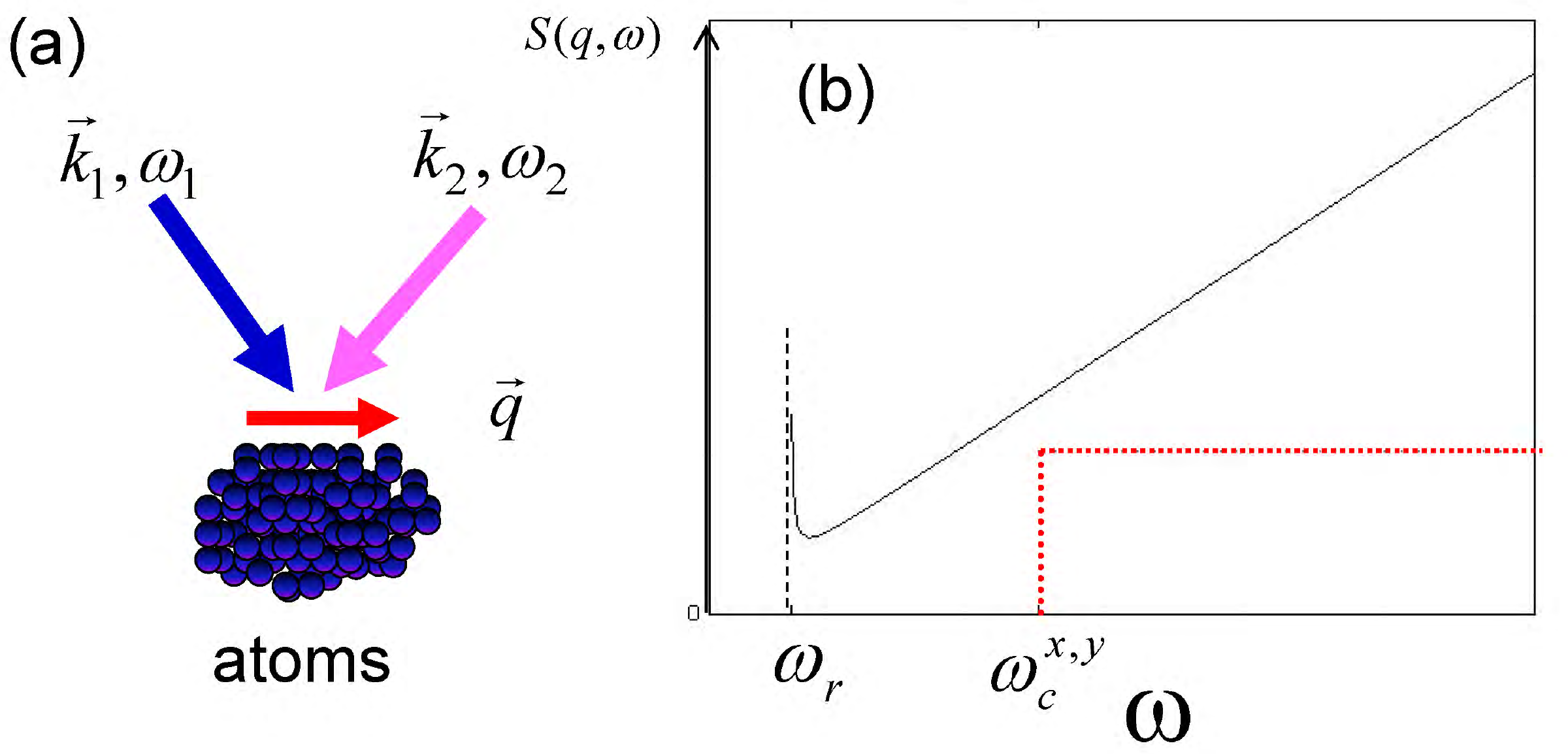}
 \caption{(Color online) Detecting Dirac fermions from Bragg spectroscopy \cite{SLZhu2007}. (a) Schematic of the Bragg scattering. (b) The dynamic structure factors $S(\mathbf{q},\omega)$ for the massless Dirac fermions (solid line) and for the massive ones (dotted line).}
\label{BraggSpe}
\end{figure}

The Bragg spectroscopy can provide an
alternative method to confirm the linear dispersion relation for
the massless Dirac fermions and the energy gap for the massive
ones \cite{SLZhu2007}. As shown in Fig. \ref{BraggSpe}(a), two
laser beams are shined on the atomic gas in the Bragg
spectroscopy, by fixing the angle between the two beams, which
gives rise to the relative momentum transfer
$\mathbf{q=k}_{2}-\mathbf{k}_{1}$, with $\mathbf{k}_{i}$ being the
wave vector of each laser beam. One can then measure the atomic
transition rate by scanning the laser frequency difference
$\omega=\omega _{2}-\omega _{1}$. From the Fermi's golden rule,
this transition rate basically measures the following dynamical
structure factor
\begin{equation}
S(\mathbf{q},\omega )=\sum_{\mathbf{k}_{1},\mathbf{k}_{2}}|\langle f_{%
\mathbf{k}_{2}}|H_{B}|i_{\mathbf{k}_{1}}\rangle |^{2}\delta
\lbrack \hbar \omega -E_{f\mathbf{k}_{2}}+E_{i\mathbf{k}_{1}}],
\end{equation}
where $H_{B}=\sum_{\mathbf{k}_{1},\mathbf{k}_{2}}\Omega
e^{i\mathbf{q\cdot r}}|i_{\mathbf{k}_{1}}\rangle \langle
f_{\mathbf{k}_{2}}|+h.c.$ is the light-atom interaction
Hamiltonian, and $|i_{\mathbf{k}_{1}}\rangle $ and
$|f_{\mathbf{k}_{2}}\rangle $ denote the initial and the final
atomic states with the energies $E_{i\mathbf{k}_{1}}$ and
$E_{f\mathbf{k}_{2}}$ and the momenta $\mathbf{k}_{1}$ and
$\mathbf{k}_{2}$, respectively. At the half filling, the
excitations are dominantly around the Dirac point, and
$S(q,\omega)$ has the expression for the  massless Dirac fermions
\cite{SLZhu2007}
\begin{equation}
S(q,\omega)=\left\{
              \begin{array}{ll}
                0, & \hbox{$\omega\leqslant\omega_r$;} \\
                \frac{\pi \Omega ^{2}}{8v_{F}}\frac{2q_{r}^{2}-q^{2}}{\sqrt{q_{r}^{2}-q^{2}}}\Upsilon(\omega -\omega _{r}), & \hbox{$\omega>\omega_r$.}
              \end{array}
            \right.
\end{equation}
where $\Upsilon$ is the unit step function, $\omega
_{r}=qv_F/\hbar $ ($q\equiv \left| \mathbf{q}\right| $) and
$q_{r}=\hbar \omega /v_F$. For massive Dirac fermions with the
dispersion $E\approx\pm (\Delta_g+\hbar ^{2}q_{x}^{2}/2m_{x}+\hbar
^{2}q_{y}^{2}/2m_{y})$ with the effective mass
$m_{x,y}=\hbar^{2}\Delta_g/v_{_{x},_{y}}^{2}$, the Fermi velocity
$q_{x,y}$ along $x,y$ axis and the energy gap $\Delta_g$, the
dynamical structure factor becomes \cite{SLZhu2007}
\begin{equation}
S(q,\omega)=\left\{
              \begin{array}{ll}
                0, & \hbox{$\omega\leqslant\omega_r$;} \\
                \frac{\pi \Omega ^{2}\Delta_g }{2v_{_{x}}v_{_{y}}}\Upsilon(\omega -\omega _{c}^{x,y}), & \hbox{$\omega>\omega_r$.}
              \end{array}
            \right.
\end{equation}
where $\omega _{c}^{x,y}=2\Delta_g +\hbar
^{2}q_{x,y}^{2}/4m_{x,y}$. As shown in Fig. \ref{BraggSpe}(b), the
dynamical structure factor for massless Dirac fermions has the
lower cutoff frequency $\omega _{r}$ that is linearly proportional
to the momentum difference $q$ and vanishes when $q$ tends to
zero. For massive ones, the lower cutoff frequency $\omega
_{c}^{x,y}$ does not vanish as the momentum transfer goes to zero.
This distinctive difference between the dynamical structure
factors can be used to distinguish the cases with massive or
massless Dirac fermions. Similar Bragg-spectroscopy methods were
proposed to probe the edge and bulk states in 2D Chern insulators
in OLs
\cite{Stanescu2010,XJLiu2010,Goldman2012,Buchhold2012}, such that
the topological phase transition from trivial insulating phase to
quantum anomalous Hall phase can be detected.

\subsection{Interferometer in momentum space}

An interferometric method for measuring Berry's phases and
topological properties of Bloch bands for  ultracold atoms in 2D
OLs was proposed in Ref.\cite{Abanin2013}. The
proposal is based on a combination of Ramsey interference and
Bloch oscillations in the BZ to measure Zak phases,
which can be used to measure $\pi$ Berry's phase of Dirac points
and the first Chern number of topological bands.

\begin{figure}\centering
\includegraphics[width=0.8\columnwidth]{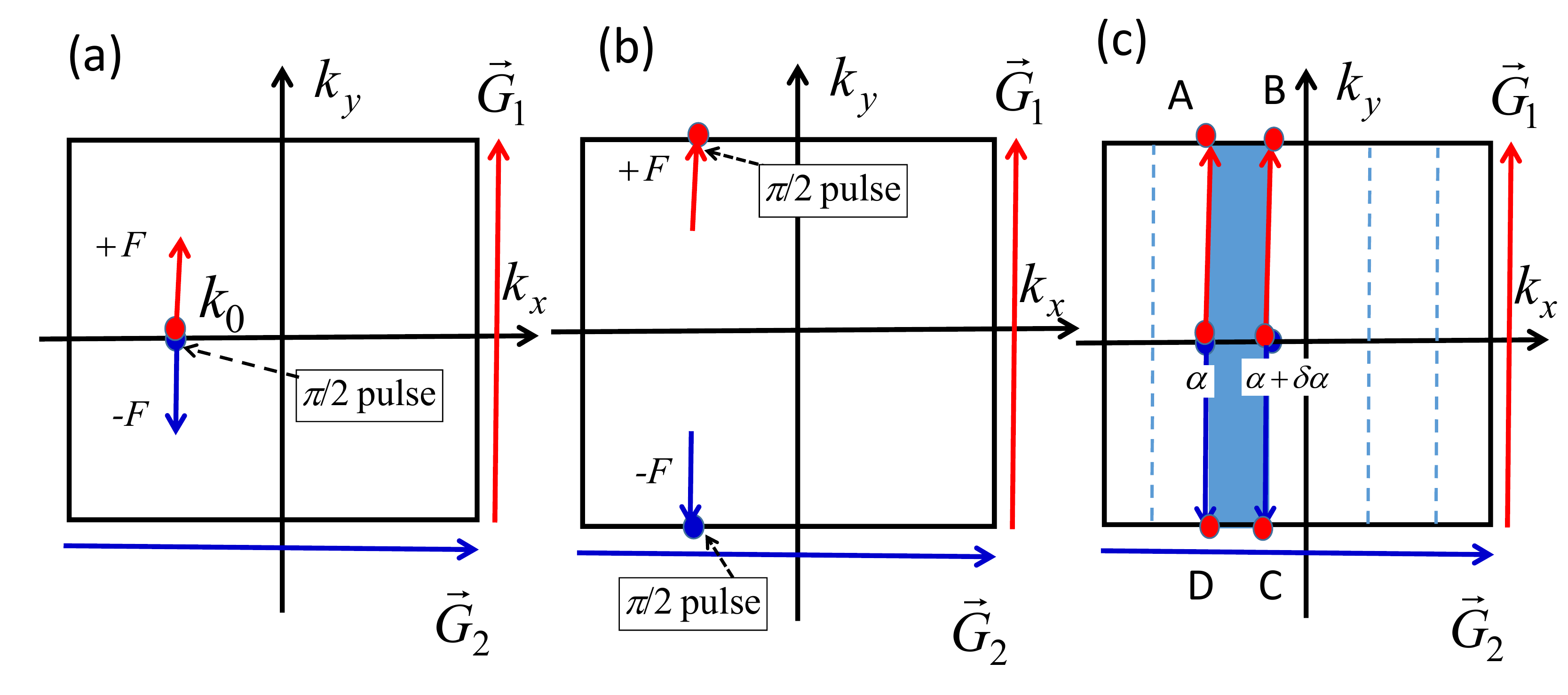}
 \caption{(Color online) An interferometer method to measure the Zak phase and Chern number \cite{Abanin2013}. A cloud of ultracold atoms with a
 well-defined quasi-momentum ${\bf k_0}$ in the spin-up state is initially loaded into a 2D OL, with ${\bf G_1}$ and ${\bf G_2}$ as reciprocal lattice vectors. (a) A $\pi/2$ pulse creates a coherent superposition of $|\!\! \uparrow \rangle$ and $|\!\! \downarrow\rangle$ states. Then spin-selective forces $\pm {\bf F}$ parallel to ${\bf G_1}$ are applied. (b) The two spins meet in the quasi-momentum space after half a period of Bloch oscillations, following by another $\pi/2$ pulse. The accumulated phase difference between the two states containing the Zak phase is measured from the resulting Ramsey fringe. (c) The Chern number of the band can be measured from the Zak phase across the primitive cell. }
\label{Interferometer}
\end{figure}

The scheme for measuring the Zak phase consists of three steps, as
shown in Figs. \ref{Interferometer} (a,b). The atoms are initially
prepared in a spin-up state with a given quasi-momentum ${\bf
k_0}$ in the $n$th band, and a first $\pi/2$ pulse is used to
create a coherent superposition of two spin states denoted by the
wave function $\psi_{{\bf k_0}n}({\bf r}) \otimes
({|\uparrow\rangle+|\downarrow\rangle})/{\sqrt{2}}$. The
eigenfunctions in the $n$th band is written as $\psi_{{\bf
k}n}({\bf r})=e^{i {\bf k r}} u_{{\bf k}n}({\bf r})$, where
$u_{{\bf k} n}$ is the cell-periodic Bloch function, satisfying
$u_{{\bf k} n} ({\bf r+G_i})=u_{{\bf k} n} ({\bf r})$ with two
primitive reciprocal lattice vectors ${\bf G_i}$ ($i=1,2$). Next,
opposite forces $\pm {\bf F}$ on the spin-up and spin-down are
applied parallel to some reciprocal lattice vector ${\bf G_1}$,
which can be created by a magnetic field gradient along the $y$
axis. The atoms then exhibit the Bloch oscillations in the
momentum space, which is assumed to be adiabatic. In this case,
the evolution under the application of the force $\pm {\bf F}$ is
described by the time-dependent wave function $(\Psi_{\uparrow}
({\bf r,t})\otimes {|
\uparrow\rangle}+\Psi_{\downarrow} ({\bf r,t})\otimes
{|\downarrow\rangle})/{\sqrt{2}}$. The wave functions
$\Psi_{\sigma} ({\bf r,t})$ ($\sigma=\uparrow,\downarrow$) obey
the Schr\"{o}dinger equation
$i\hbar{\partial\Psi_{\sigma} ({\bf r},t)}/{\partial
t}=H_{\sigma} \Psi_{\sigma}({\bf r,t}),$
where the Hamiltonian
\begin{equation}
H_{\uparrow,\downarrow}=H_0\mp {\bf F}{\bf r}\pm E_Z, \,\,
H_0=-\frac{\hbar^2}{2m}\nabla^2+V({\bf r}),
\end{equation}
with $V({\bf r})$ being the lattice potential and $E_Z$ being the
Zeeman energy. Under the adiabatic condition, the atoms remain
within the Bloch band. The wave functions take the form:
$\Psi_{\uparrow(\downarrow)} ({\bf r,t})=e^{i\xi
_{\uparrow(\downarrow)}(t)} \psi_{{\bf k_{\pm}(t)},n} ({\bf r}),$
where ${\bf k_{\pm}}(t)={\bf k_0}\pm {\bf f}t$, ${\bf f}={\bf
F}/\hbar$, and the phase $\xi_{\uparrow(\downarrow)}(t)$ is given
by:
\begin{equation}
\xi_{\uparrow(\downarrow)}(t)=i\int_{\bf k_0} ^{{\bf k_{\pm}}(t)}
\langle u_{{\bf k'}n} | \nabla_{\bf k'} u_{{\bf k'}n}\rangle d{\bf
k'}-\frac{1}{\hbar} \int _0^t \epsilon_n ({\bf
k_{\pm}}(t')) dt' \mp \frac{E_Z t}{\hbar}.
\end{equation}
After half a period of the Bloch oscillations (period is given by
$T=|\bf{G}_1 |/|{\bf f}|$), the two spins meet at the edge of the first BZ,
another $\pi/2$ pulse is applied to perform the Ramsey
interferometry, which measures the phase difference picked up by
the two spin species $\xi_{\uparrow}(T/2)-\xi_{\downarrow}(T/2)$.
During such an evolution, the up and down states pick up the
geometric Zak phase, which can be determined from the Ramsey
phase:
\begin{equation}
 \varphi_{\rm tot}=\varphi_{\rm Zak}+\varphi_{\rm dyn}+\varphi_{\rm Zeeman},
\end{equation}
where the Zak phase is given by:
\begin{equation}
\varphi_{\rm Zak}=i\int_{{\bf k_0-G}/2}^{{\bf k_0+G}/2} \langle
u_{{\bf k'}n} | \nabla_{\bf k'} u_{{\bf k'}n}\rangle d{\bf k'}
\end{equation}
and the dynamical phase and Zeeman phase are given by
$\varphi_{\rm dyn}=- \frac{1}{\hbar} \int _{-T/2}^{T/2} {\rm
sgn}(t') \epsilon_n ({\bf k_0}+{\bf f} t') dt'$ and $\varphi_{\rm
Zeeman}=-{E_ZT}/{\hbar}$, respectively. For a band structure with symmetric
dispersion relation, $\epsilon_n ({\bf k_0+{\bf f}}t')=\epsilon_n
({\bf k_0-{\bf f}}t')$, the dynamical phase vanishes. The Zeeman
phase can be also eliminated by a spin echo sequence
\cite{Abanin2013,Atala2013}. Thus, the Ramsey interferometry can
directly give the Zak phase.

The interferometer scheme can be further used to measure the Chern
number of a gapped band for cold atoms in 2D OLs. As
shown in Fig. \ref{Interferometer} (c), the initial quasi-momentum can be prepared to ${\bf k}(\alpha)=\alpha {\bf
G_2}$, where $\alpha\in [0;1)$. Then the Zak phase for a specific $\alpha$ can be measured through the
interferometer protocol. The small change of Zak phase as
$\alpha$ is increased by $\delta \alpha$ is equal to the
integral of the Berry curvature over the rectangle $\delta S$
defined by the corresponding trajectories, which is the Berry's
phase for the contour $ABCDA$. To see this, one can choose a smooth
gauge for the periodic Bloch function in $\delta S$, such that the
Berry's phase $\gamma$ can be represented as the sum of the
Berry's phases for the four sides of the rectangle,
$\gamma=\gamma_{AB}+\gamma_{BC}+\gamma_{CD}+\gamma_{DA}$. Since the sides $AB$ and $CD$ are
equivalent but traversed in the opposite direction, their
contribution vanishes, $\gamma_{AB}+\gamma_{CD}=0$, then
$\gamma_{BC}+\gamma_{DA}$ is equal to the difference of the Zak phases
for trajectories $BC$ and $DA$. Thus, the change of the Zak phase is
related to the Berry phase and is given by an integral of the
Berry curvature:
\begin{equation}
\gamma=\int_{\delta {\cal S}} d^2 k  \Omega({\bf
k})=-ie^{-i\varphi_{\rm Zak}(\alpha)} \partial_{\alpha}
e^{i\varphi_{\rm Zak}(\alpha)} \delta \alpha.
\end{equation}
As the Chern number $C=\frac{1}{2\pi} \int_{\rm BZ} d^2 k
\Omega({\bf k})$, we then obtain $C$ from the winding number of
the Zak phase \cite{Xiao2010},
\begin{equation}
C=-\frac{i}{2\pi} \int _0^1 d\alpha e^{-i\varphi_{\rm
Zak}(\alpha)} \partial_{\alpha} e^{i\varphi_{\rm
Zak}(\alpha)}.
\end{equation}
This relation implies that the Chern number can be extracted from
the interferometric measurements of the Zak phase across the
BZ. The method for a more general lattice structure is introduced in Ref. \cite{Abanin2013}. In addition, the $\pi$ Berry phase of a Dirac point can also be determined from the interferometric measurement
over a trajectory enclosing the point in the momentum space.

The proposed interferometer method has been demonstrated in
cold atom experiments \cite{Atala2013,Duca2015}. The Zak
phase of topological Bloch bands for cold atoms in a 1D dimerized
OL, which realizes the SSH/Rice-Mele Hamiltonians
(see Sec. \ref{SSHModel}), has been directly detected from the
interferometric measurements \cite{Atala2013}. Furthermore, the
atomic interferometer to measure $\pi$ Berry flux of a Dirac point
in momentum space has also been demonstrated \cite{Duca2015},
which is in analogy to an Aharonov-Bohm interferometer that
measures magnetic flux in real space (See Sec. \ref{Diracfermion}).

Based on the proposed measurements of non-Abelian generalizations
of Zak phases (the Wilson loops) or time-reversal polarizations,
the interferometric method by combining the Bloch oscillations
with Ramsey interferometry can be generalized to probe $\Z_2$
topological invariants of time-reversal-invariant topological
insulators realized in OLs \cite{Grusdt2014b}.
Moreover, by using an additional mobile impurities that bind to
quasiparticles of a host many-body system, an interferometric
scheme for detecting many-body topological invariants of
interacting states with topological order, such as the fractional
excitations in fractional quantum Hall systems, was proposed
in Ref. \cite{Grusdt2016}.

\subsection{Hall drift of accelerated wave packets}

For a wavepacket evolving on a lattice, which is centered at
position $\bs r$ with the quasimomentum $\bs k$ and driven by an
external force $\bs F$, the equations of motion are given by
\cite{Xiao2010}
\begin{eqnarray}
\dot{\bs r}_n= \frac{1}{\hbar} \frac{\partial E_n (\bs
k)}{\partial \bs k} - (\dot{\bs k} \times  \bs{e}_z) \Omega_n(\bs
k), ~~ \hbar \dot{\bs k}= \bs F ,
\end{eqnarray}
where $\Omega_n(\bs k)$ is the Berry's curvature of the $n$-band,
and  $ E_n(\bs k)$ is the corresponding band structure. Here the
equations are valid when the force $\bs{F}$ is weak enough to preclude
any inter-band transitions. Considering a 2D lattice and the force
along the $y$ direction $\bs{F}=F \bs{e}_y$, the average velocity
$\bs{v}_{n}=\dot{\bs r}_n$ along the transverse ($x$) direction
can be obtained as
\begin{eqnarray}
v_{n}^x (\bs{k}) = \frac{\partial E_{n} (\bs{k})}{\hbar \partial
k_x} - \frac{F}{\hbar} \Omega_{n} (\bs{k}).
\end{eqnarray}
The first term in the above equation describes the usual band
velocity for Bloch oscillations, and the second term related to
the Berry curvature is the so-called \emph{anomalous} velocity,
which can produce a net drift transverse to the applied force. It
was shown that the anomalous velocity can be isolated and observed
by canceling the contribution from the band velocity through
comparing trajectories for opposite forces $\pm F$
\cite{Price2012}. Following this protocol, a measurement of the
averaged velocity of the accelerated wavepacket for many
trajectories gives the Berry's curvature $\Omega_n(\bs k)$ over a
``pixelated" BZ, and thus the Chern number can be
evaluated by properly adjusting the paths. In the experiment of
realizing the Haldane model with ultracold fermions
\cite{Jotzu2014}, the drift measurement has been performed to
probe the nontrivial Berry curvature and map out the topological
regime of the model (see Sec. \ref{HaldaneModel} ).

It was further shown that the Chern number could be directly
measured by imaging the center-of-mass drift of a Fermi gas as an
effective Hall response to the external force \cite{Dauphin2013}.
Due to the periodicity of the energies in $k$-space,
\begin{equation}
\int_{\mathrm{BZ}} \bigl ( \partial E_{n} (\bs{k})/\partial k_x
\bigr )  \, d^2 k =0,
\end{equation}
so the contribution from the band velocity naturally vanishes by
averaging the velocity over the entire first BZ. By
setting the Fermi energy within a topological bulk gap, the
averaged anomalous velocity can thus be isolated by uniformly
populating the bands, and the displacement along the transverse
direction is directly proportional to the Chern number
\cite{Dauphin2013,Price2016}. Considering a general square lattice
system of size $A_{\mathrm{syst}} \!=\!L_x \!\times\! L_y$ and a
unit cell size $A_{\mathrm{cell}}$, the number of states within
each band is
$N_{\mathrm{states}}\!=\!A_{\mathrm{syst}}/A_{\mathrm{cell}}$, and
the total number of particles is $N_{\mathrm{tot}}=\sum_{n}
N^{(n)}$, where $N^{(n)}$ is the number of particles occupying the
$n$-band. By assuming that each band is populated homogeneously,
the average number of particles in a Bloch state $u_{n}(\bs{k})$
is uniform over the BZ and is given by
\begin{equation}
\rho^{(n)} (\bs{k})=\rho^{(n)}=N^{(n)} / N_{\mathrm{states}},
\end{equation}
which acts as the band filling factor. The total averaged velocity
along the direction transverse to the force is given by
\cite{Dauphin2013}
\begin{eqnarray}
v^x_{\mathrm{tot}}=- ({F A_{\mathrm{cell}}}/{h}) \sum_{n}
N^{(n)}  \, C^{(n)}  ,
\end{eqnarray}
where $C^{(n)}= \frac{1}{2 \pi} \sum_{\bs{k}} \Omega_{n}
(\bs{k}) \, \Delta k_{x}\Delta k_{y}\approx\frac{1}{2 \pi}
\int_{\mathrm{BZ}} \Omega_{n} (\bs{k}) \, d^2 k$ is the band Chern
number, with $\Delta k_{x,y}=2 \pi/L_{x,y}$. This equation reveals
that the averaged transverse velocity of the wave packet is
related to the Chern number, which is a topological invariant and
remains a constant as long as the spectral gaps to other bands do
not vanish. For a free spin-polarized Fermi gas at zero
temperature loaded in a 2D OL with topological bands,
the Fermi energy $E_{\mathrm{F}}$ within a spectral gap naturally
leads to a perfect filling of the bands located below the gap with
$\rho^{(n)}=N^{(n)}/N_{\mathrm{states}} =1$ for $E_{n}<
E_{\mathrm{F}}$. Thus one has
\begin{equation}
v^x_{\mathrm{tot}}=  - ({F A_{\mathrm{syst}}}/{h})
\sum_{E_{n}\!<\! E_{\mathrm{F}}}  C^{(n)},
\end{equation}
which indicates that the total velocity of the Fermi gas is
directly related to the sum of Chern numbers associated with
populated bands.

In cold atom experiments, one can also prepare a thermal Bose gas
filling certain Bloch bands,  such as the lowest band, to measure
such a Hall (transverse) response  \cite{Aidelsburger2014}. If
only the lowest band is filled, its Chern number $C$ can
be simply extracted from the center-of-mass displacement of the
Bose gas, as the transverse velocity of the center-of-mass is
given by
\begin{eqnarray}
v^x_{\mathrm{cm}}=v^x_{\mathrm{tot}}/N_{\mathrm{tot}}= - ({F
A_{\mathrm{cell}}}/{h}) C.
\end{eqnarray}
Since both the unit cell  area $A_{\mathrm{cell}}$ and the
strength of the applied force $F$ can be precisely determined, the
center-of-mass Hall drift $\Delta
x_{\mathrm{cm}}(t)=v^x_{\mathrm{cm}} t$ offers a direct measure of
the Chern number of the lowest band. This protocol of measuring
Chern numbers was successfully implemented with bosonic atoms in
artificially generated Hofstadter bands \cite{Aidelsburger2014}
(See Sec. \ref{HofModel}). This simple scheme is robust against perturbations and could be
applied to any cold-atom setups characterized by nontrivial Chern
numbers. Moreover, this method could be extended to detect the
$\Z_2$ topological states, where the spin Chern number could be
deduced by subtracting the center-of-mass drifts associated with
the two spin species.

\subsection{Streda formula and density profiles}

It was revealed by Str\v{e}da that the Hall conductivity and thus
the first Chern number can be represented as the number of
occupied states in the QHE \cite{Streda1982}.
Considering the Hofstadter model \cite{Hofstadter1976} on a 2D
lattice subjected to a uniform magnetic field $B$ with the
butterfly energy spectrum shown in Fig. \ref{Hofs-Butterlfy}, when
the magnetic flux per plaquette is rational $\Phi=p/q$, this
spectrum splits into $q$ sub-bands $E_{n}$ ($n=1, \dots , q$).
Each bulk band $E_{n} (\bs k)$ is associated with a Chern number
$C^{(n)}$, which remains constant as long as the bulk gaps do
not close. The quantized Hall conductivity of a 2D electron system
can be obtained from the Str\v{e}da formula (let
$h=c=e=1$)\cite{Streda1982}:
\begin{equation}
\sigma_H=\sigma_0C=\sigma_0\sum_{E_n < E_F} C^{(n)}=
\frac{\partial N(E)}{\partial B}\mid_{E=E_F}
\end{equation}
where $\sigma_0$ is the conductivity quantum, the Chern number $C$
includes the contribution of all occupied bulk bands $C^{(n)}$,
and $N$ is the number of states lying below the Fermi energy. The
Str\v{e}da formula is valid when the Fermi energy $E_F$ lies in a
gap.

For neutral atoms, by using the Str\v{e}da formula, it was
proposed to extract the quantized Hall  conductance and thus the
Chern number from the measurement of atomic density profiles
\cite{Umucalilar2008,LBShao2008}. For the OL with
Fermi atoms, this quantized quantity is related to the particle
density $\rho (r)$, which can be directly detected
in practical experiments. For OL with a smooth
confining potential $V_{c} (r)$, the spatial density profile $\rho
(r)$ in the local-density approximation is
\begin{equation}
\rho(r)= \int dE \, D (E) \, \Theta [ E_F - V_{c} (r) - E],
\end{equation}
where $D(E)$ is the homogeneous-system density of states, and the
density $\rho (r)$ actually counts the number of states below the
``local chemical potential" $\mu (r)= E_F - V_{c} (r)$. When the
local chemical potential lies in one of the gaps, we have
$\partial \rho(r)/\partial\mu(r)=0$ because of vanishing
compressibility, and thus the plateaus in the density profile
appear, which correspond to the energy gaps in the energy spectrum
\cite{Umucalilar2008}. The discernible number of plateaus is
related to the size of the energy gaps. Generally, smaller gaps
have higher values of Hall conductance, but it becomes
increasingly harder to observe these gaps, as the corresponding
plateaus will become discernible at lower temperatures and higher
particle numbers. By comparing the density plateaus $\rho_{1,2}$
obtained from two different configurations of the magnetic flux
$\Phi_{1,2}$ but corresponding to the opening of the same bulk gap
in the bulk spectrum, one can obtain the analogue of the Str\v{e}da
formula for the Hofstadter OL with the Chern number
\cite{Umucalilar2008}:
\begin{eqnarray}
C =\frac{\Delta \rho}{\Delta \Phi} = \frac{\rho_2 - \rho_1}{\Phi_2
- \Phi_1} = \sum_{n} C^{(n)}.
\end{eqnarray}
This equation is analogous to the quantized Hall conductivity of
an electronic system with the Fermi energy set within the $n$th
gap. In cold atom experiments, the Chern number can be extracted
by comparing two measurements of atomic densities at different values
of synthetic magnetic flux in the Hofstadter OL.
Notably, it was shown in Ref. \cite{Zhao2011} that the Chern
number could also be revealed in the momentum density $\rho (\bs
k)$ of the same Hofstadter OL.

A simple relation between the Chern number and  the atomic density
plateaus for the anomalous QHE in the Haldane
model (see Sec. \ref{HaldaneModel}) was also obtained from the
Str\v{e}da formula \cite{LBShao2008}. This
density-profile-measurement method for probing the band topology
has also been shown to be applicable in 1D OLs with
non-interacting Fermi atoms \cite{Lang2012a} and interacting Bose
atoms \cite{SLZhu2013}, as well as the chiral topological
insulator in 3D OLs \cite{STWang2014}. The intrinsic
anomalous Hall effect in a Fermi gas loaded in a trap without
OLs may be also observed through the response of
atomic density to the synthetic magnetic field \cite{CZhang2010}.
Thus, the measurement of the atom density plateaus with the analogous
Str\v{e}da formula for cold atoms offers a general way to identify
topological order and phase transitions
\cite{FLi2008,Bermudez2010a}. In practical experiments, one
requires low temperature to make the atomic density plateaus
visible, which is the main obstacle for demonstrating this
detection method. For $^{40}$K atoms in typical OLs,
the plateaus will become visible when the temperature is lower
than about 10 nK \cite{Umucalilar2008,Lang2012a}.

\subsection{Tomography of Bloch states}

A method for full tomography of Bloch vectors was first proposed
for a specific realization of a Haldane-Chern insulator in
spin-dependent hexagonal OLs  \cite{Alba2011}. The Haldane
model \cite{Haldane1988} and some other models of Chern insulators
can be well described by two-band Bloch Hamiltonians of the form
\begin{eqnarray}
\mathcal{H} (\bs k) = \epsilon_0 (\bs k) \mathbf{I}_{2\times2} +
\bs d (\bs k) \cdot \bs{\sigma}, \label{TwoBandHam}
\end{eqnarray}
with the Bloch vectors $\bs d (\bs k)=d_{x,y,z} (\bs k)$. The Berry
curvature $\Omega$ and the Chern number $C$ of the lowest energy
band can be expressed in terms of the normalized Bloch vector $\bs
n (\bs k) = \bs d (\bs k) / \vert \bs d (\bs k) \vert$
\cite{XLQi2008}:
\begin{equation}
\Omega(\bs k) = \frac{1}{2} \bs n \cdot (\partial_{k_x} \bs n
\times \partial_{k_y} \bs n),  \ \ C = \frac{1}{2 \pi}
\int_{\text{BZ}}\Omega(\bs k) d \bs k.
\end{equation}
Based on this equation,  when the system is in a phase $C \ne
0$, an experimental measurement of $\bs n (\bs k)$ would depict a
Skyrmion pattern on a ``pixelated" BZ, leading to an
approximate measurement of the Chern number. For a specific
Haldane-like model that could be realized with fermionic atoms of
two spin states confined on the two triangular sublattices of the
honeycomb pattern, the Bloch vector distribution $\bs n (\bs
k)\propto\langle\bs{\sigma}\rangle$ can be experimentally
determined from spin-resolved time-of-flight images
\cite{Alba2011}. In a typical experiment, after the ground state
is prepared, switching off the trap in adequate timescales
projects the atom cloud into the momentum density distributions
$\rho_{a,b}(\bs k)$, which give the pseudospin component $n_z(\bs
k)=\frac{1}{2}[\rho_a(\bs k) - \rho_b(\bs k)]/[\rho_a(\bs k) +
\rho_b(\mathbf{k})]$. A fast Raman pulse during time-of-flight
allows one to rotate the atomic states and map $n_x$ and $n_y$, which is the tomography of the whole Bloch vector field.
Actual experiments ``pixelize'' the time of flight images,
counting the number of atoms on each ``square'' of the effective
BZ and estimating the averages of $n_x,\,n_y$ or
$n_z$. Either through repetitions or through self-averaging in an
experiment with multiple copies of the lattice, a set of
normalized vectors $\{\mathbf{n}_{j}\}_{j=1}^{L\times L},$ evenly
sampled over momentum space can be obtained, which gives the Chern
number with the error $\mathcal{O}(4\pi^2/L^2)$ expected from the
discretization with the smooth integrand \cite{Alba2011}.

A different scheme for tomography of Bloch states in OLs with two sublattice states was further proposed, which is
based on the quench dynamics and thus is not restricted to a
specific system \cite{Hauke2014}. Consider spinless fermions in a
2D OL with two sublattice states $A$ and $B$, and the
system is described by the two-band Hamiltonian in Eq.
(\ref{TwoBandHam}) with $\bs{\sigma}$ acting in the sublattice
space. For every quasimomentum $\bs k$, the sublattice space
defines a Bloch sphere, with north and south poles given by $|{\bs
k} A\rangle$ and $|{\bs k} B\rangle$, respectively. The normalized
vector on the Bloch sphere is parametrized as ${\bs n} (\bs
k)=(\sin\vartheta_{\bs k}\cos\varphi_{\bs k}, \sin\vartheta_{\bs
k}\sin\varphi_{\bs k},\cos\vartheta_{\bs k})$. Then the Bloch
state of the lowest band is given by $|\bs k-\rangle=
\sin(\vartheta_{\bs k}/2)|\bs k A\rangle-\cos(\vartheta_{\bs
k}/2)e^{i\varphi_{\bs k}}|\bs k B\rangle$. In order to obtain the
full information of the Bloch state determined by $\vartheta_{\bs
k}$ and $\varphi_{\bs k}$, one can measure the momentum
distribution of the system, which is subjected to an abrupt quench
with a potential off-set $\epsilon_A-\epsilon_B\equiv\hbar\omega_{AB}$
between $A$ and $B$ sites for suppressing tunneling at the
measurement time $t_m$, leading to an observable dynamics in the
momentum distribution \cite{Hauke2014}
\begin{equation}
\rho(\bs k,t)=f(\bs k)\{1-\sin\vartheta_{\bs k}\cos[\varphi_{\bs
k}+\omega_{AB}(t-t_m)]\}.
\end{equation}
Here $f(\bs k)$ is a broad envelope function given by the momentum
distribution of the Wannier function. The oscillatory time
dependence in $\rho(\bs k,t)$ directly reveals both $\varphi_{\bs
k}$ and $\sin\vartheta_{\bs k}=1-|n_z(\bs k)|^2$. The
time-dependence of $\rho(\bs k,t)$ allows one to reconstruct
$n_{x,y,z}(\bs k)$ from the amplitude and the phase of the
oscillations. Such a complete tomography of the Bloch states is
mandatory for a measurement of the Berry curvature. It was shown
that this scheme is applicable to extract the Chern number and
topological transitions for the Hofstadter model with $\pi$
flux and the Haldane model on a honeycomb OL
\cite{Hauke2014}.

An alternative but generic scheme for measuring the Bloch
wavefunction based on the time-of-flight imaging was presented in
Ref. \cite{DLDeng2014}. For fermionic atoms with $N$ spin states
referred as $|s\rangle$ ($s=1,2,...,N$) in a generic OL, the Bloch state in the non-degenerate $n$-band can be
denoted by
$|u_{n}(\mathbf{k})\rangle=\sum_{s=1}^{N}c_{ns}(\mathbf{k})|s\rangle$,
where $c_{ns}(\mathbf{k})$ is the Bloch wavefunction with
normalization $\sum_{s}|c_{ns}(\mathbf{k})|^{2}=1$. To measure
$c_{ns}(\mathbf{k})$, one can first separate different spin
components through a magnetic field gradient and directly map out
the atomic momentum distribution
$\rho_{ns}(\textbf{k})=|c_{ns}(\textbf{k})|^{2}$ for the filled
band using the conventional time-of-flight imaging. One then
measure the phase information of $c_{ns}(\textbf{k})$ by
introducing a $\pi/2$ rotation between the two spin states denoted
by $s$ and $s'$ with an impulsive pulse light before the flight of
atoms, which induces the transition
$$c_{ns}(\textbf{k}) \rightarrow
[c_{ns}(\textbf{k})+c_{ns'}(\textbf{k})]/\sqrt{2}, \ \
c_{ns'}(\textbf{k})\rightarrow
[c_{ns}(\textbf{k})-c_{ns'}(\textbf{k})]/\sqrt{2}.$$
With this pulse, the difference between $|c_{ns}(\textbf{k})\pm
c_{ns'}(\textbf{k})|^{2}/2$ measured through time-of-flight
imaging gives the real part of the interference terms
$\text{Re}[c^*_{ns}(\textbf{k})c_{ns'}(\textbf{k})]$. The
imaginary part
$\text{Im}[c^*_{ns}(\textbf{k})c_{ns'}(\textbf{k})]$ can be
obtained by the same way with a different rotation. The
measurement of the population and interference terms determines
the Bloch wave function up to an arbitrary overall phase
$c_{ns}(\textbf{k})\rightarrow
c_{ns}(\textbf{k})e^{i\chi(\textbf{k})}$, where $\chi(\textbf{k})$
in general depends on $\textbf{k}$ instead of the spin index. This
arbitrary $\mathbf{k}$-dependent phase poses an obstacle to
measurement of the topological invariants, which can be overcame
by a gauge-invariant method to calculate the Berry curvature based
on the so-called $U(1)$-link defined for each pixel of the
discrete BZ in experiments \cite{Fukui2005}. It was shown that the proposed method is generally applicable to
probe the topological invariants in various topological bands
\cite{DLDeng2014,DWZhang2017,DLDeng2018} and robust to
typical experimental imperfections such as inhomogeneous trapping
potentials and disorders in the systems.

Recently, the tomography of Bloch states has been experimentally
demonstrated with two different approaches
\cite{Flaschner2016,TLi2016}. Based on the method proposed in Ref.
\cite{Hauke2014}, a full tomography of the Bloch states across the
entire BZ was experimentally demonstrated by observing
the quench dynamics at each momentum point \cite{Flaschner2016}.
In the experiment, a cloud of single-component fermionic $^{40}$K
atoms in a shaking hexagonal OL formed a tunable
Floquet band insulator. Even though the global of the band has
zero Chern number in the system, the measured distribution of
Berry curvature showed the rich topology, such as the phase
vortices as topological defects near Dirac points and their
chiralities as the signal of the topological transition due to the
shaking \cite{Flaschner2016}. The topological defects of the Bloch
states in the hexagonal OL were further
experimentally studied by mapping out the azimuthal phase profile
$\varphi_{\bs k}$ in the entire momentum space and by identifying
the phase windings \cite{Tarnowski2017a}.

The state tomography methods discussed above are applicable for
non-degenerate or isolated bands, where the Berry phase is merely
a number. However, some lattice systems having multiple bands with
degeneracies, such as in topological insulators and graphene, can
seldom be understood with standard Berry phases but can instead be
described using matrix-valued Wilson lines
\cite{RYu2011,Alexandradinata2014,Grusdt2014b,Wilczek1984}. Wilson
lines as non-Abelian generalization of Berry phase
\cite{Wilczek1984} provide indispensable information to identify
the topological structure of bands as they encode the geometry of
degenerate states, such as the eigenvalues of Wilson-Zak loops
(i.e., Wilson lines closed by a reciprocal lattice vector) for
formulating the $Z_2$ topological invariants
\cite{RYu2011,Alexandradinata2014,Grusdt2014b}.

\begin{figure}\centering
\includegraphics[width=0.8\columnwidth]{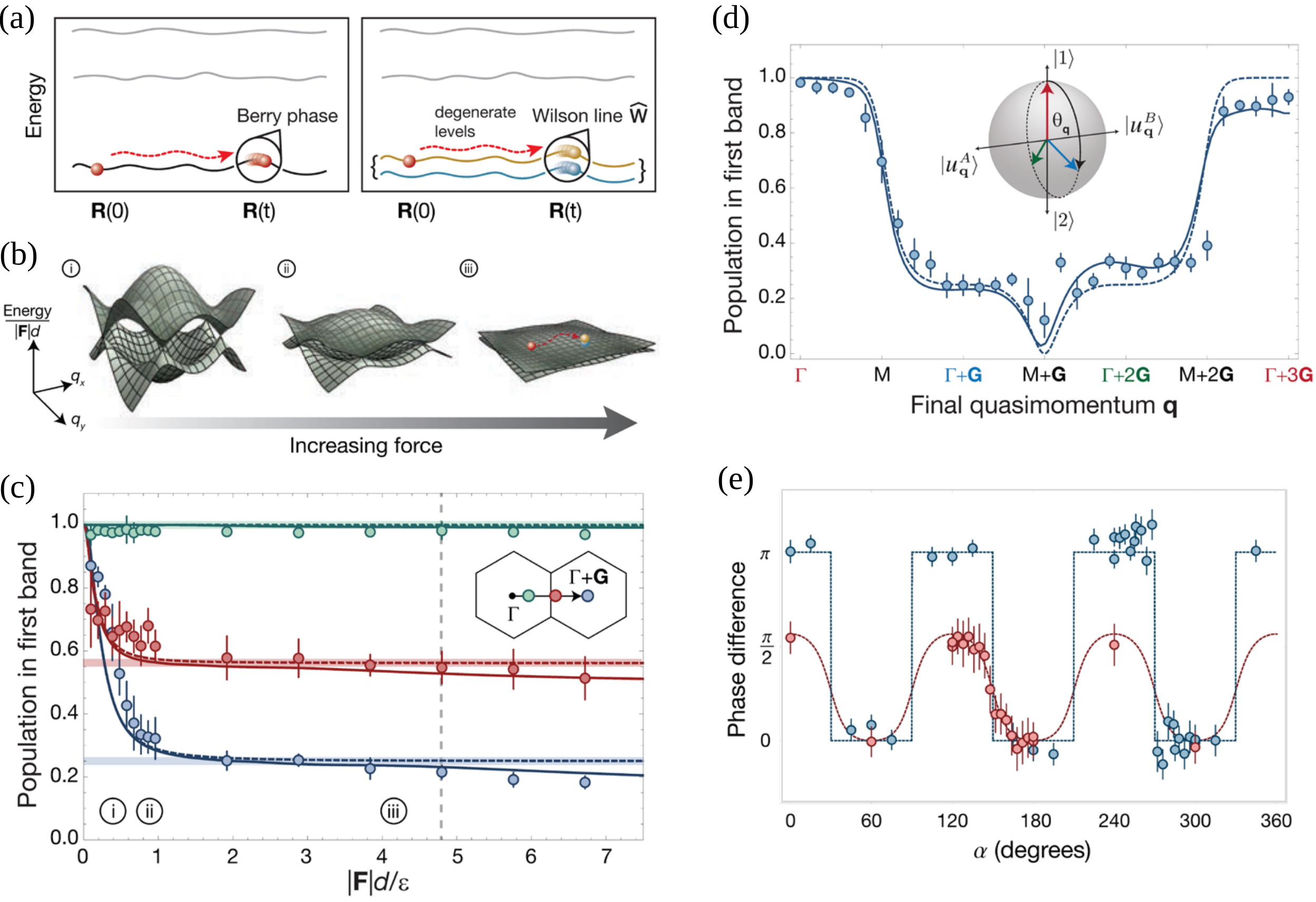}
\caption{(Color online) Realizing the Wilson lines in the honeycomb lattice. (a) In a non-degenerate system (left), adiabatic
evolution of a state through parameter space $\vec{R}$ results in
the acquisition of a Berry phase. In a degenerate system (right),
the evolution is instead governed by the matrix-valued Wilson
line, which lead to population changes between the levels. (b) The
band structure of the lowest two bands of the honeycomb lattice in
effective energy units of $|\mathbf{F}|d$. As the force $\mathbf{F}$ is
increased, the largest energy scale of the bands becomes small. At
large forces (iii), the effect of the band energies is negligible
and the system is effectively degenerate. In this regime, the
evolution is governed by the Wilson line operator. (c) The
measured population remaining in the first band for different
forces after transport to $\Gamma+0.2\mathbf{G}$ (green),
$\Gamma+0.55\mathbf{G}$ (red), and $\Gamma+\mathbf{G}$ (blue),
where inset numbers i to iii refer to band schematics in (b). (d) Measuring mixing angles $\theta_\mathbf{q}$ at different final quasimomenta $\mathbf{q}$. (e) Measuring relative phases $\phi_\mathbf{q}$  at different $\mathbf{q}$, lying at angular coordinate $\alpha$ on a circle centered at $\Gamma$. The quantized jumps of $\pi$ in the phase of the interference fringe each time $\alpha$ is swept through a Dirac point. Reprinted from Li {\it et al.}\cite{TLi2016}. Reprinted with permission from AAAS.} \label{Wilsonline}
\end{figure}

In a recent experiment \cite{TLi2016}, using an ultracold gas of
rubidium atoms loaded in a honeycomb OL, the
strong-force dynamics in Bloch bands that are described by Wilson
lines was realized and an evolution in the band populations for
revealing the band geometry was observed. This enables a full
tomography of band eigenstates using Wilson lines. This approach
can be used to determine the topological invariants in single- and
multi-band systems. As shown in Fig. \ref{Wilsonline}(a), the
Berry phase merely multiplies a state by a phase factor, while the
Wilson line is a matrix-valued operator that can mix state
populations. The Wilson line was measured by detecting changes in
the band populations under the influence of an external force
$\mathbf{F}$, such that atoms with initial quasimomentum
$\mathbf{q}(0)$ evolve to quasimomentum
$\mathbf{q}(t)=\mathbf{q}(0)+\mathbf{F}t/\hbar$ after a time $t$.
When the force is sufficiently weak and the bands are
non-degenerate, the system will remain in the lowest band and the
quantum state merely acquires a Berry phase and a dynamical phase.
Transitions to other bands occur at stronger forces, and when the
force is infinite with respect to a chosen set of bands, the
effect of the dispersion vanishes and the bands appear as
effectively degenerate, as shown in Fig. \ref{Wilsonline}(b).
The system then evolves according to the formalism for adiabatic
motion in a degenerate system \cite{Wilczek1984}, and the dynamics
is described by the unitary time-evolution operator as the Wilson
line matrix \cite{TLi2016}:
\begin{align}
\hat{\mathbf{W}}_{\mathbf{q}(0) \rightarrow
\mathbf{q}(t)}=\mathcal{\hat{P}}\text{exp}[i\int_\mathcal{C}d\mathbf{q}
\hat{\mathbf{A}}_\mathbf{q}],
\end{align}
where the path-ordered $\mathcal{\hat{P}}$ integral runs over the
path $\mathcal{C}$ in reciprocal space from $\mathbf{q}(0)$ to
$\mathbf{q}(t)$ and $\mathbf{\hat{A}_q}$ is the Wilczek-Zee
connection for local geometric properties of the state space. In
the honeycomb OL, the Wilson line operator describing
transport of a Bloch state from $\mathbf{Q}$ to $\mathbf{q}$
reduces to
$\mathbf{\hat{W}}_{\mathbf{Q}\rightarrow\mathbf{q}}=e^{i(\mathbf{q}-\mathbf{Q})\cdot\hat{\mathbf{r}}}$,
and thus the Wilson line operator simply measures the overlap
between the cell-periodic Bloch functions denoted
$|u_{\mathbf{Q}}^n\rangle$ and $|u_{\mathbf{q}}^m\rangle$ (with
the band index $n,m$) at the initial and final quasimomenta
\cite{Alexandradinata2014}:
\begin{align}
W^{mn}_{\mathbf{Q} \rightarrow \mathbf{q}}= \langle
u_{\mathbf{q}}^m|u_{\mathbf{Q}}^n\rangle.
\end{align}
This enables a tomograph of the cell-periodic Bloch functions over
the entire BZ in the basis of the states
$|u_{\mathbf{Q}}^n\rangle$. In the experiment \cite{TLi2016}, a
nearly pure BEC of $^{87}$Rb was initially
loaded into the lowest band at the center of the BZ
$\mathbf{Q}=\Gamma$, and an inertial force, created by accelerating
the lattice via linearly sweeping the frequency of the laser
beams, was used to realize the Wilson line. The Wilson line was
then verified by transporting the atoms from $\Gamma$ to different
final quasimomenta using a variable force $|\mathbf{F}|$ and
performing band mapping to measure the population remaining in the
lowest band, as shown in Fig. \ref{Wilsonline}(c). The
saturation value $|W^{11}_{\Gamma \rightarrow
\mathbf{q}}|^2=|\langle u_{\mathbf{q}}^1|u_{\Gamma}^1\rangle|^2$
of the population after transport to $\mathbf{q}$ is a measure of
the overlap between the Bloch functions of the first band at
$\Gamma$ and $\mathbf{q}$. To demonstrate the reconstruct of Bloch
states using the Wilson lines, it is convenient to represent the
state $|{u_\mathbf{q}^1}\rangle$ in the basis of
$|1\rangle=|u_\mathbf{Q}^1\rangle$ and
$|2\rangle=|u_\mathbf{Q}^2\rangle$ at a fixed reference
quasimomentum $\mathbf{Q}$ as
\begin{align}
|u_\mathbf{q}^1\rangle=\cos\frac{\theta_\mathbf{q}}{2}|1\rangle+\sin\frac{\theta_\mathbf{q}}{2}e^{i\phi_\mathbf{q}}|2\rangle.
\end{align}
Obtaining $\theta_\mathbf{q}$  and $\phi_\mathbf{q}$ for each
quasimomentum $\mathbf{q}$ will map out the geometric structure of
the lowest band \cite{Alba2011,Hauke2014}. As shown in Fig.
\ref{Wilsonline}(d,e), mixing angles $\theta_\mathbf{q}$ at
different final quasimomenta $\mathbf{q}$ was measured from the
atom population remaining in the first band after the transport,
and relative phases $\phi_\mathbf{q}$ at different $\mathbf{q}$
was measured through a procedure analogous to Ramsey or
St{\"u}ckelberg interferometry \cite{TLi2016}. Using the data, the
Bloch states in the lowest band $|u_\mathbf{q}^1\rangle$ and the
eigenvalues of Wilson-Zak loops can both be reconstructed.

\subsection{Spin polarization at high symmetry momenta}

It was proposed that for a class of Chern insulators, the
topological index can be obtained by only measuring the spin
polarization of the atomic gas at highly symmetric points of the
BZ \cite{XJLiu2013}. The two-band Bloch Hamiltonian of
the Chern insulators in square lattices  is given by
\begin{equation}
\mathcal{H}(\mathbf{k})=[m_z-2J_0\cos(k_xa)-2J_0\cos(k_ya)]\sigma_z -2J_{\rm so}\sin(k_{x}a)\sigma_x-2J_{\rm
so}\sin(k_{y}a)\sigma_y,\label{ChHam}
\end{equation}
where the $m_z$ is an effective Zeeman splitting, $J_0$ and
$J_{\rm so}$ represent the nearest-neighbor spin-conserved
and spin-flipped hopping coefficients, respectively. Notably, the
cold-atom realization of this Hamiltonian with the needed 2D
SOC has been theoretically proposed
\cite{XJLiu2014} and then experimentally achieved \cite{ZWu2016}
by a simple optical Raman lattice scheme that applies two pairs of
light beams to create the lattice and Raman potentials
simultaneously. The topology of the lowest Bloch band can be
characterized by the Chern number $C=\mbox{sgn}(m_z)$ when
$0<|m_z|<4J_0$, and otherwise $C=0$.

The lattice system has an inversion symmetry defined by the 2D
inversion transformation $\hat P\otimes\hat R_{\rm 2D}$, where
$\hat P=\sigma_z$ acting on spin space and $\hat R_{2D}$
transforms Bravais lattice vector $\bold R\rightarrow-\bold R$.
For the corresponding Bloch Hamiltonian, one has $\hat P{\cal
H}(\bold k)\hat P^{-1}={\cal H}(-\bold k)$, which follows that
$[\hat P,{\cal H}(\bold \Lambda_i)]=0$ at the four highly
symmetric points $\{\bold \Lambda_i\}=\{G(0,0), X_1(0,\pi),
X_2(\pi,0), M(\pi,\pi)\}$. Therefore the Bloch states
$|u_{\pm}(\bold \Lambda_i)\rangle$ in the two energy bands are
also eigenstates of the parity operator $\hat P$ with eigenvalues
$P_{\pm}=+1$ or $-1$. The topology of the inversion-symmetric
Chern band can also be determined by the following invariant
\cite{XJLiu2014,ZWu2016}
\begin{eqnarray}
\Theta=\prod_i \text{sgn}[P_{-}(\bold \Lambda_i)].
\end{eqnarray}\label{Theta}
It can be proven rigorously that $\Theta=-1$ for the topological
band, while $\Theta=1$ for the trivial band. Furthermore, for the
square lattice with four highly symmetric points, the
corresponding Chern number is given by
$C=-\frac{1-\Theta}{4}\sum_{i=1}^4\text{sgn}[P(\bold \Lambda_i)]$.
Since the parity eigenstates are simply the spin eigenstates (here
$\hat P=\sigma_z$), with the spin-up and spin-down corresponding
to different atomic internal states, the topological invariants
$\Theta$ and $C$ of the lowest Bloch band can be determined by
measuring the spin polarization $P_{(-)}(\mathbf{k})$ of an atomic
cloud at the four highly symmetric points:
\begin{equation}
P_{-}(\boldsymbol{\Lambda_i})=\frac{n_\uparrow(\boldsymbol{\Lambda_i})-n_\downarrow(\boldsymbol{\Lambda_i})}{n_\uparrow(\boldsymbol{\Lambda_i})+n_\downarrow(\boldsymbol{\Lambda_i})},
\end{equation}
where $n_{\uparrow,\downarrow}(\boldsymbol{\Lambda_i})$  denotes
the atomic momentum density of spin states
$|\uparrow,\downarrow\rangle$, which can be measured directly by
spin-resolved time-of-flight imaging. This method can be used to
probe other topological bands with specific symmetries by
measuring atomic spin polarization at highly symmetric points in
momentum space, such as the chiral topological insulators
\cite{XJLiu2016,BSong2018} and double-Weyl semimetals
\cite{XYMai2017}.

\begin{figure}\centering
\includegraphics[width=0.8\columnwidth]{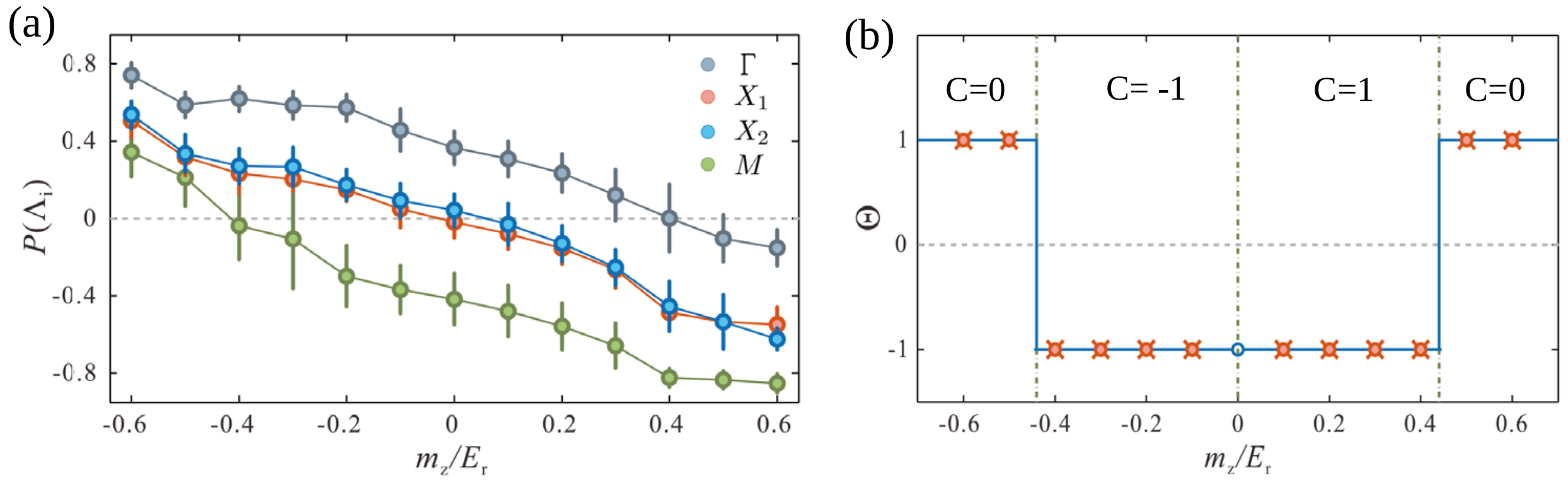}
\caption{(Color online) (a) Measured spin polarization
$P(\boldsymbol{\Lambda_i})$ at the four symmetric momenta
$\boldsymbol{\Lambda_i}$ as a function of $m_z$. (b) Obtained
invariant $\Theta$, which determines the Chern number $C$ of the
lowest band in Hamiltonian (\ref{ChHam}).
Reprinted from Wu {\it et al.}\cite{ZWu2016}. Reprinted with permission from AAAS.} \label{SpinPola}
\end{figure}

Based on the optical Raman lattice  method \cite{XJLiu2014}, the
Bloch Hamiltonian in Eq. (\ref{ChHam}) with the topological bands
has been experimentally realized for a BEC of
$^{87}$Rb atoms \cite{ZWu2016}. For the condensate in the OL, the spin polarization at the four high symmetry momenta
can be written as
\begin{equation}
\langle\sigma_z(\boldsymbol{\Lambda_i})\rangle\approx
P_-(\boldsymbol{\Lambda_i})f(E_-,T)
+P_+(\boldsymbol{\Lambda_i})f(E_+,T),
\end{equation}
where
$f(E_{\pm},T)=1/[e^{(E_{\pm}(\boldsymbol{\Lambda_i})-\mu)/k_BT}-1]$
is the BEC with  $\mu$ and $T$ respectively
being the chemical potential and temperature. Since
$P_+(\boldsymbol{\Lambda_i})=-P_-(\boldsymbol{\Lambda_i})$, one
has $\langle\sigma_z(\boldsymbol{\Lambda_i})\rangle\approx
P_-(\boldsymbol{\Lambda_i})[f(E_-,T)-f(E_+,T)]$. Thus by preparing
a cloud of bosonic atoms with the temperature satisfying
$f(E_-(\boldsymbol{\Lambda_i}),T)>f(E_+(\boldsymbol{\Lambda_i}),T)$,
one can obtain
\begin{equation}
\text{sgn}[\langle\sigma_z(\boldsymbol{\Lambda_i})\rangle]=\text{sgn}[P_-(\boldsymbol{\Lambda_i})].
\end{equation}
Thus, the spin polarization can be precisely measured with a
condensate at low temperature. In the experiment \cite{ZWu2016},
the spin polarization was measured as a function of the tunable
parameter $m_z$ to determine $\Theta$ and $C$ for the topology of
the lowest-energy band, which is topologically nontrivial when
$0<m_z<|m_z^c|$, whereas it is trivial for $m_z>|m_z^c|$, as show
in Fig. \ref{SpinPola}. The 2D SOC and the band
topology for BECs in the optical Raman
lattices have recently been further investigated
\cite{BZWang2018,WSun2017,WSun2018}

\subsection{Topological pumping approach}

As introduced in Sec. \ref{Topopuming}, the topological pumping,
geometric pumping, and spin pumping  have been recently realized
with ultracold atoms in 1D optical superlattices
\cite{Lohse2016,Nakajima2016,HILu2016,Schweizer2016}. Very
recently, a 2D topological charge pump as a dynamical version of
the 4D integer QHE was realized by using ultracold
bosonic atoms in a 2D optical superlattice \cite{Lohse2018}. The
quantized transported particle (i.e., the atomic center-of-mass
change in the experiments) in the adiabatic cyclic evolution of
these pumps indicates the underlying band topology, such as the
first and second Chern numbers characterizing the 1D and 2D
topological pumping, respectively.

Based on the pumping approach in OLs and hybrid
Wannier functions in  band theory
\cite{King-Smith1993}, it was shown that the Chern
number can be extracted from hybrid time-of-flight images
\cite{LWang2013b}. In the modern theory of polarization, a 2D
insulating lattice system can be viewed as a fictitious 1D
insulator along one direction, say along $x$, subject to an
external parameter $k_y$, where $k_y$ is the crystal-momentum
along $y$. The polarization of this 1D insulator can be defined by
means of hybrid Wannier functions
\cite{King-Smith1993}, in which the Fourier transform
from Bloch functions is carried out in the $y$ direction only. The
polarization at each $k_y$ is then given by the center of the
corresponding hybrid Wannier functions, and the change in
polarization from adiabatically changing $k_y$ by $2\pi$ is
proportional to the Chern number of the 2D insulators
\cite{King-Smith1993,Coh2009}. This is a manifestation of
topological particle pumping with $k_y$ being the adiabatic
pumping parameter. A generalization of the hybrid Wannier
functions of band theory to the hybrid particle densities in cold
atomic gases $\rho(x,k_y)$, which are the particle densities
resolved along the $x$-direction as a function of $k_{y}$,
provides a natural way to measure the Chern number in OLs. Experimentally, $\rho(x,k_y)$ can be measured by combing
\emph{in situ} imaging along $x$ and time-of-flight imaging along
the release direction $y$. In the measurement, the OL
is switched off along the $y$ direction while the system remains
unchanged in the $x$ direction. It was shown that the hybrid
particle density provides an efficient numerical reconstruction of
the Chern number in topologically-ordered OLs
\cite{LWang2013b}, such as the Hofstadter and Haldane models. This
method is general and allows the measurement of other topological
invariants in OLs, such as the $\Z_2$ topological
invariant in time-reversal symmetric insulators and the $k_z$-dependent Chern number $C_{k_z}$
in Weyl semimetals \cite{DWZhang2015,XYMai2017}.

\subsection{Detection of topological edge states}

According to the bulk-edge correspondence,  the topological index
of the bulk bands corresponds to the number of gapless edge-modes
present within the bulk gap \cite{Hasan2010,XLQi2011}. In Chern
insulators, all the gapless edge states propagate in the same
direction, such that they are chiral. In the context of the QHE, the chiral edge states are responsible for the
quantized Hall conductivity. As introduced in Sec.
\ref{sythdimention}, the chiral edge states in a quantum Hall
ribbon have been experimentally realized and detected with cold
atoms through the synthetic dimension method
\cite{Mancini2015,Stuhl2015}. In addition, the chiral currents
were also observed in an optical ladder for ultracold bosonic
atoms exposed to a uniform artificial magnetic field
\cite{Atala2014}. The high-resolution addressing technique in cold
atom gases offers the possibility of directly visualizing the
time-evolution of these edge states.

In a  2D atomic Chern insulator under an external trapping
potential, the direct detection of topological edge states is
challenging because the number of occupied edge modes within a
bulk gap and below the Fermi level contains a very small fraction
of the total number of particles. In addition, these edge states
would be washed out by the smooth harmonic trap and thus one may
not be able to distinguish them from the bulk states. To
circumvent these problems in detecting the topological edge
states, it was proposed to use a steep confining potential and to
image the edge states from optical Bragg spectroscopy
\cite{Stanescu2010,XJLiu2010,Goldman2012,Buchhold2012}. Based on a
generalization of Bragg spectroscopy sensitive to angular momentum
\cite{Goldman2012}, the Bragg probe can transfer energy and
angular momentum to atoms located in the vicinity of the Fermi
level and simultaneously changes their internal states, which
completely removes the edge states from the cloud and allows
imaging on a dark background unpolluted by the untransferred
atoms. In this scheme, the Bragg spectra can provide an
unambiguous signature of the topological edge states that
establishes their chiral nature. Another method to directly image
the propagating edge states was proposed by forcing them to
propagate in a region that is unoccupied by the bulk states after
suddenly removing the potential \cite{Goldman2013}. Other methods
to visualize the edge state currents were also proposed by
quenching the parameters of the system Hamiltonian
\cite{Killi2012,Reichl2014}. It would to interesting to extend
these cold-atom schemes to directly image the helical edge states
in $\Z_2$ topological insulators and the Fermi arc surface states
in 3D topological semimetals.

\section{Topological quantum matter in continuous form}\label{SecVI}

In this section, we move beyond topological Bloch bands in lattice systems to describe some of the quantum matter
in the continuum that have topologically nontrivial properties. Here we focus on there model systems realized with cold atoms without lattice potentials, which are the topological solitons in Jackiw-Rebbi model, various topological defects in BECs, and the atomic (quantum) spin Hall effect.

\subsection{Jackiw-Rebbi model with topological solitons}\label{JRModel&TS}

In relativistic quantum field theory, Jackiw and Rebbi introduced
a celebrated model to  generate topological soliton modes with
fractional particle numbers \cite{Jackiw1976}. The Jackiw-Rebbi
model describes a 1D Dirac field coupled to nontrivial background
fields. The relativistic Dirac Hamiltonian for 1D Dirac fermions
subjected to two static bosonic fields $\varphi_1$ and $\varphi_2$
can be written as \cite{Jackiw1976,Goldstone1981,Jackiw1983}
\begin{equation}
\label{DiracHam}
 H_D=c_x\sigma_x p_x+\varphi_1(x)\sigma_z+\varphi_2(x)\sigma_y,
\end{equation}
where $c_x$ is the (effective) speed of light and the background
field with a kink potential can be described as
\begin{equation}
\label{kink} \varphi_1(x)=\Gamma,~~
\varphi_2(x\rightarrow\pm\infty)=\pm\Delta_0,
\end{equation}
with positive constants $\Gamma$ and $\Delta_0$. The relativistic
Dirac Hamiltonian with such a topologically  nontrivial background
potential supports an nondegenerate soliton state, which gives
rise to fractionalization of particle number. In the original
Jackiw-Rebbi model with $\Gamma=0$, the nondegenerate zero-energy
soliton state is protected by the conjugation (particle-hole)
symmetry that connects each state with energy $E$ to its partner
located at the opposite energy. In the many-particle description,
there are two degenerate many-body ground states corresponding to
the soliton state being filled or empty, carrying fractional Fermi
number $\mathcal{N}_0=\pm1/2$, respectively.

\begin{figure}\centering
 \includegraphics[width=0.9\columnwidth]{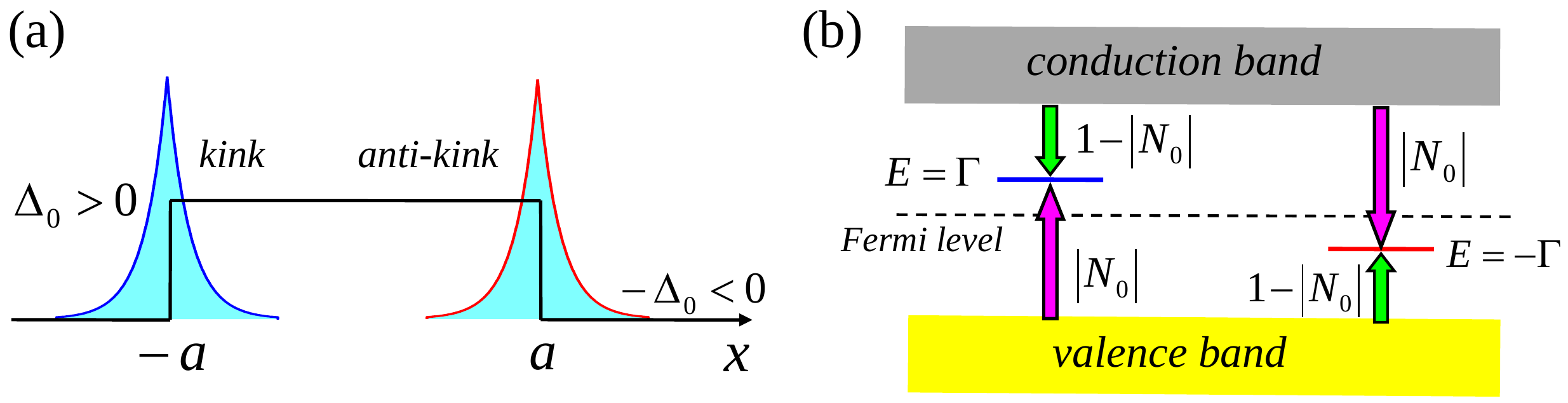}
\caption{(Color online) Fractionalization in Jackiw-Rebbi model \cite{DWZhang2012}. (a) A background
field with a pair of kink and anti-kink, both of which support a
localized soliton state. (b) The energy spectrum  and a pair of
solitons with energies $E=\pm \Gamma$ and fractional particle
numbers $\pm|\mathcal{N}_0|$. At the kink, the soliton state picks
a fractional particle number of $|\mathcal{N}_0|$ from the
effective valance band (Fermi sea) and another $1-|\mathcal{N}_0|$
from the effective conduction band, and vice versa for the
opposite case at the anti-kink. } \label{JRmodel}
\end{figure}

With the conjugation-symmetry-breaking term $\Gamma$, the soliton
mode has non-zero energy and the particle number   is generally
irrational \cite{Goldstone1981}:
\begin{equation}
\mathcal{N}_0=\frac{1}{\pi}\arctan\frac{\Delta_0}{\Gamma}.
\end{equation}
One can see that $\mathcal{N}_0$ depends only on the asymptotic
value of the kink rather than its detailed shape. In this sense,
it is topological and is insensitive to local fluctuations of the
background field. To have a better understanding of arbitrary
fractional particle number \cite{DWZhang2012}, one can consider
another kind of background field with a simple but practical
configuration, that is, a pair of kink and anti-kink both with a
step-function profile as shown in Fig. \ref{JRmodel}(a). Solving
the energy spectrum of the Dirac Hamiltonian (\ref{DiracHam}) at
the kink potential (near $x=-a$) with
$\varphi_2(x)=\Delta_{0}\text{sgn}(x+a)$ yields a localized in-gap
eigenstate in the kink at $E=\Gamma$ with the wave function
decaying as $\exp{(-\Delta_0|x+a|/\hbar c)}$ and the energy gap
$E_g=2\sqrt{\Delta^2_0+\Gamma^2}$. It is understood that the
soliton state picks up a fractional fermion number of
$|\mathcal{N}_0|$ from the effective valance band (Fermi sea) and
$(1-|\mathcal{N}_0|)$ from the effective conduction band, as shown
in Fig. \ref{JRmodel}(b). Without the soliton, one can assume $N$
fermions fully occupying the valence band acting  as a uniform
background in the state counting. For filling with $N+1$ fermions
in the presence of kink and anti-kink configuration, the
expectation value of the fermion number of the soliton modes at
the kink and anti-kink can be $|N_0|$ and $1-|N_0|$ if they are
both occupied. For an anti-kink potential (near $x=a$) with
$\varphi_2(x)=-\Delta_{0}\text{sgn}(x-a)$, the localized soliton
state is obtained at $E=-\Gamma$ with the wave function decaying
as $\exp{(-\Delta_0|x-a|/\hbar c_x)}$. It picks up
$(1-|\mathcal{N}_0)|$ from the valence band and $|\mathcal{N}_0|$
from the conduction band. There must be pairs of kink and
anti-kink in a periodic system. If both states are unoccupied, the
particle numbers are $-|\mathcal{N}_0|$ at the kink and
$|\mathcal{N}_0|-1$ at the anti-kink. When the Fermi level in this
system, is tuned up, the $E=-\Gamma$ soliton state is occupied
first and the particle numbers at kink and anti-kink are
$\mp|\mathcal{N}_0|$, respectively. When both states are occupied,
there are particles $(1-|\mathcal{N}_0|)$ and $|\mathcal{N}_0|$ at
the kink and anti-kink, respectively.

The first condensed matter realization of the Jackiw-Rebbi model
is the conducting polymers described by the SSH model for lattice
systems, wherein the low-energy effective Hamiltonian of the Bloch
Hamiltonian near Dirac points takes the Dirac form with
kink-soliton modes in the continuum. In particular, the
Jackiw-Rebbi model and the SSH model share many similar features
related to topological insulators under a suitable regularisation.
The soliton modes and related topological properties in the SSH
model have been intensively investigated with cold atoms in 1D
OLs (see Sec. \ref{SSHModel}).

The direct realization of the Jackiw-Rebbi model and the detection
of the induced soliton mode with the fractional  particle number
by using a 1D atomic Fermi gas in the continuum were proposed in
Ref. \cite{DWZhang2012}. The first procedure is to create strong
SOCs for the ultracold atoms, and the second one is
to generate a kink-like potential and a tunable
conjugation-symmetry-breaking term are properly constructed by
laser-atom interactions, leading to an effective low-energy
relativistic Dirac Hamiltonian with a topologically nontrivial
background field. The fractionalization of the particle number in
the atomic system may be detected through the soliton density and
the local density of states near the kink by using two standard
experimental detection methods for ultracold atomic gases, the in
situ absorption imaging technique and spatially resolved rf
spectroscopy \cite{DWZhang2012}. The realization of 1D homogeneous
Dirac-like Hamiltonian (particles) and related relativistic
effects (e.g. Klein tunneling and Zitterbewegung) with cold atoms
in the continuum has been studied in theories and experiments
\cite{Ruseckas2005,Juzeliunas2008,Vaishnav2008,SLZhu2009,DWZhang2012a,LeBlanc2013,Qu2013}.
See Ref. \cite{DWZhang2011} for a review on relativistic quantum
effects of Dirac particles simulated by ultracold atoms.

\subsection{Topological defects in Bose-Einstein condensates}

BECs of atomic gases in a harmonic trap without lattice potentials can host various topological
defects in real space. These topological defects include solitons
in 1D, vortices and Skyrmions in 2D, monopoles, Skyrmions and
knots in 3D. The topological defects have different physical
properties and are classified by homotopy groups of their
order-parameter space. Thus, they are distinguished by their
topological charges with discrete values \cite{Kawaguchi2012} and
robust against to external perturbations. In particular, atomic
BECs with internal spin degrees of freedom provide unique
platforms for investigating different topological objects due to
the rich structure of their superfluid order parameters, which are
vectors rather than scalar quantities. Moreover, the
well-developed manipulation techniques for atomic motion and spin
states enable one to engineer the topological defects of interest
in real space for studying their dynamics and stability in a
highly controllable manner.

In early experiments of atomic BECs, the 1D solitons in the atomic
density distributions have been  created and controlled by a phase
imprinting method \cite{Burger1999,Denschlag2000,Khaykovich2002}.
The 2D topological vortices, which are line defects in the
superfluid order parameter accompanied by a quantized phase
winding of an integer multiple of $2\pi$, have also been generated
in single- and multi-component atomic BECs by an external rotation
\cite{Madison2000,Abo-Shaeer2001} or the phase imprinting method
\cite{Matthews1999,Leanhardt2002,Leanhardt2003}.

\begin{figure}\centering
 \includegraphics[width=0.85\columnwidth]{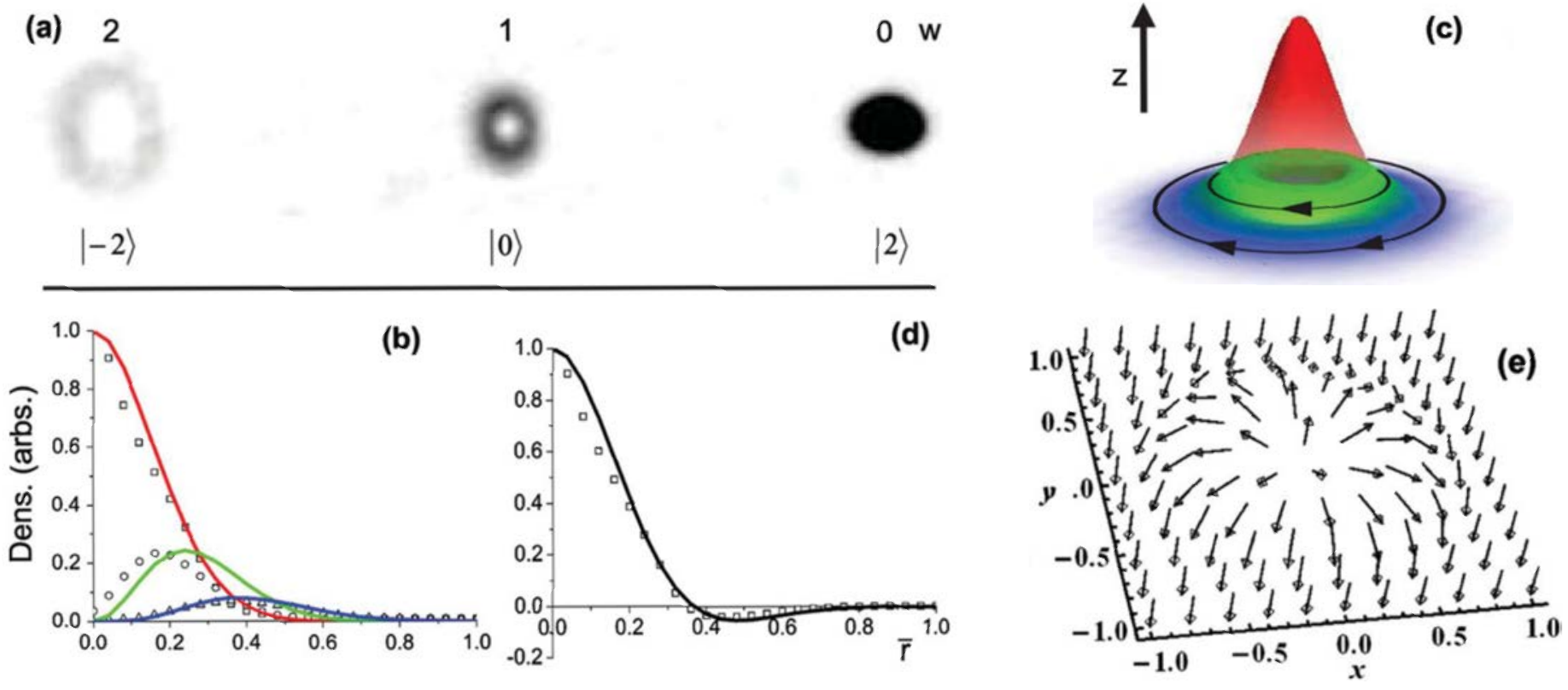}
\caption{(Color online) Creation of 2D Skyrmions in a spinor BEC. The absorption
image (a) of a 2D skyrmion  created in spin-2 $^{87}$Rb. The
winding number $\nu_\text{w}$ for each spin state is indicated. (b)
Azimuthally averaged lineouts (points) for each spin state. (c) 3D
plot of the solid lines in (b), where the colors red, green, and
blue correspond to the $\left|2\right\rangle$,
$\left|0\right\rangle$, and $\left|-2\right\rangle$ states,
respectively, with the number of arrowheads indicating the winding
number of the spin state. (d) The polarization of the skyrmion.
(e) The vector field of the Skyrmion. Reprinted with permission from Leslie {\it et al.}\cite{Leslie2009}. Copyright\copyright~(2009) by the American Physical Society. } \label{2DSkyrmion}
\end{figure}

As another kind of topological defects, Skyrmions are first
envisioned in field theory  and then extended to condensed matter
physics. A 2D Skyrmion is characterized by a local spin that
continuously rotates through an angle of $\pi$ from the center to
the boundary of the system. In terms of a unit spin vector
$\mathbf{d}$, a typical 2D Skyrmion spin texture shown in Fig.
\ref{2DSkyrmion}(e) with $z$-axis symmetry can be written as in
the polar coordinate
\begin{equation} \label{SkyField}
 \mathbf{d}(r,\phi)=\cos\beta(r) \hat{z}+\sin\beta(r)(\cos\phi\hat{x}+\sin\phi\hat{y}),
\end{equation}
where $\beta(r)$ is the bending angle characterizing the rotation
or ``bending'' of the local spin across the cloud, with the
boundary conditions $\beta(0)=0$ and $\beta(\infty)=\pi$. This
spin texture has the topological charge
\begin{equation}
\nu_\text{w}=\frac{1}{4\pi}\int dx
dy~\mathbf{d}\cdot(\partial_{x}\mathbf{d}\times\partial_{y}\mathbf{d})=1,
\end{equation}
which is the 2D winding number representing the number of times that
the spin texture encloses  the whole spin space. Such a 2D
Skyrmion spin texture was first created in a spin-2 condensate of
$^{87}$Rb atoms with coherent Raman transitions between the spin
states \cite{Leslie2009}. In the experiment, the BEC was prepared
in the $\left|F=2,m_F=2\right\rangle=\left|2\right\rangle$ state,
and then two Raman beams respectively with first-order
Laguerre-Gaussian and Gaussian intensity profiles were applied to
transfer the population to the $\left|0\right\rangle$
($\left|-2\right\rangle$) state, which acquires a $\nu_\text{w}=1$ ($\nu_\text{w}=2$)
azimuthal phase winding. The Raman interaction effectively evolves
the order parameter of the spinor BEC to \cite{Leslie2009}
\begin{equation}
\Psi(r)=\sqrt{n(r)} \left( \begin{array}{ccc} \cos^2(\beta(r)/2)
\\
0
\\
\sqrt{2} e^{i\phi} \sin(\beta(r)/2)\cos(\beta(r)/2)
\\
0
\\
e^{2i\phi}\sin^2(\beta(r)/2)
\end{array}\right).
\end{equation}
Here $n(r)$ is the density of the atomic cloud, and the
distribution $\beta(r)$ as the  form in Eq. (\ref{SkyField}) can
be engineered to generate the 2D Skyrmion spin texture. The
Skyrmion was detected from absorption image of the atomic density
profile shown in Fig. \ref{2DSkyrmion}(a-d) and further conformed
by matter-wave interference \cite{Leslie2009}. It was demonstrated
that the state 2D Skyrmions can be created in a spin-1 BEC of
$^{23}$Na atoms in a harmonic trap by using a 3D quadrupole
magnetic field \cite{Choi2012a,Choi2012b}, and moreover, an atomic
geometric Hall effect in the spinor BEC with a 2D Skyrmion spin
texture has been observed \cite{Choi2013}.

In 3D, Skyrmion is a particlelike soliton hypothetically
introduced by Skyrme \cite{Skyrme1961}. A 3D Skyrmion has a
nonsingular texture that can be topologically characterized by a
3D winding number. The stability of a 3D Skyrmion in two-component
BECs, which can be simply viewed as a vortex ring containing a
superflow, has been theoretically studied and found as metastable
solutions of the energy functional
\cite{Khawaja2001a,Khawaja2001b}. Several schemes have been
proposed to create and stabilize the metastable 3D Skyrmions in
multi-component BECs
\cite{Ruostekoski2001,Battye2002,Savage2003,Wuester2005,Herbut2006}.
Recently, it was shown that a fully stable 3D Skyrmion can
spontaneously emerge as the ground state of a two-component BECs
coupled with a synthetic non-Abelian gauge field
\cite{Kawakami2012}. The 3D Skyrmion spin texture is elusive in
experiments until recently, and it was realized  within a
spin-polarized ferromagnetic $^{87}$Rb BEC that is exposed to an
externally controlled magnetic field \cite{Lee2018}.

On the other hand, knots are another 3D topological objects, which are
characterized by a linking number or a Hopf invariant
\cite{Battye1998}. The existence of a stable knot soliton was
first discussed in the context of a two-component BEC
\cite{Babaev2002}. An experimental scheme for generation of knot
spin textures in a spin-1 BEC was proposed \cite{Kawaguchi2008}.
Based on this theoretical proposal, the creation and observation
of knot solitons in the spinor BEC has recently been demonstrated
\cite{Hall2016}.

\begin{figure}\centering
 \includegraphics[width=0.85\columnwidth]{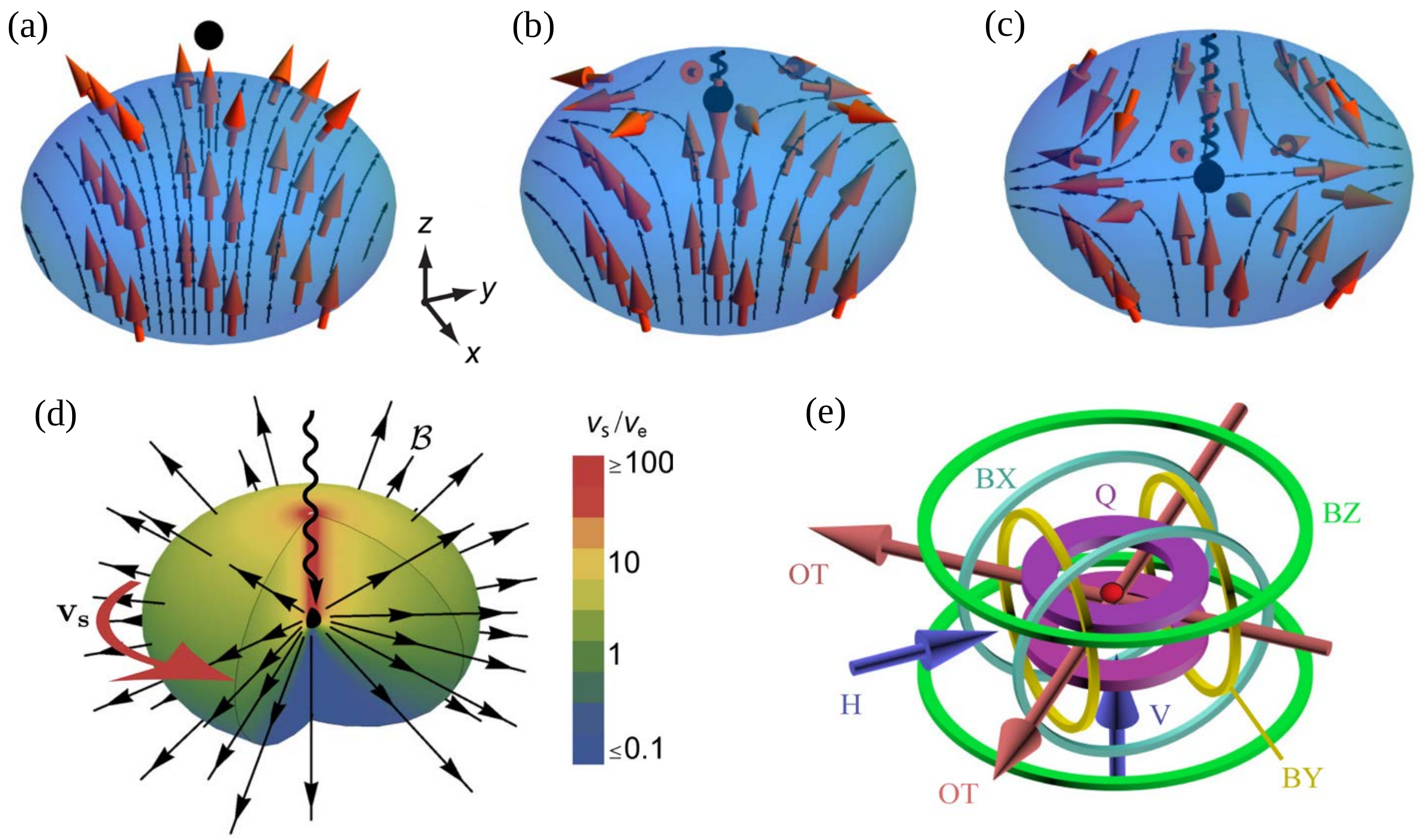}
\caption{(Color online) Creation of Dirac monopoles in a spinor BEC. (a-c) Spin
orientation (red arrows) in the  condensate when the magnetic
field zero (black dot) is above (a), entering (b) and in the
middle of (c) the condensate. The helix represents the
singularity in the vorticity. (d) Azimuthal superfluid velocity
$v_s$ (red arrow and colour scale by equatorial velocity $v_e$).
Black arrows depict the synthetic magnetic field $B^*$. (e)
Experimental setup with a controlled 3D magnetic quadrupole. Reprinted by permission from Macmillan Publishers Ltd: Ray {\it et al.}\cite{Ray2014}, copyright\copyright~(2014).} \label{Dirac-monopole}
\end{figure}

A fundamentally important and interesting topological defect is
monopole,  following Dirac's theory of magnetic monopoles which
are consistent with both quantum mechanics and the gauge
invariance of the electromagnetic field~\cite{Dirac1931}. The
experimental evidence of magnetic monopoles as fundamental
constituents of matter is still absent, however, they can emerge
as quasiparticle excitations or other analogies in condensed
matter systems, such as topological insulators \cite{XLQi2009}. It
has been proposed that the light-induced gauge potentials for
neural atoms in proper Raman laser fields can provide the
realization of synthetic magnetic monopoles and even non-Abelian
monopoles \cite{PZhang2005,Ruseckas2005,Pietila2009b,Sonner2009}.
Alternatively, it was theoretically demonstrated that a
topological defect as the Dirac magnetic monopole can be imprint
on the spin texture of an atomic BEC by using external magnetic
fields \cite{Pietila2009a}. Due to the spin of the condensate
aligning with the local magnetic field with nontrivial 3D
structures, one can create a pointlike defect to the spin texture
of the condensate giving rise to a vorticity equivalent to the
magnetic field of a magnetic monopole. A synthetic monopole field on a sphere with exact flat Landau levels on curved spherical geometry in a system of spinful cold atoms could be realized by engineering of a magnetic quadrupole field \cite{XFZhou2018}.

Following the method introduced in Ref.~\cite{Pietila2009a}, the
Dirac monopoles have been experimentally created in the synthetic
electromagnetic field that arises in the order parameter of a
ferromagnetic spin-1 $^{87}$Rb BEC in a tailored excited state
\cite{Ray2014}. The order parameter $\mathit{\Psi}({r},t) =
\psi({r},t) \zeta({r},t)$ is the product of a scalar order
parameter $\psi$, and a spinor $\zeta =
(\zeta_{+1},\zeta_0,\zeta_{-1})^\mathrm{T}\equiv|\zeta\rangle$,
where $\zeta_m=\langle m|\zeta\rangle$ represents the $m$th spinor
component along $z$, with $\zeta = (1,0,0)^\mathrm{T}$ at the
beginning. The spin texture
$\mathbf{S}=\mathit{\Psi}^{\dagger}\mathbf{F}\mathit{\Psi}$ are
given by the condensate order parameter and the spin-$1$ matrices
$\mathbf{F}$.  The spinor order parameter corresponding to the
Dirac monopole was generated by an adiabatic spin rotation in
response to a time-varying magnetic field \cite{Ray2014}
\begin{equation}
\mathbf{B}(r,t)=b_q(x\hat{x}+y\hat{y}-2z\hat{z})+B_z(t)\hat{z},
\end{equation}
where $b_q>0$ is the strength of a quadrupole field gradient and
$B_z(t)$ is  a uniform bias field. As shown in Fig.
\ref{Dirac-monopole}, the magnetic field zero is initially located
on the $z$~axis at $z=B_z(0)/2b_q$, and the spin rotation occurs
as $B_z$ is reduced, drawing the magnetic field zero into the
region occupied by the superfluid. In the experiment, the
condensate spin nearly adiabatically follows the local direction
of the field, as shown in Fig.~\ref{Dirac-monopole}(a-c). Using a
scaled and shifted coordinate system with $x'=x$, $y'=y$,
$z'=2z-B_z/b_q$, corresponding spherical coordinates
$(r',\theta',\phi')$, the applied magnetic field is then
$\mathbf{B} = b_q(x'\hat{x}' + y'\hat{y}' - z'\hat{z}')$. As $B_z$
is reduced, each spin rotates by an angle $\pi-\theta'$ about an
axis defined by the unit vector $\hat{n}(r',\theta',\phi')=
-\hat{x}'\sin\phi' + \hat{y}'\cos\phi'$. In the adiabatic limit,
the condensate order parameter corresponds to the local eigenstate
of the linear Zeeman operator
$g_{F}\mu_{B}\mathbf{B}\cdot\mathbf{F}$, and this
spatially-dependent rotation leads to a superfluid velocity
\cite{Pietila2009a,Ray2014}
\begin{align}
\mathbf{v}_s =
\frac{\hbar}{Mr'}\frac{1+\cos\theta'}{\sin\theta'}{{\hat{\varphi'}}},
\end{align}
and vorticity
\begin{align}
\mathbf{\Omega}_s= \nabla' \times \mathbf{v}_s = -\frac{\hbar}{M
r'^2}\hat{r'} + \frac{4 \pi \hbar}{M} \delta(x') \delta(y')
\mathit{\Theta}(z') \hat{r'},
\end{align}
where $M$ is the atomic mass, $\delta$ is the Dirac delta function
and  $\mathit{\Theta}$ is the Heaviside step function. The
vorticity is  a monopole attached to a semi-infinite vortex
line singularity as an analog of Dirac string, with phase winding
$4\pi$, extending along the positive $z'$ axis. The synthetic
vector potential arising from the spin rotation can be written as
$\mathbf{A}^{*}= - M {\mathbf{v}_s}/\hbar$, and the synthetic
magnetic field of the monopole is
\begin{align}
\mathbf{B}^{*}= \frac{\hbar}{r'^2}\hat{r}'.
\end{align}
The fields $\mathbf{v}_s$ and $\mathbf{B}^{*}$ are depicted in
Fig.~\ref{Dirac-monopole}(d). The  created Dirac monopoles were
then experimentally identified at the termini of vortex lines
within the condensate by directly imaging such a vortex line in
real space \cite{Ray2014}. Based on this method, a topological
point defect as an isolated monopole without terminating nodal
lines (the Dirac string) was also created and observed in the
order parameter of the spin-1 BEC \cite{Ray2015}.

\subsection{Spin Hall effect in atomic gases}

Spin Hall effects \cite{Sinova2015} are a class of SOC phenomena where flowing particles experience
orthogonally directed spin-dependent Lorentz-like forces and give
rise spin currents. This is analogous to the conventional Lorentz
force for the Hall effect, but opposite in sign for two spin
states. A quantized spin Hall effect is closely related to the
$\Z_2$ topological insulators which preserves time-reversal
symmetry (see Sec. \ref{KMModel} and \ref{3DZ2TI}). The spin Hall
effects have been observed for electrons in spin-orbit coupled
materials \cite{Kato2004,Wunderlich2005} and circularly polarized
photons passing through certain surfaces
\cite{Hosten2008,XYin2013}. It was proposed that the spin Hall
effects can be realized for neutral atoms with spin-dependent
Lorentz forces \cite{SLZhu2006,XJLiu2007}, which can be achieved
by the synthetic gauge potentials as discussed in Sec.
\ref{AMFSOC}. Following the proposal of Ref. \cite{SLZhu2006}, the
spin Hall effect was observed in a pseudospin-1/2 $^{87}$Rb
BEC subjected to spin- and space-dependent vector potentials
\cite{Beeler2013}.

\begin{figure}\centering
 \includegraphics[width=0.9\columnwidth]{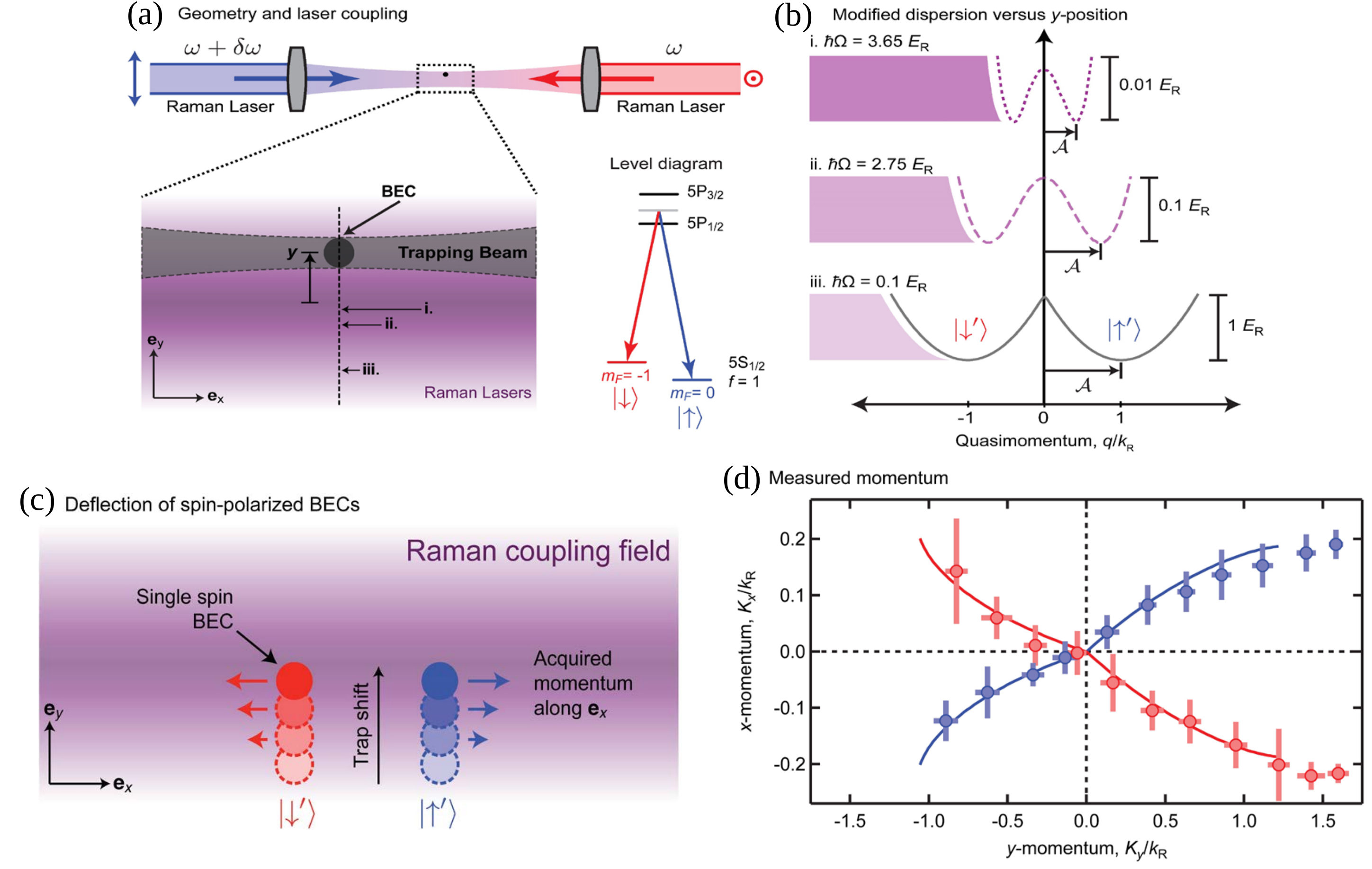}
\caption{(Color online) Atomic spin Hall effect.  (a) Experiment schematic of two
controlled Raman beams propagating along $\hat{x}$ coupled two
ground states in $^{87}$Rb atoms. (b) The induced double-well
dispersion $E(q)$ for the three different $y$-positions marked in
(a), with the synthetic vector potentials $\mathcal{A}$. (c) Spin
Hall effect with spin-dependent forces along $\hat{x}$ from motion
along $\hat{y}$. (d) Acquired momentum along $\hat{x}$ versus
final momentum along $\hat{x}$. Reprinted by permission from Macmillan Publishers Ltd: Beeler {\it et al.}\cite{Beeler2013}, copyright\copyright~(2013).} \label{SHE}
\end{figure}

Experimentally, two Raman lasers counterpropagating with wave
number $k_R$ along $\hat{x}$ were used to couple the internal
states $\left|f =1; m_F = 0, -1\right> = |\uparrow,
\downarrow\rangle$, which comprise pseudospin-$1/2$ atomic system
with strength $\Omega$, as shown in Fig. \ref{SHE}(a). For $\hbar
\Omega < 4E_{\rm{R}}$ with the single-photon recoil energy
$E_{\rm{R}}$, the 2D effective pseudospin Hamiltonian (ignoring
the trap, the light shift and the zero-energy shift from the Raman
dressing) can be written as \cite{Beeler2013}
\begin{equation}
\label{pseudoham} \hat{H} = \frac{1}{2 m}\left(\mathbf{\hat{p}}-
\mathcal{A} \hat{\sigma}_z \mathbf{e}_x \right)^2,
\end{equation}
where $\mathcal{A}=\hbar k_{\rm{R}}  \left[   1-\left(  \hbar
\Omega  /  4 E_{\rm{R}}\right)^2  \right]^{1/2}$ is a
light-induced spin-dependent vector potential along $\hat{x}$. As
shown in Fig. \ref{SHE}(b), the modified dispersion features two
degenerate wells and each of them displaces from zero by an amount
$\mathcal{A}$. Here $\mathcal{A}$ has the spatial dependence of
the Raman lasers' Gaussian intensity profile since $\Omega$
depends on the laser intensity. The spatial dependence of
$\mathcal{A}$ gives rise to a spin Hall effect in the atomic gas
\cite{SLZhu2006,XJLiu2007}. In the experiment \cite{Beeler2013},
the mechanism underlying the spin Hall effect was first probed
from observing spin-dependent shear in the atomic density
distribution by abruptly changing $\mathcal{A}$, which gives rise
to a spin-dependent ``electric'' force $-\partial
\mathcal{A}/\partial t$ on the atom cloud. Then for a
time-independent $\mathcal{A}$, the resulting spin Hall effect was
further observed by detecting a spin-dependent Lorentz-like
response along $\pm \hat{x}$ with atoms propelled in either spin
state along $\hat{y}$ and realizing an atomic spin transistor
using a mixture of both spins. As shown in Fig \ref{SHE}(c,d),
each spin-polarized BEC acquired a momentum along $\hat{x}$ that
was detected oppositely for the two spins and related to its final
momentum along $\hat{y}$, which demonstrates an intrinsic spin
Hall effect.

Since the (pseudo)spin here is a good quantum number, the system
can be thought of as two independent subsystems that respond
oppositely to temporal and spatial gradients of the light-induced
gauge potential $\mathcal{A}$. By introducing a large non-zero
curl for $\mathcal{A}$, each spin state could be separately driven
to the regime of integer QHE
\cite{SLZhu2006,Beeler2013}. Therefore, one can create a quantum
spin Hall effect in this atomic system composed of an equal
mixture of both spins.

\section{Topological matter with interactions}\label{SecVII}

Until now we have mainly reviewed cold-atom realizations of essentially non-interacting topological phases. But interactions can lead to new topological phases including intrinsic topological phases and symmetry-protected topological phases for both bosonic and fermionic systems in all dimensions. While tremendous theoretical efforts have been recently paid with fruitful achievements usually by advanced mathematics, the experimental realization for most theoretically predicted phases hardly has any solid progress yet except several classic examples. This section is intended not to give a systematical review of interacting topological phases, which is beyond the scope of this review article, but much more modestly to introduce a number of interesting models with interactions, which can exhibit interaction-intrinsic topological phases. In particular we focus on how to realize them by recent advances of cold-atom techniques, and wish to convey the expectation that the high-tunability of cold-atom systems would enable us to explore this open and deep field further.

\subsection{Spin chains}

\subsubsection{Spin-1/2 chain}

Ultracold atoms in OLs is a promising platform to
realize some spin-1/2 models. We here consider the well-know
anisotropic Heisenberg model (XXZ model) in 1D, which
arises in the context of various condensed matter systems. The
Hamiltonian of XXZ spin model is given by
\begin{equation}
\label{HXXZ}
 H_{\text{XXZ}}=-\sum_{j=1}^N \left[ \lambda _{z}\sigma
_{j}^{z}\sigma _{j+1}^{z} + \lambda _{ \perp }\left( \sigma
_{j}^{x}\sigma _{j+1}^{x}+\sigma _{j}^{y}\sigma _{j+1}^{y}\right)
\right],
\end{equation}
where $\lambda_{z}$ ($\lambda_{ \perp })$ denotes the nearest
neighbor interaction along $z$-direction ($x-$ and $y$-direction),
and $\sigma^{x,y,z}_j$ are the Pauli matrices for the $j$th spin.
The phase diagram of this Hamiltonian is pretty rich. In addition,
the geometric phase of the ground state in this spin model is
quantized in certain parameters and it obeys scaling behavior in
the vicinity of a quantum phase
transition\cite{Carollo2005,SLZhu2006a}.

When $\lambda_{\perp}=0$ and in the presence of an applied
magnetic field along $x$ direction, the model in Eq.(\ref{HXXZ})
becomes the transverse Ising model, which Hamiltonian is given by
\begin{equation}
   H_{\text{Ising}} = -\lambda_z \sum_{j=1}^{N-1} \sigma_j^z \sigma_{j+1}^z - h_x
   \sum_{j=1}^N \sigma_j^x,\label{HIsing}
\end{equation}
where the parameter $h_x$  is the intensity of the magnetic field
applied in the $x$ direction. 
Consider the projection onto the $x$-axis of
the spin with the fermionic occupation number $n$:
$|\uparrow\rangle \leftrightarrow n = 0,~~|\downarrow\rangle
\leftrightarrow n = 1$, one has $\sigma_j^x = (-1)^{a_j^\dagger
a_j}$. Employing the string-like annihilation and creation
operators (the Jordan-Wigner transformation):
\begin{eqnarray}
a_j=\left(\prod_{k=1}^{j-1} \sigma_k^x\right) \sigma_j^+,~~
a_j^\dagger=\left(\prod_{k=1}^{j-1} \sigma_k^x\right) \sigma_j^-,
\end{eqnarray}
where $\sigma^+$ and $\sigma^-$ are the spin raising and lower
operators, $H_{\text{Ising}}$ can be rewritten as
\begin{equation}
H_{\text{Ising}} = J \sum_{j=1}^{N-1} (a_j - a_j^\dagger) (a_{j+1} +
a_{j+1}^\dagger) +2h_z \sum_{j=1}^N\left(a_j^\dagger a_j -
1/2\right).
\end{equation}
This shows that the 1D transverse Ising chain is mathematically
equivalent to the Kiteav's chain \cite{Kitaev2009} of p-wave superconductor (see the
next section) and thus exhibits the same topological phase, in
which the ground state degeneracy is dependent on the boundary
conditions of the chain. For the superconductors, the $\Z_2$
symmetry of fermionic parity cannot be lifted by any local
physical operators, because such operators must contain an even
number of fermion operators. However, the $\Z_2$ symmetry in the
Ising model is given by a global spin flip in the $\sigma_z$
basis: $P_S = \prod_{j=1}^N \sigma_j^x$, such that its degeneracy
can be lifted by a simple longitudinal magnetic field $h_z \sum_j
\sigma_j^z$. This indicates that the topological phase in the
transverse Ising chain is much weaker, and thus  the creation and
manipulation of Majorana edge modes in this system are more
difficult.

We now turn to address a scheme proposed in Ref. \cite{LMDuan2003}
to realize the spin models with ultracold atoms in OLs. Consider an ensemble of ultracold bosonic or fermionic
atoms confined in an OL. We are interested in the Mott insulator regime,
and the atomic density of roughly one atom per lattice site. Each
atom is assumed to have two relevant internal states, which are
denoted with the effective spin index $\sigma
=\uparrow ,\downarrow $, respectively. We assume that the atoms with spins $%
\sigma =\uparrow ,\downarrow $ are trapped by independent standing
wave laser beams through polarization (or frequency) selection.
Each laser beam
creates a periodic potential $V_{\mu \sigma }\sin^{2}(\vec{k}_\mu\cdot \vec{r}%
)$ in a certain direction $\mu $, where $\vec{k}_\mu$ is the
wavevector of light. For sufficiently strong periodic potential
and low temperatures, the atoms will be confined to the lowest
Bloch band, and the low energy Hamiltonian is then given by the
Boson- or Fermi-Hubbard Hamiltonian
\begin{equation}
H =-\sum_{\langle ij\rangle \sigma }\left( J_{\mu \sigma
}a_{i\sigma
}^{\dagger }a_{j\sigma }+H.c.\right)  +\frac{1}{2}\sum_{i,\sigma }U_{\sigma }n_{i\sigma }\left(
n_{i\sigma }-1\right) +U_{\uparrow \downarrow
}\sum_{i}n_{i\uparrow }n_{i\downarrow }, \label{HspinAtom}
\end{equation}
where $\left\langle i,j\right\rangle $ denotes the near neighbor
sites, $a_{i\sigma }$ are bosonic (or
fermionic) annihilation operators respectively for bosonic (or
fermionic) atoms of spin $\sigma $ localized on site $i$, and
$n_{i\sigma }=a_{i_{\sigma }}^{\dagger }a_{i_{\sigma }}$.

For the cubic lattice ($\mu =x$,$y$,$z$) and using a harmonic
approximation around the minima of the potential, the
spin-dependent tunneling energies and the on-site interaction
energies are given by

\begin{eqnarray*}
&J_{\mu \sigma } \approx \frac{4E_{R}^{1/4}\left( V_{\mu \sigma }\right) ^{3/4}}{\sqrt{\pi }}
e^{-2(V_{\mu
\sigma }/E_{R})^{1/2}},\ U_{\uparrow
\downarrow }\approx \left(\frac{8}{\pi}\right)^{1/2}(ka_{s\uparrow \downarrow })(E_{R}%
\overline{V}_{1\uparrow \downarrow }\overline{V}_{2\uparrow \downarrow }%
\overline{V}_{3\uparrow \downarrow })^{1/4},\\
& U_{\sigma } \approx \left(\frac{8}{\pi} \right)^{1/2}\left( ka_{s\sigma }\right)
\left( E_{R}V_{1\sigma }V_{2\sigma }V_{3\sigma }\right) ^{1/4} \ (\text{for bosons}),
U_{\sigma } \approx  2\sqrt{V_{\mu \sigma }E_{R}} \ (\text{for fermions}),
 \end{eqnarray*}
 where
$\overline{V}_{\mu \uparrow \downarrow }=4V_{\mu \uparrow }V_{\mu
\downarrow }/(V_{\mu \uparrow }^{1/2}+V_{\mu \downarrow
}^{1/2})^{2}$ is the spin average potential in
each direction, $E_{R}=\hbar ^{2}k^{2}/2m$ is the atomic recoil energy, and $%
a_{s\uparrow \downarrow }$ is the scattering length between the
atoms of different spins.  For
fermionic atoms, $U_{\sigma }$ is on the order of Bloch band separation $%
\sim 2\sqrt{V_{\mu \sigma }E_{R}}$, which is typically much larger than $%
U_{\uparrow \downarrow }$ and can be taken to be infinite. In
writing Eq.(\ref{HspinAtom}), overall energy shifts $\sum_{i\mu }\left( \sqrt{%
E_{R}V_{\mu \uparrow }}-\sqrt{E_{R}V_{\mu \downarrow }}\right)
\left( n_{i\uparrow }-n_{i\downarrow }\right) /2$  have been neglected, which can be
easily compensated by a homogeneous external magnetic field
applied in the $z$ direction.

From the above expressions, we observe that $J_{\mu \sigma }$
depend sensitively (exponentially) upon the ratios $V_{\mu \sigma }/E_{R}$ while $%
U_{\uparrow \downarrow }$ and $U_{\sigma }$ exhibit only weak
dependence. So we can easily introduce spin-dependent tunneling
$J_{\mu \sigma }$ by varying the potential depth $V_{\mu \uparrow
}$ and $V_{\mu \downarrow }$ with control of the intensity of the
trapping laser. This simple experimental method provides us a
powerful tool to engineer many-body Hamiltonians.  In the regime
where $J_{\mu \sigma }\ll U_{\sigma },U_{\uparrow \downarrow }$
and $\left\langle n_{i\uparrow }\right\rangle +\left\langle
n_{i\downarrow }\right\rangle \simeq 1$, which corresponds to an
insulating phase, the terms proportional to tunneling $J_{\mu
\sigma }$ can be considered via perturbation theory.  To the leading order in $%
J_{\mu \sigma }/U_{\uparrow \downarrow }$, Eq. (\ref{HspinAtom})
is equivalent to the following effective Hamiltonian
\begin{equation}\label{HXXZAtom}
 H=-\sum_{j=1}^N \left[ \lambda _{\mu z}\sigma
_{j}^{z}\sigma _{j+1}^{z} + \lambda _{\mu \perp }\left( \sigma
_{j}^{x}\sigma _{j+1}^{x}+\sigma _{j}^{y}\sigma _{j+1}^{y}\right)
\right],
\end{equation}
where $\sigma _{i}^{z}=n_{i\uparrow }-n_{i\downarrow }$, $\sigma
_{i}^{x}=a_{i\uparrow }^{\dagger }a_{i\downarrow }+a_{i\downarrow
}^{\dagger }a_{i\uparrow }$, and $\sigma _{i}^{y}=-i\left(
a_{i\uparrow }^{\dagger }a_{i\downarrow }-a_{i\downarrow
}^{\dagger }a_{i\uparrow }\right) $ are the usual spin operators.
The $+$ and $-$ signs before $\lambda _{\mu \perp }$ correspond,
respectively, to the cases of fermionic and bosonic atoms.
 The parameters $\lambda _{\mu z}$ and $\lambda _{\mu \perp
}$ for the bosonic atoms are given by
\begin{equation}
\lambda _{\mu z}=\frac{J_{\mu \uparrow }^{2}+J_{\mu \downarrow }^{2}}{%
2U_{\uparrow \downarrow }}-\frac{J_{\mu \uparrow }^{2}}{U_{\uparrow }}-\frac{%
J_{\mu \downarrow }^{2}}{U_{\downarrow }},\;\;\lambda _{\mu \perp }=\frac{%
J_{\mu \uparrow }J_{\mu \downarrow }}{U_{\uparrow \downarrow }}.
\label{lambda}
\end{equation}
For fermionic atoms the expression for $\lambda _{\perp }$ is the
same as in (\ref{lambda}), but in the expression for $\lambda
_{z}$ the last two terms vanish since $U_{\sigma }\gg U_{\uparrow
\downarrow }$. In writing Eq. (\ref{HXXZAtom}), the term $\sum_{i\mu }4\left(
J_{\mu \uparrow }^{2}/U_{\uparrow }-J_{\mu \downarrow
}^{2}/U_{\downarrow }\right) \sigma _{i}^{z}$ is neglected, since
it can be easily compensated by an applied external magnetic
field. When we set $V_{\mu \downarrow }/V_{\mu
\uparrow }\gg 1$, so that $J_{\mu \downarrow }$ becomes negligible while $J_{\mu \uparrow }$ remains finite. In this case, the Hamiltonian
(\ref{HXXZAtom}) reduces to the Ising model $H=\sum_{\left\langle
i,j\right\rangle }\lambda _{\mu z}\sigma _{i}^{z}\sigma _{j}^{z}$,
with $\lambda _{\mu z}=J_{\mu \uparrow }^{2}/\left( 1/2U_{\uparrow
\downarrow }-1/U_{\uparrow }\right) $. The transverse field term
in Eq.(\ref{HIsing}) can be easily achieved with an applied
external magnetic field along the $x$ direction.
 The Ising model has been realized experimentally with atoms in OLs
\cite{Meinert2013,Simon2011}.

The approach using ultracold atoms to realize the spin models has
a unique advantage in that the parameters $\lambda _{\mu z}$ and
$\lambda _{\mu \perp }$ can be easily controlled by adjusting the
intensity of the trapping laser beams. They can also be changed
within a broad range by tuning the ratio between the scattering
lengths $a_{s\uparrow \downarrow }$ and $a_{s\sigma }$ $\left(
\sigma =\uparrow ,\downarrow \right) $ by adjusting an external
magnetic field through Feshbach resonance. Therefore, even with
bosonic atoms alone, it is possible to realize the entire class of
Hamiltonians in the general form (\ref{HXXZAtom}) with an
arbitrary ratio $\lambda _{\mu z}/\lambda _{\mu \perp }$. This is
important since bosonic atoms are generally easier to cool.
We estimate the typical energy scales for the realized Hamiltonian. For Rb atoms with a lattice constant $%
\pi /\left| \vec{k}\right| \sim 426$nm, the typical tunnelling
rate $J/\hbar
$ can be chosen from zero to a few kHz. The on-site interaction $%
U/\hbar $ corresponds to a few kHz at zero magnetic field, but can
be much larger near the Feshbach resonance. The energy scale for
magnetic interaction is about $J^{2}/\hbar U\sim 0.1$kHz
(corresponding to a time scale of $10$ms) with a conservative
choice of $U\sim 2$kHz and $\left( J/U\right) ^{2}\sim 1/20$.

\begin{figure}[ptb]\centering
\includegraphics[width=0.7\columnwidth]{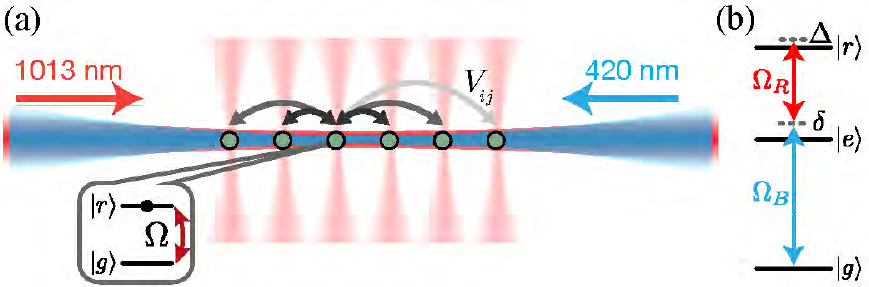}
 \caption{(Color online) Experimental platform for realization of the Ising model. (a) Individual $^{87}$Rb atoms are
trapped using optical tweezers (vertical red beams) and arranged
into defect-free arrays. Coherent interactions $V_{ij}$ between
the atoms are enabled by exciting them to a Rydberg state with strength $\Omega$ and
detuning $\Delta$. (b) A two-photon process couples the ground
state $|g\rangle=|5S_{1/2},F=2,m_F=-2\rangle$ to the Rydberg state
$|r\rangle=|70S_{1/2},J=1/2,m_J=-1/2\rangle$ via an intermediate
state $|e\rangle=|6P_{3/2},F=3,m_F=-3\rangle$ with detuning
$\delta$, using circularly polarized 420 nm and 1013 nm lasers
with single-photon Rabi frequencies of $\Omega_B$ and $\Omega_R$,
respectively. Reprinted by permission from Macmillan Publishers Ltd: Bernien {\it et al.} \cite{Bernien2017}, copyright\copyright~(2017). }
\label{IsingRydberg}
\end{figure}

Furthermore, the Ising model with long range interactions can be
realized with the Rydberg gas in OLs
\cite{Schaub2012,Schaub2015,Zeiher2017,Bernien2017}. A promising
avenue for realizing strongly interacting quantum matter involves
coherent coupling of neutral atoms to highly excited Rydberg
states. This results in repulsive van der Waals interactions of
strength $V_{ij}=C/R_{ij}^6$ between Rydberg atom pairs at a
distance $R_{ij}$, where $C$
 is the van der Waals coefficient. Such interactions have
recently been used to  explore quantum many-body physics of Ising
spin systems in OLs \cite{Schaub2012,Schaub2015,Zeiher2017,Bernien2017}. The
achieved Hamiltonian $(\hbar=1)$ is given by
\begin{equation}
\tilde{H}_{\text{Ising}}=\sum_{i}\frac{\Omega_i}{2}\sigma_i^x-\sum_{i}\Delta_i
n_i+\sum_{i<j}V_{ij}n_i n_j, \label{HIsRyd}
\end{equation}
where $\Delta_i$ are the detunings of the driving lasers from the
Rydberg state, $\Omega_i$ is the Rabi frequency describing the coupling between the ground state
$|g_i\rangle$ and the Rydberg state $|r_i\rangle$ of an atom at
position $i$, and $n_i\rangle=|r_i\rangle\langle r_i|$. The Hamilton with the first
two terms in Eq. (\ref{HIsRyd}) is equivalent to the Ising
Hamiltonian in Eq. (\ref{HIsing}), while the last term represents
the long range interactions induced by the van der Waals
interactions. Recently a programmable Ising-type quantum spin
model with tunable interactions and system sizes of up to 51
qubits was experimentally realized \cite{Bernien2017}. Their
approach combines these strong, controllable interactions with
atom-by-atom assembly of arrays of cold neutral $^{87}$Rb atoms.
The experimental platform is shown in  Fig. \ref{IsingRydberg}, each
$^{87}$Rb atoms can be controlled by optical tweezers, the
parameters ($|\Omega_i|=\Omega$ and $\Delta_i=\Delta$) can be
controlled by changing laser intensities and detunings in time.
The interaction strength $V_{ij}$ can be tuned either by varying
the distance between the atoms or by coupling them to a different
Rydberg state.

\subsubsection{Spin-1 chain and Haldane phase}

Another topological spin model is the spin-1 quantum Heisenberg chain, with the ground state belonging to the Haldane phase \cite{Haldane1983a,Haldane1983b}.
In the topological aspect, the Haldane phase is protected by the lattice inversion symmetry and can be classified as a symmetry-protected topological phase \cite{Gu2009,Pollmann2012}. According to Haldane's seminal work \cite{Haldane1983a,Haldane1983b}, the 1D integer-spin Heisenberg antiferromagnets have an exotic unordered ground state with unbroken rotational symmetry and with a finite excitation gap in the spectrum, while half-integer antiferromagnets are gapless. There are two topological features of the Haldane insulator phase: the existence of hidden non-local string order \cite{Nijs1989} and emergent fractional spin-1/2 edge states at the boundaries of
open chains \cite{Hagiwara1990}. For integer spin $S=1$, the following quadratic-biquadratic Hamiltonian contains the Haldane phase in a rich phase structure \cite{Affleck1987,Affleck1988,Garcia-Ripoll2004}:
\begin{equation}
H_{\rm QB} = \alpha \sum_{j=1}^{N-1}\left[
  \mathbf{S}_{j}\cdot \mathbf{S}_{j+1} +
  \beta (\mathbf{S}_{j}\cdot \mathbf{S}_{j+1})^2 \right],
\label{H-spin}
\end{equation}
where $\beta$ is a relative coupling constant, the sign of $\alpha$ determines the ferro or antiferromagnetic regimes, and $\mathbf{S}_{j}=(S_j^x,S_j^y,S_j^z)$ are spins at lattice site $j$ with
\begin{equation}
\begin{aligned}
S_j^x=\frac{1}{\sqrt{2}}\left(
                          \begin{array}{ccc}
                            0 & 1 & 0 \\
                            1 & 0 & 1 \\
                            0 & 1 & 0 \\
                          \end{array}
                        \right), \ \
S_j^y=\frac{1}{\sqrt{2}}\left(
                          \begin{array}{ccc}
                            0 & -i & 0 \\
                           i & 0 & -i \\
                            0 & i & 0 \\
                          \end{array}
                        \right), \ \
S_j^z=\left(
                          \begin{array}{ccc}
                            1 & 0 & 0 \\
                            0 & 0 & 0 \\
                            0 & 0 & -1 \\
                          \end{array}
                        \right).
\end{aligned}
\end{equation}
The properties of the ground state of $H_{\rm QB}$ are entirely
determined by an angle, $\theta$, such that $\alpha =
|a|\cos\theta$ and  $\alpha \beta = |a|\sin\theta$. The ground
state belongs to the Haldane phase when $\theta \in
[-\frac{\pi}{4},\frac{\pi}{4}]$, with $\theta =0$ being the
Heisenberg point and $\theta =\arctan(\frac{1}{3})$ the
Affleck-Lieb-Kennedy-Tasaki point \cite{Affleck1987,Affleck1988},
which can be described with an exact valence-bond wavefunction.
There are two critical points $\theta =\pm\frac{\pi}{4}$, at which
a phase transition occurs into a gapless phase for $\theta \in
[\frac{\pi}{4},\frac{\pi}{2}]$ and into a gapped dimerized phase
for $\theta \in [-\frac{3\pi}{4},-\frac{\pi}{4}]$, respectively.
For other values of $\theta$, the ground state belongs to the
ferromagnetic phase.

Although the Haldane phase of integer-spin Heisenberg
antiferromagnets has been extensively studied in theories in the
last decades, to the best of our knowledge, there is no
experimentally convenient system for realzing this phase.
Recently, it has been proposed that the Haldane phase can be
explored with ultracold atoms in OLs thanks to the
high tunability in these systems. A first possible direction is to
use spin-1 bosonic atoms in a deep 1D OL with one
atom per site \cite{Garcia-Ripoll2004}, which can be described by
the following extended Bose-Hubbard Hamiltonian:
\begin{equation} \label{HBH_haldane}
  H_{BH} = -J \sum_{\langle{j,l}\rangle,\alpha} (a^\dagger_{j\alpha} a_{l\alpha}+
  a^\dagger_{l\alpha} a_{j\alpha})   +\sum_{S=0,2} \frac{U_S}{2} \sum_{j,\alpha,\beta,\gamma,\delta}
  (\Psi^{(S)}_{\gamma\delta}a_{j\gamma}a_{j\delta})^\dagger
  (\Psi^{(S)}_{\alpha\beta}a_{j\alpha}a_{j\beta}).
\end{equation}
Here the indices $j$ and $l$ run over the lattice sites, the Greek
letters label the three spin-component of an atom ($\alpha, \beta,
\gamma, \delta=-1,0,+1$). Then the first term in the Hamiltonian
is the single-particle hopping with amplitude $J$, and the second
term  describes the interaction between bosons within a site. Two
bosons interact only when their total spin is either $0$ or $2$
because the state $S=1$ is antisymmetric, and the interaction may
be different for each value of the total spin. This is taken into
account by the spin-dependent interaction constants $U_S$ and the
tensors $\Psi^{(S)}_{\gamma\beta} = \langle{S|s,\gamma;s,
\beta}\rangle$, which are the Clebsch-Gordan coefficients between
the states $|s=1,\gamma\rangle\otimes|s=1,\beta\rangle$ and
$|S=0,2\rangle$. For the case of one atom per site, one can define
the effective spin-1 operators
$$S_j^z = a^\dagger_{j,+1} a_{j,+1}-a^\dagger_{j,-1} a_{j,-1},\ \
S_j^+=(S_j^-)^{\dag}=a^\dagger_{j,+1} a_{j,0}+a^\dagger_{j,0}
a_{j,-1},$$
where $S_j^{\pm}=S_j^{x}\pm S_j^{y}$. Using a perturbative
calculation for $J \ll U_S$ and around states with unit occupation, one can obtain the Hamiltonian $H_{QB}$
in Eq. (\ref{H-spin}) from the Hamiltonian $H_{BH}$ in
Eq.(\ref{HBH_haldane}) with the parameters $\alpha =
\tfrac{1}{2}C_2$, $\beta = \tfrac{1}{3}(2C_0/C_2+1)$, where
$C_S=J^2/U_S$ is tunable in the OL. The ways for
adiabatically preparing the Haldane phase and detecting its
intrinsic properties (the energy gap, end-chain spins effects and
the string order parameter) in the OL have also been
proposed \cite{Garcia-Ripoll2004}.

Sequent studies to realize the Haldane phase have
been made on spinless bosons with long-range dipole interactions
in 1D OLs \cite{Torre2006,Brennen2007,Berg2008,Amico2010,Dalmonte2011,Ruhman2012},
where the effective spin-1 chain can be obtained by truncating the
Hilbert space of the Bose system to three occupation states per
site. The Haldane insulators were also predicted for two-component
\cite{Ho2006,Kobayashi2012,Kobayashi2014,Fazzini2017} and
multi-component \cite{Nonne2010a,Nonne2011,Nonne2013} Fermi gases
in 1D OLs. A two-leg spin-1/2 ladder, which can be
realized with cold atoms in a ladder-like OL, was also
shown to host a Haldane-like phase
\cite{Nonne2010b,Cardarelli2017,Xu2017a}. A gapless topological
Haldane liquid phase in a 1D cold polar molecular lattice and an
exotic topological Haldane superfluid phase of soft-core bosons in
1D OLs were predicted \cite{Kestner2011,Lv2016}. Recently, the
direct observation of hidden non-local string order via quantum gas microscopy of doped ultracold Fermi-Hubbard chains was reported \cite{Hilker2017}, which is an important step toward experimental studies of emergent topological order in integer spin chains.

\begin{figure}[htbp]\centering
\includegraphics[width=0.9\columnwidth]{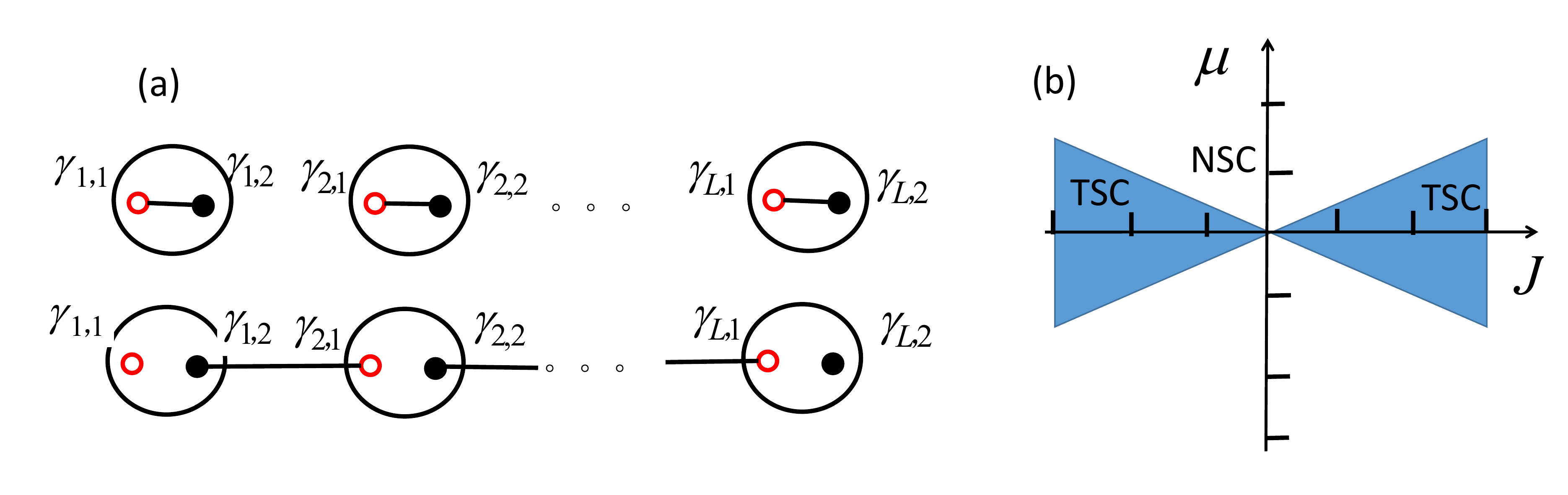}
 \caption{(Color online) Two phases of the Kitaev chain. (a) Upper: In the
topological trivial phase Majorana fermions on each lattice site can be
thought of as bound into ordinary fermions. Lower: In the topological nontrivial
phase Majorana fermions on neighboring sites are bound leaving two
unpaired Majorana fermions at the two ends of the chain. (b) The
phase diagram of the Kitaev chain in the $\mu-J$ plane, showing
the topological superconducting phase and the normal superconducting phase.} \label{Kitaev1}
\end{figure}

\subsection{Kitaev chain model}

The well-known Kitaev chain was proposed by Alexei Kitaev \cite{Kitaev2001}, which is the simplest model system that
shows unpaired Majorana zero modes. The Kitaev model is a toy model but can be exactly solved, which provides an extremely useful paradigm for Majorana zero modes at the two ends of a quantum wire of $p$-wave superconductor. The Kitaev chain is the spinless fermion model with nearest-neighbor hopping and pairing between the sites of a 1D lattice described by the Hamiltonian
\begin{equation}\label{Kitaevchain}
H=\sum_j[-J(c^\dagger_jc_{j+1}+c^\dagger_{j+1}c_j)-\mu(c^\dagger_jc_j-\frac{1}{2})
+\Delta{c_jc_{j+1}}+\Delta^*c^\dagger_{j+1}c^\dagger_j],
\end{equation}
where  $\mu$ is a chemical potential, and
$\Delta=|\Delta|e^{i\theta}$ is a superconducting gap.
Consider a chain with $L$ sites and open boundary conditions, as shown in Fig.(\ref{Kitaev1}),  we
can rewrite this Hamiltonian in the Majorana representation
by using the Majorana operators as
\begin{equation}\label{Kitaevm}
H=\frac{i}{2}\sum_j[-\mu\gamma_{j,1}\gamma_{j,2}+(J+|\Delta|)\gamma_{j,2}\gamma_{j+1,1}
+(-J+|\Delta|)\gamma_{j,1}\gamma_{j+1,2}].
\end{equation}
Here the Majorana operators are defined as
\begin{equation}
\gamma_{j,1}=e^{i\frac{\theta}{2}}c_j+e^{-i\frac{\theta}{2}}c_j^{\dagger},~~~~\gamma_{j,2}=(e^{i\frac{\theta}{2}}c_j-e^{-i\frac{\theta}{2}}c_{j}^{\dagger})/i.
\end{equation}
which satisfy the relations
\begin{equation}
\gamma^\dagger_{j,\alpha}=\gamma_{j,\alpha}, ~~~\{\gamma_{j,\alpha},\gamma_{k,\beta}\}=2\delta_{jk}\delta_{\alpha\beta},
\end{equation}
for $j,k=1,2,...,L$ and $\alpha,\beta=1,2$.

Now we discuss two specific cases. The topological trivial case for $|\Delta|=J=0$ is considered first. The Hamiltonian becomes
\begin{equation}
H=-\mu\sum_{j}(c^\dagger_jc_j-\frac{1}{2})=\frac{i}{2}(-\mu)\sum_{j}\gamma_{j,1}\gamma_{j,2}.
\end{equation}
The Majorana operators $\gamma_{j,1}$, $\gamma_{j,2}$ from the
same site $j$ are paired together, as shown in Fig. \ref{Kitaev1}(a), to form
a ground state with the occupation number $0$ ($\mu<0$) or $1$
($\mu>0$). Secondly, we consider the cases of $|\Delta|=J>0$ and $\mu=0$,
and we have
\begin{equation}\label{kitaev}
H=iJ\sum^{L-1}_{j}\gamma_{j,2}\gamma_{j+1,1}.
\end{equation}
Now the Majorana operators $\gamma_{j,2}$ and $\gamma_{j+1,1}$ from
different sites are paired together, as illustrated in Fig.
\ref{Kitaev1}(a). The ground state of this Hamiltonian is easily
found by defining new annihilation and creation operators
\begin{equation}
a_j=\frac{1}{2}(\gamma_{j,2}+i\gamma_{j+1,1}),~~~a^\dagger_j=\frac{1}{2}(\gamma_{j,2}-i\gamma_{j+1,1}),
\end{equation}
with $i\gamma_{j,2}\gamma_{j+1,1}=2a^\dagger_ja_j-1$ for $j=1,2,...,L-1$. Subsequently, the Hamiltonian (\ref{kitaev}) can be rewritten in a canonical form
\begin{equation}
H=2J\sum^{L-1}_{j}a^\dagger_ja_j-J(L-1).
\end{equation}
As we can see that Hamiltonian (\ref{kitaev}) does not contain
operators $\gamma_{1,1}$ and $\gamma_{L,2}$, i.e.,
$[\gamma_{1,1},H]=[\gamma_{L,2},H]=0$, while all pairs of
($\gamma_{j,2}$, $\gamma_{j+1,1}$) for $j=1,2,...,L-1$ form new
fermions. The ground states with twofold degeneracy for $J>0$
satisfy the condition $a_j{|g\rangle}=0$ for all $j$, and
\begin{equation}
H|g\rangle=-J(L-1)|g\rangle.
\end{equation}
These represent zero-energy Majorana modes localized at the
two ends of the chain. Since $[\gamma_{1,1},H]=[\gamma_{L,2},H]=0$, the two orthogonal ground
states of the Kitaev chain model can be constructed as $|g\rangle$ and
$a^\dagger|g\rangle$, where
$a=\frac{1}{2}(\gamma_{1,1}+i\gamma_{L,2})$ is an ordinary
zero-energy fermion operator. These states have
different fermionic parities: one is even and the other is
odd (i.e., it is a superposition of states with even or odd number
of electrons). Note that the ground states with double degeneracies
or not reveal that the system is topologically nontrivial or
trivial, respectively. Similarly, the considerations will also
yield unpaired Majorana zero modes for the special case
$|\Delta|=-J$ and $\mu=0$. These two specific cases represent two distinct phases of the
Kitaev chain: topologically trivial or nontrivial, corresponding
to different pairing methods without or with unpaired Majorana
zero modes localized at the ends of the chain.

To study the general properties of the Hamiltonian (\ref{Kitaevchain}) at
arbitrary values of $J$, $\mu$ and $\Delta$, we diagonalize the
Kitaev Hamiltonian under periodic boundary condition. After the
Fourier transformation with
$c^\dagger_j=1/\sqrt{L}\sum_{k}c^\dagger_ke^{i\mathbf{k}\cdot\mathbf{r}_j}$,
the Hamiltonian in momentum space can be written in Bogoliubov-de
Gennes form
\begin{equation}
\begin{aligned}
H=\frac{1}{2}\sum_k\begin{pmatrix} c^\dagger_k& c_{-k}\end{pmatrix}\mathcal{H}_{\text{BdG}}(k)\begin{pmatrix} c_k\\c^\dagger_{-k}\end{pmatrix},
\end{aligned}
\end{equation}
where the Bogoliubov-de
Gennes Hamiltonian is written in terms of Pauli matrices $\vec{\tau}$ as
\begin{equation}\label{BdG}
\mathcal{H}_{\text{BdG}}(k)=\epsilon(k)\tau_z +\Delta(k)\text{cos}\theta\tau_y+\Delta(k)\text{sin}\theta\tau_x,
\end{equation}
with $\epsilon(k)=-2J\text{cos}k-\mu$, $\Delta(k)=2|\Delta|\text{sin}k$. The energy spectrum is given by
\begin{equation}
E(k)=\pm\sqrt{\epsilon(k)^2+|\Delta(k)|^2}.
\end{equation}
For $\Delta\neq0$, the system is in superconducting states. The energy spectrum always fully gapped except
when $2J=\pm\mu$. As shown in Fig. \ref{Kitaev1}(b), two lines
represent gap closing are defined, which mark the phase transition
between the two distinct phases of the model. We can identify that
the system in the region $|J|>|\mu|/2$ is a topological
superconductor. In the other region, the
system is a normal superconductor (topologically trivial).

Since the two distinct phases of the model have the same
symmetries but different topological features, we can distinguish
these two phases by calculating the topological invariants. As is
known, $\mathcal{H}_{\text{BdG}}$ preserves intrinsic
particle-hole symmetry. One can check that the Hamiltonian
(\ref{BdG}) satisfies the relation
\begin{equation}
\hat{C}\mathcal{H}_{\text{BdG}}(k)\hat{C}^{-1}=-\mathcal{H}_{\text{BdG}}(-k),
\end{equation}
where the particle-hole operator $\hat{C}=\tau_x\hat{K}$ satisfies $\hat{C}^2=+1$.
According to the topological classifications, Hamiltonian
(\ref{BdG}) belongs to  the symmetry class D $(d=1)$ and thus has
a $\Z_2$-type topological number. The relevant topological
invariant of the system described by the Hamiltonian
(\ref{Kitaevchain}) is the so-called Majorana number
$\mathcal{M}=\pm1$, which is actually the $\Z_2$ index, first
formulated by Kitaev. In Kitaev's paper \cite{Kitaev2001}, it was
shown that all 1D fermionic systems with superconducting order
fall into two categories distinguished by $\mathcal{M}$. One is
topologically trivial with $\mathcal{M}=+1$ and the other is
nontrivial with $\mathcal{M}=-1$ and the existence of unpaired
Majorana zero modes.

To calculate $\mathcal{M}$, we consider the Hamiltonian that can be written in the Majorana representation as
\begin{equation}
H=\frac{i}{4}\sum_{lm\alpha\beta}B_{\alpha\beta}(l-m)\gamma_{l\alpha}\gamma_{m\beta},
\end{equation}
where $l$ and $m$ label the lattice sites while $\alpha$ and $\beta$ denote all other quantum numbers. Then $\mathcal{M}$ is defined as
\begin{equation}\label{MN}
\mathcal{M}=\text{sgn}\{\text{Pf}[\tilde{B}(0)]\}\text{sgn}\{\text{Pf}[\tilde{B}(\pi)]\}
\end{equation}
where $\tilde{B}(k)$ denotes the spatial Fourier transform of $B(l-m)$ regarded as a matrix in indices $\alpha$, $\beta$ and Pf$[A]$ denotes the Pfaffian where Pf$[A]^2$=det$[A]$, with $A$ being an antisymmetric matrix. Thus, we can calculate the Majorana number of the Kitaev model by using Eq. (\ref{MN}). In momentum space, the Kitaev Hamiltonian (\ref{Kitaevm}) can be written in the following form
\begin{equation}\label{Majornum}
\begin{aligned}
H=\frac{i}{4}\sum_{k}\begin{pmatrix} \gamma_{k,1}& \gamma_{k,2}\end{pmatrix}\begin{pmatrix} 0& D(k)\\ -D^*(k)& 0\end{pmatrix}\begin{pmatrix} \gamma_{-k,1}\\ \gamma_{-k,2}\end{pmatrix},
\end{aligned}
\end{equation}
with $D(k)=-2J\cos k-2i|\Delta|{\sin k}-\mu$. The operators
$\gamma_{k,1}$ and $\gamma_{k,2}$ are defined as
\begin{equation}
\gamma_{k,1}=c^\dagger_{-k}+c_k,~~~\gamma_{k,2}=i(c^\dagger_{-k}-c_k),
\end{equation}
which satisfy the relations
\begin{equation}
\{\gamma^\dagger_{k,\alpha},\gamma_{k',\beta}\}=2\delta_{kk'}\delta_{\alpha\beta},~~~\gamma^\dagger_{k,\alpha}=\gamma_{-k,\alpha}.
\end{equation}
Note that $\gamma_{k,1}$ and $\gamma_{k,2}$ are not the Majorana
operators except when $k=0$. As we can see that the matrix
$\tilde{B}(k)$ here for $k=0,\pi$ is antisymmetric, the Pfaffian
of $2\times2$ antisymmetric matrix can be simply given by its
upper off-diagonal component $D(k)_{0,\pi}$. It yields the
Majorana number
\begin{equation}
\mathcal{M}=\text{sgn}(D(0))\text{sgn}(D(\pi))=\text{sgn}(\mu^2-4J^2).
\end{equation}
One can check that the topological superconducting phase occurs
when $\mathcal{M}=-1$ for $|\mu|<2|J|$, the other phase is trivial
when $\mathcal{M}=+1$ for $|\mu|>2|J|$, as we have discussed
above.

The Kitaev's model describes a 1D system of spinless fermions but
electron spectra are usually degenerate with respect to spin in
real system. For this reason it has been initially viewed as a
somewhat unphysical toy model because the physical realization of
a quantum wire $\mathcal{M}=-1$ in condensed matter systems is very
difficult. This problem can be avoided in higher dimensional
space, which involve various combinations of the SOC and magnetic interactions that the produced normal metal
is effectively spinless \cite{LFu2008,Elliott2015,XLQi2011}.  There are many efforts to search for $p$-wave superconductors and Majorana fermions in condensed matter systems, but their unambiguous
detection (realization) remains an outstanding challenge.


\begin{figure}[htbp]\centering
\includegraphics[width=0.85\columnwidth]{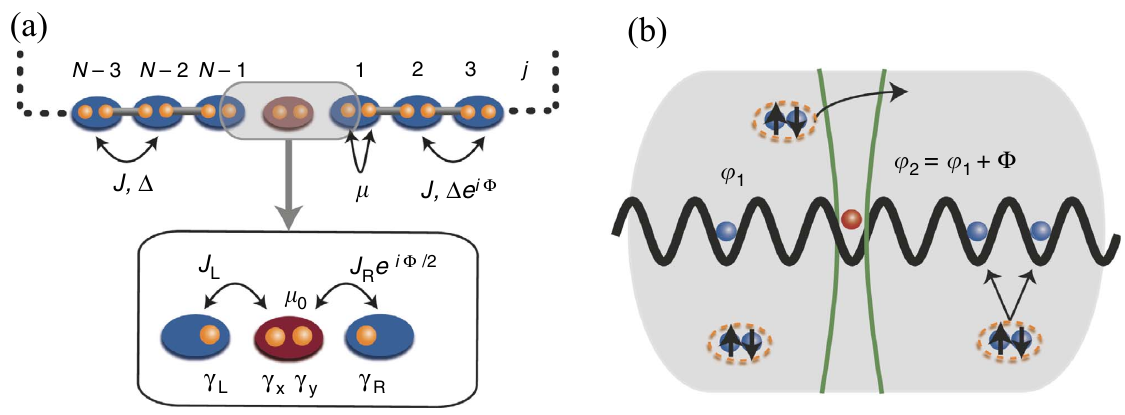}
 \caption{(Color online) (a) Schematics of the model Hamiltonian: the central part of the system is magnified in
 the box at the bottom, where the Majorana degrees of freedom included in the simplified model are highlighted. (b) Realization
in a cold atom system. A 1D OL is coupled to a BEC reservoir that gives rise to the Kitaev Hamiltonian in the chain. An optical barrier acts both to create the impurity site (red) and triggers the Josephson effect in the reservoir itself. The phase difference across the barrier in the reservoir then
acts as the phase $\Phi$ for the OL. Rrprinted by permission from Laflamme {\it et al.}\cite{Laflamme2016}.} \label{atomicKitaev}
\end{figure}

Cold atom systems may provide a platform with high controllability to simulate and study the Kitaev chain model based
on the remarkable advances in recent experiments. Several experimental proposals for realization of the Kitaev wires with
cold atoms in the past years have been presented \cite{LJiang2011,Diehl2011,Kraus2013,Nascimbene2013,Laflamme2014,Laflamme2016},
where the two key ingredients to induce the unpaired Majorana zero modes can be created. In these proposals, a single-piece gas
of cold fermionic atoms which can be regarded as the spinless
fermions are considered. Furthermore, the effective $p$-wave
pairing term can be realized by a Raman induced dissociation of
Cooper pairs \cite{LJiang2011} or Feshbach molecules \cite{Nascimbene2013} forming an atomic BCS (or BEC)
reservoir. The Kitaev chain model is usually considered as a noninteracting system; however, the pairing in the superfluid should be formed in an interacting atomic system.

Without loss of generality, we describes a recent work for realizing the Kitaev wires with cold atoms \cite{Laflamme2016}. In this experimental scheme, the model Hamiltonian
describes the spinless fermions with field operator $\psi_j$ in a ring OL reads as \cite{Laflamme2016}
\begin{equation}
\begin{aligned}
H(\Phi)&=\sum^{N-1}_{j=1}[-J\psi^\dagger_j\psi_{j+1}+\Delta\psi_j\psi_{j+1}-\frac{\mu}{2}(\psi^\dagger_j\psi_j-\frac{1}{2})]\\
&+J_L\psi^\dagger_{N-1}\psi_0+J_R\psi^\dagger_0\psi_1e^{i\Phi/2}+\frac{\mu_0}{2}\psi^\dagger_0\psi_0+h.c.,
\end{aligned}
\end{equation}
where $j=0,1...N-1$ labels the lattice sites. As shown in Fig. \ref{atomicKitaev}(a), it describes a
proximity-induced $p$-wave superconductor with pairing $\Delta$,
interrupted by an extra site at $j=0$ (assumed to be not
affected by the pairing), $J$ and $\mu$ denote the normal
nearest-neighbour hopping and the chemical potential relative to
half-filling, respectively. The site at $j=0$ is connected to its
neighbours by the hopping amplitudes $J_L$ and $J_R$, respectively, and has an energy offset $\mu_0$. The phase factor
$e^{i\Phi/2}$ on the hopping between $j=0$ and $j=1$ represents the phase of a Cooper pair by $\Phi$ when moving around
the ring. For $J_L=J_R=\mu_0=0$, the model Hamiltonian returns to
the Eq. (\ref{Kitaevchain}), which describes the original Kitaev
chain model. Thus, for the case $|\mu|<2J$ and $|\Delta|>0$, the
system hosts unpaired Majorana zero modes, $\gamma_L$ and
$\gamma_R$, which are localized around $j=N-1$ and $j=1$,
respectively.

In their proposed setup, three points to realize the model
Hamiltonian are required: the realization of a 1D Kitaev chain,
the additional single site separating the two ends of the chain,
and the time control of the phase $\Phi$. A system of fermionic alkaline earth atoms prepared in their
$^1S_0$ ground state in a 1D ring lattice was considered. The choice of alkaline
earth atoms allows one to independently trap the $^1S_0$
ground state $|g\rangle$ and the $^3P_0$ metastable excited state $|e\rangle$. To address the first issue, the hopping terms $J$ in the lattice
naturally arise and the pairing term $\Delta$ can be induced by
coupling the fermions in the lattice to a BEC reservoir. Here a
radio-frequency field is used to break up Cooper pairs into
neighbouring sites directly in the lattice \cite{LJiang2011}, as
depicted in Fig. \ref{atomicKitaev}(b). The second step is to
interrupt the chain with a single site. Following this idea, a
barrier is engineered to inhibit $|g\rangle$ atoms from being at
site $j=0$, which splits the Kitaev wire into two wires. It can be done
by using a highly focused beam at the so-called anti-magic wavelength, which acts as a sink for $|e\rangle$, and as a source
for $|g\rangle$. Consequently, the $|e\rangle$ atom only being
trapped at site $j=0$ acts as the additional site coupling the two
ends of the chain. Although the natural tunnelling into and out of this
site is deterred by this barrier, the hopping $J_L$ and $J_R$
are then can be reintroduced with Raman processes involving a clock
transition \cite{Jaksch2003,Gerbier2010,Wall2016}. Finally, the
realization of the time control of phase $\Phi$ is related to the
Josephson effect where the additional site $j=0$ and its
nearest-neighbour form a Josephson-like knot \cite{Laflamme2016}. Within these setups, this system will occur a
non-equilibrium Josephson effect with a characteristic $8\pi$
periodicity of the Josephson current. At this point, the system is
pumped to an excited state after slowly increasing $\Phi$ by
$4\pi$, and returns to the ground state after a second $4\pi$ cycle.

\subsection{1D Anyon-Hubbard model}

The Hubbard model of 1D lattice anyons with on-site interactions, called Anyon-Hubbard model, takes the form
\begin{equation}
\label{HA} H_A=-J\sum_{j=1}^N (a_j^\dagger
a_{j+1}+h.c.)+U\sum_{j=1}^N n_j (n_j-1).
\end{equation}
Here $n_j=a_j^\dagger a_j$ is the number operator for anyons and
the operators $a_j$ and $a_j^\dagger$ annihilate or create an
anyon on site $j$, and they are defined by the commutation relations
\begin{equation}
\label{Anyon1} a_j a_k^\dagger-e^{-i\theta
\text{sgn}(j-k)}a_k^\dagger a_j = \delta_{jk},\ \ \ a_j a_k =
e^{i\theta \text{sgn}(j-k)} a_k a_j,
\end{equation}
which are parameterized by the statistical angle $\theta$. The sign
function in the above equations is $\text{sgn}(j-k)=-1,0,1$ for $j<k$, $j=k$, $j>k$, respectively. Thus, two particles on the same site behave as ordinary bosons. Consequently, even for $\theta=\pi$, these lattice anyons are just
pseudofermions: they are bosons on-site and fermions off-site, since many of them are allowed to occupy the same site.

\begin{figure}[htbp]
\centering
\includegraphics[width=0.7\columnwidth]{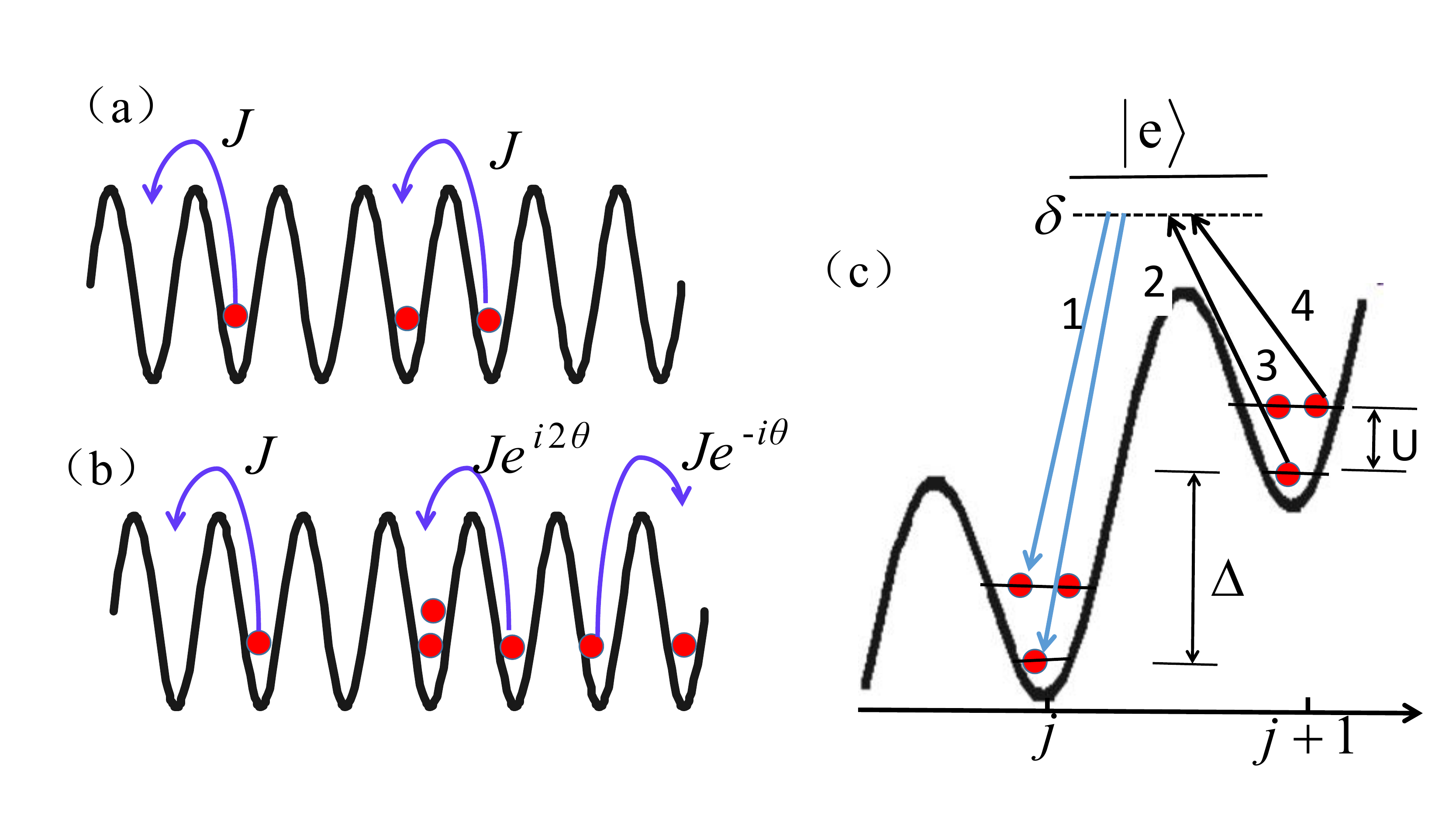}
\caption{(Color online) The mapping between Anyon- and Boson-Hubbard models and
a scheme to realize occupation-dependent gauge potential \cite{Keilmann2011}. (a)
Anyons in 1D can be mapped onto bosons featuring
occupation-dependent gauge potential. (b) Assisted Raman
tunnelling proposed for the realization of the Anyon-Hubbard
model.}
\label{Anyon-Boson}
\end{figure}

There exists an exact mapping between anyons and bosons in 1D. Define the fractional version of a Jordan-Wigner transformation,
\begin{equation}
\label{Mapping}
a_j=b_j \exp\left( i\theta\sum_{k=1}^{j-1}
n_k\right),
\end{equation}
with $n_k=a_k^\dagger a_k=b_k^\dagger b_k$ the number operator for
both particle types. One can check that the mapped operators $a_j$
and $a^\dagger_j$ indeed obey the anyonic commutation relations
in Eq. (\ref{Anyon1}), provided that the particles of
type $b$ are bosons with the bosonic commutation relations: $[b_j,b_k^\dagger]=\delta_{jk}$ and
$[b_j,b_k]=0$. By inserting the anyon-boson mapping
(\ref{Mapping}), the Hamiltonian (\ref{HA}) can be rewritten in
terms of bosons \cite{Keilmann2011},
\begin{equation}
\label{HB} H_B=-J\sum_{j=1}^N (b_j^\dagger b_{j+1} e^{i\theta n_j}
+h.c.)+U\sum_{j=1}^N n_j (n_j-1).
\end{equation}
Therefore, the anyonic exchange phase has been translated to an
occupation-dependent Peierls phase: when tunneling from right to
left $(j+1,j)$, a boson picks up a phase given by $\theta$ times
the number of particles occupying the site that it jumps to. Under this
condition, the many-body wave function picks up a phase of
$\theta\ (-\theta)$ if two particles pass each other via two
subsequent tunneling processes to the right (left). The proposed
conditional-hopping scheme is depicted in Fig. \ref{Anyon-Boson}.
Interestingly, the non-local mapping between
anyons and bosons in Eq. (\ref{Mapping}) leads to a purely local and thus realizable Hamiltonian (\ref{HB}).

The occupation-dependent gauge potential can be implemented in OLs with a laser-assisted Raman tunneling scheme \cite{Keilmann2011}, generalized the idea proposed in Ref. \cite{Jaksch2003} to create an artificial gauge potential, see the Sec.\ref{Laser-assisted-tunnelings}. Figure (\ref{Anyon-Boson}) displays the basic concept. A non-zero on-site interaction $U$ is required to distinguish between different local occupational states.
The OL is tilted, with an energy offset $\Delta$
between neighbouring sites. For simplicity, we consider lattice
site occupations that are restricted to $n_j=0,1,2$, but higher
local truncations are also possible. Two different occupational
states in either of the two sites form in total a 4D atomic ground state manifold, which are coupled to an excited
state $|e\rangle$ via four external driving fields, labelled
$L_1$, $L_2$, $L_3$ and $L_4$ in Fig. \ref{Anyon-Boson}(b).

The excited state can be experimentally realized in at least two
alternative ways. First, two spin-dependent lattices can be used.
We take $^{87}$Rb as an example. One lattice traps atoms in the $F
= 1$, $m_F = -1$ hyperfine state, which is the ground state
manifold. The $F = 1$, $m_F = 0$ hyperfine state is chosen as the
excited state $|e\rangle$ and trapped in a second lattice. Atoms in
the excited state would then be localized between the left and
right wells of the $F = 1$, $m_F = -1$ lattice, but not
necessarily in their center. This implementation has the advantage
of external driving fields in the radio-frequency regime. Such
frequencies could resolve the typical energies $U$ and $\Delta$,
both of the order of a few kHz,  which requires a laser with a
linewidth $\delta \ll ªÞU,\Delta$. This is a necessary
condition for selectively coupling the four different states in
the ground state manifold. Second, one can use two optical
lattices, and trap ground state manifold atoms in the red-detuned
lattice, while the excited state would live in the blue-detuned
one. The driving fields required in this case would be typically
in the THz frequency regime, making a precise resolution of $U$ and
$\Delta$ more challenging in experiments.

The effective tunnelling rates $J_{ab}$ ($a\in {1, 2}$, $b\in{3,
4})$ between the four different levels are obtained in terms of
the effective Rabi frequencies
$J_{ab}=\Omega_a^{*}\Omega_b/\Delta$, where an overlap integral
should be included in the Rabi frequencies $\Omega_{a,b}$ since
ground and excited states feel different lattices. It is
demonstrated that the model in Eq. (\ref{HB}) can be implemented if
the conditions $J_{23}=J_{24} \equiv J$ and $J_{13}=J_{14} \equiv
J e^{i\theta}$ were satisfied \cite{Keilmann2011}. It is notable
that the tilt energy $\Delta$ disappears in the effective
Hamiltonian after rotating out time-dependent phase factors, this
energy offset is absorbed by the external radiation field,
yielding a total Hamiltonian without a tilt term.

To realize the model in Eq. (\ref{HB}), one has the parameters to
satisfy the following conditions. (i) The lasers linewidth
$\delta\ll\Delta,U$, so that the external driving fields can
resolve the different levels of the ground state manifold. (ii) A
short-lived excited state and the validity of the adiabatic
elimination require large detunings $\Delta \gg |\Omega_{1-4}|$.
(iii) $\Delta$ and $U$ can be in the same frequency regime (a
few kHz), but their difference should be much larger than the
lasers linewidth $\delta$. As an example, $\Delta \approx 2$
kHz, $U\approx 3$ kHz, $|J_{ab}| = J \approx 5$ kHz and
$|\Omega_{ab}|\approx 20$ kHz would be sufficient if the
linewidth of the radiation field were $\delta\approx 50$ Hz,
which is a realistic assumption for typical radio-frequency
driving fields. However, it was shown in Ref. \cite{Greschner2015}
that a further condition $U,\Delta \gg\delta$ is also required
in the above scheme. For typical experimental parameters, it would
lead to large heating. This drawback was solved by a scheme
proposed in Ref. \cite{Greschner2015}, where one ground-state
component bosonic gas is replaced by two ground-state components
atomic gas.

An alternative scheme for the experimental realization of 1D
Anyon-Hubbard model, based on time-periodic forcing, was proposed
in Ref. \cite{Strater2016}. The occupation-dependent Peierls phase
can be engineered by means of coherent lattice-shaking-assisted
tunneling against potential offsets created by a combination of a
static potential tilt and strong on-site interaction. The
potential tilt $\Delta$ is added in the Hamiltonian with the
term $\Delta j b_j^\dagger b_j$. By shaking the lattice, a
similar term $F(t) j b_j^\dagger b_j$ can be further added in the
Hamiltonian, where $F(t)=F(t+T)$ incorporates a homogeneous
time-periodic force of angular frequency $\omega=2\pi/T$ with
vanishing cycle average $1/T\int_0^T dt F(t)=0$ and the resonance
condition $\Delta=\hbar\omega$. It can be implemented as an
inertial force $F(t)/a =-m\ddot{x} (t)$, with lattice constant
$a$, by shaking the lattice $x(t)$ back and forth.

A fully 1D Anyon-Hubbard model introduced here has not yet been
experimentally realized. However, some relevant ingredients have been achieved, such as the experimental implementation of tunable
occupation-dependent tunneling with Floquet engineering of the
on-site interaction energy \cite{Meinert2016} and the realization
of the occupation-number sensitivity of the tunneling \cite{RMa2011}. These techniques
may immediately applied to generating low-dimensional anyons.



\subsection{Bosonic quantum Hall states}

The integer and fractional QHE are among the
most important discoveries in condensed matter physics in 1980s.
It is a quantum-mechanical version of the Hall effect, observed in
2D electron systems subjected to low temperatures and
strong magnetic fields, in which the Hall conductance $\sigma_H$
undergoes quantum Hall transitions to take on the quantized values
$\sigma_H= \nu \frac {e^2}{h}$, where $e$ is the elementary charge
and $h$ is Planck's constant. The prefactor $\nu$ is known as the
filling factor, and can take on either integer or fractional
values. The QHE is referred to as the integer or
fractional QHE depending on whether $\nu$ is an
integer or fraction, respectively. Until now, the QHE
has been observed only in electron systems. Can we
experimentally observe such important quantum properties in other
systems is still a long-standing open question. Recently, there has been
considerable progress towards their realization in cold-atom
systems. In this section, we introduce several theoretical
proposals for realization of the QHE with cold atoms.
In principle, both bosonic and fermionic atoms can be used in the
experiments; however, the preparation of topological states of
matter relies on quick thermalization and cooling below the
many-body gap, which is hard to achieve in cold atom systems. Since
bosonic atoms are generally easier to cool, we focus on the
realization of the QHE with bosonic atoms. We will
mainly focus on the realizations of bosonic integer QHE with a Chern number $C=2$, and the fractional quantum Hall
state with the filling factor $\nu=1/2$, since they will be the
most experimentally accessible conditions.

Compared with the QHE of fermions, non-interacting boson phases are topologically trivial, and
integer QHE with bosons can only occur under the
strong interactions. The needed strong interactions for creating
bosonic quantum Hall states makes them harder to study than their
fermionic cousins. As a smoking gun of the realization of quantum
Hall state, one can compute the many-body Chern number of the
ground state $|\Psi\rangle$. In the theory of the QHE, it is well understood that the conductance quantization is
due to the existence of certain topological invariants, the so-called
Chern numbers. The Chern numbers with the single-particle problem
and Bloch waves have been introduced in the previous sections.
For fermions, the Chern number is defined as an integration over
the occupied states in momentum space \cite{Thouless1982}.
This definition cannot be applied to the bosonic system as many
bosons can occupy the same momentum state. The generalization
to many-body systems has been proposed by Niu {\sl et al.}
\cite{QNiu1985} by manipulating the phases under the closed
boundary conditions on a torus for both the integer and fractional quantum
Hall systems. Suppose the ground
state $|\Psi\rangle$ has a gap to the excited state and depends on the parameters $%
\theta_x ,\theta_y $ through the generalized periodic boundary
conditions:
$$|\Psi (m+M,l )\rangle
=e^{i\theta_x }|\Psi (m,l )\rangle, \ \ \  |\Psi (m,l+L )\rangle
=e^{i\theta_y }|\Psi (m,l )\rangle,$$
where $M\times L$ denotes the system size, and $(\theta_x,\theta_y)$ are
the twist angles vary on the torus. Under this boundary condition,
we numerically diagonalize the Hamiltonian of the system and
derive the ground state $|\Psi (\theta_x ,\theta_y )\rangle$, and
then one can define the many-body Chern
number $C_{\text{MB}}$ as a topological invariant by the following formula \cite{QNiu1985}
\begin{equation}
C_{\text{MB}}=\frac{1}{2\pi }\int_{0}^{2\pi }d\theta_x \int_{0}^{2\pi
}d\theta_y (\partial _{\theta_x }A_{\theta_y }-\partial _{\theta_y
}A_{\theta_x }),  \label{CMB}
\end{equation}%
where the Berry connection $A_{\mu }\equiv i\langle \Psi (\theta_x
,\theta_y )|\partial _{\mu }|\Psi (\theta_x ,\theta_y )\rangle $
$(\mu =\theta_x ,\theta_y )$.

As for fractional quantum Hall state, one can also calculate the
overlap between the ground state and the Laughlin wavefunction. If
$N$ is the number of particles in the system and $N_\phi$ is the
number of magnetic fluxes measured in units of the fundamental
flux quanta $\Phi_0=2\pi\hbar/e$, we can define the filling factor
$\nu=N/N_\phi$. In the simplest form the fractional QHE occurs if the number of magnetic fluxes is an integer
$1/\nu$. At this value of the magnetic field, the ground state of
the system is an incompressible quantum liquid which is separated
from all other states by an energy gap and is well described by
the Laughlin wavefunction
\begin{equation}
\label{Laughlin} \Psi({z}_1,{z}_2,\dots
{z}_N)=e^{-\sum_j|{z}_j|^2/4}\prod_{j<k}({z}_j-{z}_k)^{1/\nu},
\end{equation}
where $z=x+iy$. Due to the Pauli principle, only the states with
odd (even) $1/\nu$ is applicable to fermions (bosons). In the following, we will address bosonic QHE in
both single- and two-component Bose-Hubbard models.

\subsubsection{Single-component Bose-Hubbard model}

We consider single-component bosonic atoms at zero temperature
confined in a 2D square OL with the lattice
constant $a$ and a background harmonic potential in the presence
of artificial magnetic field. In the Landau gauge, the system is
well described by the Bose-Hubbard model with Peierls substitution
term in the nearest-neighbor hopping,
\begin{equation}
\label{HBH}  H=\sum_{m,l}\left[-J\left(e^{i2\pi\alpha
l}b^\dagger_{m+1,l}b_{m,l}+b^\dagger_{m,l+1}b_{m,l}+h.c. \right)
+\frac{U}{2}
\hat{n}_{m,l}(\hat{n}_{m,l}-1)-(\mu-V_{m,l})\hat{n}_{m,l} \right],
\end{equation}
where $J$ is the hopping energy between two neighboring sites, $U$
is the on-site interaction energy, and the phase $2\pi\alpha$
arises from the artificial magnetic field and $0\le \alpha \le 1$
is the flux quanta per plaquette. $\mu-V_{m,l}$ is the local
chemical potential with $V_{m,l}$ being the trapping potential. In
the Bose-Hubbard Hamiltonian, the hopping and on-site interaction
are the two competing terms, and both of them can be tuned over a
wide range of values by changing the depth of the lattice
potential and employing Feshbach resonance. The phase diagram of
the Bose-Hubbard model for $\alpha=0$ is well-known: the
Mott-insulator phase corresponds to the strong on-site interaction
limit $J/U \ll 1$, and superfluid phase corresponds to the
opposite limit $J/U \gg 1$.

It was first proposed in Ref. \cite{Sorensen2005} that fractional
quantum Hall states may occur in the single-component Bose-Hubbard
model. They argued that the interactions of atoms localized in the
lattices are strongly enhanced compared to the interaction of
atoms in free space, so the created states of the quantum Hall
type in OLs are characterized by large energy gaps.
It is a clear advantage from an experimental point of view because
the state is more robust against external perturbations. There are two
energy scales for the system in the presence of an artificial
magnetic field: the first is the magnetic tunneling term,
$J\alpha$, which is related to the cyclotron energy in the
continuum limit $\hbar\omega_c = 4\pi J\alpha$, and the second is
the on-site interaction energy $U$. By using the method of exact
diagonalization \cite{Sorensen2005}, it was shown that the overlap
of the ground state wave function $|\Psi\rangle$ of the Hamiltonian
in Eq. (\ref{HBH}) with the Laughlin wave function is very good when
$\nu=1/2$ for $\alpha \le 0.3$, but the overlap start to fall off
for $\alpha \ge 0.3$. Furthermore, the Chern numbers for fixed
$\nu=1/2$ and different $\alpha$'s were calculated in
Ref. \cite{Hafezi2007}. The results show that, for higher $\alpha$,
the lattice structure becomes more apparent and the overlap with
the Laughlin state decreases. However, the ground state remains
twofold degenerate and the ground state Chern number tends to
remain equal to 1 before reaching some critical $\alpha_c\approx
0.4$.

\begin{figure}[ptb]\centering
\includegraphics[width=0.8\columnwidth]{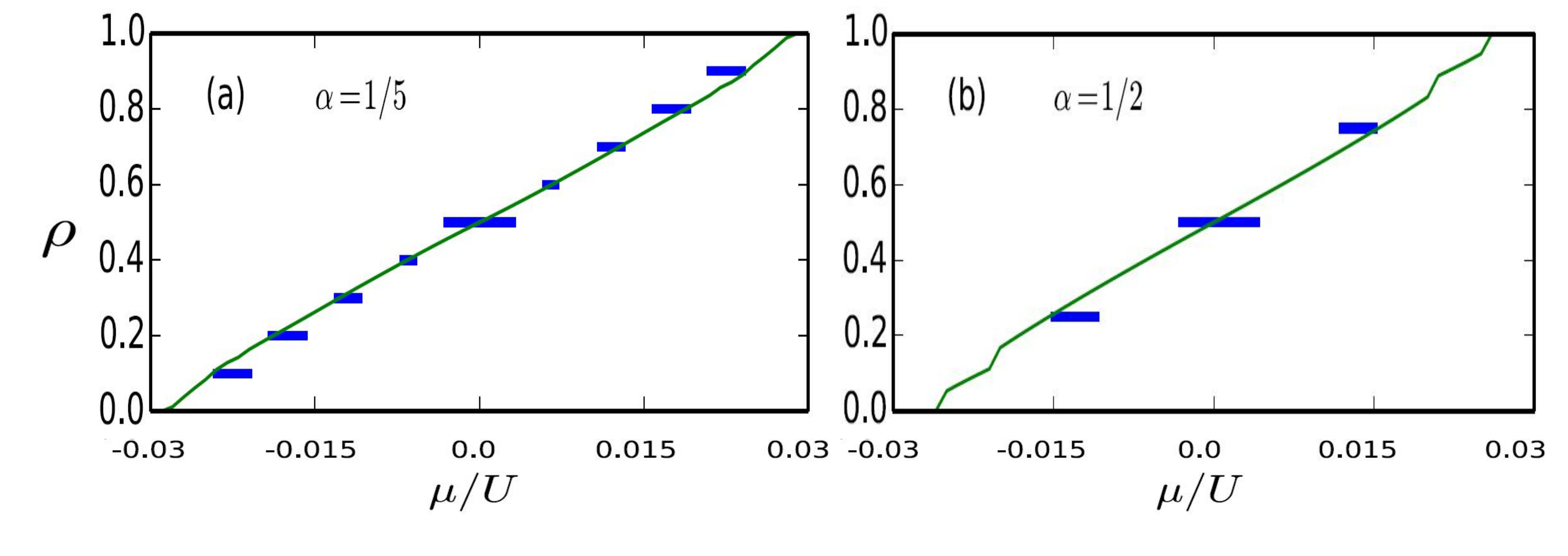}
 \caption{(Color online) The variation in the number density $\rho$ in the presence of an artificial magnetic field with $\alpha=1/5$ in (a) and for
 $\alpha=1/2$ in (b). The states in superfluid are compressible and $\rho$ varies linearly with $\mu$, as shown with solid black lines.
 For specific values of filling factor $\nu$ there are states with constant $\rho$, represented by the blue lines, and these incompressible states correspond to the existence of quantum Hall states. Reprinted with permission from Bai {\it et al.}\cite{Bai2018}. Copyright\copyright~(2018) by the American Physical Society.}
\label{QHEcomp}
\end{figure}

The states in the Bose-Hubbard model can be classified based on
the compressibility defined by $\kappa=\partial \rho/\partial
\mu$, where the density $\rho=\sum_{m,l}\langle
\Psi|\hat{n}_{m,l}|\Psi\rangle/(M L)$. It is incompressible
($\kappa=0$) for the quantum Hall states, and finite for the
superfluid states. The compressibility of the Hamiltonian in Eq.
(\ref{HBH}) was calculated by using the cluster Gutzwiller mean
field theory in Ref. \cite{Bai2018}. The results for $\alpha=1/5$
and $\alpha=1/2$ are plotted in Fig. \ref{QHEcomp}. The states in
superfluid phase are compressible, as a result, the density $\rho$
varies linearly with the chemical potential $\mu$. However, for
specific values of filling factor $\nu$ there are states with
constant $\rho$, represented by the blue horizontal lines, and
these incompressible positions correspond to the existence of
quantum Hall states. In Fig. \ref{QHEcomp}(a), the plateaus or the
constant $\rho$ values correspond to $\nu=1/2, 1, 3/2, 2, 5/2, 3,
7/2, 4, 9/2$ and the corresponding $\rho$ values are $\nu\alpha$.
In Fig. \ref{QHEcomp}(b), the plateaus correspond to
$\nu=1/2,1,3/2$.

Whether a state with integer filling factor in Fig.
\ref{QHEcomp} is an integer quantum Hall state should be carefully
analyzed. It was demonstrated that the conductivity $\sigma_{xy}$
must be even for any bosonic quantum Hall state without fractional
quasiparticle excitations \cite{Senthil2013}. To have a basic idea
about this issue, we consider some excitations created in a
general bosonic quantum Hall state. Each of them can be considered
as a bosonic particle attaching with $2\pi$ flux and has charge
$\sigma_{xy}$. If we braid one excitation around another, the
statistical phase follows from the Aharonov-Bohm effect:
$\theta=2\pi\sigma_{xy}$. Similarly, if we exchange two
excitations, the associated phase is $\theta/2=\pi\sigma_{xy}$.
On the other hand, if the state does not support fractional quasiparticles, then these excitations
must be bosons. Therefore, we conclude that $\sigma_{xy}$ must be even for any bosonic quantum Hall state without fractional quasiparticle excitations. Based on this argument, the $\nu=1$ state in Fig. \ref{QHEcomp}(a) cannot be a stable integer quantum Hall state.

Notably, the integer quantum Hall state for single component bosons can
occur in some lattice structures with Chern number $C_{\text{MB}}=2$
\cite{TSZeng2016,YCHe2015,Sterdyniak2015,Moller2009}. Recently,
two different lattice versions of bosonic quantum Hall states have
been proposed at integer filling $\nu=1$ of the lowest topological
flat-band with $C_{\text{MB}}=2$. The optical flux lattice has
been studied by exact diagonalization of the projected Hamiltonian
in momentum space \cite{Sterdyniak2015} and correlated Haldane
honeycomb lattice has been studied by infinite density matrix
renormalization group of hardcore boson in real space. The authors
in Ref. \cite{YCHe2015} established the existences of the bosonic
quantum Hall phase in their model by providing numerical evidence:
(i) a quantized Hall conductance with $\sigma_{xy}=2$; (ii) two
counter propagating gapless edge modes. On the other hand, it was
demonstrated that bosonic integer quantum Hall state emerges in
integer boson filling factor $\nu=1$ of the lowest band in a
generalized Hofstadter lattice (including the nearest neighbor
hopping) with $C_{\text{MB}}=2$ \cite{TSZeng2016}.


\subsubsection{Two-component Bose-Hubbard model}

\begin{figure}[ptb]\centering
\includegraphics[width=0.7\columnwidth]{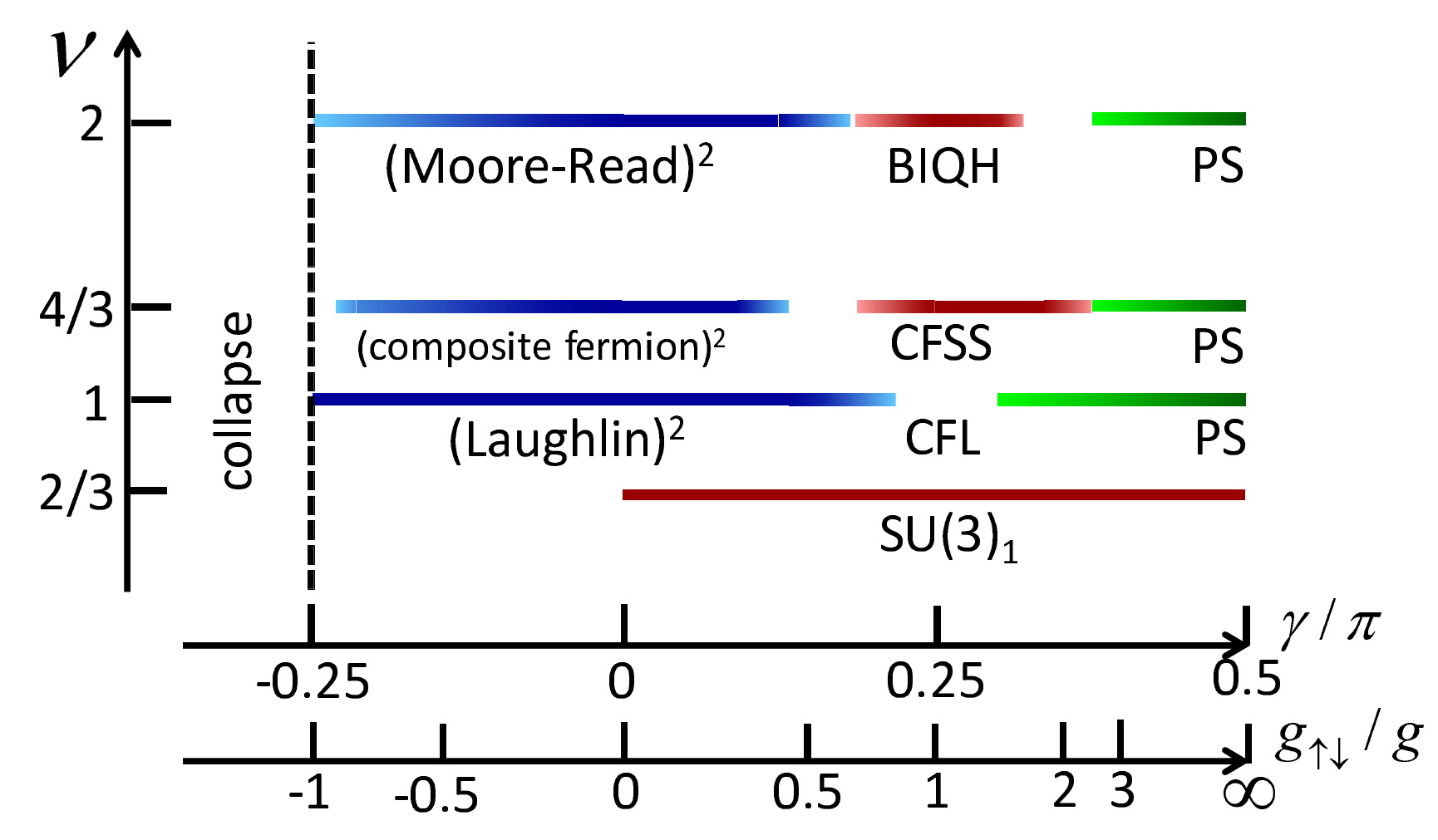}
 \caption{(Color online) Ground-state phase diagram in the space of the total filling
factor $\nu$ and the ratio $U_{\uparrow\downarrow}/U = \tan\gamma$
between the intercomponent coupling constant and the
intracomponent one. The product states of a pair of nearly
uncorrelated quantum Hall states (Laughlin, composite fermion, and
Moore-Read states) appear when $U_{\uparrow\downarrow}<0$. BIQH:
bosonic integer quantum Hall state; PS: phase separation;
CFSS:composite fermion spin-singlet state; CFL: composite fermion
liquid. $SU(3)_1$: the Halperin state with an $SU(3)_1$ symmetry. Reprinted with permission from Furukawa {\it et al.} \cite{Furukawa2017}. Copyright\copyright~(2017) by the American Physical Society.} \label{Phase_2CBHM}
\end{figure}

We further address the quantum Hall states in a two-component
Bose-Hubbard model. We consider a system of a 2D pseudospin-1/2
bosonic gas (in the $xy$ plane) subject to the same magnetic
fields $B$ along the $z$ axis for both spin states. In the
second-quantized form, the interaction Hamiltonian is written as
$$
H_{\text{int}}=\sum_{\alpha\beta}\frac{U_{\alpha\beta}}{2}\int
d^2\mathbf{r}\hat{\Psi}_\alpha^\dagger(\mathbf{r})\hat{\Psi}_\beta^\dagger
(\mathbf{r}) \hat{\Psi}_\alpha
(\mathbf{r})\hat{\Psi}_\beta(\mathbf{r}),
$$
where $\hat{\Psi}_\alpha
(\mathbf{r})$ is the bosonic field operator for the spin state
$\alpha$ ($=\uparrow$ or $\downarrow$). We set the strengths of
the intracomponent contact interactions
$U_{\uparrow\uparrow}=U_{\downarrow\downarrow}=U
>0 $ and the strengths of the intercomponent contact interactions
$U_{\uparrow\downarrow}=U_{\downarrow\uparrow}$. For a 2D system
of area A, the number of magnetic flux quanta piercing each
component is given by $N_\phi = |\phi|/(2\pi\hbar) =
A/(2¦Ð\ell^2)$, where $\ell= \sqrt{\hbar A/|\phi|}$ is the
magnetic length. Strongly correlated physics is expected to emerge
when $N_\phi$ becomes comparable with or larger than the total
number of particles, $N = N_\uparrow + N_\downarrow$, where
$N_\uparrow$ and $N_\downarrow$ are the numbers of
pseudospin-$\uparrow$ and $\downarrow$ bosons, respectively.

The ground-state phase diagram of pseudospin-1/2 bosonic gases in
a uniform artificial magnetic field in the space of the total
filling factor $\nu=N/N_\phi$ and the coupling ratio
$U_{\uparrow\downarrow}/U$ were numerically calculated by
performing an extensive exact diagonalization analysis in the
lowest-Landau level based on spherical and torus geometries
\cite{Furukawa2017}. The main results are summarized in Fig.
\ref{Phase_2CBHM}. In the figure, the two coupling constants are
parametrized as
$$
(U,U_{\uparrow\downarrow}) = G \ell^2 (\cos\gamma, \sin\gamma),
$$
where $G > 0$, and $\gamma\in[-\pi/2,\pi/2]$. As can be seen in
this diagram, when the intercomponent coupling is attractive
($U_{\uparrow\downarrow} < 0$), doubled quantum Hall states are
remarkably robust and persist even when $|U_{\uparrow\downarrow}|$
is comparable to the intracomponent coupling $U$. This sharply
contrasts with the case of an intercomponent repulsion
($U_{\uparrow\downarrow} < 0$), where a variety of spin-singlet
quantum Hall states with high intercomponent entanglement emerge
for $U_{\uparrow\downarrow} \approx U$. This remarkable dependence
on the sign of $U_{\uparrow\downarrow} $ can be interpreted in
light of Haldane¡¯s pseudopotentials on a sphere. More
specifically, the stability of the doubled quantum Hall states for
$U_{\uparrow\downarrow} < 0$ can be understood from the
¡°ferromagnetic¡± nature of the intercomponent interaction in
terms of (modified) angular momenta of particles. Meanwhile,
various spin-singlet quantum Hall states with a finite excitation gap emerge for pseudospin-independent
[$SU(2)$-symmetric] interactions with $U_{\uparrow\downarrow}=U$.
Among those states, relatively large gaps are found for the
Halperin state with an $SU(3)_1$ symmetry at $\nu = 2/3$
\cite{Paredes2002} and a bosonic integer quantum Hall state
protected by a $U(1)$ symmetry at $\nu = 2$
\cite{Senthil2013,Furukawa2013}. At $\nu=4/3$, two types of
spin-singlet quantum Hall states compete in finite-size systems: a
non-Abelian $SU(3)_2$ state  and a composite fermion spin-singlet
state. Furthermore, a gapless spin-singlet composite Fermi liquid
has been shown to appear at $\nu = 1$ \cite{YHWu2013}. In all these
spin-singlet states, the two components are highly entangled. For
small $U_{\uparrow\downarrow}/U$, in contrast, the system can be
viewed as two weakly coupled scalar bosonic gases, and the product
states of nearly independent quantum Hall states (doubled quantum
Hall states) are expected to appear.

We can compare the phase diagram in Fig. \ref{Phase_2CBHM} with
that of the two-component bosonic gases in antiparallel magnetic
fields \cite{Furukawa2014}. In the latter case, the pseudospin
$\uparrow$ ($\downarrow$) component is subject to the magnetic
field $B$ $(-B)$ in the direction perpendicular to the 2D gas, and
the system possesses the TRS
\cite{SLZhu2006,XJLiu2007}. In the regime with $\nu = O(1)$,
(fractional) quantum spin Hall states \cite{Bernevig2006a} composed
of a pair of quantum Hall states with opposite chiralities are
robust for an intercomponent repulsion $U_{\uparrow\downarrow}> 0$
and persist for $U_{\uparrow\downarrow}$ as large as $U$. Similar
results have also been found in the stability of two coupled
bosonic Laughlin states in lattice models. These results suggest
that the case of $U_{\uparrow\downarrow}> 0$ for antiparallel
fields essentially corresponds to the case of
$U_{\uparrow\downarrow}< 0$ for parallel fields
\cite{Furukawa2017}.


\subsection{Kitaev honeycomb model}

Generalized the previous idea on realization of the unpaired Majorana
zero modes in a 1D system, in 2006, Kitaev further
proposed another model that unpaired zero-energy Majorana modes
can appear in a 2D spin-1/2 system on a honeycomb lattice, where
nearest-neighbor interactions can be reduced to a problem of
non-interacting Majorana fermions \cite{Kitaev2006}. It is one of
the rare examples where a complex system is described by an
exactly solvable 2D spin Hamiltonian. Its quantum-mechanical ground state is a quantum spin liquid and supports
exotic excitations which obey Abelian or non-Abelian statistics.

\begin{figure}[tbph]\centering
\includegraphics[width=0.7\columnwidth]{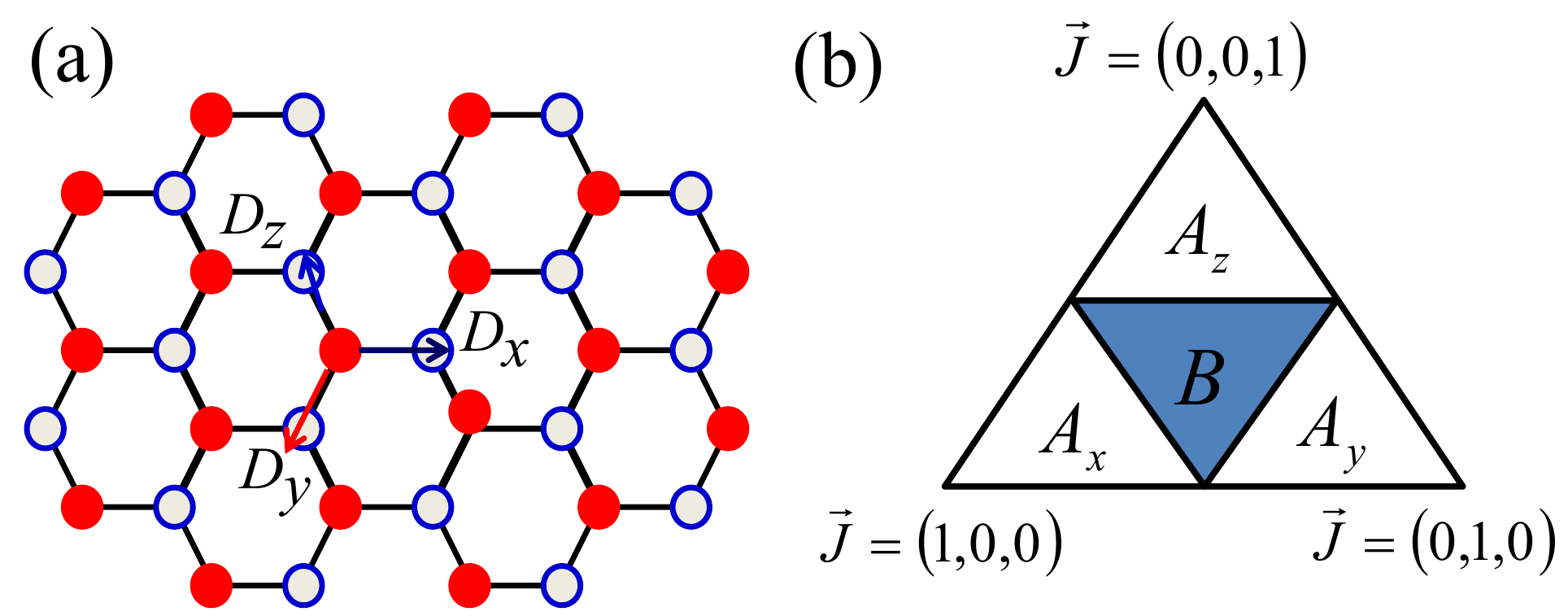}
 \caption{(Color online) (a) Kitaev model on the honeycomb
 lattice where interactions between nearest neighbors are $J_\nu$,
 depending on the direction of the link $D_\nu$.
 (b) The phase diagram of the Kitaev honeycomb model in the
$J_x + J_y + J_z = 1$ plane in the parameter space. In the three
unshaded areas labeled $A_x$, $A_y$, and $A_z$, the system is
gapped with Abelian anyon excitations, and in the shaded area
labeled $B$ the system is gapless with non-Abelian excitations. }
\label{Kitaev2D}
\end{figure}

The Kitaev honeycomb model is a spin-1/2 system in which spins are
located at the vertices of a honeycomb lattice with a spatially
anisotropic interaction between neighboring spins, as shown in Fig.
\ref{Kitaev2D}(a). This lattice consists of two equivalent
sublattices labelled `A' and `B', which are shown by open and
filled circles. A unit cell of the lattice contains both of them. The Hamiltonian is given by  \cite{Kitaev2006}
\begin{equation}\label{KitaevH}
H_\text{KHM}=-\sum_{\nu,\langle j,l\rangle\in
D_\nu}J_\nu\sigma^\nu_{j}\sigma^\nu_{l},
\end{equation}
where $\sigma^\nu_{j}$ are the Pauli matrices at the site $j$,
$J_\nu$ ($\nu=x,y,z$) are interaction parameters, and the symbol
$\langle j,l\rangle \in D_\nu $ denotes the neighboring spins in
the $D_\nu$ directions. Neighboring spins in Heisenberg models
normally interact isotropically so that the spin-spin interaction
does not depend on the spatial direction between neighbors. In the
above model, however, neighboring spins along links pointing in
different directions interact differently.

The ground state of the Kitaev honeycomb model has two distinct
phases in the parameter space and the phase diagram can be shown
in terms of points in an equilateral triangle satisfying
$J_x+J_y+J_z=1$ (the value of $J_\nu$ is the distance from the
opposite side), as shown in Fig. \ref{Kitaev2D}(b)
\cite{Kitaev2006}. If $J_x<J_y+J_z$, $J_y<J_z+J_x$ and
$J_z<J_x+J_y$, the system is gapless with non-Abelian excitations
corresponding to $B$ phase. For all other values of
$(J_x,J_y,J_z)$, the system is gapped with Abelian anyon
excitations, labelled, $A_x$ where $J_x>J_y+J_z$, $A_y$ where
$J_y>J_z+J_x$, $A_z$ where $J_z>J_x+J_y$. The gapped phases,
$A_x$, $A_y$, and $A_z$, are algebraically distinct, though
related to each other by rotational symmetry. They differ in the
way lattice translations act on anyonic states, and thus a
continuous transition from one gapped phase to another is
impossible. The two phases $A$ and $B$ are separated by three
transition lines, i.e., $J_x= 1/2$, $J_y= 1/2$, and $J_z= 1/2$,
which form a small triangle in the $B$ phase.

Precise proposals to realize an artificial Kitaev model using
atomic OLs have been made in the
literature \cite{LMDuan2003,CWZhang2007}, where the
well-controllable OLs offer the possibility of
designing such anisotropic spin lattice models. The main idea is
that the two-component Bose-Hubbard model in a honeycomb lattice
can reduce to the Kitaev spin model at half filling and large
on-site repulsion. For completeness, we here focus on the proposal
in Ref. \cite{LMDuan2003} and its modified implementation scheme
for $^{87}$Rb atoms proposed in Ref. \cite{CWZhang2007}. To
implement the Kitaev honeycomb model using ultracold atoms, we first get an
effective 2D configuration with a set of independent identical 2D
lattice in the $xy$ plane by raising the potential barriers along
the vertical direction $z$ in the 3D OL
so that the tunneling and the spin exchange interactions in $z$
direction are completely suppressed. And then a honeycomb lattice
can be constructed with three trapping potentials of the forms
\begin{equation}
\label{Vhoneycomb} V_j(x, y)= V_0\sin^2[k_\parallel(x \cos\theta_j
+ y \sin\theta_j+\varphi_0),
\end{equation}
where $j=1, 2, 3$, and $\theta_1=\pi/6$, $\theta_2=\pi/2$,
$\theta_3=-\pi/6$. Each of the potentials is formed by two
blue-detuned interfering traveling laser beams above the $xy$ plane
with an angle $\varphi_\parallel$, so that the wave vector
$k_\parallel$ projected onto the $xy$ plane has the value
$k_\parallel=k\sin (\varphi_\parallel/2)=k/\sqrt{3}$. The relative
phase $\varphi_0$ is chosen as $\pi/2$ in Eq. (\ref{Vhoneycomb})
so that the maxima of the three potentials overlap. In this case,
the atoms are trapped at the minima of the potentials and the
potential barrier between neighboring atoms is around $V_b \approx
V_0/4$. Actually, the honeycomb (hexagonal) lattice and its
topologically equivalent configuration, the brick wall lattice, have
been experimentally realized by several
groups \cite{Tarruell2012,Jotzu2014,Duca2015}. We consider a
$^{87}$Rb BEC and two hyperfine ground states
$|\uparrow\rangle=|F=2,m_F=-2\rangle$ and
$|\downarrow\rangle=|F=1,m_F=-1\rangle$ are defined as the
effective atomic spin. The potential barrier between neighboring
atoms in the honeycomb lattice is adiabatically ramped up to
approximately $V_b=14 E_R$ to obtain a Mott insulator state with
one atom per lattice site, where $E_R=\hbar^2 k^2/(2m)$ is the
recoil energy for Rb atoms.

In this hexagonal lattice, one can engineer the anisotropic
spin-spin interactions $J_\nu\sigma_j^\nu\sigma_m^\nu$ in
Eq. (\ref{KitaevH}) using additional spin-dependent standing wave
laser beams in the $xy$ plane. To this end, one can apply three
blue-detuned standing-wave laser beams in the $xy$ plane along the
tunneling directions denoted by $D_x$, $D_y$, and $D_z$,
respectively:
\begin{equation}
\label{Vspin} V_{\nu\sigma}(x, y)= V_{\nu\sigma}\sin^2[k(x
\cos\chi_\nu + y \sin\chi_\nu),
\end{equation}
where $\chi_x=-\pi/3$, $\chi_y=\pi$, $\chi_z=\pi/3$. With properly chosen laser configurations, a spin-dependent potential
\begin{equation}
\label{Vspin1} V_{\nu\sigma}= V_{\nu +}|+\rangle_\nu\langle +|
+V_{\nu -}|-\rangle_\nu \langle -|
\end{equation}
along different tunneling directions $\nu$ can be generated, where
$|+\rangle_\nu$ ($|-\rangle_\nu$) is the eigenstate of the
corresponding Pauli operator $\sigma^\nu$ with the eigenvalue $+1
(-1)$. One can adjust $V_{+}^\nu$ and $V_{-}^\nu$ by varying the
laser intensity in the $D_\nu$ direction so that atoms can virtually
tunnel with a rate $t_{+\nu}$ only when it is in the eigenstate
$|+\rangle_\nu$, which yields the effective spin-spin exchange
interactions $J_\nu\sigma_j^\nu\sigma_m^\nu$ with the interaction
strength $J_\nu\approx-t^2_{+\nu}/(2U)$. Here the on-site
interactions $U_{\uparrow\downarrow}\approx U_\uparrow\approx
U_\downarrow\approx U$.

As for a typical example, we introduce more detailed on how to
generate the spin-spin interaction $J_z\sigma_j^z\sigma_m^z$ in the
Hamiltonian Eq. (\ref{KitaevH}) following the proposal in
Ref. \cite{CWZhang2007}, and the other spin-spin interaction terms
can be created using a similar procedure \cite{LMDuan2003}. The
potentials (\ref{Vspin}) and (\ref{Vspin1}) do not have influence
on the equilibrium positions of the atoms, but they change the
potential barrier between the neighboring atoms in the $D_z$
direction from $V_b\approx V_0/4$ to
$V_{z\sigma}^\prime=V_b+V_{z\sigma}$. The standing wave laser beam
used for generating spin-dependent tunneling is along the $z$-link
direction and has a detuning $\Delta_0=2\pi\times 3600$ GHz to the
$5^2 P_{3/2}$ state (corresponding to a wavelength $787.6$ nm). This
laser beam forms a blue-detuning potential for atoms with spin
$|\uparrow\rangle$, but a red-detuning potential for
$|\downarrow\rangle$ atoms. For instance, with a properly chosen
laser intensity, the spin-dependent potential barriers may be set
as $V_\downarrow= 8 E_R$ and $V_\uparrow=-4 E_R$, which, combined
with the spin-independent lattice potential barrier $V_0=14E_R$,
yield the total effective spin-dependent lattice potential
barriers $V_\downarrow=22E_R$ and $V_\uparrow=10E_R$ for
neighboring atoms in the honeycomb lattice. Therefore, the
tunneling rates for two spin states satisfy
$t_\uparrow/t_\downarrow\gg1$, which leads to the spin-spin
interaction $J_z\sigma_j^z\sigma_m^z$ with $J_z
 \approx t^2_\uparrow /U$, as shown in Ref. \cite{LMDuan2003}. For $^{87}$Rb
atoms, the time scale for the spin-spin interaction $h/J_z \approx$
10 ms. By carefully tuning the spin-dependent lattice depth in
different directions, one can in principle access all phases of
the Kitaev honeycomb model. However, to observe the properties in this
model, the temperature of the system needs to be much
lower than the spin-spin interaction strength $T \ll
J_z/k_B\approx  1$ nK, which sets a strict requirement for
experiments. Higher temperatures will populate the system with an
excess of unwanted excitations.

The Kitaev honeycomb model is an exactly soluble spin model that
carries excitations with both Abelian and non-Abelian anyonic
braiding statistics, which are the hallmarks of topological
quantum matter. The advantage of the realization of this model with
OLs is that both the Abelian and non-Abelian phases
can be accessed just by varying the OL parameters.
On the other hand, how to create, braid, and detect these anyons in
this spin model defined on a honeycomb OL have also been proposed in
Ref. \cite{CWZhang2007}, which is an important first step towards
topological quantum computation. Furthermore, even the simple
observation of Abelian anyonic properties in an OL
and the subsequent read-out them will be a break-through
achievement in itself, because anyonic statistic has never been
directly demonstrated in any experimental system.




%
%
%
%
%
%

\section{Conclusion and outlook}

In the previous sections, we have reviewed the recent theoretical and experimental advances on exploring topological quantum matter with cold atoms. The cold atom systems provide many interesting possibilities of searching for exotic topological states that are currently absent or unrealizable in real materials. These include some unconventional topological insulators and semimetals, the topological phases with many-body interactions, non-equilibrium dynamics, and non-Hermitian or dissipation perturbations. So far, the theoretical understanding of these phases is limited and most experimental studies are at the single-particle level, but progress is already being made. In this final section of the review, we aim to discuss some promising developments for the near future.

\subsection{Unconventional topological bands}

Ultracold atomic gases in OLs with fully engineered geometries and atomic hopping forms provide a promising platform for exploring certain unconventional topological bands. These include some unconventional topological insulators and semimetals that are difficult to realize in solid-state materials, such as the chiral topological insulators protected by the chiral symmetry \cite{Essin2012,STWang2014,Velasco2017}, topological nodal-line semimetals protected by the combined space-time symmetry \cite{DWZhang2016a,WChen2017}, and the topological bands with unconventional relativistic quasiparticles \cite{Bradlyn2016,YQZhu2017,HPHu2017}. The chiral symmetry played by certain sublattice symmetry is typically broken by disorder potential in real materials; however, it naturally arises for cold atoms in OLs with negligible disorder \cite{STWang2014}. The topological semimetals or metals with tunable structures of nodal points or lines could be experimentally implemented by varying the atom-laser interaction configuration \cite{YQZhu2017,DWZhang2016}.

Furthermore, some theoretically predicted topological insulators beyond the ten-fold classification could be implemented in OLs. These include the Hopf insulators \cite{Moore2008,DLDeng2013,DLDeng2018}, the 3D quantum Hall states \cite{DWZhang2017}, the topological crystalline insulators stabilized by crystalline lattice symmetries \cite{LFu2011}, the topological Anderson insulators \cite{JLi2009,Groth2009,HJiang2009}, and the so-called higher-order topological insulators and semimetals \cite{Benalcazar2017a,Benalcazar2017b,Langbehn2017,ZSong2017,Schindler2018,MLin2017}. The topological Anderson insulator is a disorder-driven topological phase, where the static disorder induces nontrivial topology when added to a trivial band structure \cite{JLi2009,Groth2009,HJiang2009}. Although there are many theoretical studies, the topological Anderson insulator has so far evaded experimental realization due to the lack of precise control over disorder and topology in real materials. In a recent experiment, evidence for the topological Anderson insulator phase in synthetic cold-atomic wires with controllable disorder has been found \cite{Meier2018}. As an extension of the topological insulator family, the recently proposed higher-order ($n$-th order) topological insulators can host quantized multipole moments in the bulk bands, such as quadrupole and octupole, and has robust gapless states at the intersection of $n$ crystal faces (but is gapped otherwise). For instance, a bulk 2D topological quadrupole insulator described by ¡°nested¡± Wilson loops hosts protected corner states with fractional charges \cite{Benalcazar2017a}, and its extension to a layered 3D system can give rise to a topological quadrupolar semimetal \cite{MLin2017}. It was also proposed to realize topological quadrupole insulators using ultracold atoms in an optical superlattice \cite{Benalcazar2017a}.

Cold gases also allow the exploration of band topology in $D=d_r+d_s$ dimensions higher than the real dimension $d_r=3$, through the use of $d_s$ synthetic dimensions. A recent experiment has demonstrated the topological response of an effective 4D system by using 2D pumping in an OL \cite{Lohse2018}. Other topological states in 4D would be studied in the near future, such as 4D intriguing fractional phases induced by interactions \cite{ZhangSC2001} and time-reversal-symmetric 4D QHE. In addition, the 5D generalization of the topological Weyl semimetals with Yang monopoles and linked Weyl surfaces in the BZ \cite{BLian2016,BLian2017} would be similarly simulated with cold atoms. Very recently, the quantum simulation of a Yang monopole in a 5D parameter space built
from the internal states of an atomic quantum gas was reported \cite{Sugawa2018}. Moreover, its topological charges (the second Chern numbers) were measured by experimentally characterizing the associated non-Abelian Berry curvatures in the parameter space.

\subsection{Other interacting topological phases}

\emph{Topological superfluids with Majorana bound states.} It has been theoretically shown that the $p+ip$-wave or $p$-wave topological superconductors/superfluids can be effectively induced in conventional $s$-wave superconductors/superfluids by combining the SOC and Zeeman splitting \cite{Sau2010,Tewari2007,CZhang2008,Sato2009}. The zero-energy Majorana bound states with non-Abelian statistics can emerge in these systems, but they have not yet been experimentally confirmed. With the recent advances, all the individual ingredients including the synthetic SOC and effective Zeeman fields for topological superfluids in fermionic quantum gases are in place. Several concrete proposals for realizing exotic topological superfluids with Majorana bound states for cold atomic gases have been proposed \cite{Tewari2007,CZhang2008,Sato2009,Massignan2010,SLZhu2011,LJiang2011,FWu2013,YCao2014}.
Recently, a 2D SOC and a perpendicular Zeeman field have been simultaneously generated in ultracold Fermi gases \cite{LHuang2016,ZMeng2016}, which paved the way for future exploration of topological superfluids in ultracold atoms. Once the systems are experimentally realized, the high degree of experimental control over these cold atom systems will enable new approaches for the direct observation and manipulation of Majorana bound states, such as non-Abelian braiding.

\emph{Topological Mott insulators.} In general, a strong interaction will open a trivial energy gap and break the band topology. However, there exists a class of topological insulators called the topological Mott insulators \cite{Raghu2008}, where the many-body interactions are responsible for topological insulator behaviors. Although the topological Mott insulator phase was first reveled in an extended Fermi-Hubbard model on a 2D honeycomb lattice \cite{Raghu2008}, it has now been known as a class of interaction-induced topological insulators for interacting fermions or bosons. The topological Mott insulators in 1D and 3D fermion systems have also been investigated \cite{Yoshida2014,Herbut2014,Amaricci2016}. Whereas many studies provide growing evidence for the existence of the topological Mott insulating phase, its experimental observation in electron systems is still outstanding. The schemes for realizing the topological Mott insulator with Rydberg-dressed fermionic atoms in 2D OLs were put forward \cite{Dauphin2012,Dauphin2016}. Furthermore, several works have  theoretically and numerically shown that the topological Mott insulator phase can occur for interacting bosonic atoms \cite{SLZhu2013,XDeng2014,TLi2015,Kuno2017}, atomic mixtures \cite{ZXu2013}, and fermionic atoms \cite{HHu2017} in 1D optical superlattices. Due to the tunable atomic interactions in OLs (similar as the engineered Bose- or Fermi-Hubbard models for cold atoms), these artificial systems may provide the first realization of topological Mott insulators in the near future.

\emph{Topological Kondo insulators.} A class of topological insulators called topological Kondo insulators for few heavy-fermion materials were recently predicted \cite{Dzero2010,FLu2013,Alexandrov2013}, which originates from the hybridization between itinerant conduction bands and correlated electrons. Topological Kondo insulators are essentially induced by the strongly correlated Kondo effect that leads to the insulating gap, although they share the same topological properties with conventional topological insulators. Little evidence for a topological Kondo insulator state in SmB$_6$ has been reported \cite{NXu2014,Kim2014}, and more theoretical and experimental works are needed to fully understand it. On the other hand, a scheme has recently been proposed to realize and observe topological (Chern) Kondo insulators in a 2D optical superlattice with laser-assisted $s$ and $p$ orbital hybridization and a synthetic gauge field \cite{HChen2016}. The topological Kondo insulator phase was also predicted on interacting ``sp-ladder" models \cite{Lisandrini2017,SNiu2018}, which could be experimentally realized in OLs with higher orbitals loaded with ultracold fermionic atoms. Motivated by experimental advances on ultracold atoms coupled to a pumped optical cavity \cite{Ritsch2013}, a scheme for synthesizing and observing the topological Kondo insulator in Fermi gases trapped in OLs was also proposed \cite{ZZheng2018}.

\subsection{Non-equilibrium dynamics and band topology}

In a recent experiment \cite{Flaeschner2017}, the dynamical evolution of the Bloch
wavefunction was studied by using time- and momentum-resolved
state tomography for spinless fermionic atoms in the driven
hexagonal OL. In particular, the appearance, movement
and annihilation of dynamical vortices in momentum space after
sudden quenches close to the topological phase transition were
observed. Furthermore, it was theoretically proposed \cite{CWang2017} and experimentally
demonstrated \cite{Tarnowski2017b} that the topological Chern
number of a static Hamiltonian can be measured from a dynamical
quench process through a rigorous mapping between the band
topology and the quench dynamics, i.e., the mapping of the Chern
number to the linking number of dynamical vortex trajectories
appearing after a quench to the Hamiltonian. It was also predicted that a topologically quantized Hall response can be dynamically built up from nontopological states \cite{YHu2016}.

Very recently, a different dynamical approach with high precision has
been experimentally demonstrated to reveal topology through the unitary evolution after a quench from a topological trivial
initial state to a 2D Chern band realized in an ultracold $^{87}$Rb atom gas \cite{WSun2018}. The emerging
ring structure in the spin dynamics uniquely determines the Chern number for the post-quench
band. The dynamical quantum phase transition and the topological properties in the quench dynamics
have been theoretically studied for various topological systems \cite{ZGong2017,Kruckenhauser2017,XQiu2018,CYang2018,PChang2018,LZhang2018,LZhang2018b}. These studies have shown that the cold atom systems provide a natural and promising platform to explore the connection between topological phases and non-equilibrium dynamics.

\subsection{Topological states in open or dissipative systems}

\emph{Topological states in open systems.} The topological phases discussed so far are ground state phases in isolated systems. With an additional dissipative coupling between atoms and an environment, one may consider the possibility of engineering topological states using the concept of dissipative state preparation \cite{Diehl2008,Verstraete2009}. For simplicity, here we take the environment temperature to be $T=0$ and restrict our discussion to the cold atom implementations \cite{Diehl2011,Bardyn2012,Bardyn2013}. For a weak coupling to a Markovian bath, which is a good approximation of atoms coupled to a continuum of radiation modes, the master equation takes the Lindblad form \cite{Lindblad1976}
\begin{align}
\text{d} \hat \rho/\text{d} t  = i\left[\hat \rho, \hat H\right]+\sum_j\left(\hat L_j \hat \rho \hat L_j^\dag -\frac{1}{2}\left\{\hat L_j^\dag \hat L_j, \hat\rho\right\}\right),
\label{eqn:Lindblad}
\end{align}
where $\hat \rho$ denotes the reduced density matrix of the system, and the incoherently acting Lindblad operators $\hat L_j$ (also called jump operators) account for the system-bath coupling with the dissipative channels being labelled by $j$. In the open systems, the steady states $\hat \rho_s$ are defined by $\text{d} \hat \rho_s/\text{d} t=0$, and the counterpart to an energy gap is provided by a damping gap defined as the smallest rate at which deviations from $\hat \rho_s$ are damped out. When the coupling is engineered so that the system ends up after some relaxation time into a pure state (called a dark state), ``topology by dissipation" is achieved where the pure state has nontrivial topological properties \cite{Diehl2011,Bardyn2012}. The specific system studied in Ref. \cite{Diehl2011} is a quantum wire of spinless atomic fermions in an OL coupled to a bath. The key feature of the dissipative dynamics described by the Lindblad master equation is the existence of Majorana edge modes, and their topological protection is granted by a non-trivial winding number of the system density matrix. Such a concept of topology by dissipation has formally been extended to higher spatial dimensions and various symmetry classes \cite{Bardyn2013}. Furthermore, it was shown that the dissipation can lead to a novel manifestation of topological states with no Hamiltonian counterpart \cite{Bardyn2012}, such as spatially separated Majorana zero modes in the dissipation-induced $p$-wave paired phase of 2D spin-polarized fermions with zero Chern number.

\emph{Topological superradiant states.} The experimental advances on ultracold atomic gases coupled to an optical cavity have shown that the interplay between the atomic motion and the light fields can give rise to rich dynamical processes and exotic many-body collective phenomena \cite{Ritsch2013}, such as the Dicke superradiant state \cite{Baumann2010}. Recently, a topological superradiant state in a 1D spin-1/2 degenerate Fermi gas in a cavity with cavity-assisted Raman processes was predicted \cite{JSPan2015,DYu2017}. This novel steady-state topological phase of a driven-dissipative system is characterized simultaneously by a local order parameter and a global topological invariant (the winding number of momentum-space spin texture) with a superradiance-induced bulk gap. It was also suggested to detect the topological phase transition between normal and topological superradiant states from its signatures in the momentum distribution of the atoms or the variation of the cavity photon occupation, due to the nontrivial feedback of the atoms on the cavity field \cite{JSPan2015}. A superradiant topological Peierls insulator involving transversely laser-driven atoms coupled to a single mode of an
optical resonator in the dispersive regime was also predicted \cite{Mivehvar2017}. A fermionic quantum gas in a 2D OL coupled to an optical cavity can self-organize into a state in which the cavity mode is occupied and an artificial magnetic field dynamically emerges, such that the fermionic atoms can form steady-state chiral insulators \cite{Kollath2016} or topological Hofstadter insulators \cite{Sheikhan2016}.

\emph{Topological states in non-Hermitian systems.} Recently, the search for topological states of matter in non-Hermitian systems has attracted increasing interest (see Ref. \cite{Alvarez2018} for a review). For a dissipative cold atom system with particle gain and loss, a new type of topological ring characterized by both a quantized Chern number and a quantized Berry phased (defined via the Riemann surfaces) was revealed \cite{YXu2017b}, dubbed a Weyl exceptional ring consisting of exceptional points at which two eigenstates coalesce. Realizing the Weyl exceptional ring requires a non-Hermitian term associated with particle loss for spin-down atoms. Without this term, the system is in the Weyl semimetal phase that can be realized with cold atoms in 3D OLs (see Sec. \ref{WeylSM}). To generate the decay term representing an atom loss for spin-down atoms, one may consider using a resonant
optical beam to kick the spin-down atoms out of a weak trap, or alternatively, using a radio frequency pulse to excite the spin-down atoms to another irrelevant internal state. A possible approach to measure the Weyl exceptional ring is probing the dynamics of atom numbers of each spin component after a quench  \cite{YXu2017b}. The non-Hermitian Hamiltonian was recently realized in a noninteracting $^6$Li Fermi gas via generating state-dependent atom loss, and the non-Hermitian term was achieved by an optical beam resonant with the atomic decay coupling \cite{JLi2016}.

\section*{Acknowledgements}

For figures with copyright from the American Physical Society: Readers may view, browse, and/or download material for temporary copying purposes only, provided these uses are for noncommercial personal purposes. Except as provided by law, this material may not be further reproduced, distributed, transmitted, modified, adapted, performed, displayed, published, or sold in whole or part, without prior written permission from the American Physical Society.

\section*{Disclosure statement}
No potential conflict of interest was reported by the authors.

\section*{Funding}
This work was supported by the NKRDP of China [grant no. 2016YFA0301803], the NSFC
[grant nos. 11604103, 11474153, 91636218, and 11874201], the NSAF [grant nos. U1830111
and U1801661], the NSF of Guangdong Province [grant no. 2016A030313436], the KPST of
Guangzhou [grant no. 201804020055], and the Startup Foundation of SCNU.

\begin{appendix}

\section{Formulas of topological invariants} \label{App}
The purpose of this Appendix is to provide general discussions and more details of derivations of topological invariants referenced in this review. Mathematically, these topological invariants are defined for vector or principal bundles to characterize the topological (isomorphism) classes of the bundles and have applications wherever the bundles find their manifestations in physical systems. In the course of adapting this mathematical subject into physics, condensed matter and high energy physics communities made tremendous endeavors. To serve our main subject of topological cold-atom systems, we do not intend a complete review but focus on the Berry bundle of band theories for an insulator of non-interacting fermions. The formulas of topological invariants are applicable to other suitable bundles of physical systems.

\subsection{Flattened Hamiltonians and Berry Bundles} \label{Berry Bundles}
We denote the momentum-space Hamiltonian of the insulator as $\H(\k)$ with $\k$ in the first Brillouin zone (BZ), and assume finite number of bands, namely $\H(\k)$ is a $(M+N)$-dimensional matrix at each $\k$, where $M$ and $N$ are numbers of conduction and valence bands, respectively. At each $\k$, $\H(\k)$ can be diagonalized and the conduction and valence eigenpairs are $(E_{+,a},~|+,\k,a \rangle)$ and $(E_{-,b},~|-,\k,b\rangle)$, respectively, with $a=1,\cdots,M$ and $b=1,\cdots,N$. Therefore the Hamiltonian is now expressed as $\H(\k)=\sum_a E_{+,a}|+,\k,a\rangle \langle +,\k,a|+\sum_b E_{-,b}|-,\k,b\rangle \langle -,\k,b|$. We further introduce the projectors onto conduction and valence spaces as $\Pi_{\pm}(\k)=\sum_a|\pm,\k,a\rangle \langle \pm,\k,a|$, which satisfy the following relations,
\begin{equation}
1=\Pi_{+}+\Pi_{-},\quad \Pi^2_{\pm}=\Pi_{\pm},\quad \Pi_{+}\Pi_{-}=\Pi_{-}\Pi_{+}=0.
\end{equation}
Then, it is clear that $\H(\k)$ can be adiabatically deformed to be the flattened Hamiltonian
\begin{equation}
\widetilde{\H}(\k)=\Pi_{+}(\k)-\Pi_{-}(\k)\label{Flattened-Hamiltonian}
\end{equation}
without closing the energy gap by smoothly regulating positive and negative energies converging to $\pm1$, respectively. Since the topological properties of an insulator do not change under gap-preserving continuous deformations, it is sufficient and more convenient to adopt the flattened $\widetilde{\H}(\k)$ for studying topological properties.

At each $\k$, valence states $|-,\k,b\rangle$ span an $N$ dimensional vector space that is the image of $\Pi_{-}(\k)$, and these vector spaces spread smoothly over the whole BZ, forming an $N$D vector bundle, which is called the Berry bundle of valence bands of an insulator. Since the Berry bundle is generated by the projector $\Pi_{-}(\k)$, there exists a canonical Levi-Civita connection,
called the Berry connection, which is given by
\begin{equation}
\A_{b,b'}^\mu(\k)=\langle-,\k,b|\frac{\partial}{\partial k_\mu}|-,\k,b'\rangle \label{Berry-connection}
\end{equation}
with $\mu=1,2,\cdots,d$ labeling momentum coordinates.
To see that Eq.~\eqref{Berry-connection} is indeed a connection, one may check that under a gauge transformation
\begin{equation}
|-,\k,b\rangle \longrightarrow |-,\k,b'\rangle U_{b'b}(\k) \label{Gauge-Trans}
\end{equation}
with $U(\k)$ being a field of unitary matrices globally defined in the whole BZ, the Berry connection transforms as
\begin{equation}
\A^\mu(\k)\longrightarrow U^\dagger \A^\mu U+U^\dagger\partial^\mu U. \label{A-Gauge-Trans}
\end{equation}
Accordingly, the Berry curvature of the Berry bundle is
\begin{equation}
\F^{\mu\nu}=\partial^\mu \A^\nu-\partial^\nu\A^\mu+[\A^\mu,\A^\nu], \label{F-components}
\end{equation}
whose gauge transformation is given by
\begin{equation}
\F^{\mu\nu}\longrightarrow U^\dagger\F^{\mu\nu}U. \label{F-Gauge-Trans}
\end{equation}

As an example we discuss the general two-band model for insulators,
\begin{equation}\label{2-b-model}
\mathcal{H}_{2b}(\k)=\mathbf{d}(\k)\cdot\sigma,
\end{equation}
where $\sigma_i$ with $i=1,2,3$ are the Pauli matrices. The $\sigma_0$ term with $\sigma_0$ the $2\times 2$ identity matrix has been ignored for it only shifts the energy spectrum and does not affect eigenstates. For insulator $|\mathbf{d}(\k)|$ is not equal to zero for all $\k$, since the spectrum is given by $E_{\pm}(\k)=\pm |\mathbf{d}(\k)|$. The Hamiltonian of Eq.~\eqref{2-b-model} can be flattened as
\begin{equation}\label{2-band-flattened}
\widetilde{\H}_{2b}(\k)=\hat{\mathbf{d}}(\k)\cdot\sigma
\end{equation}
with $\hat{\mathbf{d}}(\k)$ being the unit vector $\mathbf{d}(\k)/|\mathbf{d}(\k)|$. Accordingly the projectors for valence and conduction bands are
\begin{equation}
\Pi^{2b}_{\pm}(\k)=\frac{1}{2}[\sigma_0\pm \hat{\mathbf{d}}(\k)\cdot\sigma].
\end{equation}
The valence eigenstates can be represented by $|-,\k\rangle=e^{-i\sigma_3\phi(\k)/2}e^{-i\sigma_2\theta(\k)/2}|\downarrow\rangle$, where $\theta(\k)$ and $\phi(\k)$ are the standard spherical coordinates of $\hat{\mathbf{d}}(\k)$, and $|\downarrow\rangle$ is the negative eigenstate of $\sigma_3$. The Berry connection can be straightforwardly derived as
\begin{equation}
\A^\mu(\k)=\frac{i}{2}\cos\theta(\k)\,\partial^{\mu}\phi(\k).
\end{equation}
Note that for two-band case the Berry connection is Abelian. Under the $U(1)$ gauge transformation $|-,\k\rangle\rightarrow e^{i\varphi(\k)}|-,\k\rangle$, the Berry connection $\A^{\mu}(\k)$ is transformed to be $\A^\mu(\k)+i\partial_{k_\mu}\varphi(\k)$. But the Berry curvature is invariant under gauge transformations, and is given from Eq.~\eqref{F-components} by $\F^{\mu\nu}(\k)=-\frac{i}{2}\sin\theta(\k)(\partial^\mu\theta(\k)\partial^\nu\phi(\k)-\partial^\nu\theta(\k)\partial^\mu\phi(\k))$, which can be recast in terms of $\hat{\mathbf{d}}(\k)$ as
\begin{equation}\label{F-2b}
\F^{\mu\nu}(\k)=\frac{1}{2i}\hat{\mathbf{d}}\cdot(\partial^\mu \hat{\mathbf{d}}\times\partial^\nu\hat{\mathbf{d}}).
\end{equation}

Equations \eqref{A-Gauge-Trans}, \eqref{F-components} and \eqref{F-Gauge-Trans} also appear in gauge theory. More specifically the Berry connection $\A^\mu(\k)$ and curvature $\F^{\mu\nu}(\k)$ corresponds to the gauge field or potential and field strength tensor, respectively, according to the terminologies of $U(N)$ gauge theory, for which the base space is the spacetime. Conventionally the connection of a vector bundle is not unique (but usually forming a space), and actually the definition of the Berry connection, namely Eq.~\eqref{Berry-connection}, is just a canonical way to assign a vector bundle with Hermitian metric  connection. Therefore a band theory can be regarded as a $U(N)$ gauge theory with a given connection, the Berry connection.
Once a connection is assigned for a vector bundle, we can compare two vectors at two separate points $y$ and $z$ through parallel transport of the one at $y$ along a path $\mathcal{C}$ to $z$. Parameterizing the path $\mathcal{C}$ as $x(\tau)$, where $\tau\in[\tau_i,\tau_f]$, $x(\tau_i)=y$ and $x(\tau_f)=z$, the mutually parallel vectors $|\psi(x(\tau))\rangle$ along $\mathcal{C}$ satisfy the equation,
\begin{equation}
\frac{dx_\mu}{d\tau}D^\mu\psi=0,
\end{equation}
with the covariant derivative $D^{\mu}=\frac{\partial}{\partial{x_{\mu}}}+\A^{\mu}(x)$.
The solution is $|\psi(x(\tau))\rangle=U_P(z,y)|\psi(y)\rangle$, and the parallel-transport operator $U_P$ is given by
\begin{equation}
U_P(z,y)=\hat{P}\exp\left[-\int_{\mathcal{C}}d\tau~\frac{dx_\mu}{d\tau}\A^\mu(x(\tau)) \right],
\end{equation}
where $\hat{P}$ indicates that the integral is path ordered. If $\mathcal{C}$ is a closed path, the parallel-transport operator is called the Wilson-loop operator in the context of gauge theory, or the holonomy along $\mathcal{C}$ in mathematics.
If $\tau\in[\tau_i,\tau_f]$ is divided into $N$ equal intervals, the parallel-transport operator can be approximated by
\begin{equation}
U_P(z,y)\approx U_NU_{N-1}\cdots U_1,
\end{equation}
where $U_j=\exp[-\A^y((\tau_{j-1}+\tau_j)/2)\Delta\tau/N]$ with $\tau_j=\tau_i+j\Delta\tau/N$ and $\Delta\tau=(\tau_f-\tau_i)/N$, so that the equality is recovered taking $N$ to infinity.

Performing a gauge transformation $|\psi(x(\tau))\rangle\rightarrow V(x(\tau))|\psi(x(\tau))\rangle$ along the path $\mathcal{C}$, the parallel-transport operator is transformed as
\begin{equation}
U_P(z,y)\longrightarrow V(z)U_P(z,y)V^\dagger(y).
\end{equation}
For a Wilson loop $\mathcal{C}$, which is a closed circle, it is a unitary transformation given by the reference point $y$, namely $U_P(y,y)\rightarrow V(y)U_P(y,y)V^\dagger(y)$. In particular, for Abelian connection $\A^\mu$, the path order $\hat{P}$ is not important, and therefore the Wilson loop is gauge invariant, and is given by the flux inserted over the area surrounded by $\mathcal{C}$,
\begin{equation}
U_\mathcal{C}=\mathrm{exp}\left[-\int_{\mathcal{D}}d^2x~\F^{12}(x)\right],
\end{equation}
where $\mathcal{D}$ is any smooth surface with $\mathcal{C}$ being its boundary, and $~\F^{12}$ is the corresponding Abelian Berry curvature. In this case, the phase factor $U_\mathcal{C}$ is called the geometric Berry phase along $\mathcal{C}$ as well.

\subsection{Chern Number and Chern-Simons Term}
The Chern number can be formulated for any even-dimensional integral domain. For $2n$ dimensions, the corresponding Chern number is called the $n$th Chern number, and the corresponding integrand is called the $n$th Chern character.
When $n=1$, the first Chern number for a $2$D insulator is explicitly given as
\begin{equation}
{C}=\frac{i}{2\pi}\int_{\mathbb{T}^2} d^2k~\mathrm{tr}\F^{12}. \label{First-Chern}
\end{equation}
Here and hereafter $\mathbb{T}^{2n}$ represents a $2n$ dimensional torus. Noting that the trace over the commutator in Eq.~\eqref{F-components} vanishes, the first Chern number essentially comes from the Abelian connection
$
a^\mu=\mathrm{tr}\A^\mu,
$
and can be accordingly recast as
$
C=(i/2\pi)\int_{\mathbb{T}^2}d^2k~f^{12} \label{First-Chern-Num}
$
with $f^{\mu\nu}=\partial^\mu a^\nu-\partial^\nu a^\mu$.
If $n=2$, the second Chern number for a $4$D insulator is
\begin{equation}
{C}_2=-\frac{1}{32\pi^2}\int_{\mathbb{T}^4}d^4k~\epsilon_{\mu\nu\lambda\sigma}\mathrm{tr} \F^{\mu\nu}\F^{\lambda\sigma}, \label{Second-Chern}
\end{equation}
which is essentially non-Abelian.

For $2$D insulators, the first Chern number of Eq.~\eqref{First-Chern} is also called the Thouless-Kohmoto-Nightingale-den Nijs (TKNN) invariant, and was shown to be the transverse conductance in the unit of $e^2/h$ by using the Kubo formula~\cite{Thouless1982}. A nonvanishing transverse conductance requires the breaking of TRS. This is consistent with Eq.~\eqref{First-Chern}, because the first Chern number has to be vanishing in order to preserve TRS, since $i\F$ is odd under time-reversal, which shall be clear when we discuss TRS.
In other words, a $2$D Chern insulator cannot have TRS. In contrast, the second Chern number of Eq.~\eqref{Second-Chern} is time reversal symmetric, namely there exist time-reversal-symmetric $4$D Chern insulators. The meaning of the second Chern number for electromagnetic response can be found in Refs.~\cite{Golterman1993,XLQi2008}.

If a $2n$ dimensional sphere $\mathbb{S}^{2n}$ is chosen to enclose a singular point, where the bundle is not well-defined, in a $(2n+1)$D space, the Chern number may be calculated on the $\mathbb{S}^{2n}$, and is referred to as the monopole charge of the singular point. For monopoles in $3$D space, the monopole charge can be calculated by the Abelian connection $a^\mu=\mathrm{tr}\A^\mu$, and therefore are termed as Abelian monopoles. For instance the Weyl points for $\H_{W}(\k)=\pm\k\cdot\sigma$ can be interpreted as unit Abelian monopoles in momentum space for the respective Abelian Berry bundles of valence band restricted on $\mathbb{S}^2$ surrounding the origin.  Monopoles in $5$D space are categorized into non-Abelian monopoles, the monopole charge must be calculated by non-Abelian connections. Accordingly the $5$D Weyl points $\H^{5D}_{W}(\k)=\pm\k_\mu\Gamma^\mu$ with $\mu=1,2,\cdots,5$, where $\Gamma^\mu$ are $4\times 4$ Dirac matrices.

It is noteworthy that if the Chern number is nontrivial (nonzero), it is impossible to find a complete set of globally well-defined valence eigenstates in the whole BZ. Otherwise,  the Chern character is a total derivative, and the Chern number must be trivial, which shall be clear when we discuss Chern-Simons forms.
Due to the lack of global wavefunctions in general, it is usually technically difficult to directly calculate the Chern number via Eq.~\eqref{First-Chern} or \eqref{Second-Chern}. To avoid this difficulty, we can reformulate the Chern number in terms of the Green's function of imaginary time, $G(\omega,\k)=1/[i\omega-\H(\k)]$~\cite{volovik2003,YXZhao2013,XLQi2008,ZWang2010,ZWang2012b}, which for Eq.~\eqref{First-Chern} is explicitly given by
\begin{equation}
C[\k]=-\frac{1}{24\pi^2}\int_{-\infty}^{\infty}d\omega \int_{\mathbb{T}^2} d^2k\epsilon_{\mu\nu\lambda}\mathrm{tr}~ G\partial^\mu G^{-1}G\partial^\nu G^{-1}G\partial^\lambda G^{-1}. \label{2D-G-Chern}
\end{equation}
Note that $G(\omega,\k)$ is an invertible matrix for each $(\omega,\k)$, namely $G(\omega,\k)\in GL(N+M,\mathbb{C})$, because $\H(\k)$ is invertible.

Although there is no global Berry connection $\A$ over the base manifold if the Berry bundle is nontrivial, if the base manifold is trivially a disk $D^{2n}$, $\A$ can be given over the whole $D^{2n}$, and furthermore the Chern character can be expressed as a total derivative of the Chern-Simons form. If $n=1$, it is obvious that the first Chern character $C(\F)=dQ_1(\A)$ with
\begin{equation}
Q_1(\A)=\frac{i}{2\pi}\mathrm{tr}\A. \label{Q1}
\end{equation}
For $n=2$, the Chern-Simons form is a third form
\begin{equation}
Q_3(\A,\F)=\frac{1}{2}\left(\frac{i}{2\pi}\right)^2\mathrm{tr}(\A d\A+\frac{2}{3}\A^3), \label{Q3}
\end{equation}
and it is straightforward to check that $dQ_3(\A,\F)=\mathrm{C}_2(\F)$.  A general formula $Q_{2n-1}(\mathcal{A},\mathcal{F})$ for any $n$ is
\begin{equation}
Q_{2n-1}(\mathcal{A},\mathcal{F})=\frac{1}{(n-1)!}\left(\frac{i}{2\pi}\right)^{n}\int_{0}^{1}dt\:\mathrm{tr}\left(\mathcal{A}\mathcal{F}_{t}^{(n-1)}\right)
\end{equation}
with $\mathcal{F}_{t}=t\mathcal{F}+t(t-1)\mathcal{A}^{2}$. The integration of $Q_{2n-1}$ over a $(2n-1)$ dimensional manifold, for instance $\mathbb{S}^{2n-1}$, is called the Chern-Simons term,
\begin{equation}
\nu_{CS}^{2n-1}[\A]=\int_{\mathbb{S}^{2n-1}} Q_{2n-1}(\A,\F).
\end{equation}

A significant difference of the Chern-Simons term from Chern number is that it is not gauge invariant. For instance,
\begin{equation}
Q_1(\A^{U})=Q_1(\A)+\frac{i}{2\pi}\mathrm{tr} U^\dagger \partial_k U,
\end{equation}
and therefore the change of $\nu_{CS}^{1}$ is just the winding number of $U(k)$. Analogous calculations for $Q_3(\A,\F)$, although are a little more complicated, can also be made straightforwardly. It is actually a general conclusion that
a gauge transformation of $U$ changes the Chern-Simons term over $\mathbb{S}^{2n-1}$ by a winding number $\nu_\text{w}^{2n-1}[U]$ of $U$, namely
\begin{equation}
\nu_{CS}^{2n-1}[\A^U]-\nu_{CS}^{2n-1}[\A]=\nu_\text{w}^{2n-1}[U]. \label{CS-gauge-trans}
\end{equation}

We now make a classic application of the mathematics introduced in this subsection, which is of fundamental importance. Consider a Berry bundle on a $2$D sphere $\mathbb{S}^{2}$. If the bundle is nontrivial, there is no globally well-defined Berry connection $\A$. But one can always have Berry connections on the north hemisphere $D^{2}_N$ and the south as $\A_N$ and $\A_S$, respectively, and glue the wave functions along the equator $\mathbb{S}^{1}$, which is given by the transition function from the south hemisphere to the north, namely $U(\k)\in U(N)$ with $\k\in \mathbb{S}^{1}$. Then, the Chern number is calculated as
\begin{eqnarray}
{C} &=&\int_{D^{2}_N} dQ_{1}(\A_N,\F_N)+\int_{D^{2}_S}dQ_{1}(\A_S,\F_S)
=\int_{\mathbb{S}^{1}}Q_{1}(\A_N,\F_N)-\int_{\mathbb{S}^{1}}Q_{1}(\A_S,\F_S) \nonumber \\
&=&\nu_\text{w}^1[U],\label{Chern-Winding}
\end{eqnarray}
%
%
%
where the minus sign in the second equality is due to the opposite orientations of $\mathbb{S}^{2n-1}$ with respect to the north hemisphere and the south, and the third equality has used Eq.~\eqref{CS-gauge-trans}. It is concluded that the Chern number of the bundle on the sphere is just the winding number of the transition function along the equator.

{\it 1st Chern Number in Terms of Wilson Loop.} The topological invariant for $2$D Chern insulators is the first Chern number of Eq.~\eqref{First-Chern}. There does not exist a complete set of globally well-defined valence eigenstates in the whole torus $\mathbb{T}^2$. But we can find it over the cylinder, which does not require periodic boundary condition over $k_y\in(-\pi,\pi]$, and then use a transition
function from the bundle on the circle $S^{1}_{-}$ at $k_y=-\pi$ to that on $S^{1}_{+}$ at $k_y=\pi$. The Chern number is just the winding number of the transition function as a mapping from $S^{1}$ to $U(N)$, which can be inferred from the discussions about Eq.~\eqref{Chern-Winding}. Given such a set of eigenstates over the cylinder,
we can use the method of Wilson loops over $k_y$, which are parametrized by $k_x\in(-\pi,\pi]$, to obtain the transition function. At each $k_x$, the Wilson-loop operator is given by
\begin{equation}
U(k_x)=\hat{P}\exp\int_{-\pi}^{\pi}dk_y~\A^y(k_x,k_y)\in U(N).
\end{equation}
After working out the transition function by Wilson loops, the Chern number is calculated by the winding number
\begin{equation}
\nu_\text{w}=\frac{i}{2\pi}\oint dk_x~ \mathrm{tr}U(k_x)\partial_{ k_x}U^\dagger(k_x).
\end{equation}
Note that $\mathrm{tr}\F=\mathrm{tr}(d\A+\A_\wedge \A)=\mathrm{tr}d\A=\sum_a d\A_{aa}$, it implies the fact that whether $\A$ is non-Abelian is not important in two dimensions, namely each valence band can be treated individually. In practice, one can always add appropriate perturbations to separate valence bands, and for the $a$th band the corresponding Berry connection $\A_a$ is just an Abelian connection.
Thus the Chern number is just the summation of winding numbers for each valence band,
\begin{equation}
{C}=\frac{i}{2\pi}\sum_a \oint dk_x~ U_a(k_x)\partial_{ k_x}U_a^\dagger(k_x).
\end{equation}

{\it Chern Number of Two-band Model.} For the two-band model of Eq.~\eqref{2-b-model} as an example, the Chern number can be expressed explicitly by
\begin{equation}
C=\frac{1}{4\pi}\int_{\mathbb{T}^2}d^2k~\hat{\mathbf{d}}\cdot(\partial_{k_x}\hat{\mathbf{d}}\times\partial_{k_y}\hat{\mathbf{d}}), \label{Simplified-Chern}
\end{equation}
which can be derived by directly substituting Eq.~\eqref{F-2b} into Eq.~\eqref{First-Chern}, or alternatively by
substituting the Green' function with imaginary frequency, $G(\omega,\k)=1/[i\omega-\mathbf{d}(\k)\cdot\sigma]$, into Eq.~\eqref{2D-G-Chern}.
The simplified formula of Eq.~\eqref{Simplified-Chern} is just  the winding number of the vector field $\hat{\mathbf{d}}(\k)$ as a mapping from $\mathbb{T}^2$ to $\mathbb{S}^2$.

\subsection{Topological Invariants for Topological Insulators}\label{TIZ2}

\subsubsection{3D Topological Insulators.}
The Chern insulators do not require any symmetry for the momentum-space Hamiltonian, but the $3$D topological insulator requires TRS. The TRS is represented in momentum space (even in real space) by $\hat{T}=U_T\K$, where $U_T$ is a unitary matrix, and $\K$ is the complex conjugate operator, satisfying $\hat{T}^2=-1$. Note that for electronic systems, $U_T=-i\sigma_2$ with $\sigma_2$ acting in the spin space. If a system has TRS, then $\T^{\dagger}\H(-\k)\T=\H(\k)$, which may be explicitly expressed as
\begin{equation}
U_T^\dagger \H(-\k) U_T=\H^*(\k).\label{time-reversal}
\end{equation}
For a valence eigenstate $|a,\k\rangle$ with $\H(\k)|a,\k\rangle=E_a(\k)|a,\k\rangle$, $U_T|a,\k\rangle^*$ is an eigenstate of $\H(-\k)$ with the same energy $E_a(\k)$, which can be deduced from Eq.~\eqref{time-reversal}. Here we abbreviate $|-,a,\k\rangle$ to be $|a,\k\rangle$ for simplicity. Thus, the spectrum is inversion symmetric in momentum space, and $U_T|a,\k\rangle^*$ can be expanded by the basis at $-\k$ as
\begin{equation}
U_T|a,\k\rangle^*=\sum_b \mathcal{U}^*_{ab}(-\k)|b,-\k\rangle, \label{T-band-transf}
\end{equation}
where $U(\k)$ is a unitary matrix for each $\k$. Due to the constraint of Eq.~\eqref{T-band-transf} exerted by the TRS in the valence bands, the Berry connection satisfies the relation,
\begin{equation}
\A^*(\k)=\mathcal{U}(-\k)\A(-\k)\mathcal{U}^\dagger(-\k)+\mathcal{U}(-\k)d\mathcal{U}^\dagger(-\k). \label{T-A-transf}
\end{equation}
In general the Chern-Simons term can take any real number, but symmetry could lead to quantization of the Chern-Simons term. For $3$D topological insulators with TRS, the relation of Eq.~\eqref{T-A-transf} can be applied to quantize the Chern-Simons term. Observing that $\A^*(\k)$ is just the gauge transformed $\A(-\k)$ by $\mathcal{U}^\dagger(-\k)$ from Eq.~\eqref{T-A-transf}, and the Chern-Simons term is a real number odd under inversion, we can deduce
\begin{equation}
2\nu_{CS}^3[\A]=\nu_\text{w}^3[\mathcal{U}], \label{CS-quantization}
\end{equation}
in the light of Eq.~\eqref{CS-gauge-trans}. The right hand of Eq.~\eqref{CS-quantization} is the winding number of $\mathcal{U}(\k)$ over the $3$D BZ. Because of the gauge ambiguity described by Eq.~\eqref{CS-gauge-trans}, $2\nu_{CS}^3[\A]$ can be regarded as a $\Z_2$ topological invariant for $3$D topological insulators, which is explicitly given by~\cite{XLQi2008}
\begin{equation}
\nu_{\Z_2}^{(1)}=-\frac{1}{4\pi^2}\int_{\mathbb{T}^3}d^3k~\\\epsilon_{\mu\nu\lambda}\mathrm{tr}(\A^\mu\partial^\nu\A^\lambda+\frac{2}{3}\A^\mu\A^\nu\A^\lambda)\mod 2. \label{CS-Z2}
\end{equation}

It is essential important for $3$D topological insulators that $\hat{T}^2=-1$, which can readily be seen from another topological invariant given by Fu-Kane-Mele~\cite{LFu2007a}. The fact that $\hat{T}^2=-1$ implies $U_T$ is anti-symmetric, namely $U_T^t=-U_T$, which further implies $\mathcal{U}^t(\k)=-\mathcal{U}(-\k)$.  In the $3$D BZ, there are eight inversion invariant points $\Gamma_i$ with $i=1,2,\cdots,8$, where $\mathcal{U}(\Gamma_i)$ are anti-symmetric. A significant consequence of TRS is the presence of the two-fold Kramers degeneracy for energy eigenstates, and therefore the valence-state number is even at $\Gamma_i$. For an anti-symmetric even-dimensional matrix $\mathcal{U}$, the Pfaffian $\mathrm{Pf}(\mathcal{U})$ can be defined as a polynomial of entries of $\mathcal{U}$, and the topological invariant is given by
\begin{equation}
(-1)^{\nu_{\Z_2}^{(1)}}=\prod_{i=1}^{8}\frac{\mathrm{Pf}(\mathcal{U}(\Gamma_i))}{\sqrt{\mathrm{Det}(\mathcal{U}(\Gamma_i))}}, \label{FKM-invariant}
\end{equation}
which is called the Fu-Kane-Mele invariant~\cite{LFu2007a}. Note that the determinants of $\mathcal{U}(\Gamma_i)$ as unitary matrices are all definitely positive. Although, the expression of Eq.~\eqref{FKM-invariant} is local at $\Gamma_j$, the global information is acquired by the requirement that the valence wave functions are globally well-defined in the whole BZ. A proof for the equality of Eqs.~\eqref{CS-Z2} and \eqref{FKM-invariant} can be found in Ref. \cite{ZhongWang-NJP}. Despite of using the unfamiliar Pfaffian, Eq.~\eqref{FKM-invariant} can be radically simplified in the presence of inversion symmetry $\hat{P}$. Since each $\Gamma_i$ is inversion invariant, each valence state $|-,\Gamma_i,a\rangle$ at $\Gamma_i$ with $a=1,2,\cdots, 2N$ is also an eigenstate of $\hat{P}$ with eigenvalue (parity) $\xi_{a}(\Gamma_i)=\pm 1$. Assuming that $(2m-1)$th and $2m$th states form Kramers pairs with $m=1,2,\cdots, N$, we define $\delta_i=\prod_{m=1}^{N}\xi_{2m-1}(\Gamma_i)$ noticing that two states of each Kramers pair have the same parity because of $[\hat{P},\hat{T}]=0$. Then Eq.~\eqref{FKM-invariant} takes the simple expression~\cite{LFu2007b},
\begin{equation}\label{Inversion-Z2}
(-1)^{\nu_{\Z_2}^{(1)}}=\prod_{i}\delta_i.
\end{equation}
The convenience of Eq.~\eqref{Inversion-Z2} lies in that it is entirely determined by the representation of $\hat{P}$, which can be derived from local eigenstates at $\Gamma_i$, and hence its practice does not require global valence eigenstates, which as aforementioned are usually technically difficult to obtain.

\subsubsection{2D Topological Insulators}

There exists no Chern insulator without TRS, since the first Chern number is odd under time-reversal. From Eq.~\eqref{A-Gauge-Trans}, it is found that $\F^*(\k)=\mathcal{U}(-\k)\F(-\k)\mathcal{U}^\dagger(-\k)$, but $\F^*(\k)=-\F^t(\k)$. However, new time reversal invariant topological insulators arise with $\Z_2$ classification, and the corresponding topological properties requires and are protected by TRS~\cite{Kane2005a}. The $\Z_2$ topology can be characterized by a topological invariant quite similar to Eq.~\eqref{FKM-invariant}, but now the product is over four inversion invariant points in the $2$D BZ~\cite{LFu2006},
\begin{equation}\label{Pf-2D-Z2}
(-1)^{\nu_{\Z_2}^{(2)}}=\prod_{i=1}^{4}\frac{\mathrm{Pf}(\mathcal{U}(\Gamma_i))}{\sqrt{\mathrm{Det}(\mathcal{U}(\Gamma_i))}},
\end{equation}
which again requires globally well-defined wave functions. If the inversion symmetry is present, Eq.~\eqref{Inversion-Z2} is also a usually more convenient alternative for Eq.~\eqref{Pf-2D-Z2}, where the product is now over four inversion invariant points.

A topological invariant in terms of the Berry connection may also be formulated for the time-reversal-symmetric topological phase. For each $\k$, let  $|I,\alpha,\k\rangle$ and $|II,\alpha,-\k\rangle$ be a pair of states labelled by $\alpha$, which are related by TRS by
$$|I,\alpha,\k\rangle=U_T|II,\alpha,-\k\rangle^*,\ \ \  |II,\alpha,\k\rangle=-U_T|I,\alpha,-\k\rangle^*.$$
Locally the state label $a$ is further specified to be $(s,\alpha)$ with $s=I$ or $II$. However, if the topological insulator has nontrivial topological invariant, such a basis $|s,\alpha,\k\rangle$ does not exist globally. Instead, we can choose such a basis for the $1$D subsystem with $k_y=-\pi$, and also for the one with $k_y=0$. Then the topological invariant is given by
\begin{equation}
\begin{split}
\nu_{\Z_2}^{(2)}&=\frac{i}{2\pi}\int_{-\pi}^{\pi} dk_x~\mathrm{tr}[\A^1(k_x,-\pi)-\A^1(k_x,0)]
-\frac{i}{2\pi}\int_{-\pi}^{\pi} dk_x\int_{-\pi}^{0}dk_y~\mathrm{tr}\F^{12}\mod 2, \label{T-Z2-Chern}
\end{split}
\end{equation}
noting that the $1$D subsystems with $k_y=-\pi$ and $0$ are the boundary of the integration domain of the second term, which is a cylinder. The gauge freedom of the first term justifies the $\Z_2$ nature of the topological invariant, recalling Eq.~\eqref{CS-gauge-trans}. The equivalence of the two topological invariants Eqs.~\eqref{Pf-2D-Z2} and \eqref{T-Z2-Chern} can be found in Ref.~\cite{LFu2006}.

\subsection{Winding Numbers for Chiral Classes}
In this section, we consider a Hamiltonian $\H(\k)$ with chiral symmetry $\Gamma$, namely $\H(k)$ anti-commutes with $\Gamma$,
\begin{equation}
\{\H(\k),\Gamma\}=0,
\end{equation}
and assume that $\Gamma^2=1$ and $\Gamma^\dagger=\Gamma$. The chiral symmetry implies at each $\k$ the valence states of the insulator $\H(\k)$ have a one-to-one correspondence to conduction states, noting that if $\H(\k)|\psi\rangle= E|\psi\rangle$, then $\H(\k)\Gamma|\psi\rangle= -E\Gamma|\psi\rangle$. One can always choose a basis, for which $\Gamma=\mathrm{diag}(1_N,-1_N)$ with $1_N$ being the $N\times N$ identity matrix, and then the Hamiltonian takes the anti-diagonal form,
\begin{equation}
\H(\k)=\begin{pmatrix}
0 & q^\dagger(\k)\\
q(\k) & 0
\end{pmatrix}, \label{antidiagonal-H}
\end{equation}
where $q(\k)$ is an $N\times N$ invertible matrix for each $\k$, since $\H(\k)$ describing an insulator is invertible. Thus, the winding number of $q$ from the BZ to $GL(N,\mathbb{C})$ is a topological invariant for this symmetry class, which is given by
\begin{equation}\label{Un-winding-number}
\nu_\text{w}^{2n+1}[q]=\eta_n\int_{\mathbb{T}^{2n+1}} \mathrm{tr}(qdq^{-1})^{2n+1}
\end{equation}
with $\eta_n=n!/[(2n+1)!(2\pi i)^{n+1}]$ for a $(2n+1)$D BZ~\cite{Schnyder2008}. Note that even-dimensional insulators in this symmetry class are all trivial.
In physical dimensions, for $1$D insulators with $n=0$, the topological invariant is explicitly given by
\begin{equation}
\nu^1_\text{w}[q]=\frac{1}{2\pi i}\oint dk~ \mathrm{tr}~qdq^{-1}, \label{winding-1d}
\end{equation}
and for $3$D insulators with $n=1$,
\begin{equation}
\nu^3_\text{w}[q]=-\frac{1}{24\pi^2}\int d^3k\epsilon_{\mu\nu\lambda}\mathrm{tr}~ q\partial^\mu q^{-1}q\partial^\nu q^{-1}q\partial^\lambda q^{-1}.
\end{equation}
The definition of $q(\k)$ or the anti-diagonal form of Eq.~\eqref{antidiagonal-H} is based on the particular representation of $\Gamma=\mathrm{diag}(1_N,-1_N)$, and therefore Eq.~\eqref{Un-winding-number} is gauge dependent. A gauge independent expression of the topological invariant can be given for hermitian chiral symmetry $\Gamma$ with the normalization $\Gamma^2=1$ as
\begin{equation}
\nu_\text{w}^{2n+1}[\H]=\frac{\eta_n}{2}\int_{\mathbb{T}^{2n+1}}\mathrm{tr}\Gamma (\H d\H^{-1})^{2n+1}.
\end{equation}

\subsection{Quantized Zak Phase}
In general the Berry phase of valence bands in the unit of $\pi$ for a $1$D gapped system is given by
\begin{equation}
\nu=\frac{i}{\pi}\oint dk~ \mathrm{tr}\A \label{Berry-phase-1D}
\end{equation}
with $\A$ given by Eq.~\eqref{Berry-connection}, may be any real number, and thus cannot be a topological invariant.
However, certain symmetries can quantize it into integers, which is similar to that the quantized Chern-Simons term in three dimensions, recalling from Eq.~\eqref{Q1} that Eq.~\eqref{Berry-phase-1D} is just the Chern-Simons term in one dimension. The quantization of Eq.~\eqref{Berry-phase-1D} was first discussed in 1D band theory by Zak taking into account inversion symmetry~\cite{Zak1989}, and therefore the Berry phase in band theory is also called the Zak phase. Only the parity of the quantized Berry phase of Eq.~\eqref{Berry-phase-1D} is gauge invariant, since a large gauge transformation, $|k,b\rangle\rightarrow \sum_cU_{bc}(k)|k,c\rangle$, can change Eq.~\eqref{Berry-phase-1D} by two times of the winding number of $u(k)=\mathrm{Det}[U(k)]\in U(1)$, namely
\begin{equation}
\nu\rightarrow \nu+\frac{1}{\pi i}\int dk~u(k)\partial_{k} u^\dagger(k),
\end{equation}
which is just Eq.~\eqref{CS-gauge-trans} specialized to $n=1$.  These results are consistent with the physical meaning of the Zak phase. The Zak phase of Eq.~\eqref{Berry-phase-1D} is just the center of the Wannier functions with the lattice constant normalized to be $2$. For a periodic system, of course the center of the Wannier functions should be a position modulo the lattice constant. It is also clear that in order to preserve inversion symmetry the center has to be concentrated at lattice sites or at the midpoints of lattice sites, namely it is an integer for the lattice constant $2$.

If a gapped system has chiral symmetry, casting the Hamiltonian into the form of Eq.~\eqref{antidiagonal-H}, globally well-defined valence states are given readily as
\begin{equation}
|k,b\rangle=\frac{1}{\sqrt{2}}\begin{pmatrix}
-v_b\\ q(k)v_b
\end{pmatrix} \label{Chiral-States}
\end{equation}
where $q=1,\cdots,N$, and $v_q$ is a $N$-vector with all entries being zero except the $b$th being $1$. In Eq.~\eqref{Chiral-States}, we have assumed $\widetilde{\mathcal{H}}$ with $q(k)\in U(N)$. Using this set of valence states, one may check that Eq.~\eqref{Berry-phase-1D} is equal to Eq.~\eqref{winding-1d}. Thus, the Berry phase of Eq.~\eqref{Berry-phase-1D} is quantized into integers by chiral symmetry.  Since the Zak phase is only a $\mathbb{Z}_2$ invariant, it counts only the parity of the topological invariant of Eq.~\eqref{winding-1d} for the symmetry class.

The Zak phase can also be quantized by charge-conjugate or particle-hole symmetry. In momentum space charge-conjugate symmetry is represented by $\C=U_C\K\hat{I}$ with $U_C$ being a unitary matrix, and is required to satisfy $\C^2=1$, which implies $U_CU^*_C=1$ and $U_C=U_C^t$. The momentum-space Hamiltonian is transformed under charge-conjugate symmetry as
\begin{equation}
U_C^\dagger\H(-k)U_C=-\H^*(k). \label{C-symmetry}
\end{equation}
The charge-conjugate symmetric 1D gapped systems have a $\Z_2$ topological classification, and the Zak phase is just the topological invariant. Noting that the Bogoliubov-de Gennes Hamiltonians of superconductors are naturally charge-conjugate symmetric, thus the Zak phase is the topological invariant for 1D topological superconductors. Similar to time-reversal-symmetric topological insulators, the topological invariant can also be equivalently expressed as a product of Pfaffians at two inversion invariant points, which is given by the Majorana representation of free fermionic systems in Ref.~\cite{Kitaev2001}.

\subsection{Skyrmions in two and three dimensions}
As discussed in Sec.~\ref{Berry Bundles}, a two-band model gives a field of unit vectors $\mathbf{n}(\k)$ over the 2D BZ, and the Chern number is just the winding number of the unit-vector field as a mapping from $\mathbb{T}^2$ to $\mathbb{S}^2$. If a similar unit-vector texture $\mathbf{n}(\mathbf{x})$ occurs in $2$D real space $\mathbb{R}^2$, the winding number is also called the topological charge $\nu_\text{w}^{2D}$ of skyrmions in the vector field $\mathbf{n}(\mathbf{x})$, which is explicitly given by
\begin{equation}
\nu_\text{w}^{2D}=\frac{1}{4\pi}\int_{\mathbb{R}^2}d^2x~\mathbf{n}\cdot(\partial_{x}\mathbf{n}\times\partial_{y}\mathbf{n}).\label{skyrmion-charge}
\end{equation}
Vectors at infinity are usually assumed to be oriented toward the same direction, namely the plane $\mathbb{R}^2$ is effectively compactified to be $\mathbb{S}^2$, and therefore the topological charge of skyrmions is an integer.

In some physical systems, there may exist a field of four-component unit vectors in $3$D space. For example, $SU(2)$ order parameters after condensation, or two-component normalized quantum states. A group element $U(x)$ of $SU(2)$ can be expressed as $U(x)=\hat{d}_0(x)\sigma_0+i\hat{d}_i(x)\sigma^i$ with $d^\mu=(\hat{d}_0,\hat{\mathbf{d}})$ being a unit vector in $4$D Euclidean space, and a two-component normalized quantum state $\psi(x)$ can be represented as $\psi(x)=(\hat{d}_0+i\hat{d}_1,~\hat{d}_2+i\hat{d}_3)^t$. If vectors are constant at infinity, the $3$D space $\mathbb{R}^3$ is topologically identical to $\mathbb{S}^3$, and the vector fields are mappings from $\mathbb{S}^3$ to $\mathbb{S}^3$. Therefore, there exist $3$D skyrmions as counterpart of $2$D ones can exist in such systems, whose topological charge is just the corresponding winding number. For $SU(2)$ field $U(x)$, as afore-mentioned, the winding number is given by
\begin{equation}
\nu_\text{w}^{3D}=-\frac{1}{24\pi^2}\int d^3x~\epsilon_{\mu\nu\lambda}\mathrm{tr}~ U(x)\partial^\mu U^{\dagger}(x)U(x)\partial^\nu U^{\dagger}(x)U(x)\partial^\lambda U^{\dagger}(x).
\end{equation}
Substituting $U(x)=\hat{d}_0(x)\sigma_0+i\hat{d}_i(x)\sigma^i$, the formula in terms of the unit vectors is
\begin{equation}
\nu_\text{w}^{3D}=\frac{1}{2\pi^2}\int_{\mathbb{R}^3} d^3x~\epsilon_{\mu\nu\lambda\rho}\hat{d}^\mu \partial_1 \hat{d}^\nu\partial_2 \hat{d}^\lambda \partial_3 \hat{d}^\rho.
\end{equation}

The notion of skyrmion can be readily generalized to any dimension $n$, where the skyrmion charge is just the winding number of the unit vector field $d^\mu(x)$ as a mapping from $\mathbb{S}^n$ to $\mathbb{S}^n$. The formula for topological charge is explicitly given  as
\begin{equation}
\nu^{nD}_\text{w}=\frac{1}{\Omega_n}\int_{\mathbb{R}^n}d^nx~\epsilon_{\mu_0\mu_1\cdots\mu_n}
\hat{d}^{\mu_0}\partial_1 \hat{d}^{\mu_1}\cdots\partial_{n} \hat{d}^{\mu_n},
\end{equation}
where $\Omega_n$ is the geometric angle of $(n+1)$ dimensional Euclidean space equal to $2\pi^{d/2}/\Gamma(d/2)$, and the integrand is just the volume element of the unit-vector-valued function $\hat{d}^\mu(x)\in \mathbb{S}^n$.

\subsection{Hopf Invariant}\label{HopfIndex}
So far, we have mainly concerned with band theories of sufficiently many bands. We now consider a topological insulator, which has only two bands and occurs only in three dimensions. For a two-band insulator, the flattened Hamiltonian can always be written as Eq.~\eqref{2-band-flattened}. Hence $\widetilde{\H}(\k)$ at each $\k$ can be regarded as a point on the unit sphere $\mathbb{S}^2$, and $\widetilde{\mathcal{H}}$ gives a mapping from the $3$D momentum space to $\mathbb{S}^2$.
Because of the homotopy group $\pi_{3}(\mathbb{S}^2)\cong \Z$, there exist (strong) $3$D two-band topological insulators with $\Z$ classification, which is termed Hopf insulators. The topological invariant is called the Hopf invariant~\cite{Wilczek1983,Moore2008}, and is given by
\begin{equation}
\nu_{H}=-\frac{1}{4\pi^2}\int_{\mathbb{T}^3}d^3k~\epsilon_{\mu\nu\lambda} \A^{\mu}\partial^{\nu}\A^{\lambda}, \label{Hopf-invariant}
\end{equation}
where $\A^\mu=\langle-,\k|\partial_{ k_\mu}|-,\k\rangle$ is the Berry connection of the valence band defined in Eq.~\eqref{Berry-connection}.  It is worth noting that we have ignored all cases of weak topological insulators, i.e., the Chern number over any 2D sub BZ has been assumed to be zero, such that the valence wave function $|-,\k\rangle$ can be globally well-defined in the whole 3D BZ.

We now develop another expression of the Hopf invariant, which also gives an explanation of the homotopy group $\pi_3(\mathbb{S}^2)\cong\Z$ and Eq.~\eqref{Hopf-invariant}. At each $\k$, $\hat{d}_i(\k)\sigma^i$ can be obtained from $\sigma^3$ by a $SU(2)$ rotation, namely
\begin{equation}
\widetilde{\H}(\k)=U(\k)\sigma^3 U^{-1}(\k),
\end{equation}
with $U(\k)\in SU(2)$. Then the valence eigenstates are $|-,\k\rangle=U(\k)|\downarrow\rangle$. Note that the rotation to the $z$-axis, $e^{-i\sigma^3\phi/2}$, does not change the orientation of the state $|\downarrow\rangle$, namely $U(\k)$ and $U(\k)e^{-i\sigma^3\phi(\k)/2}$ give the same $\hat{\mathbf{d}}(\k)$. Due to the $U(1)$ gauge freedom in the presentation of $\hat{\mathbf{d}}(\k)$ by an element of $SU(2)$, $U(\k)$ can be made globally well-defined in the whole BZ. Then, the Hopf invariant is equal to the winding number of $U(\k)$ as a mapping from the $3$D BZ to $SU(2)$~\cite{Volovik1989},
\begin{equation} \label{Hopf-Winding}
\nu_\text{w}[U]=-\frac{1}{24\pi^2}\int_{\mathbb{T}^3} d^3k~\epsilon_{\mu\nu\lambda}\mathrm{tr}U\partial^\mu U^{-1}U\partial^\nu U^{-1}U\partial^\lambda U^{-1}.
\end{equation}

\end{appendix}


\bibliographystyle{tADP}
\bibliography{reference}

\begin{thebibliography}{100}
\newcommand{\noopsort}[1]{}
\newcommand{\printfirst}[2]{#1}
\newcommand{\singleletter}[1]{#1}
\newcommand{\switchargs}[2]{#2#1}
\providecommand{\url}[1]{\normalfont{#1}}
\providecommand{\urlprefix}{Available at }

\bibitem{Dirac1931}
P.A.M. Dirac, Proc. R. Soc. Lond. A 133 (1931), pp. 60--72.

\bibitem{Penrose1965}
R.~Penrose, Phys. Rev. Lett. 14 (1965), pp. 57--59.

\bibitem{WPSu1979}
W.P. Su, J.R. Schrieffer, and A.J. Heeger, Phys. Rev. Lett. 42 (1979), p. 1698.

\bibitem{Heeger1988}
A.J. Heeger, S.~Kivelson, J.R. Schrieffer, and W.P. Su, Rev. Mod. Phys. 60
  (1988), pp. 781--850.

\bibitem{LYu1988}
L.~Yu, \emph{Solitons and Polarons in Conducting Polymers}, World Scientific,
  1988 Aug.

\bibitem{Haldane2017}
F.D.M. Haldane, Rev. Mod. Phys. 89 (2017), p. 040502.

\bibitem{Klitzing1980}
K.V. Klitzing, G.~Dorda, and M.~Pepper, Phys. Rev. Lett. 45 (1980), pp.
  494--497.

\bibitem{Tsui1982}
D.C. Tsui, H.L. Stormer, and A.C. Gossard, Phys. Rev. Lett. 48 (1982), p. 1559.

\bibitem{Haldane1983a}
F.D.M. Haldane, Phys. Rev. Lett. 50 (1983), p. 1153.

\bibitem{Thouless1982}
D.J. Thouless, M.~Kohmoto, M.P. Nightingale, and M.~den Nijs, Phys. Rev. Lett.
  49 (1982), p. 405.

\bibitem{Simon1983}
B.~Simon, Phys. Rev. Lett. 51 (1983), pp. 2167--2170.

\bibitem{Berry1984}
M.V. Berry, Proc. R. Soc. Lond. A 392 (1984), pp. 45--57.

\bibitem{Haldane1988}
F.D.M. Haldane, Phys. Rev. Lett. 61 (1988), p. 2015.

\bibitem{Hasan2010}
M.Z. Hasan and C.L. Kane, Rev. Mod. Phys. 82 (2010), pp. 3045--3067.

\bibitem{XLQi2011}
X.L. Qi and S.C. Zhang, Rev. Mod. Phys. 83 (2011), pp. 1057--1110.

\bibitem{ZhangSC2001}
S.C. Zhang and J.~Hu, Science 294 (2001), pp. 823--828.

\bibitem{Lohse2018}
M.~Lohse, C.~Schweizer, H.M. Price, O.~Zilberberg, and I.~Bloch, Nature 553
  (2018), p.~55.

\bibitem{Kane2005a}
C.L. Kane and E.J. Mele, Phys. Rev. Lett. 95 (2005), p. 146802.

\bibitem{Kane2005b}
C.L. Kane and E.J. Mele, Phys. Rev. Lett. 95 (2005), p. 226801.

\bibitem{Bernevig2006b}
B.A. Bernevig, T.L. Hughes, and S.C. Zhang, Science 314 (2006), pp. 1757--1761.

\bibitem{Bernevig2006a}
B.A. Bernevig and S.C. Zhang, Phys. Rev. Lett. 96 (2006), p. 106802.

\bibitem{Konig2007}
M.~König~\emph{et al}., Science 318 (2007), p. 766.

\bibitem{Wehling2014}
T.~Wehling, A.~Black-Schaffer, and A.~Balatsky, Adv. Phys. 63 (2014), pp.
  1--76.

\bibitem{Armitage2018}
N.P. Armitage, E.J. Mele, and A.~Vishwanath, Rev. Mod. Phys. 90 (2018), p.
  015001.

\bibitem{Windpassinger2013}
P.~Windpassinger and K.~Sengstock, Rep. Prog. Phys. 76 (2013), p. 086401.

\bibitem{Dalibard2011}
J.~Dalibard, F.~Gerbier, G.~Juzeli\ifmmode~\bar{u}\else \={u}\fi{}nas, and
  P.~\"Ohberg, Rev. Mod. Phys. 83 (2011), pp. 1523--1543.

\bibitem{Goldman2014}
N.~Goldman, G.~Juzeliūnas, P.~Öhberg, and I.B. Spielman, Rep. Prog. Phys. 77
  (2014), p. 126401.

\bibitem{Jotzu2014}
G.~Jotzu~\emph{et al}., Nature 515 (2014), pp. 237--240.

\bibitem{Miyake2013}
H.~Miyake, G.A. Siviloglou, C.J. Kennedy, W.C. Burton, and W.~Ketterle, Phys.
  Rev. Lett. 111 (2013), p. 185302.

\bibitem{Aidelsburger2013}
M.~Aidelsburger~\emph{et al}., Phys. Rev. Lett. 111 (2013), p. 185301.

\bibitem{YJLin2011b}
Y.J. Lin, K.~Jimenez-Garcia, and I.B. Spielman, Nature 471 (2011), pp. 83--86.

\bibitem{JYZhang2012}
J.Y. Zhang~\emph{et al}., Phys. Rev. Lett. 109 (2012), p. 115301.

\bibitem{PWang2012}
P.~Wang~\emph{et al}., Phys. Rev. Lett. 109 (2012), p. 095301.

\bibitem{Cheuk2012}
L.W. Cheuk~\emph{et al}., Phys. Rev. Lett. 109 (2012), p. 095302.

\bibitem{LHuang2016}
L.~Huang~\emph{et al}., Nat. Phys. 12 (2016), pp. 540--544.

\bibitem{ZWu2016}
Z.~Wu~\emph{et al}., Science 354 (2016), pp. 83--88.

\bibitem{Mancini2015}
M.~Mancini~\emph{et al}., Science 349 (2015), pp. 1510--1513.

\bibitem{Stuhl2015}
B.K. Stuhl, H.I. Lu, L.M. Aycock, D.~Genkina, and I.B. Spielman, Science 349
  (2015), pp. 1514--1518.

\bibitem{Livi2016}
L.~Livi~\emph{et al}., Phys. Rev. Lett. 117 (2016), p. 220401.

\bibitem{YLi2013a}
Y.~Li and C.~Wu, Phys. Rev. Lett. 110 (2013), p. 216802.

\bibitem{Price2015}
H.M. Price, O.~Zilberberg, T.~Ozawa, I.~Carusotto, and N.~Goldman, Phys. Rev.
  Lett. 115 (2015), p. 195303.

\bibitem{Chin2010}
C.~Chin, R.~Grimm, P.~Julienne, and E.~Tiesinga, Rev. Mod. Phys. 82 (2010), pp.
  1225--1286.

\bibitem{Baranov2008}
M.~Baranov, Phys. Rep. 464 (2008), pp. 71--111.

\bibitem{Lahaye2009}
T.~Lahaye, C.~Menotti, L.~Santos, M.~Lewenstein, and T.~Pfau, Rep. Prog. Phys.
  72 (2009), p. 126401.

\bibitem{TLi2016}
T.~Li~\emph{et al}., Science 352 (2016), pp. 1094--1097.

\bibitem{Flaschner2016}
N.~Flaschner~\emph{et al}., Science 352 (2016), p. 1091.

\bibitem{Flaeschner2017}
N.~Fläschner~\emph{et al}., Nat. Phys. 14 (2017), pp. 265--268.

\bibitem{Aidelsburger2014}
M.~Aidelsburger~\emph{et al}., Nat. Phys. 11 (2014), p. 162.

\bibitem{Lewenstein2007}
M.~Lewenstein~\emph{et al}., Adv. Phys. 56 (2007), pp. 243--379.

\bibitem{Bloch2008}
I.~Bloch, J.~Dalibard, and W.~Zwerger, Rev. Mod. Phys. 80 (2008), pp. 885--964.

\bibitem{Cooper2008}
N.~Cooper, Adv. Phys. 57 (2008), pp. 539--616.

\bibitem{DWZhang2011}
D.W. Zhang, Z.D. Wang, and S.L. Zhu, Front. Phys. 7 (2011), pp. 31--53.

\bibitem{Bloch2012}
I.~Bloch, J.~Dalibard, and S.~Nascimb{\`{e}}ne, Nat. Phys. 8 (2012), pp.
  267--276.

\bibitem{Galitski2013}
V.~Galitski and I.B. Spielman, Nature 494 (2013), p.~49.

\bibitem{HZhai2015}
H.~Zhai, Rep. Prog. Phys. 78 (2015), p. 026001.

\bibitem{XZhou2015}
X.~Zhou, Y.~Li, Z.~Cai, and C.~Wu, Journal of Physics B: Atomic, Molecular and
  Optical Physics 48 (2015), p. 249501.

\bibitem{Goldman2016}
N.~Goldman, J.C. Budich, and P.~Zoller, Nat. Phys. 12 (2016), pp. 639--645.

\bibitem{Gross2017}
C.~Gross and I.~Bloch, Science 357 (2017), pp. 995--1001.

\bibitem{Cooper2018}
N.R. Cooper, J.~Dalibard, and I.B. Spielman, arXiv:1803.00249v1  (2018).

\bibitem{Bansil2016}
A.~Bansil, H.~Lin, and T.~Das, Rev. Mod. Phys. 88 (2016).

\bibitem{Chiu2016}
C.K. Chiu, J.C.Y. Teo, A.P. Schnyder, and S.~Ryu, Rev. Mod. Phys. 88 (2016), p.
  035005.

\bibitem{Zak1989}
J.~Zak, Phys. Rev. Lett. 62 (1989), pp. 2747--2750.

\bibitem{Golterman1993}
M.F.L. Golterman, K.~Jansen, and D.B. Kaplan, Phys. Lett. B 301 (1993), p. 219.

\bibitem{XLQi2008}
X.L. Qi, T.L. Hughes, and S.C. Zhang, Phys. Rev. B 78 (2008), p. 195424.

\bibitem{DNSheng2006}
D.N. Sheng, Z.Y. Weng, L.~Sheng, and F.D.M. Haldane, Phys. Rev. Lett. 97
  (2006), p. 036808.

\bibitem{Moore2008}
J.E. Moore, Y.~Ran, and X.G. Wen, Phys. Rev. Lett. 101 (2008), p. 186805.

\bibitem{Wilczek1983}
F.~Wilczek and A.~Zee, Phys. Rev. Lett. 51 (1983), pp. 2250--2252.

\bibitem{Hansch1975}
T.W. Hänsch and A.L. Schawlow, Opt. Commun. 13 (1975), pp. 68--69.

\bibitem{Leggett2001}
A.J. Leggett, Rev. Mod. Phys. 73 (2001), pp. 307--356.

\bibitem{Moerdijk1995}
A.J. Moerdijk, B.J. Verhaar, and A.~Axelsson, Phys. Rev. A 51 (1995), pp.
  4852--4861.

\bibitem{Bloch2005}
I.~Bloch, Nat. Phys. 1 (2005), p.~23.

\bibitem{Fallani2007}
L.~Fallani, J.E. Lye, V.~Guarrera, C.~Fort, and M.~Inguscio, Phys. Rev. Lett.
  98 (2007), p. 130404.

\bibitem{Roati2008}
G.~Roati~\emph{et al}., Nature 453 (2008), p. 895.

\bibitem{Billy2008}
J.~Billy~\emph{et al}., Nature 453 (2008), p. 891.

\bibitem{Blakie2004}
P.B. Blakie and J.V. Porto, Phys. Rev. A 69 (2004), p. 013603.

\bibitem{SLZhu2007}
S.L. Zhu, B.~Wang, and L.M. Duan, Phys. Rev. Lett. 98 (2007), p. 260402.

\bibitem{Tarruell2012}
L.~Tarruell, D.~Greif, T.~Uehlinger, G.~Jotzu, and T.~Esslinger, Nature 483
  (2012), pp. 302--305.

\bibitem{Jaksch2005}
D.~Jaksch and P.~Zoller, Ann. Phys. 315 (2005), pp. 52 -- 79, special Issue.

\bibitem{Jaksch1998}
D.~Jaksch, C.~Bruder, J.I. Cirac, C.W. Gardiner, and P.~Zoller, Phys. Rev.
  Lett. 81 (1998), pp. 3108--3111.

\bibitem{Fisher1989}
M.P.A. Fisher, P.B. Weichman, G.~Grinstein, and D.S. Fisher, Phys. Rev. B 40
  (1989), pp. 546--570.

\bibitem{Esslinger2010}
T.~Esslinger, Ann. Rev. Condens. Matter Phys. 1 (2010), pp. 129--152.

\bibitem{Mead1980}
C.A. Mead, Chem. Phys. 49 (1980), pp. 23--32.

\bibitem{Aharonov1959}
Y.~Aharonov and D.~Bohm, Phys. Rev. 115 (1959), pp. 485--491.

\bibitem{SLZhu2006}
S.L. Zhu, H.~Fu, C.J. Wu, S.C. Zhang, and L.M. Duan, Phys. Rev. Lett. 97
  (2006), p. 240401.

\bibitem{Juzeliunas2006}
G.~Juzeli\ifmmode~\bar{u}\else \={u}\fi{}nas, J.~Ruseckas, P.~\"Ohberg, and
  M.~Fleischhauer, Phys. Rev. A 73 (2006), p. 025602.

\bibitem{Juzeliunas2004}
G.~Juzeli\ifmmode\bar{u}\else\={u}\fi{}nas and P.~\"Ohberg, Phys. Rev. Lett. 93
  (2004), p. 033602.

\bibitem{Spielman2009}
I.B. Spielman, Phys. Rev. A 79 (2009), p. 063613.

\bibitem{Gunter2009}
K.J. G\"unter, M.~Cheneau, T.~Yefsah, S.P. Rath, and J.~Dalibard, Phys. Rev. A
  79 (2009), p. 011604.

\bibitem{YJLin2009a}
Y.J. Lin~\emph{et al}., Phys. Rev. Lett. 102 (2009), p. 130401.

\bibitem{YJLin2009b}
Y.J. Lin, R.L. Compton, K.~Jimenez-Garcia, J.V. Porto, and I.B. Spielman,
  Nature 462 (2009), pp. 628--632.

\bibitem{ZFu2011}
Z.~Fu, P.~Wang, S.~Chai, L.~Huang, and J.~Zhang, Phys. Rev. A 84 (2011), p.
  043609.

\bibitem{Ruseckas2005}
J.~Ruseckas, G.~Juzeli\ifmmode~\bar{u}\else \={u}\fi{}nas, P.~\"Ohberg, and
  M.~Fleischhauer, Phys. Rev. Lett. 95 (2005), p. 010404.

\bibitem{Juzeliunas2008}
G.~Juzeli\ifmmode\bar{u}\else\={u}\fi{}nas~\emph{et al}., Phys. Rev. A 77
  (2008), p. 011802.

\bibitem{Stanescu2007}
T.D. Stanescu, C.~Zhang, and V.~Galitski, Phys. Rev. Lett. 99 (2007), p.
  110403.

\bibitem{Vaishnav2008}
J.Y. Vaishnav and C.W. Clark, Phys. Rev. Lett. 100 (2008), p. 153002.

\bibitem{SLZhu2009}
S.L. Zhu, D.W. Zhang, and Z.D. Wang, Phys. Rev. Lett. 102 (2009), p. 210403.

\bibitem{CWu2008}
C.~Wu, I.~Mondragon-Shem, and X.F. Zhou, arXiv:0809.3532; Chinese. Phys. Lett.
  28 (2011), p.097102 .

\bibitem{Stanescu2008}
D.~Stanescu Tudor, B.~Anderson, and V.~Galitski, Phys. Rev. A 78 (2008), p.
  023616 .

\bibitem{CWang2010}
C.~Wang, C.~Gao, C.M. Jian, and H.~Zhai, Phys. Rev. Lett. 105 (2010), p.
  160403.

\bibitem{Ho2011}
T.L. Ho and S.~Zhang, Phys. Rev. Lett. 107 (2011), p. 150403.

\bibitem{HHu2012}
H.~Hu, B.~Ramachandhran, H.~Pu, and X.J. Liu, Phys. Rev. Lett. 108 (2012), p.
  010402 .

\bibitem{HHu2011}
H.~Hu, L.~Jiang, X.J. Liu, and H.~Pu, Phys. Rev. Lett. 107 (2011), P. 195304 .

\bibitem{SLZhu2011}
S.L. Zhu, L.B. Shao, Z.D. Wang, and L.M. Duan, Phys. Rev. Lett. 106 (2011), p.
  100404.

\bibitem{Beeler2013}
M.C. Beeler~\emph{et al}., Nature 498 (2013), pp. 201--204.

\bibitem{Qu2013}
C.~Qu, C.~Hamner, M.~Gong, C.~Zhang, and P.~Engels, Phys. Rev. A 88 (2013), p.
  021604.

\bibitem{Hamner2014}
C.~Hamner~\emph{et al}., Nat. Commun. 5 (2014), p. 4023.

\bibitem{Olson2014}
A.J. Olson~\emph{et al}., Phys. Rev. A 90 (2014), p. 013616.

\bibitem{XJLiu2007}
X.J. Liu, X.~Liu, L.C. Kwek, and C.H. Oh, Phys. Rev. Lett. 98 (2007), p.
  026602.

\bibitem{Wilczek1984}
F.~Wilczek and A.~Zee, Phys. Rev. Lett. 52 (1984), pp. 2111--2114.

\bibitem{YZhang2012}
Y.~Zhang, L.~Mao, and C.~Zhang, Phys. Rev. Lett. 108 (2012), p. 035302.

\bibitem{Campbell2011}
D.L. Campbell, G.~Juzeli\ifmmode~\bar{u}\else \={u}\fi{}nas, and I.B. Spielman,
  Phys. Rev. A 84 (2011), p. 025602.

\bibitem{CZhang2010}
C.~Zhang, Phys. Rev. A 82 (2010), p. 021607.

\bibitem{ZMeng2016}
Z.~Meng~\emph{et al}., Phys. Rev. Lett. 117 (2016), p. 235304.

\bibitem{Jaksch2003}
D.~Jaksch and P.~Zoller, New J. Phys. 5 (2003), p.~56.

\bibitem{Bermudez2010b}
A.~Bermudez~\emph{et al}., Phys. Rev. Lett. 105 (2010), p. 190404.

\bibitem{Goldman2010}
N.~Goldman~\emph{et al}., Phys. Rev. Lett. 105 (2010), p. 255302.

\bibitem{XJLiu2014}
X.J. Liu, K.T. Law, and T.K. Ng, Phys. Rev. Lett. 112 (2014), p. 086401.

\bibitem{Aidelsburger2011}
M.~Aidelsburger~\emph{et al}., Phys. Rev. Lett. 107 (2011), p. 255301.

\bibitem{FMei2014}
F.~Mei~\emph{et al}., Phys. Rev. A 90 (2014), p. 063638.

\bibitem{WZheng2014}
W.~Zheng and H.~Zhai, Phys. Rev. A 89 (2014), p. 061603.

\bibitem{Goldman2014b}
N.~Goldman and J.~Dalibard, Phys. Rev. X 4 (2014), p. 031027.

\bibitem{Sorensen2005}
A.S. S\o{}rensen, E.~Demler, and M.D. Lukin, Phys. Rev. Lett. 94 (2005), p.
  086803.

\bibitem{Schnyder2008}
A.P. Schnyder, S.~Ryu, A.~Furusaki, and A.W.W. Ludwig, Phys. Rev. B 78 (2008),
  p. 195125.

\bibitem{Jackiw1976}
R.~Jackiw and C.~Rebbi, Phys. Rev. D 13 (1976), pp. 3398--3409.

\bibitem{Atala2013}
M.~Atala~\emph{et al}., Nat. Phys. 9 (2013), pp. 795--800.

\bibitem{Rice1982}
M.J. Rice and E.J. Mele, Phys. Rev. Lett. 49 (1982), pp. 1455--1459.

\bibitem{Goldstone1981}
J.~Goldstone and F.~Wilczek, Phys. Rev. Lett. 47 (1981), pp. 986--989.

\bibitem{Jackiw1983}
R.~Jackiw and G.~Semenoff, Phys. Rev. Lett. 50 (1983), pp. 439--442.

\bibitem{Thouless1983}
D.J. Thouless, Phys. Rev. B 27 (1983), p. 6083.

\bibitem{Niu1984}
Q.~Niu and D.J. Thouless, J. Phys. A: Math. Gen. 17 (1984), p. 2453.

\bibitem{Xiao2010}
D.~Xiao, M.C. Chang, and Q.~Niu, Rev. Mod. Phys. 82 (2010), pp. 1959--2007.

\bibitem{Ruostekoski2002}
J.~Ruostekoski, G.V. Dunne, and J.~Javanainen, Phys. Rev. Lett. 88 (2002), p.
  180401.

\bibitem{Javanainen2003}
J.~Javanainen and J.~Ruostekoski, Phys. Rev. Lett. 91 (2003), p. 150404.

\bibitem{Ruostekoski2008}
J.~Ruostekoski, J.~Javanainen, and G.V. Dunne, Phys. Rev. A 77 (2008), p.
  013603.

\bibitem{Zheng2017}
Z.~Zheng, H.~Pu, X.~Zou, and G.~Guo, Phys. Rev. A 95 (2017), p. 013616.

\bibitem{DWZhang2015}
D.W. Zhang, F.~Mei, Z.Y. Xue, S.L. Zhu, and Z.D. Wang, Phys. Rev. A 92 (2015),
  p. 013612.

\bibitem{Li2013}
X.~Li, E.~Zhao, and W.~Vincent~Liu, Nat. Commun. 4 (2013), p. 1523.

\bibitem{Przysiezna2015}
A.~Przysiężna, O.~Dutta, and J.~Zakrzewski, New J. Phys. 17 (2015), p.
  013018.

\bibitem{SLZhang2017}
S.L. Zhang and Q.~Zhou, Phys. Rev. A 95 (2017), p. 061601.

\bibitem{Duca2015}
L.~Duca~\emph{et al}., Science 347 (2015), pp. 288--292.

\bibitem{Leder2016}
M.~Leder~\emph{et al}., Nat. Commun. 7 (2016), p. 13112.

\bibitem{Meier2016}
E.J. Meier, F.A. An, and B.~Gadway, Nat. Commun. 7 (2016), p. 13986.

\bibitem{Gadway2015}
B.~Gadway, Phys. Rev. A 92 (2015), p. 043606.

\bibitem{Meier2016a}
E.J. Meier, F.A. An, and B.~Gadway, Phys. Rev. A 93 (2016), p. 051602.

\bibitem{FAn2017a}
F.A. An, E.J. Meier, and B.~Gadway, Sci. Adv. 3 (2017), p. e1602685.

\bibitem{Nakajima2016}
S.~Nakajima~\emph{et al}., Nat. Phys. 12 (2016), pp. 296--300.

\bibitem{Qian2011}
Y.~Qian, M.~Gong, and C.~Zhang, Phys. Rev. A 84 (2011), p. 013608.

\bibitem{Wang2013}
L.~Wang, M.~Troyer, and X.~Dai, Phys. Rev. Lett. 111 (2013), p. 026802.

\bibitem{Grusdt2014}
F.~Grusdt and M.~H\"oning, Phys. Rev. A 90 (2014), p. 053623.

\bibitem{Marra2015}
P.~Marra, R.~Citro, and C.~Ortix, Phys. Rev. B 91 (2015), p. 125411.

\bibitem{Zeng2015}
T.S. Zeng, C.~Wang, and H.~Zhai, Phys. Rev. Lett. 115 (2015), p. 095302.

\bibitem{Zeng2016}
T.S. Zeng, W.~Zhu, and D.N. Sheng, Phys. Rev. B 94 (2016), p. 235139.

\bibitem{Taddia2017}
L.~Taddia~\emph{et al}., Phys. Rev. Lett. 118 (2017), p. 230402.

\bibitem{Xu2017}
Z.~Xu, Y.~Zhang, and S.~Chen, Phys. Rev. A 96 (2017), p. 013606.

\bibitem{Sun2017}
N.~Sun and L.K. Lim, Phys. Rev. B 96 (2017), p. 035139.

\bibitem{YKe2017}
Y.~Ke, X.~Qin, Y.S. Kivshar, and C.~Lee, Phys. Rev. A 95 (2017), p. 063630.

\bibitem{Lohse2016}
M.~Lohse, C.~Schweizer, O.~Zilberberg, M.~Aidelsburger, and I.~Bloch, Nat.
  Phys. 12 (2016), pp. 350--354.

\bibitem{HILu2016}
H.I. Lu~\emph{et al}., Phys. Rev. Lett. 116 (2016), p. 200402.

\bibitem{Schweizer2016}
C.~Schweizer, M.~Lohse, R.~Citro, and I.~Bloch, Phys. Rev. Lett. 117 (2016), p.
  170405.

\bibitem{LFu2006}
L.~Fu and C.L. Kane, Phys. Rev. B 74 (2006), p. 195312.

\bibitem{Ryu2010}
S.~Ryu, A.P. Schnyder, A.~Furusaki, and A.W.W. Ludwig, New J. Phys. 12 (2010),
  p. 065010.

\bibitem{Liu2013}
X.J. Liu, Z.X. Liu, and M.~Cheng, Phys. Rev. Lett. 110 (2013), p. 076401.

\bibitem{Velasco2017}
C.G. Velasco and B.~Paredes, Phys. Rev. Lett. 119 (2017), p. 115301.

\bibitem{Zhou2017}
X.~Zhou~\emph{et al}., Phys. Rev. Lett. 119 (2017), p. 185701.

\bibitem{BSong2018}
B.~Song~\emph{et al}., Sci. Adv. 4 (2018), p. 4748.

\bibitem{Creutz1999}
M.~Creutz, Phys. Rev. Lett. 83 (1999), p. 2636.

\bibitem{Sticlet2014}
D.~Sticlet, L.~Seabra, F.~Pollmann, and J.~Cayssol, Phys. Rev. B 89 (2014), p.
  115430.

\bibitem{Mazza2012}
L.~Mazza~\emph{et al}., New J. Phys. 14 (2012), p. 015007.

\bibitem{Huegel2014}
D.~H\"ugel and B.~Paredes, Phys. Rev. A 89 (2014), p. 023619.

\bibitem{Mazza2015}
L.~Mazza, M.~Aidelsburger, H.H. Tu, N.~Goldman, and M.~Burrello, New J. Phys.
  17 (2015), p. 105001.

\bibitem{Mugel2016}
S.~Mugel, Phys. Rev. A 94 (2016), p. 023631.

\bibitem{Juenemann2017}
J.~J\"unemann~\emph{et al}., Phys. Rev. X 7 (2017), p. 031057.

\bibitem{Barbarino2018}
S.~Barbarino, M.~Dalmonte, R.~Fazio, and G.E. Santoro, Phys. Rev. A 97 (2018),
  p. 013634.

\bibitem{Atala2014}
M.~Atala~\emph{et al}., Nat. Phys. 10 (2014), p. 588.

\bibitem{JHKang2018}
J.H. Kang, J.H. Han, and Y.~Shin, Phys. Rev. Lett. 121 (2018), p. 150403.

\bibitem{Lang2012a}
L.J. Lang, X.~Cai, and S.~Chen, Phys. Rev. Lett. 108 (2012), p. 220401.

\bibitem{Kraus2012a}
Y.E. Kraus, Y.~Lahini, Z.~Ringel, M.~Verbin, and O.~Zilberberg, Phys. Rev.
  Lett. 109 (2012), p. 106402.

\bibitem{Aubry1980}
S.~Aubry and G.~Andre, Ann. Israel Phys. Soc. 3 (1980), p. 133.

\bibitem{Harper1955}
P.G. Harper, Proc. Phys. Soc. Sect. A 68 (1955), pp. 874--878.

\bibitem{Schreiber2015}
M.~Schreiber~\emph{et al}., Science 349 (2015), pp. 842--845.

\bibitem{Singh2015}
K.~Singh, K.~Saha, S.A. Parameswaran, and D.M. Weld, Phys. Rev. A 92 (2015), p.
  063426.

\bibitem{Ganeshan2013}
S.~Ganeshan, K.~Sun, and S.~Das~Sarma, Phys. Rev. Lett. 110 (2013), p. 180403.

\bibitem{Mei2012}
F.~Mei, S.L. Zhu, Z.M. Zhang, C.H. Oh, and N.~Goldman, Phys. Rev. A 85 (2012),
  p. 013638.

\bibitem{Ray2017}
S.~Ray, B.~Mukherjee, S.~Sinha, and K.~Sengupta, Phys. Rev. A 96 (2017), p.
  023607.

\bibitem{Lang2012b}
L.J. Lang and S.~Chen, Phys. Rev. B 86 (2012), p. 205135.

\bibitem{Cai2013}
X.~Cai, L.J. Lang, S.~Chen, and Y.~Wang, Phys. Rev. Lett. 110 (2013), p.
  176403.

\bibitem{DeGottardi2013}
W.~DeGottardi, D.~Sen, and S.~Vishveshwara, Phys. Rev. Lett. 110 (2013), p.
  146404.

\bibitem{Xu2013}
Z.~Xu, L.~Li, and S.~Chen, Phys. Rev. Lett. 110 (2013), p. 215301.

\bibitem{SLZhu2013}
S.L. Zhu, Z.D. Wang, Y.H. Chan, and L.M. Duan, Phys. Rev. Lett. 110 (2013), p.
  075303.

\bibitem{Novoselov2005}
K.S. Novoselov~\emph{et al}., Nature 438 (2005), p. 197.

\bibitem{YZhang2005}
Y.~Zhang, Y.W. Tan, H.L. Stormer, and P.~Kim, Nature 438 (2005), p. 201.

\bibitem{DXiao2007}
D.~Xiao, W.~Yao, and Q.~Niu, Phys. Rev. Lett. 99 (2007), p. 236809.

\bibitem{CWu2008b}
C.~Wu and S.~Das~Sarma, Phys. Rev. B 77 (2008), p. 235107.

\bibitem{Wunsch2008}
B.~Wunsch, F.~Guinea, and F.~Sols, New J. Phys. 10 (2008), p. 103027.

\bibitem{KLLee2009}
K.L. Lee, B.~Gr\'emaud, R.~Han, B.G. Englert, and C.~Miniatura, Phys. Rev. A 80
  (2009), p. 043411.

\bibitem{Poletti2011}
D.~Poletti, C.~Miniatura, and B.~Grémaud, Europhys. Lett. 93 (2011), p. 37008.

\bibitem{Bercioux2009}
D.~Bercioux, D.F. Urban, H.~Grabert, and W.~H\"ausler, Phys. Rev. A 80 (2009),
  p. 063603.

\bibitem{RShen2010}
R.~Shen, L.B. Shao, B.~Wang, and D.Y. Xing, Phys. Rev. B 81 (2010), p. 041410.

\bibitem{Satija2008}
I.I. Satija, D.C. Dakin, J.Y. Vaishnav, and C.W. Clark, Phys. Rev. A 77 (2008),
  p. 043410.

\bibitem{JMHou2009}
J.M. Hou, W.X. Yang, and X.J. Liu, Phys. Rev. A 79 (2009), p. 043621.

\bibitem{Lim2008}
L.K. Lim, C.M. Smith, and A.~Hemmerich, Phys. Rev. Lett. 100 (2008), p. 130402.

\bibitem{Goldman2009}
N.~Goldman~\emph{et al}., Phys. Rev. Lett. 103 (2009), p. 035301.

\bibitem{XJLiu2010}
X.J. Liu, X.~Liu, C.~Wu, and J.~Sinova, Phys. Rev. A 81 (2010), p. 033622.

\bibitem{Kennett2011}
M.P. Kennett, N.~Komeilizadeh, K.~Kaveh, and P.M. Smith, Phys. Rev. A 83
  (2011), p. 053636.

\bibitem{Goldman2011}
N.~Goldman, D.F. Urban, and D.~Bercioux, Phys. Rev. A 83 (2011), p. 063601.

\bibitem{Semenoff1984}
G.W. Semenoff, Phys. Rev. Lett. 53 (1984), pp. 2449--2452.

\bibitem{XGWen2004}
X.G. Wen, \emph{Quantum Field Theory of Many-body Systems: From the Origin of
  Sound to an Origin of Light and Electrons}, Oxford: Oxford University, 2004.

\bibitem{Bermudez2010a}
A.~Bermudez, N.~Goldman, A.~Kubasiak, M.~Lewenstein, and M.A. Martin-Delgado,
  New J. Phys. 12 (2010), p. 033041.

\bibitem{Block2010}
J.K. Block and N.~Nygaard, Phys. Rev. A 81 (2010), p. 053421.

\bibitem{EZhao2006}
E.~Zhao and A.~Paramekanti, Phys. Rev. Lett. 97 (2006), p. 230404.

\bibitem{CWu2007}
C.~Wu, D.~Bergman, L.~Balents, and S.~Das~Sarma, Phys. Rev. Lett. 99 (2007), p.
  070401.

\bibitem{Soltan-Panahi2011}
P.~Soltan-Panahi~\emph{et al}., Nat. Phys. 7 (2011), pp. 434--440.

\bibitem{Soltan-Panahi2012}
P.~Soltan-Panahi, D.S. Luhmann, J.~Struck, P.~Windpassinger, and K.~Sengstock,
  Nat. Phys. 8 (2012), pp. 71--75.

\bibitem{Uehlinger2013a}
T.~Uehlinger, G.~Jotzu, M.~Messer, D.~Greif, W.~Hofstetter, U.~Bissbort, and
  T.~Esslinger, Phys. Rev. Lett. 111 (2013), p. 185307.

\bibitem{Uehlinger2013}
T.~Uehlinger, D.~Greif, G.~Jotzu, L.~Tarruell, T.~Esslinger, L.~Wang, and
  M.~Troyer, Eur. Phys. J. Special Topics 217 (2013), pp. 121--133.

\bibitem{Lim2012}
L.K. Lim, J.N. Fuchs, and G.~Montambaux, Phys. Rev. Lett. 108 (2012), p.
  175303.

\bibitem{Montambaux2009}
G.~Montambaux, F.~Pi\'echon, J.N. Fuchs, and M.O. Goerbig, Phys. Rev. B 80
  (2009), p. 153412.

\bibitem{Hofstadter1976}
D.R. Hofstadter, Phys. Rev. B 14 (1976), pp. 2239--2249.

\bibitem{Dean2013}
C.R. Dean~\emph{et al}., Nature 497 (2013), pp. 598--602.

\bibitem{Hunt2013}
B.~Hunt~\emph{et al}., Science 340 (2013), pp. 1427--1430.

\bibitem{Mueller2004}
E.J. Mueller, Phys. Rev. A 70 (2004), p. 041603.

\bibitem{Osterloh2005}
K.~Osterloh, M.~Baig, L.~Santos, P.~Zoller, and M.~Lewenstein, Phys. Rev. Lett.
  95 (2005), p. 010403.

\bibitem{Goldman2009b}
N.~Goldman, A.~Kubasiak, P.~Gaspard, and M.~Lewenstein, Phys. Rev. A 79 (2009),
  p. 023624.

\bibitem{Gerbier2010}
F.~Gerbier and J.~Dalibard, New J. Phys. 12 (2010), p. 033007.

\bibitem{Bilitewski2015}
T.~Bilitewski and N.R. Cooper, Phys. Rev. A 91 (2015), p. 063611.

\bibitem{Bukov2014}
M.~Bukov and A.~Polkovnikov, Phys. Rev. A 90 (2014), p. 043613.

\bibitem{Zhou2014b}
Z.~Zhou, I.I. Satija, and E.~Zhao, Phys. Rev. B 90 (2014), p. 205108.

\bibitem{Creffield2016}
C.E. Creffield, G.~Pieplow, F.~Sols, and N.~Goldman, New J. Phys. 18 (2016), p.
  093013.

\bibitem{Umucalilar2008}
R.O. Umucal\ifmmode\imath\else\i\fi{}lar, H.~Zhai, and M.O. Oktel, Phys. Rev.
  Lett. 100 (2008), p. 070402.

\bibitem{Yilmaz2015}
F.~Y{\i}lmaz, F.N. Ünal, and M.. Oktel, Phys. Rev. A 91 (2015), p. 063628.

\bibitem{Yilmaz2017}
F.~Y{\i}lmaz and M.. Oktel, Phys. Rev. A 95 (2017), p. 063628.

\bibitem{Zhao2011}
E.~Zhao, N.~Bray-Ali, C.J. Williams, I.B. Spielman, and I.I. Satija, Phys. Rev.
  A 84 (2011), p. 063629.

\bibitem{Wang2014b}
L.~Wang and M.~Troyer, Phys. Rev. A 89 (2014), p. 011603.

\bibitem{Goldman2012}
N.~Goldman, J.~Beugnon, and F.~Gerbier, Phys. Rev. Lett. 108 (2012), p. 255303.

\bibitem{Goldman2013}
N.~Goldman~\emph{et al}., Proc. Natl Acad. Sci. 110 (2013), pp. 6736--6741.

\bibitem{Kolovsky2011}
A.R. Kolovsky, Europhys. Lett. 93 (2011), p. 20003.

\bibitem{Kennedy2013}
C.J. Kennedy, G.A. Siviloglou, H.~Miyake, W.C. Burton, and W.~Ketterle, Phys.
  Rev. Lett. 111 (2013), p. 225301.

\bibitem{Kennedy2015}
C.J. Kennedy, W.C. Burton, W.C. Chung, and W.~Ketterle, Nat. Phys. 11 (2015),
  pp. 859--864.

\bibitem{Tai2017}
M.E. Tai~\emph{et al}., Nature 546 (2017), pp. 519--523.

\bibitem{Zhai2010}
H.~Zhai, R.O. Umucal{\i}lar, and M.. Oktel, Phys. Rev. Lett. 104 (2010), p.
  145301.

\bibitem{Lacki2016}
M.~Lacki~\emph{et al}., Phys. Rev. A 93 (2016), p. 013604.

\bibitem{LBShao2008}
L.B. Shao, S.L. Zhu, L.~Sheng, D.Y. Xing, and Z.D. Wang, Phys. Rev. Lett. 101
  (2008), p. 246810.

\bibitem{CWu2008a}
C.~Wu, Phys. Rev. Lett. 101 (2008), p. 186807.

\bibitem{Alba2011}
E.~Alba, X.~Fernandez-Gonzalvo, J.~Mur-Petit, J.K. Pachos, and J.J.
  Garcia-Ripoll, Phys. Rev. Lett. 107 (2011), p. 235301.

\bibitem{GCLiu2010}
G.~Liu, S.L. Zhu, S.~Jiang, F.~Sun, and W.M. Liu, Phys. Rev. A 82 (2010), p.
  053605.

\bibitem{Cocks2012}
D.~Cocks~\emph{et al}., Phys. Rev. Lett. 109 (2012), p. 205303.

\bibitem{Young2012}
S.M. Young~\emph{et al}., Phys. Rev. Lett. 108 (2012), p. 140405.

\bibitem{ZWang2012}
Z.~Wang~\emph{et al}., Phys. Rev. B 85 (2012), p. 195320.

\bibitem{ZWang2013}
Z.~Wang, H.~Weng, Q.~Wu, X.~Dai, and Z.~Fang, Phys. Rev. B 88 (2013), p.
  125427.

\bibitem{MYang2010}
M.~Yang and S.L. Zhu, Phys. Rev. A 82 (2010), p. 064102.

\bibitem{Lepori2010}
L.~Lepori, G.~Mussardo, and A.~Trombettoni, Europhys. Lett. 92 (2010), p.
  50003.

\bibitem{Wilson1977}
K.~Wilson, \emph{New Phenomena in Subnuclear Physics}, Plenum, New York, 1977.

\bibitem{XWan2011}
X.~Wan, A.M. Turner, A.~Vishwanath, and S.Y. Savrasov, Phys. Rev. B 83 (2011),
  p. 205101.

\bibitem{Burkov2011}
A.A. Burkov and L.~Balents, Phys. Rev. Lett. 107 (2011), p. 127205.

\bibitem{Burkov2011b}
A.A. Burkov, M.D. Hook, and L.~Balents, Phys. Rev. B 84 (2011), p. 235126.

\bibitem{Ganeshan2015}
S.~Ganeshan and S.~Das~Sarma, Phys. Rev. B 91 (2015), p. 125438.

\bibitem{JHJiang2012}
J.H. Jiang, Phys. Rev. A 85 (2012), p. 033640.

\bibitem{WYHe2016}
W.Y. He, S.~Zhang, and K.T. Law, Phys. Rev. A 94 (2016), p. 013606.

\bibitem{JMHou2016}
J.M. Hou and W.~Chen, Sci. Rep. 6 (2016), p. 33512.

\bibitem{BZWang2018}
B.Z. Wang~\emph{et al}., Phys. Rev. A 97 (2018), p. 011605.

\bibitem{XLi2015}
X.~Li and S.D. Sarma, Nat. Commun. 6 (2015), p. 7137.

\bibitem{Syzranov2016}
S.V. Syzranov, M.L. Wall, B.~Zhu, V.~Gurarie, and A.M. Rey, Nat. Commun. 7
  (2016), p. 13543.

\bibitem{LJLang2017}
L.J. Lang, S.L. Zhang, K.T. Law, and Q.~Zhou, Phys. Rev. B 96 (2017), p.
  035145.

\bibitem{Dubcek2015}
T.~Dub\ifmmode\check{c}\else\v{c}\fi{}ek~\emph{et al}., Phys. Rev. Lett. 114
  (2015), p. 225301.

\bibitem{DWZhang2017}
D.W. Zhang, R.B. Liu, and S.L. Zhu, Phys. Rev. A 95 (2017), p. 043619.

\bibitem{ZLi2016}
Z.~Li, H.Q. Wang, D.W. Zhang, S.L. Zhu, and D.Y. Xing, Phys. Rev. A 94 (2016),
  p. 043617.

\bibitem{Soluyanov2015}
A.A. Soluyanov~\emph{et al}., Nature 527 (2015), pp. 495--498.

\bibitem{YXu2016b}
Y.~Xu and L.M. Duan, Phys. Rev. A 94 (2016), p. 053619.

\bibitem{Shastri2017}
K.~Shastri, Z.~Yang, and B.~Zhang, Phys. Rev. B 95 (2017), p. 014306.

\bibitem{XKong2017}
X.~Kong, J.~He, Y.~Liang, and S.P. Kou, Phys. Rev. A 95 (2017), p. 033629.

\bibitem{GXu2011}
G.~Xu, H.~Weng, Z.~Wang, X.~Dai, and Z.~Fang, Phys. Rev. Lett. 107 (2011), p.
  186806.

\bibitem{CFang2012}
C.~Fang, M.J. Gilbert, X.~Dai, and B.A. Bernevig, Phys. Rev. Lett. 108 (2012),
  p. 266802.

\bibitem{SHuang2016}
S.M. Huang~\emph{et al}., Proc. Natl. Acad. Sci. 113 (2016), pp. 1180--1185.

\bibitem{Lepori2016}
L.~Lepori, I.C. Fulga, A.~Trombettoni, and M.~Burrello, Phys. Rev. A 94 (2016),
  p. 053633.

\bibitem{XYMai2017}
X.Y. Mai, D.W. Zhang, Z.~Li, and S.L. Zhu, Phys. Rev. A 95 (2017), p. 063616.

\bibitem{YXu2014}
Y.~Xu, R.L. Chu, and C.~Zhang, Phys. Rev. Lett. 112 (2014), p. 136402.

\bibitem{BLiu2015}
B.~Liu, X.~Li, L.~Yin, and W.V. Liu, Phys. Rev. Lett. 114 (2015), p. 045302.

\bibitem{YXu2015}
Y.~Xu, F.~Zhang, and C.~Zhang, Phys. Rev. Lett. 115 (2015), p. 265304.

\bibitem{YJWu2017}
Y.J. Wu, W.Y. Zhou, and S.P. Kou, Phys. Rev. A 95 (2017), p. 023620.

\bibitem{CFang2016}
C.~Fang, H.~Weng, X.~Dai, and Z.~Fang, Chin. Phys. B 25 (2016), p. 117106.

\bibitem{RYu2017}
R.~Yu, Z.~Fang, X.~Dai, and H.~Weng, Front. Phys. 12 (2017), p. 127202.

\bibitem{GBian2016}
G.~Bian~\emph{et al}., Nat. Commun. 7 (2016), p. 10556.

\bibitem{JHu2016}
J.~Hu~\emph{et al}., Phys. Rev. Lett. 117 (2016), p. 016602.

\bibitem{Pezzini2017}
S.~Pezzini~\emph{et al}., Nat. Phys. 14 (2017), pp. 178--183.

\bibitem{DWZhang2016a}
D.W. Zhang~\emph{et al}., Phys. Rev. A 93 (2016), p. 043617.

\bibitem{YXZhao2016}
Y.X. Zhao, A.P. Schnyder, and Z.D. Wang, Phys. Rev. Lett. 116 (2016), p.
  156402.

\bibitem{YXu2016}
Y.~Xu and C.~Zhang, Phys. Rev. A 93 (2016), p. 063606.

\bibitem{WYHe2018}
W.Y. He, D.H. Xu, B.T. Zhou, Q.~Zhou, and K.T. Law, arXiv:1801.05182v1  (2018).

\bibitem{Lim2017}
L.K. Lim and R.~Moessner, Phys. Rev. Lett. 118 (2017), p. 016401.

\bibitem{YXu2017b}
Y.~Xu, S.T. Wang, and L.M. Duan, Phys. Rev. Lett. 118 (2017), p. 045701.

\bibitem{Bzdusek2016}
T.~Bzdu{\v{s}}ek, Q.~Wu, A.~Rüegg, M.~Sigrist, and A.A. Soluyanov, Nature 538
  (2016), pp. 75--78.

\bibitem{WChen2017}
W.~Chen, H.Z. Lu, and J.M. Hou, Phys. Rev. B 96 (2017), p. 041102.

\bibitem{ZYan2017}
Z.~Yan~\emph{et al}., Phys. Rev. B 96 (2017), p. 041103.

\bibitem{Ezawa2017}
M.~Ezawa, Phys. Rev. B 96 (2017), p. 041202.

\bibitem{BSong2018b}
B.~Song~et al., arXiv:1808.07428(2018) .

\bibitem{LFu2007a}
L.~Fu, C.L. Kane, and E.J. Mele, Phys. Rev. Lett. 98 (2007), p. 106803.

\bibitem{Moore2007}
J.E. Moore and L.~Balents, Phys. Rev. B 75 (2007), p. 121306.

\bibitem{Roy2009}
R.~Roy, Phys. Rev. B 79 (2009), p. 195322.

\bibitem{LFu2007b}
L.~Fu and C.L. Kane, Phys. Rev. B 76 (2007), p. 045302.

\bibitem{Hsieh2008}
D.~Hsieh~\emph{et al}., Nature 452 (2008), p. 970.

\bibitem{HJZhang2009}
H.~Zhang~\emph{et al}., Nat. Phys. 5 (2009), p. 438.

\bibitem{YXia2009}
Y.~Xia~\emph{et al}., Nat. Phys. 5 (2009), p. 398.

\bibitem{Cooper2011a}
N.R. Cooper, Phys. Rev. Lett. 106 (2011), p. 175301.

\bibitem{Beri2011}
B.~B\'eri and N.R. Cooper, Phys. Rev. Lett. 107 (2011), p. 145301.

\bibitem{Tannoudji1992}
C.~Cohen-Tannoudji, J.~Dupont-Roc, and G.~Grynberg, \emph{Atom-Photon
  Interactions}, Wiley, New York, 1992.

\bibitem{BLi2017}
B.~Li and A.A. Kovalev, arXiv:1712.08612v2  (2017).

\bibitem{Hosur2010}
P.~Hosur, S.~Ryu, and A.~Vishwanath, Phys. Rev. B 81 (2010), p. 045120.

\bibitem{Essin2012}
A.M. Essin and V.~Gurarie, Phys. Rev. B 85 (2012), p. 195116.

\bibitem{STWang2014}
S.T. Wang, D.L. Deng, and L.M. Duan, Phys. Rev. Lett. 113 (2014), p. 033002.

\bibitem{Silaev2010}
M.A. Silaev and G.E. Volovik, J. Low Temp. Phys. 161 (2010), pp. 460--473.

\bibitem{Essin2011}
A.M. Essin and V.~Gurarie, Phys. Rev. B 84 (2011), p. 125132.

\bibitem{Neupert2012}
T.~Neupert, L.~Santos, S.~Ryu, C.~Chamon, and C.~Mudry, Phys. Rev. B 86 (2012),
  p. 035125.

\bibitem{DLDeng2014}
D.L. Deng, S.T. Wang, and L.M. Duan, Phys. Rev. A 90 (2014), p. 041601.

\bibitem{DLDeng2013}
D.L. Deng, S.T. Wang, C.~Shen, and L.M. Duan, Phys. Rev. B 88 (2013), p.
  201105.

\bibitem{CYWang2015}
C.Y. Wang and Y.~He, J. Phys.: Condens. Matter 27 (2015), p. 075603.

\bibitem{CXLiu2017}
C.~Liu, F.~Vafa, and C.~Xu, Phys. Rev. B 95 (2017), p. 161116.

\bibitem{DLDeng2018}
D.L. Deng, S.T. Wang, K.~Sun, and L.M. Duan, Chin. Phys. Lett. 35 (2018),
  013701.

\bibitem{XXYuan2017}
X.X. Yuan~\emph{et al}., Chin. Phys. Lett. 34 (2017), p. 060302.

\bibitem{Whitehead1947}
J.H.C. Whitehead, Proc. Natl Acad. Sci. USA 33 (1947), p. 117.

\bibitem{Lyons2003}
D.W. Lyons, Math. Mag. 76 (2003), pp. 87--98.

\bibitem{Hatcher2002}
A.~Hatcher, \emph{Algebraic Topology}, Cambridge University Press, 2002.

\bibitem{Kauffman2013}
L.H. Kauffman, \emph{Knots and physics, vol. 53}, World scientific, 2013.

\bibitem{Faddeev1997}
L.~Faddeev and A.J. Niemi, Nature 387 (1997), p.~58.

\bibitem{Halperin1987}
B.I. Halperin, Jpn. J. Appl. Phys. 26 (1987), p. 1913.

\bibitem{Montambaux1990}
G.~Montambaux and M.~Kohmoto, Phys. Rev. B 41 (1990), p. 11417.

\bibitem{Kohmoto1992}
M.~Kohmoto, B.I. Halperin, and Y.S. Wu, Phys. Rev. B 45 (1992), p. 13488.

\bibitem{Koshino2001}
M.~Koshino, H.~Aoki, K.~Kuroki, S.~Kagoshima, and T.~Osada, Phys. Rev. Lett. 86
  (2001), p. 1062.

\bibitem{Balicas1995}
L.~Balicas, G.~Kriza, and F.I.B. Williams, Phys. Rev. Lett. 75 (1995), p. 2000.

\bibitem{McKernan1995}
S.K. McKernan, S.T. Hannahs, U.M. Scheven, G.M. Danner, and P.M. Chaikin, Phys.
  Rev. Lett. 75 (1995), p. 1630.

\bibitem{Bernevig2007}
B.A. Bernevig, T.L. Hughes, S.~Raghu, and D.P. Arovas, Phys. Rev. Lett. 99
  (2007), p. 146804.

\bibitem{Mullen2015}
K.~Mullen, B.~Uchoa, and D.T. Glatzhofer, Phys. Rev. Lett. 115 (2015), p.
  026403.

\bibitem{Avron1983}
J.E. Avron, R.~Seiler, and B.~Simon, Phys. Rev. Lett. 51 (1983), p.~51.

\bibitem{Kraus2013b}
Y.E. Kraus, Z.~Ringel, and O.~Zilberberg, Phys. Rev. Lett. 111 (2013), p.
  226401.

\bibitem{Zilberberg2018}
O.~Zilberberg~\emph{et al}., Nature 553 (2018), p.~59.

\bibitem{Boada2012}
O.~Boada, A.~Celi, J.I. Latorre, and M.~Lewenstein, Phys. Rev. Lett. 108
  (2012), p. 133001.

\bibitem{Celi2014}
A.~Celi~\emph{et al}., Phys. Rev. Lett. 112 (2014), p. 043001.

\bibitem{Barbarino2015}
S.~Barbarino, L.~Taddia, D.~Rossini, L.~Mazza, and R.~Fazio, Nat. Commun. 6
  (2015), p. 8134.

\bibitem{Ghosh2015}
S.K. Ghosh, U.K. Yadav, and V.B. Shenoy, Phys. Rev. A 92 (2015), p. 051602.

\bibitem{ZYan2015}
Z.~Yan, S.~Wan, and Z.~Wang, Sci. Rep. 5 (2015), p. 15927.

\bibitem{Barbarino2016}
S.~Barbarino, L.~Taddia, D.~Rossini, L.~Mazza, and R.~Fazio, New J. Phys. 18
  (2016), p. 035010.

\bibitem{TSZeng2015}
T.S. Zeng, C.~Wang, and H.~Zhai, Phys. Rev. Lett. 115 (2015), p. 095302.

\bibitem{YLi2013}
Y.~Li, S.C. Zhang, and C.~Wu, Phys. Rev. Lett. 111 (2013), p. 186803.

\bibitem{Edge2012}
J.M. Edge, J.~Tworzyd\l{}o, and C.W.J. Beenakker, Phys. Rev. Lett. 109 (2012),
  p. 135701.

\bibitem{Price2016}
H.M. Price, O.~Zilberberg, T.~Ozawa, I.~Carusotto, and N.~Goldman, Phys. Rev. B
  93 (2016), p. 245113.

\bibitem{XWLuo2015}
X.W. Luo~\emph{et al}., Nat. Commun. 6 (2015), p. 7704.

\bibitem{Price2017}
H.M. Price, T.~Ozawa, and N.~Goldman, Phys. Rev. A 95 (2017), p. 023607.

\bibitem{FAn2017b}
F.A. An, E.J. Meier, and B.~Gadway, Nat. Commun. 8 (2017), p. 325.

\bibitem{FAn2018}
F.A. An, E.J. Meier, J.~Ang'ong'a, and B.~Gadway, Phys. Rev. Lett. 120 (2018),
  p. 040407.

\bibitem{Meier2018}
E.J. Meier~\emph{et al}., Science 362 (2018), p. 929.

\bibitem{Bradlyn2016}
B.~Bradlyn~\emph{et al}., Science 353 (2016), p. 5039.

\bibitem{ZLan2011}
Z.~Lan, N.~Goldman, A.~Bermudez, W.~Lu, and P.~\"Ohberg, Phys. Rev. B 84
  (2011), p. 165115.

\bibitem{LLiang2016}
L.~Liang and Y.~Yu, Phys. Rev. B 93 (2016), p. 045113.

\bibitem{HMWeng2016}
H.~Weng, C.~Fang, Z.~Fang, and X.~Dai, Phys. Rev. B 94 (2016), p. 165201.

\bibitem{BQLv2017}
B.Q. Lv~\emph{et al}., Nature 546 (2017), p. 627.

\bibitem{YQZhu2017}
Y.Q. Zhu, D.W. Zhang, H.~Yan, D.Y. Xing, and S.L. Zhu, Phys. Rev. A 96 (2017),
  p. 033634.

\bibitem{YXu2017a}
Y.~Xu and L.M. Duan, Phys. Rev. B 96 (2017), p. 155301.

\bibitem{XSTan2018}
X.~Tan~\emph{et al}., Phys. Rev. Lett. 120 (2018), p. 130503.

\bibitem{HPHu2017}
H.~Hu, J.~Hou, F.~Zhang, and C.~Zhang, Phys. Rev. Lett. 120 (2018), p. 240401.

\bibitem{Fulga2017}
I.C. Fulga, L.~Fallani, and M.~Burrello, Phys. Rev. B 97 (2018), p. 121402.

\bibitem{Fulga2017b}
I.C. Fulga and A.~Stern, Phys. Rev. B 95 (2017), p. 241116.

\bibitem{HPHu2018}
H.~Hu and C.~Zhang, arXiv:1802.08222v1  (2018).

\bibitem{Anderson2012}
B.M. Anderson, G.~Juzeli\ifmmode~\bar{u}\else \={u}\fi{}nas, V.M. Galitski, and
  I.B. Spielman, Phys. Rev. Lett. 108 (2012), p. 235301.

\bibitem{Campbell2016}
D.L. Campbell~\emph{et al}., Nat. Commun. 7 (2016), p. 10897.

\bibitem{XWLuo2017}
X.W. Luo, K.~Sun, and C.~Zhang, Phys. Rev. Lett. 119 (2017), p. 193001.

\bibitem{Zhang2015}
D.W. Zhang, S.L. Zhu, and Z.D. Wang, Phys. Rev. A 92 (2015), p. 013632.

\bibitem{Stanescu2010}
T.D. Stanescu, V.~Galitski, and S.~Das~Sarma, Phys. Rev. A 82 (2010), p.
  013608.

\bibitem{Buchhold2012}
M.~Buchhold, D.~Cocks, and W.~Hofstetter, Phys. Rev. A 85 (2012), p. 063614.

\bibitem{Abanin2013}
D.A. Abanin, T.~Kitagawa, I.~Bloch, and E.~Demler, Phys. Rev. Lett. 110 (2013),
  p. 165304.

\bibitem{Grusdt2014b}
F.~Grusdt, D.~Abanin, and E.~Demler, Phys. Rev. A 89 (2014), p. 043621.

\bibitem{Grusdt2016}
F.~Grusdt, N.Y. Yao, D.~Abanin, M.~Fleischhauer, and E.~Demler, Nat. Commun. 7
  (2016), p. 11994.

\bibitem{Price2012}
H.M. Price and N.R. Cooper, Phys. Rev. A 85 (2012), p. 033620.

\bibitem{Dauphin2013}
A.~Dauphin and N.~Goldman, Phys. Rev. Lett. 111 (2013), p. 135302.

\bibitem{Streda1982}
P.~Streda, J. Phys. C: Solid State Phys. 15 (1982), p. L717.

\bibitem{FLi2008}
F.~Li, L.B. Shao, L.~Sheng, and D.Y. Xing, Phys. Rev. A 78 (2008), p. 053617.

\bibitem{Hauke2014}
P.~Hauke, M.~Lewenstein, and A.~Eckardt, Phys. Rev. Lett. 113 (2014), p.
  045303.

\bibitem{Fukui2005}
T.~Fukui, Y.~Hatsugai, and H.~Suzuki, J. Phys. Soc. Jpn. 74 (2005), pp.
  1674--1677.

\bibitem{Tarnowski2017a}
M.~Tarnowski~\emph{et al}., Phys. Rev. Lett. 118 (2017), p. 240403.

\bibitem{RYu2011}
R.~Yu, X.L. Qi, A.~Bernevig, Z.~Fang, and X.~Dai, Phys. Rev. B 84 (2011), p.
  075119.

\bibitem{Alexandradinata2014}
A.~Alexandradinata, X.~Dai, and B.A. Bernevig, Phys. Rev. B 89 (2014), p.
  155114.

\bibitem{XJLiu2013}
X.J. Liu, K.T. Law, T.K. Ng, and P.A. Lee, Phys. Rev. Lett. 111 (2013), p.
  120402.

\bibitem{XJLiu2016}
X.J. Liu, Z.X. Liu, K.T. Law, W.V. Liu, and T.K. Ng, New J. Phys. 18 (2016), p.
  035004.

\bibitem{WSun2017}
W.~Sun~\emph{et al}., Phys. Rev. Lett. 121 (2018), p. 150401.

\bibitem{WSun2018}
W.~Sun~\emph{et al}., Phys. Rev. Lett. 121 (2018), p. 25403.

\bibitem{King-Smith1993}
R.D. King-Smith and D.~Vanderbilt, Phys. Rev. B 47 (1993), p. 1651.

\bibitem{LWang2013b}
L.~Wang, A.A. Soluyanov, and M.~Troyer, Phys. Rev. Lett. 110 (2013), p. 166802.

\bibitem{Coh2009}
S.~Coh and D.~Vanderbilt, Phys. Rev. Lett. 102 (2009), p. 107603.

\bibitem{Killi2012}
M.~Killi and A.~Paramekanti, Phys. Rev. A 85 (2012), p. 061606.

\bibitem{Reichl2014}
M.D. Reichl and E.J. Mueller, Phys. Rev. A 89 (2014), p. 063628.

\bibitem{DWZhang2012}
D.W. Zhang~\emph{et al}., Phys. Rev. A 86 (2012), p. 063616.

\bibitem{DWZhang2012a}
D.W. Zhang, Z.Y. Xue, H.~Yan, Z.D. Wang, and S.L. Zhu, Phys. Rev. A 85 (2012),
  p. 013628.

\bibitem{LeBlanc2013}
L.J. LeBlanc, M.C. Beeler, K.~Jim{\'{e}}nez-Garc{\'{\i}}a, A.R. Perry,
  S.~Sugawa, R.A. Williams, and I.B. Spielman, New J. Phys. 15 (2013), p.
  073011.

\bibitem{Kawaguchi2012}
Y.~Kawaguchi and M.~Ueda, Phys. Rep. 520 (2012), pp. 253--381.

\bibitem{Burger1999}
S.~Burger~\emph{et al}., Phys. Rev. Lett. 83 (1999), p. 5198.

\bibitem{Denschlag2000}
J.~Denschlag~\emph{et al}., Science 287 (2000), pp. 97--101.

\bibitem{Khaykovich2002}
L.~Khaykovich~\emph{et al}., Science 296 (2002), pp. 1290--1293.

\bibitem{Madison2000}
K.W. Madison, F.~Chevy, W.~Wohlleben, and J.~Dalibard, Phys. Rev. Lett. 84
  (2000), p. 806.

\bibitem{Abo-Shaeer2001}
J.R. Abo-Shaeer, C.~Raman, J.M. Vogels, and W.~Ketterle, Science 292 (2001),
  pp. 476--479.

\bibitem{Matthews1999}
M.R. Matthews~\emph{et al}., Phys. Rev. Lett. 83 (1999), pp. 2498--2501.

\bibitem{Leanhardt2002}
A.E. Leanhardt~\emph{et al}., Phys. Rev. Lett. 89 (2002), p. 190403.

\bibitem{Leanhardt2003}
A.E. Leanhardt, Y.~Shin, D.~Kielpinski, D.E. Pritchard, and W.~Ketterle, Phys.
  Rev. Lett. 90 (2003), p. 140403.

\bibitem{Leslie2009}
L.S. Leslie, A.~Hansen, K.C. Wright, B.M. Deutsch, and N.P. Bigelow, Phys. Rev.
  Lett. 103 (2009), p. 250401.

\bibitem{Choi2012a}
J.y. Choi, W.J. Kwon, and Y.i. Shin, Phys. Rev. Lett. 108 (2012), p. 035301.

\bibitem{Choi2012b}
J.~yoon Choi, W.J. Kwon, M.~Lee, H.~Jeong, K.~An, and Y.~il~Shin, New J. Phys.
  14 (2012), p. 053013.

\bibitem{Choi2013}
J.y. Choi, S.~Kang, S.W. Seo, W.J. Kwon, and Y.i. Shin, Phys. Rev. Lett. 111
  (2013), p. 245301.

\bibitem{Skyrme1961}
T.H.R. Skyrme, Proc. R. Soc. A: Math., Phys. Eng. Sci. 260 (1961), pp.
  127--138.

\bibitem{Khawaja2001a}
U.A. Khawaja and H.~Stoof, Nature 411 (2001), pp. 918--920.

\bibitem{Khawaja2001b}
U.A. Khawaja and H.T.C. Stoof, Phys. Rev. A 64 (2001), p. 043612.

\bibitem{Ruostekoski2001}
J.~Ruostekoski and J.R. Anglin, Phys. Rev. Lett. 86 (2001), p. 3934.

\bibitem{Battye2002}
R.A. Battye, N.R. Cooper, and P.M. Sutcliffe, Phys. Rev. Lett. 88 (2002), p.
  080401.

\bibitem{Savage2003}
C.M. Savage and J.~Ruostekoski, Phys. Rev. Lett. 91 (2003), p. 010403.

\bibitem{Wuester2005}
S.~Wüster, T.E. Argue, and C.M. Savage, Phys. Rev. A 72 (2005), p. 043616.

\bibitem{Herbut2006}
I.F. Herbut and M.~Oshikawa, Phys. Rev. Lett. 97 (2006), p. 080403.

\bibitem{Kawakami2012}
T.~Kawakami, T.~Mizushima, M.~Nitta, and K.~Machida, Phys. Rev. Lett. 109
  (2012), p. 015301.

\bibitem{Lee2018}
W.~Lee~\emph{et al}., Sci. Adv. 4 (2018), p. eaao3820.

\bibitem{Battye1998}
R.A. Battye and P.M. Sutcliffe, Phys. Rev. Lett. 81 (1998), p. 4798.

\bibitem{Babaev2002}
E.~Babaev, L.D. Faddeev, and A.J. Niemi, Phys. Rev. B 65 (2002), p. 100512.

\bibitem{Kawaguchi2008}
Y.~Kawaguchi, M.~Nitta, and M.~Ueda, Phys. Rev. Lett. 100 (2008), p. 180403.

\bibitem{Hall2016}
D.S. Hall~\emph{et al}., Nat. Phys. 12 (2016), pp. 478--483.

\bibitem{Ray2014}
M.W. Ray, E.~Ruokokoski, S.~Kandel, M.~Möttönen, and D.S. Hall, Nature 505
  (2014), pp. 657--660.

\bibitem{XLQi2009}
X.L. Qi, R.~Li, J.~Zang, and S.C. Zhang, Science 323 (2009), pp. 1184--1187.

\bibitem{PZhang2005}
P.~Zhang, Y.~Li, and C.P. Sun, Eur. Phys. J. D - At., Mol., Opt. Plasma Phys.
  36 (2005), pp. 229--233.

\bibitem{Pietila2009b}
V.~Pietil\"a and M.~M\"ott\"onen, Phys. Rev. Lett. 102 (2009), p. 080403.

\bibitem{Sonner2009}
J.~Sonner and D.~Tong, Phys. Rev. Lett. 102 (2009), p. 191801.

\bibitem{Pietila2009a}
V.~Pietil\"a and M.~M\"ott\"onen, Phys. Rev. Lett. 103 (2009), p. 030401.

\bibitem{XFZhou2018}
X.F. Zhou, C.~Wu, G.C. Guo, R.~Wang, H.~Pu, and Z.W. Zhou, Phys. Rev. Lett. 120
  (2018), p. 130402.

\bibitem{Ray2015}
M.W. Ray, E.~Ruokokoski, K.~Tiurev, M.~Mottonen, and D.S. Hall, Science 348
  (2015), pp. 544--547.

\bibitem{Sinova2015}
J.~Sinova, S.O. Valenzuela, J.~Wunderlich, C.~Back, and T.~Jungwirth, Rev. Mod.
  Phys. 87 (2015), pp. 1213--1260.

\bibitem{Kato2004}
Y.K. Kato, R.C. Myers, A.C. Gossard, and D.D. Awschalom, Science 306 (2004),
  pp. 1910--1913.

\bibitem{Wunderlich2005}
J.~Wunderlich, B.~Kaestner, J.~Sinova, and T.~Jungwirth, Phys. Rev. Lett. 94
  (2005), p. 047204.

\bibitem{Hosten2008}
O.~Hosten and P.~Kwiat, Science 319 (2008), pp. 787--790.

\bibitem{XYin2013}
X.~Yin, Z.~Ye, J.~Rho, Y.~Wang, and X.~Zhang, Science 339 (2013), pp.
  1405--1407.

\bibitem{Carollo2005}
A.C.M. Carollo and J.K. Pachos, Phys. Rev. Lett. 95 (2005), p. 157203.

\bibitem{SLZhu2006a}
S.L. Zhu, Phys. Rev. Lett. 96 (2006), p. 077206.

\bibitem{Kitaev2009}
A.~Kitaev and C.~Laumann, arXiv:0904.2771v1  (2009).

\bibitem{LMDuan2003}
L.M. Duan, E.~Demler, and M.D. Lukin, Phys. Rev. Lett. 91 (2003), p. 090402.

\bibitem{Meinert2013}
F.~Meinert~\emph{et al}., Phys. Rev. Lett. 111 (2013), p. 053003.

\bibitem{Simon2011}
J.~Simon~\emph{et al}., Nature 472 (2011), p. 307.

\bibitem{Bernien2017}
H.~Bernien~\emph{et al}., Nature 551 (2017), p. 579.

\bibitem{Schaub2012}
P.~Schauß~\emph{et al}., Nature 491 (2012), p.~87.

\bibitem{Schaub2015}
P.~Schauß~\emph{et al}., Science 347 (2015), p. 1455.

\bibitem{Zeiher2017}
J.~Zeiher~\emph{et al}., Phys. Rev. X 7 (2017), p. 041063.

\bibitem{Haldane1983b}
F.~Haldane, Phys. Lett. A 93 (1983), p. 464.

\bibitem{Gu2009}
Z.C. Gu and X.G. Wen, Phys. Rev. B 80 (2009), p. 155131.

\bibitem{Pollmann2012}
F.~Pollmann, E.~Berg, A.M. Turner, and M.~Oshikawa, Phys. Rev. B 85 (2012), p.
  075125.

\bibitem{Nijs1989}
M.~den Nijs and K.~Rommelse, Phys. Rev. B 40 (1989), pp. 4709--4734.

\bibitem{Hagiwara1990}
M.~Hagiwara, K.~Katsumata, I.~Affleck, B.I. Halperin, and J.P. Renard, Phys.
  Rev. Lett. 65 (1990), p. 3181.

\bibitem{Affleck1987}
I.~Affleck, T.~Kennedy, E.H. Lieb, and H.~Tasaki, Phys. Rev. Lett. 59 (1987),
  p. 799.

\bibitem{Affleck1988}
I.~Affleck, T.~Kennedy, E.H. Lieb, and H.~Tasaki, \emph{Valence bond ground
  states in isotropic quantum antiferromagnets}, in \emph{Condensed Matter
  Physics and Exactly Soluble Models}, Springer Berlin Heidelberg,  1988, pp.
  253--304.

\bibitem{Garcia-Ripoll2004}
J.J. Garc\'{\i}a-Ripoll, M.A. Martin-Delgado, and J.I. Cirac, Phys. Rev. Lett.
  93 (2004), p. 250405.

\bibitem{Torre2006}
E.G. Dalla~Torre, E.~Berg, and E.~Altman, Phys. Rev. Lett. 97 (2006), p.
  260401.

\bibitem{Brennen2007}
G.K. Brennen, A.~Micheli, and P.~Zoller, New J. Phys. 9 (2007), p. 138.

\bibitem{Berg2008}
E.~Berg, E.G. Dalla~Torre, T.~Giamarchi, and E.~Altman, Phys. Rev. B 77 (2008),
  p. 245119.

\bibitem{Amico2010}
L.~Amico, G.~Mazzarella, S.~Pasini, and F.S. Cataliotti, New J. Phys. 12
  (2010), p. 013002.

\bibitem{Dalmonte2011}
M.~Dalmonte, M.~Di~Dio, L.~Barbiero, and F.~Ortolani, Phys. Rev. B 83 (2011),
  p. 155110.

\bibitem{Ruhman2012}
J.~Ruhman, E.G. Dalla~Torre, S.D. Huber, and E.~Altman, Phys. Rev. B 85 (2012),
  p. 125121.

\bibitem{Ho2006}
A.F. Ho, Phys. Rev. A 73 (2006), p. 061601.

\bibitem{Kobayashi2012}
K.~Kobayashi, M.~Okumura, Y.~Ota, S.~Yamada, and M.~Machida, Phys. Rev. Lett.
  109 (2012), p. 235302.

\bibitem{Kobayashi2014}
K.~Kobayashi, Y.~Ota, M.~Okumura, S.~Yamada, and M.~Machida, Phys. Rev. A 89
  (2014), p. 023625.

\bibitem{Fazzini2017}
S.~Fazzini, A.~Montorsi, M.~Roncaglia, and L.~Barbiero, New J. Phys. 19 (2017),
  p. 123008.

\bibitem{Nonne2010a}
H.~Nonne, P.~Lecheminant, S.~Capponi, G.~Roux, and E.~Boulat, Phys. Rev. B 81
  (2010), p. 020408.

\bibitem{Nonne2011}
H.~Nonne, P.~Lecheminant, S.~Capponi, G.~Roux, and E.~Boulat, Phys. Rev. B 84
  (2011), p. 125123.

\bibitem{Nonne2013}
H.~Nonne, M.~Moliner, S.~Capponi, P.~Lecheminant, and K.~Totsuka, Europhys.
  Lett. 102 (2013), p. 37008.

\bibitem{Nonne2010b}
H.~Nonne, E.~Boulat, S.~Capponi, and P.~Lecheminant, Phys. Rev. B 82 (2010), p.
  155134.

\bibitem{Cardarelli2017}
L.~Cardarelli, S.~Greschner, and L.~Santos, Phys. Rev. Lett. 119 (2017), p.
  180402.

\bibitem{Xu2017a}
J.~Xu, Q.~Gu, and E.J. Mueller, arXiv:1703.05842v1  (2017).

\bibitem{Kestner2011}
J.P. Kestner, B.~Wang, J.D. Sau, and S.~Das~Sarma, Phys. Rev. B 83 (2011), p.
  174409.

\bibitem{Lv2016}
J.P. Lv and Z.D. Wang, Phys. Rev. B 93 (2016), p. 174507.

\bibitem{Hilker2017}
T.A. Hilker, G.~Salomon, F.~Grusdt, A.~Omran, M.~Boll, E.~Demler, I.~Bloch, and
  C.~Gross, Science 357 (2017), pp. 484--487.

\bibitem{Kitaev2001}
A.Y. Kitaev, Phys.-Usp. 44 (2001), p. 131.

\bibitem{LFu2008}
L.~Fu and C.L. Kane, Phys. Rev. Lett. 100 (2008), p. 096407.

\bibitem{Elliott2015}
S.R. Elliott and M.~Franz, Rev. Mod. Phys. 87 (2015), pp. 137--163.

\bibitem{Laflamme2016}
C.~Laflamme, J.C. Budich, P.~Zoller, and M.~Dalmonte, Nat. Commun. 7 (2016), p.
  12280.

\bibitem{LJiang2011}
L.~Jiang~\emph{et al}., Phys. Rev. Lett. 106 (2011), p. 220402.

\bibitem{Diehl2011}
S.~Diehl, E.~Rico, M.A. Baranov, and P.~Zoller, Nat. Phys. 7 (2011), p. 971.

\bibitem{Kraus2013}
C.V. Kraus, P.~Zoller, and M.A. Baranov, Phys. Rev. Lett. 111 (2013), p.
  203001.

\bibitem{Nascimbene2013}
S.~Nascimbène, J. Phys. B: At., Mol. Opt. Phys. 46 (2013), p. 134005.

\bibitem{Laflamme2014}
C.~Laflamme, M.A. Baranov, P.~Zoller, and C.V. Kraus, Phys. Rev. A 89 (2014),
  p. 022319.

\bibitem{Wall2016}
M.L. Wall~\emph{et al}., Phys. Rev. Lett. 116 (2016), p. 035301.

\bibitem{Keilmann2011}
T.~Keilmann, S.~Lanzmich, I.~McCulloch, and M.~Roncaglia, Nat. Commun. 2
  (2011), p. 361.

\bibitem{Greschner2015}
S.~Greschner and L.~Santos, Phys. Rev. Lett. 115 (2015), p. 053002.

\bibitem{Strater2016}
C.~Sträter, S.C.L. Srivastava, and A.~Eckardt, Phys. Rev. Lett. 117 (2016), p.
  205303.

\bibitem{Meinert2016}
F.~Meinert, M.J. Mark, K.~Lauber, A.J. Daley, and H.C. Nägerl, Phys. Rev.
  Lett. 116 (2016), p. 205301.

\bibitem{RMa2011}
R.~Ma~\emph{et al}., Phys. Rev. Lett. 107 (2011), p. 095301.

\bibitem{QNiu1985}
Q.~Niu, D.J. Thouless, and Y.S. Wu, Phys. Rev. B 31 (1985), p. 3372.

\bibitem{Hafezi2007}
M.~Hafezi, A.S. S\o{}rensen, E.~Demler, and M.D. Lukin, Phys. Rev. A 76 (2007),
  p. 023613.

\bibitem{Bai2018}
R.~Bai, S.~Bandyopadhyay, S.~Pal, K.~Suthar, and D.~Angom, Phys. Rev. A 98
  (2018), p. 023606.

\bibitem{Senthil2013}
T.~Senthil and M.~Levin, Phys. Rev. Lett. 110 (2013), p. 046801.

\bibitem{TSZeng2016}
T.S. Zeng, W.~Zhu, and D.N. Sheng, Phys. Rev. B 93 (2016), p. 195121.

\bibitem{YCHe2015}
Y.C. He, S.~Bhattacharjee, R.~Moessner, and F.~Pollmann, Phys. Rev. Lett. 115
  (2015), p. 116803.

\bibitem{Sterdyniak2015}
A.~Sterdyniak, N.R. Cooper, and N.~Regnault, Phys. Rev. Lett. 115 (2015), p.
  116802.

\bibitem{Moller2009}
G.~M\"oller and N.R. Cooper, Phys. Rev. Lett. 103 (2009), p. 105303.

\bibitem{Furukawa2017}
S.~Furukawa and M.~Ueda, Phys. Rev. A 96 (2017), p. 053626.

\bibitem{Paredes2002}
B.~Paredes, P.~Zoller, and J.I. Cirac, Phys. Rev. A 66 (2002), p. 033609.

\bibitem{Furukawa2013}
S.~Furukawa and M.~Ueda, Phys. Rev. Lett. 111 (2013), p. 090401.

\bibitem{YHWu2013}
Y.H. Wu and J.K. Jain, Phys. Rev. B 87 (2013), p. 245123.

\bibitem{Furukawa2014}
S.~Furukawa and M.~Ueda, Phys. Rev. A 90 (2014), p. 033602.

\bibitem{Kitaev2006}
A.~Kitaev, Ann. Phys. 321 (2006), pp. 2 --111.

\bibitem{CWZhang2007}
C.~Zhang, V.W. Scarola, S.~Tewari, and S.~Das~Sarma, Proc. Natl. Acad. Sci. 104
  (2007), pp. 18415--18420.

\bibitem{DWZhang2016}
D.W. Zhang, Quantum Inf. Process. 15 (2016), pp. 4477--4487.

\bibitem{LFu2011}
L.~Fu, Phys. Rev. Lett. 106 (2011), p. 106802.

\bibitem{JLi2009}
J.~Li, R.L. Chu, J.K. Jain, and S.Q. Shen, Phys. Rev. Lett. 102 (2009), p.
  136806.

\bibitem{Groth2009}
C.W. Groth, M.~Wimmer, A.R. Akhmerov, J.~Tworzyd\l{}o, and C.W.J. Beenakker,
  Phys. Rev. Lett. 103 (2009), p. 196805.

\bibitem{HJiang2009}
H.~Jiang, L.~Wang, Q.F. Sun, and X.C. Xie, Phys. Rev. B 80 (2009), p. 165316.

\bibitem{Benalcazar2017a}
W.A. Benalcazar, B.A. Bernevig, and T.L. Hughes, Science 357 (2017), pp.
  61--66.

\bibitem{Benalcazar2017b}
W.A. Benalcazar, B.A. Bernevig, and T.L. Hughes, Phys. Rev. B 96 (2017), p.
  245115.

\bibitem{Langbehn2017}
J.~Langbehn, Y.~Peng, L.~Trifunovic, F.~von Oppen, and P.W. Brouwer, Phys. Rev.
  Lett. 119 (2017), p. 246401.

\bibitem{ZSong2017}
Z.~Song, Z.~Fang, and C.~Fang, Phys. Rev. Lett. 119 (2017), p. 246402.

\bibitem{Schindler2018}
F.~Schindler~\emph{et al}., Sci. Adv. 4 (2018), p. eaat0346.

\bibitem{MLin2017}
M.~Lin and T.L. Hughes, arXiv:1708.08457v1  (2017).

\bibitem{BLian2016}
B.~Lian and S.C. Zhang, Phys. Rev. B 94 (2016), p. 041105.

\bibitem{BLian2017}
B.~Lian and S.C. Zhang, Phys. Rev. B 95 (2017), p. 235106.

\bibitem{Sugawa2018}
S.~Sugawa, F.~Salces-Carcoba, A.R. Perry, Y.~Yue, and I.B. Spielman, Science
  360 (2018), pp. 1429--1434.

\bibitem{Sau2010}
J.D. Sau, R.M. Lutchyn, S.~Tewari, and S.D. Sarma, Phys. Rev. Lett. 104 (2010),
  p. 040502.

\bibitem{Tewari2007}
S.~Tewari, S.D. Sarma, C.~Nayak, C.~Zhang, and P.~Zoller, Phys. Rev. Lett. 98
  (2007), p. 010506.

\bibitem{CZhang2008}
C.~Zhang, S.~Tewari, R.M. Lutchyn, and S.D. Sarma, Phys. Rev. Lett. 101 (2008),
  p. 160401.

\bibitem{Sato2009}
M.~Sato, Y.~Takahashi, and S.~Fujimoto, Phys. Rev. Lett. 103 (2009), p. 020401.

\bibitem{Massignan2010}
P.~Massignan, A.~Sanpera, and M.~Lewenstein, Phys. Rev. A 81 (2010), p. 031607.

\bibitem{FWu2013}
F.~Wu, G.C. Guo, W.~Zhang, and W.~Yi, Phys. Rev. Lett. 110 (2013), p. 110401.

\bibitem{YCao2014}
Y.~Cao, S.H. Zou, X.J. Liu, S.~Yi, G.L. Long, and H.~Hu, Phys. Rev. Lett. 113
  (2014), p. 115302.

\bibitem{Raghu2008}
S.~Raghu, X.L. Qi, C.~Honerkamp, and S.C. Zhang, Phys. Rev. Lett. 100 (2008),
  p. 156401.

\bibitem{Yoshida2014}
T.~Yoshida, R.~Peters, S.~Fujimoto, and N.~Kawakami, Phys. Rev. Lett. 112
  (2014), p. 196404.

\bibitem{Herbut2014}
I.F. Herbut and L.~Janssen, Phys. Rev. Lett. 113 (2014), p. 106401.

\bibitem{Amaricci2016}
A.~Amaricci, J.C. Budich, M.~Capone, B.~Trauzettel, and G.~Sangiovanni, Phys.
  Rev. B 93 (2016), p. 235112.

\bibitem{Dauphin2012}
A.~Dauphin, M.~M\"uller, and M.A. Martin-Delgado, Phys. Rev. A 86 (2012), p.
  053618.

\bibitem{Dauphin2016}
A.~Dauphin, M.~M\"uller, and M.A. Martin-Delgado, Phys. Rev. A 93 (2016), p.
  043611.

\bibitem{XDeng2014}
X.~Deng and L.~Santos, Phys. Rev. A 89 (2014), p. 033632.

\bibitem{TLi2015}
T.~Li, H.~Guo, S.~Chen, and S.Q. Shen, Phys. Rev. B 91 (2015), p. 134101.

\bibitem{Kuno2017}
Y.~Kuno, K.~Shimizu, and I.~Ichinose, New J. Phys. 19 (2017), p. 123025.

\bibitem{ZXu2013}
Z.~Xu and S.~Chen, Phys. Rev. B 88 (2013), p. 045110.

\bibitem{HHu2017}
H.~Hu, S.~Chen, T.S. Zeng, and C.~Zhang, arXiv:1712.03671v1  (2017).

\bibitem{Dzero2010}
M.~Dzero, K.~Sun, V.~Galitski, and P.~Coleman, Phys. Rev. Lett. 104 (2010), p.
  106408.

\bibitem{FLu2013}
F.~Lu, J.~Zhao, H.~Weng, Z.~Fang, and X.~Dai, Phys. Rev. Lett. 110 (2013), p.
  096401.

\bibitem{Alexandrov2013}
V.~Alexandrov, M.~Dzero, and P.~Coleman, Phys. Rev. Lett. 111 (2013), p.
  226403.

\bibitem{NXu2014}
N.~Xu~\emph{et al}., Nat. Commun. 5 (2014), p. 4566.

\bibitem{Kim2014}
D.J. Kim, J.~Xia, and Z.~Fisk, Nat. Mater. 13 (2014), p. 466.

\bibitem{HChen2016}
H.~Chen, X.J. Liu, and X.C. Xie, Phys. Rev. Lett. 116 (2016), p. 046401.

\bibitem{Lisandrini2017}
F.T. Lisandrini, A.M. Lobos, A.O. Dobry, and C.J. Gazza, Phys. Rev. B 96
  (2017), p. 075124.

\bibitem{SNiu2018}
S.~Niu and X.J. Liu, Phys. Rev. B 98 (2018), p. 125141.

\bibitem{Ritsch2013}
H.~Ritsch, P.~Domokos, F.~Brennecke, and T.~Esslinger, Rev. Mod. Phys. 85
  (2013), pp. 553--601.

\bibitem{ZZheng2018}
Z.~Zheng, X.B. Zou, and G.C. Guo, New J. Phys. 20 (2018), p. 023039.

\bibitem{CWang2017}
C.~Wang, P.~Zhang, X.~Chen, J.~Yu, and H.~Zhai, Phys. Rev. Lett. 118 (2017), p.
  185701.

\bibitem{Tarnowski2017b}
M.~Tarnowski~\emph{et al}., arXiv:1709.01046  (2017).

\bibitem{YHu2016}
Y.~Hu, P.~Zoller, and J.C. Budich, Phys. Rev. Lett. 117 (2016), p. 126803.

\bibitem{ZGong2017}
Z.~Gong and M.~Ueda, arXiv:1710.05289v4  (2017).

\bibitem{Kruckenhauser2017}
A.~Kruckenhauser and J.C. Budich, arXiv:1712.02440v1  (2017).

\bibitem{XQiu2018}
X.~Qiu, T.S. Deng, G.C. Guo, and W.~Yi, arXiv:1804.09032v2  (2018).

\bibitem{CYang2018}
C.~Yang, L.~Li, and S.~Chen, Phys. Rev. B 97 (2018), p. 060304.

\bibitem{PChang2018}
P.Y. Chang, Phys. Rev. B 97 (2018), p. 224304.

\bibitem{LZhang2018}
L.~Zhang, L.~Zhang, S.~Niu, and X.J. Liu, arXiv:1802.10061 (2018) .

\bibitem{LZhang2018b}
L.~Zhang, L.~Zhang, and X.J. Liu, arXiv:1807.10782  (2018).

\bibitem{Diehl2008}
S.~Diehl~\emph{et al}., Nat. Phys. 4 (2008), p. 878.

\bibitem{Verstraete2009}
F.~Verstraete, M.M. Wolf, and J.~Ignacio~Cirac, Nat. Phys. 5 (2009), p. 633.

\bibitem{Bardyn2012}
C.E. Bardyn~\emph{et al}., Phys. Rev. Lett. 109 (2012), p. 130402.

\bibitem{Bardyn2013}
C.E. Bardyn~\emph{et al}., New J. Phys. 15 (2013), p. 085001.

\bibitem{Lindblad1976}
G.~Lindblad, Commun. Math. Phys. 48 (1976), pp. 119--130.

\bibitem{Baumann2010}
K.~Baumann, C.~Guerlin, F.~Brennecke, and T.~Esslinger, Nature 464 (2010), p.
  1301.

\bibitem{JSPan2015}
J.S. Pan, X.J. Liu, W.~Zhang, W.~Yi, and G.C. Guo, Phys. Rev. Lett. 115 (2015),
  p. 045303.

\bibitem{DYu2017}
D.~Yu, J.S. Pan, X.J. Liu, W.~Zhang, and W.~Yi, Front. Phys. 13 (2017), p.
  136701.

\bibitem{Mivehvar2017}
F.~Mivehvar, H.~Ritsch, and F.~Piazza, Phys. Rev. Lett. 118 (2017), p. 073602.

\bibitem{Kollath2016}
C.~Kollath, A.~Sheikhan, S.~Wolff, and F.~Brennecke, Phys. Rev. Lett. 116
  (2016), p. 060401.

\bibitem{Sheikhan2016}
A.~Sheikhan, F.~Brennecke, and C.~Kollath, Phys. Rev. A 94 (2016), p. 061603.

\bibitem{Alvarez2018}
V.M.M. Alvarez, J.E.B. Vargas, M.~Berdakin, and L.E.F.F. Torres,
  arXiv:1805.08200v1  (2018).

\bibitem{JLi2016}
J.~Li~\emph{et al}., Nat. Commu. 10 (2019), p. 855.

\bibitem{volovik2003}
G.E. Volovik, \emph{The Universe in a Helium Droplet}, International Series of
  Monographs on Physics, Clarendon Press, 2003.

\bibitem{YXZhao2013}
Y.X. Zhao and Z.D. Wang, Phys. Rev. Lett. 110 (2013), p. 240404.

\bibitem{ZWang2010}
Z.~Wang, X.L. Qi, and S.C. Zhang, Phys. Rev. Lett. 105 (2010), p. 256803.

\bibitem{ZWang2012b}
Z.~Wang and S.C. Zhang, Phys. Rev. X 2 (2012), p. 031008.

\bibitem{ZhongWang-NJP}
Z.~Wang, X.L. Qi, and S.C. Zhang, New J. Phys. 12 (2010), p. 065007.

\bibitem{Volovik1989}
G.E. Volovik and V.M. Yakovenko, J. Phys.: Condens. Matter 1 (1989), p. 5263.

\end{thebibliography}

%

\end{document}